\documentclass[defaultstyle,10pt]{thesis_svk}

\usepackage[utf8]{inputenc}
\usepackage{amsfonts}
\usepackage{amssymb}
\usepackage{amsmath}
\usepackage{graphicx}
\usepackage{tikz}
\usetikzlibrary{decorations.pathreplacing}
\usepackage{caption}
\usepackage{hyperref}
\usepackage{xcolor}
\usepackage{aas_macros}
\usepackage{bm}
\usepackage{fnpct}
\hypersetup{
    colorlinks,
    allcolors=[rgb]{0 0 .5},
    linktoc=all
}
\usepackage{natbib}

\DeclareUnicodeCharacter{0306}{\u{s}}

\usepackage{subfiles} 

\def\realbiblio{

}
\def\biblio{

}

\newcommand{\RF}{ROUGH}
\newcommand{\muram}{MURaM}
\newcommand{\Al}{Alfv\'en}
\newcommand{\BP}{Bright Point}
\newcommand{\Bp}{Bright point}
\newcommand{\bp}{bright point}
\newcommand{\bhp}{bright-point}
\newcommand{\Bhp}{Bright-point}
\newcommand{\BHP}{Bright-Point}
\newcommand{\bz}{\ensuremath{B_z}}
\newcommand{\vz}{\ensuremath{V_z}}
\newcommand{\vx}{\ensuremath{V_x}}
\newcommand{\vy}{\ensuremath{V_y}}
\newcommand{\vr}{\ensuremath{V_r}}
\renewcommand{\max}{\ensuremath{_\text{max}}}
\renewcommand{\min}{\ensuremath{_\text{min}}}
\renewcommand{\vec}[1]{\ensuremath{\boldsymbol{\mathbf{#1}}}}

\newcommand{\abs}{\ensuremath{_\text{abs}}}
\newcommand{\rel}{\ensuremath{_\text{rel}}}
\newcommand{\rms}{\text{rms}}
\newcommand{\sd}{\sigma}

\newcommand{\fe}[1][]{\ensuremath{F_\text{8#1}}}
\newcommand{\feobs}{\fe[,obs]}
\newcommand{\femod}{\fe[,mod]}
\newcommand{\RMS}{\ensuremath{_\text{RMS}{}}}
\newcommand{\eff}{\ensuremath{_\text{eff}{}}}
\newcommand{\Teff}{\ensuremath{T\eff}}
\newcommand{\logg}{\ensuremath{\log g}}
\newcommand{\feh}{[\text{Fe/H}]}
\newcommand{\Ma}{\ensuremath{\mathcal{M}}}
\newcommand{\CBP}{\ensuremath{C_\text{BP}}}
\newcommand{\kep}{\textit{Kepler}}
\newcommand{\sun}{\ensuremath{_\odot}}
\newcommand{\grad}{\ensuremath{\nabla}}
\newcommand{\arcsec}{\mbox{$^{\prime\prime}$}}%

\let\realtableofcontents\tableofcontents
\renewcommand{\tableofcontents}{\setstretch{1.98}\hypersetup{linkcolor=black}\realtableofcontents}
\let\reallistoffigures\listoffigures
\renewcommand{\listoffigures}{\hypersetup{linkcolor=black}\reallistoffigures}
\let\reallistoftables\listoftables
\renewcommand{\listoftables}{\hypersetup{linkcolor=black}\reallistoftables}

\title{Cause and Effect: Stellar Convection Studied Through Flickering Brightness, and the Convectively-Driven Motions of Solar Bright Points}

\author{Samuel J.}{Van Kooten}

\otherdegrees{B.S., Calvin College, 2014 \\
M.S., University of Colorado, Boulder, 2016}

\degree{Doctor of Philosophy}		
       {Ph.D., Astrophysical and Planetary Sciences}		

\dept{Department of}			
     {Astrophysical and Planetary Sciences}		

\advisor{Prof.}				
        {Steven R. Cranmer}			

\reader{Mark P. Rast}
\readerThree{Benjamin P. Brown}
\readerFour{Zachory K. Berta-Thompson}
\readerFive{Ivan Milić}

\abstract{ \OnePageChapter
Magnetic bright points on the solar photosphere, prominent in the G band but also visible in the continuum, mark the footpoints of kilogauss magnetic flux tubes extending toward the corona.
Convective buffeting of these tubes is believed to excite MHD waves, which can propagate to the corona and there deposit heat.
Measuring wave excitation via bright-point motion can thus constrain coronal and heliospheric models, and this has been done extensively with centroid tracking, which can estimate kink-mode wave excitation.
DKIST will be the first telescope to provide well-resolved observations of bright points, allowing shape and size measurements to probe the excitation of other wave modes that have been difficult, if not impossible, to study to date.
I develop two complementary ways to take the first step in such an investigation, which I demonstrate on bright points in MURaM-simulated images of DKIST-like resolution, as a proof-of-concept in preparation for future DKIST observations.
This demonstration shows that accounting for these additional wave modes may increase the energy budget of this wave-heating model by a factor of two.

I also investigate the convection which drives bright-point motion.
I present a simplified model of solar granulation, which I use alongside MURaM to explore how bright-point motion depends on properties of convection, and I show the importance of turbulence to high-frequency motion.

Separately, I investigate high-frequency, stochastic brightness fluctuations (``flicker'' or \fe) in \kep\ light curves, which are the signature of stellar convection (as well as a source of noise for exoplanetary studies).
I confront a physical model of this flicker with measured values across the H-R diagram.
I revise this model to improve its agreement with observations by including the effect of the \kep\ bandpass on measured flicker, by incorporating metallicity in determining convective Mach numbers, and by using scaling relations from a wider set of numerical simulations.
I also explore what future lines of research might improve the model further.
In doing so, I help to establish this flicker as a source of stellar constraints on simulations of convection, which may support future advances in understanding both stellar and solar convection.

	}

	\dedication[Dedication]{	
	\begin{center}To everyone who's helped and supported me along the way\end{center}
	}

\acknowledgements{	\OnePageChapter	
	\vspace{-0.1cm}
This section of a dissertation is usually about expressing gratitude.
Considering the decade's worth of events that have occurred over the past year, I'm perhaps most thankful for that which I've liked least about writing this thesis: writing it alone at home.
During this pandemic I've been able to stay safe and isolated, a privilege that has been unavailable to so many.
Difficult as it was to complete this work and write this thesis through a year of isolation, I'm so thankful it was possible.

Thank you to Steve for being the very model of a good advisor.
Your patience, encouragement, advice and support have done so much to give me a positive experience in grad school and bring me to where I am today.
I'm so glad you came to CU when you did.

Thank you to my fellow grad students, especially those in my year (Evan, Nick, Piyush, Jhett, Jessica and Tristan), for being such a positive, supportive group of classmates and friends.
Never once have I felt in competition with anyone, and plenty of times have I completed something only with your help.

Thank you to my parents for countless opportunities to develop and pursue an interest in science, and for all your support through the years.

Thank you to Evan and Sami for sharing your dog Finn with me.

Thank you to all my previous research advisors---Larry Molnar, Derek Lamb, and Than Putzig---for helping me grow as a scientist and get to where I am today.

Thank you to my committee for reading this thesis and for your ideas and suggestions.

Thank you to Matthias Rempel for letting me use your simulation outputs, which have been the backbone of most of this thesis.

And now, with an eye toward remembering the work I did in 2020 but not too much else from that year, I present to you my dissertation.
	
	}

\degreeyear{2021}

\LoTisShort	

\begin{document}

\def\biblio{}

\chapter{Introduction}
\label{chap:intro}

The Sun is very hot; this much is clear.
Just where and to what degree is generally well-known.
But the question of \textit{why} the Sun is so hot---why, the favorite question of scientists and five-year-olds alike---is less settled.
Ultimately, the Sun is hot because of nuclear fusion in its core.
That heat is carried to the surface of the Sun by radiation and then convection.
These processes all have some open questions (e.g. the ``convective conundrum'' discussed later), but the most famously open question lies in the next step, how some of that heat crosses the temperature minimum just above the photosphere and concentrates in the corona, reaching temperatures orders of magnitude above that of the photosphere.
This question will motivate the bulk of this thesis.

\section{The Corona and the Photosphere}

\begin{figure}[t]
	\centering
	\includegraphics[width=6.5in]{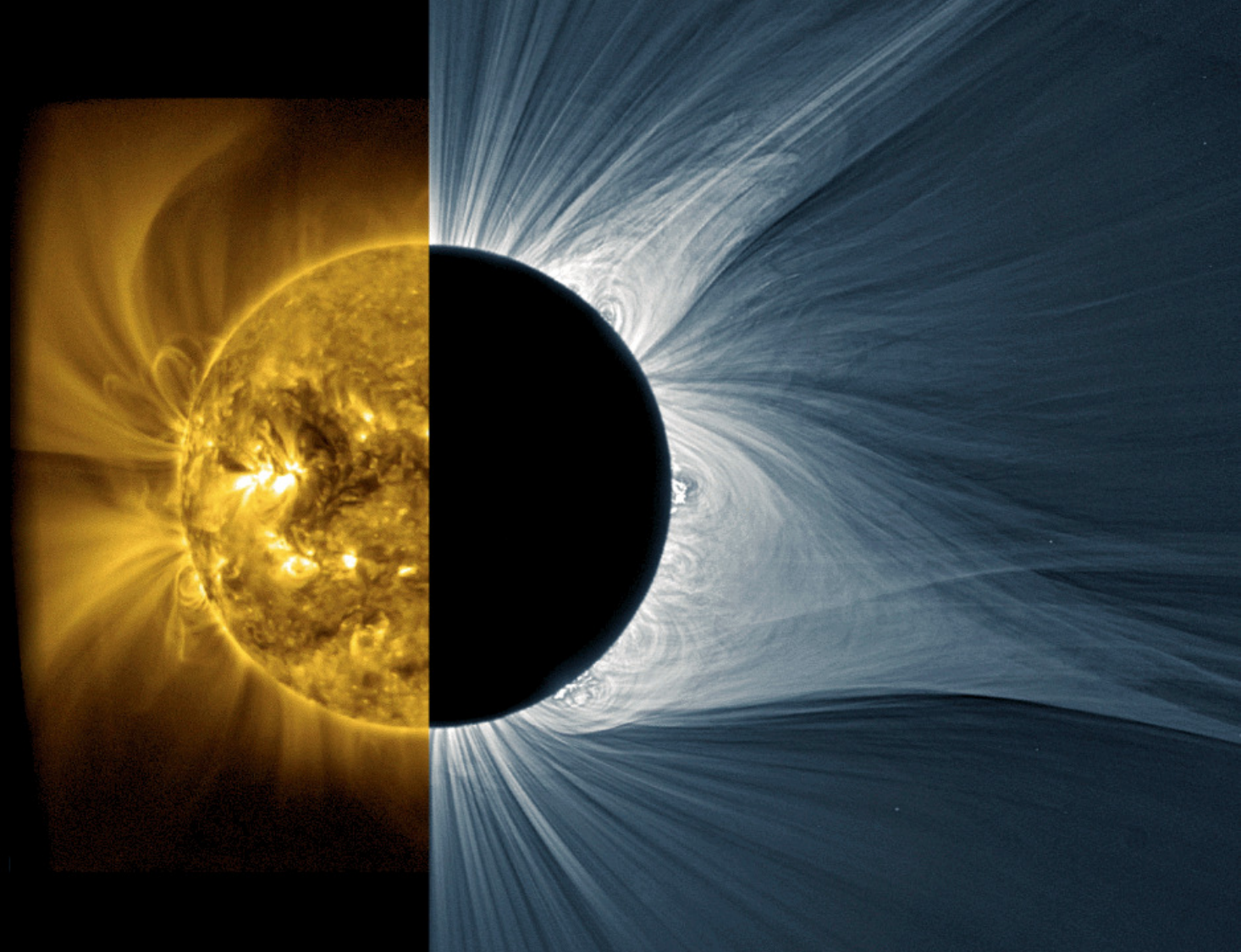}
	\caption[Two views of the solar corona]{Two views of the solar corona. \textit{(left)} An extreme-ultraviolet image from the SWAP telescope \citep{Seaton2013} on the PROBA2 satellite. \textit{(right)} A white-light image from the August 21, 2017, total solar eclipse, obtained and processed by M. Druckm\"uller, P. Aniol, and S. Habbal \citep[see][]{Druckmuller2006}. In both images, the fine magnetic structure and the large spatial extent of the corona can be clearly seen, filling the region above the solar surface. (This combined image was prepared by \citealp{Cranmer2019}.)}
	\label{fig:intro-corona}
\end{figure}

\begin{figure}[t]
	\centering
	\includegraphics[width=6in]{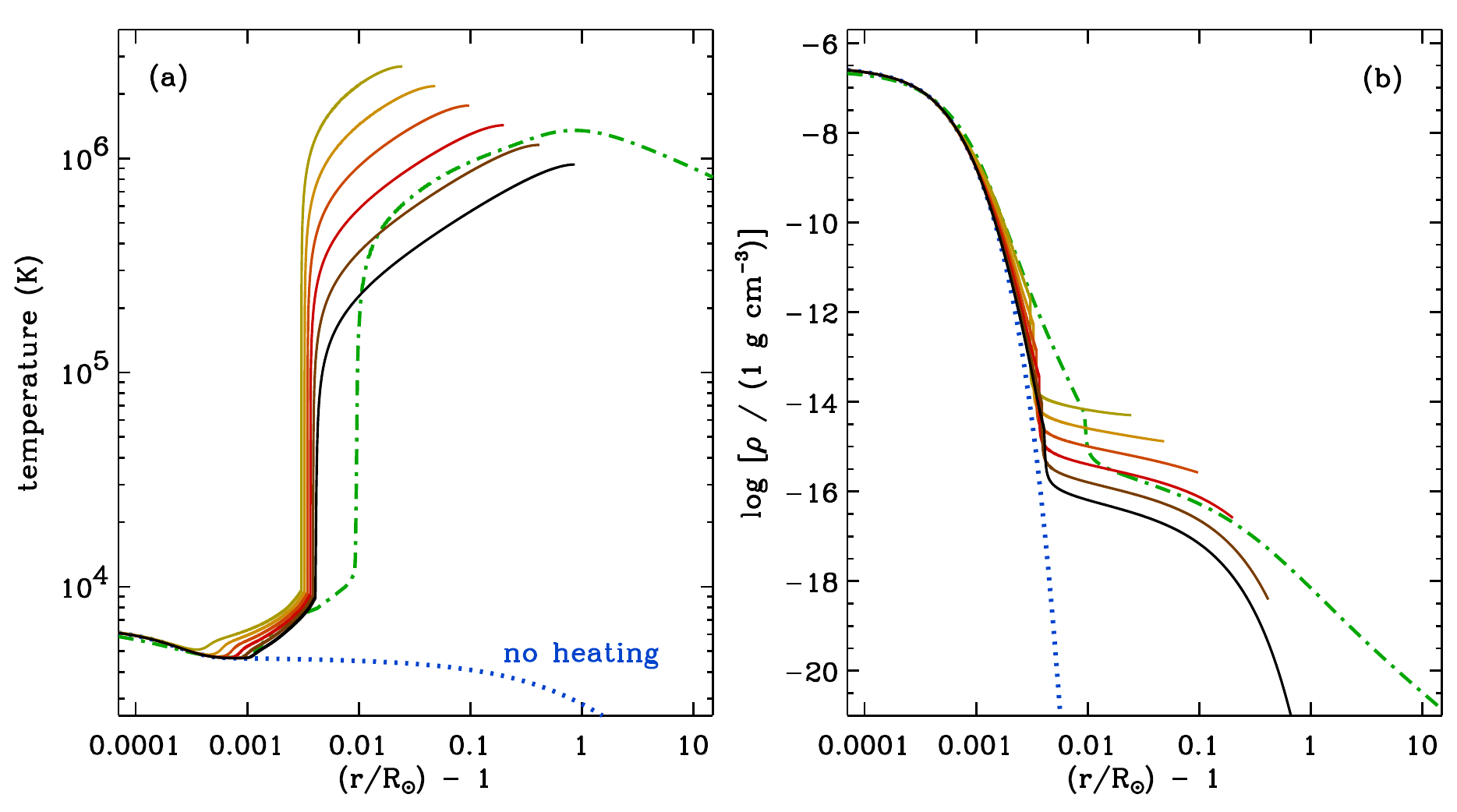}
	\caption[Radial variation of temperature and density in the solar atmosphere]{Radial variation of temperature \textit{(a)} and density \textit{(b)} with height in the solar atmosphere. Heights are in units of the solar radius $R_\odot$. The solid curves represent models of closed loops with lengths between 30 and 1,200~Mm, while the dash-dotted line represents a coronal hole model from \citet{Cranmer2007}. The dotted, blue curve indicates expected behavior in the absence of any coronal heating mechanism. (This figure is from \citet{Cranmer2019}.)}
	\label{fig:intro-temp-profile}
\end{figure}

The corona is the outer layer of the solar atmosphere (and by volume, it is almost the entire solar atmosphere), extending from a few megameters above the solar surface out to a variable and little-studied boundary with the heliosphere, of order ten solar radii from the Sun, where outflowing coronal plasma becomes the solar wind \citep[driven, in part, by the coronal temperautre;][]{Parker1958}.
The defining features of the corona are fine structuring by a dominant magnetic field (as seen in Figure~\ref{fig:intro-corona}), low plasma densities, and surprisingly high temperatures.
These latter two properties are illustrated in Figure~\ref{fig:intro-temp-profile}, where it can be seen that the plasma density drops by anywhere from 7 to 13 or more orders of magnitude over a distance of one solar radius from the photosphere (i.e., well within the frame of Figure~\ref{fig:intro-corona}).
Over this same distance, the plasma temperature dips slightly from its surface value of $\sim5,770$~K before rising extremely sharply to over a million Kelvin.
(This narrow region of the most rapidly-increasing temperature is called the transition region.)
The cause of this dramatic and counter-intuitive increase in the coronal temperature with height is the defining question of what is called the coronal heating problem.

\begin{figure}[t]
	\centering
	\includegraphics[width=6.5in]{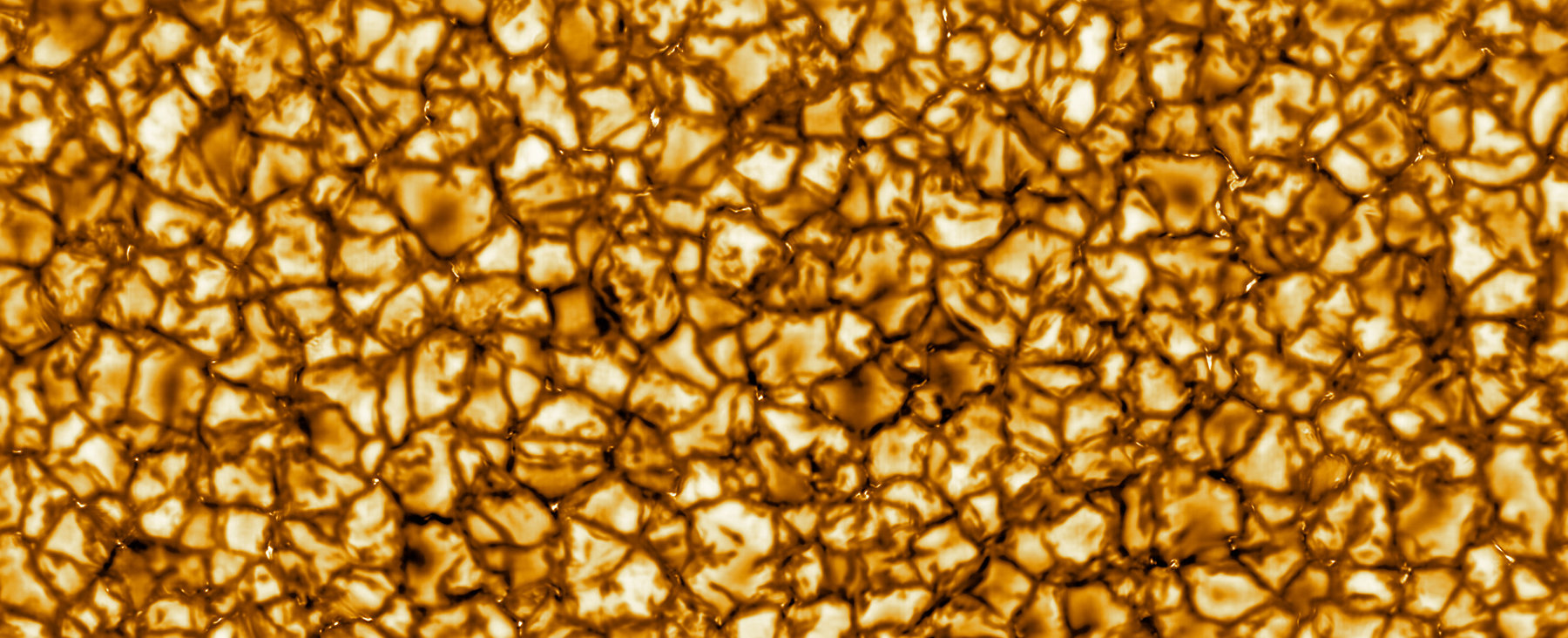}
	\caption[Granules in a DKIST first-light image]{Cropped region of a DKIST first-light image, illustrating the extent of the granulation pattern. (A zoomed-in view in shown in Figure~\ref{fig:intro-dkist-first-light-full}.) The field of view is approximately 40,000~km wide. (Image credit: NSO/AURA/NSF.)}
	\label{fig:intro-dkist-first-light}
\end{figure}

It is known that, in part, the corona remains so hot because it \textit{is} so hot: the atoms are very thoroughly ionized at these million-Kelvin temperatures, and so they are very restricted in their ability to radiate away heat \citep[e.g.,][]{Rosner1978}.
But some heat does travel away, whether through radiation, advection by the solar wind, or conduction down to he chromosphere, and a counterbalancing source of energy must replace that heat.
It is difficult to image this energy flux coming from empty space beyond the corona, and so solar physicists universally look downward to the photosphere for the source of this flux.

The photosphere is often taken to be the surface of the Sun, as it is a thin layer that is the source of almost all photons that escape into the solar system.
Outside of regions featuring sunspots, the photosphere is covered by granulation, shown in Figure~\ref{fig:intro-dkist-first-light-full}, which is caused by ubiquitous convection.
Each bright patch---a \textit{granule}--- is a region of warm, upflowing plasma.
At the surface, this rising plasma overturns and flows horizontally to the relatively dark lanes surrounding each granule, where plasma that has cooled somewhat flows back down.
(Despite these lanes being presented as dark in nearly every image, they are only a few hundred Kelvin cooler than the granules, meaning they would be considered bright if not for the contrast with the surrounding plasma.)
Typical speeds for these flows are a few km~s$^{-1}$ \citep{Oba2017a}, and the visible pattern of granulation is constantly changing on timescales of 5--10~min \citep{Nesis2002}.
This constant plasma motion is a ready reservoir of energy, far more than the energy required to maintain the coronal temperature.
The question is how this kinetic energy in the photosphere is transformed into thermal energy in the corona.

The answer lies in the prominent magnetic field of the Sun, though the exact mechanism is still unclear.
Theorists have proposed many ways that such an energy flux might be driven opposite the direction expected from the thermal gradient (see, e.g., the 19 models across four categories summarized in Table 1 of \citealp{Cranmer2019}), and somewhere within that theoretical arsenal is quite possibly the correct answer (or answers).
These models largely fall into two leading classes: turbulent dissipation of waves traveling along field lines \citep{Alfven1947,Osterbrock1961,Cranmer2007,Oran2015} and reconnection triggered by photospheric motion yielding increasingly-complex magnetic field topologies \citep{Parker1988,Fisk1999,Priest2018}.
Unfortunately, the proposed mechanisms all occur quickly on small spatial scales, making them difficult to observe, and so catching any one of them in action, to support and constrain models, is quite difficult.
Compounding this difficulty is that they occur in an optically-thin corona in close proximity to the photosphere which is, in many wavelengths, orders of magnitude brighter, meaning that careful observational techniques are required even for easier-to-observe phenomena.
Because of these observational difficulties, models of coronal heating have too few constraints and entirely too much freedom, and strong answers are yet unavailable.

\section{\BP s}

\begin{figure}[t]
	\centering
	\includegraphics[width=6.5in]{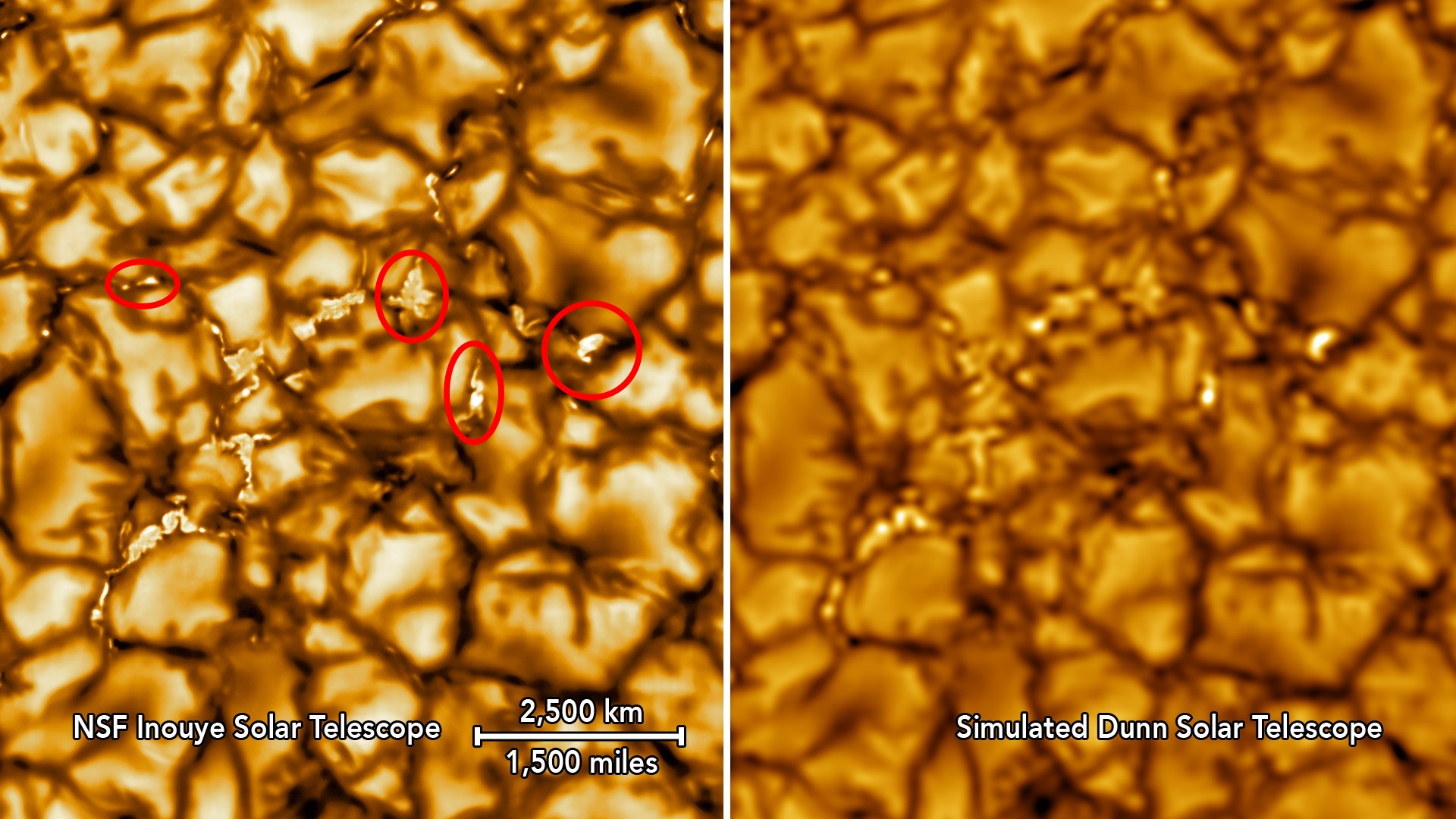}
	\caption[\Bp s in a DKIST first-light image]{DKIST first-light image. \textit{(left)} \Bp s and granules can be seen at full resolution. A selection of \bp s have been circled, and many more can be seen. \textit{(right)} The same image has been downgraded to simulate the resolution of the Dunn solar telescope, which has a 76~cm aperture diameter (compare to DKIST's 4~m diameter). (Both images are from NSO's first-light press release and are not calibrated for science use; credit: NSO/AURA/NSF.)}
	\label{fig:intro-dkist-first-light-full}
\end{figure}

In this thesis, I will not address coronal heating itself, but I will work to better understand the potential energy budget of one proposed heating mechanism through studies of \bp s.

\begin{figure}[t]
	\centering
	\includegraphics{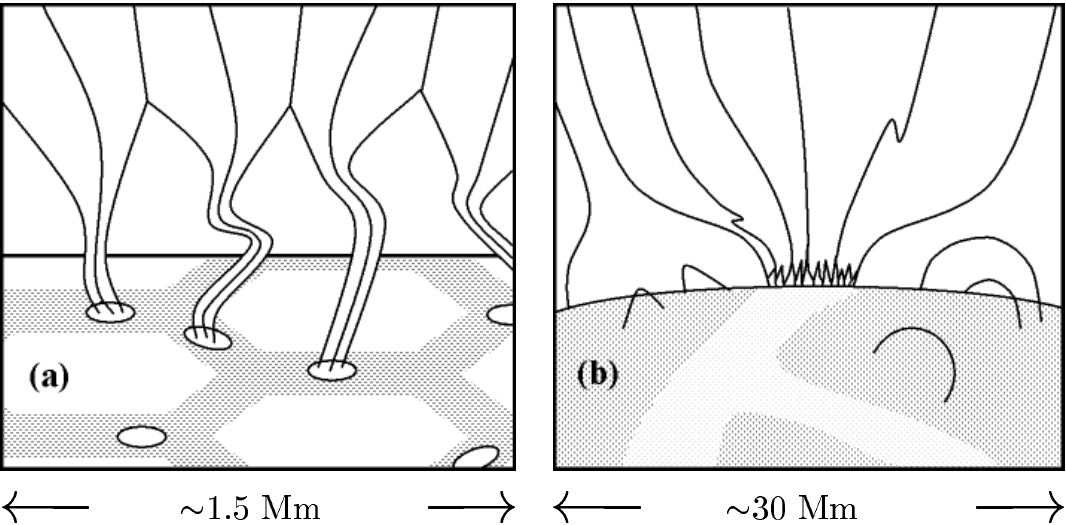}
	\caption[Illustration of expanding flux tubes]{Illustration of expanding flux tubes. \textit{(a)} Individual flux tubes arise from \bp s in the dark, intergranular lanes, expanding horizontally due to decreased plasma pressure and merging together at a height of approximately 0.5~Mm. \textit{(b)} As it rises, the merged flux continues to expand horizontally until it meets another concentration of flux (which may be rooted in the other side of a supergranular cell). Above a height of approximately 1~Mm, the volume has been filled by these expanded flux concentrations, which form a magnetic canopy. (This figure originally appeared in \citealp{Cranmer2005}.)}
	\label{fig:intro-expanding-tubes}
\end{figure}

Photospheric bright points,\footnote{Also called intergranular bright points, G-band bright points, magnetic bright points, or network bright points; not to be confused with coronal or x-ray bright points \citep{Madjarska2019} or chromospheric bright points (also called sequential chromospheric brightenings, \citealp{Kirk2016a}; distinct from ``chromospheric bright points'' as the chromospheric manifestation of the flux tube associated with a photospheric bright point, e.g. \citealp{Xiong2017,Liu2018a,Narang2019}). This region of solar nomenclature is truly unfortunate.} shown in Figure~\ref{fig:intro-dkist-first-light-full}, are intensity enhancements of order 100~km in size, seen in the intergranular lanes of the quiet-Sun photosphere, with a greater abundance along the supergranular network.
\Bp s usually correspond with kilogauss concentrations of vertical magnetic flux \citep{Utz2013}, sometimes called \textit{flux elements}, which are believed to be flux tubes that may rise to the corona (illustrated in Figure~\ref{fig:intro-expanding-tubes}), making them the source of much of the open magnetic flux in coronal holes \citep{Hofmeister2019}.
The rise of these flux tubes up to the low chromosphere has been observed as brightenings in the chromospheric line Ca II H coinciding with photospheric \bp s \citep{Xiong2017,Liu2018a,Narang2019}.
As they rise, these flux tubes expand rapidly \citep{Kuckein2019} due to the decreasing plasma pressure \citep{Spruit1976}.

\Bp s appear bright because of reduced gas pressure and density within the \bp, required to maintain pressure balance with their surroundings given their high magnetic pressure, and this lowers the optical surface by a few hundred kilometers \citep{Carlsson2004}.
At these slightly lower depths, heating from the surrounding granular plasma allows enhanced continuum emission, but the heating and reduced density also reduces the abundance of molecules such as CH, which allows a significant increase in emission escaping through molecular lines in spectral regions such as the G-band \citep{Steiner2001,Schussler2003,Uitenbroek2006}.
\Bp s are thus frequently observed in these molecularly-dominated bands where they show a greatly enhanced contrast.

\Bp s are generally believed to form via the process of convective collapse \citep{Parker1978,Spruit1979,Nagata2008,Utz2014}, in which a concentration of vertical flux restricts horizontal plasma motions and thereby suppresses convection within a small region, causing the plasma within that region to cool and form a downdraft.
This evacuates much of the plasma within the proto-\bhp\ region, allowing further concentration of the magnetic flux by the external plasma pressure.
This results in the high magnetic field and low plasma density typical of \bp s.
Some observational evidence, however, suggests that convective collapse is not the sole phenomenon responsible for \bp s reaching their kilogauss field strengths \citep{Keys2019}.

\Bp s are seen to be in near-continuous motion, buffeted by the constantly-evolving granular pattern \citep{VanBallegooijen1998,Nisenson2003,Utz2010,Chitta2012}, and this motion is believed to excite flux tube waves which propagate up the tube to the corona, where they deposit heat through turbulent dissipation\footnote{Many mechanisms have been proposed for the details of this dissipation---I again refer to Table 1 of \citet{Cranmer2019}.} \citep{Cranmer2005,Soler2019}.
(Evidence of oscillations in the chromosphere, serving as evidence of wave propagating up from the photosphere through magnetized regions, has been observed above magnetic elements, \citealp{Jafarzadeh2017a,Stangalini2017}, and in chromospheric plage, \citealp{GuevaraGomez2021}.)
In addition to waves, the dynamics of \bp s have also been hypothesized to produce jet-like upflows, a possible origin for at least some spicules \citep{Oxley2020}.
The motions of \bp s are complex \citep{Keys2020}, to say the least.

It is important to note that, while \bp s usually correspond to magnetic flux tubes, it is not always a one-to-one correlation.
Weaker flux elements may not possess the magnetic pressure necessary to reduce the plasma density sufficiently for an intensity enhancement to be visible, flux elements of any strength may not produce \bp s if their geometry is unfavorable relative to the viewing angle, and some observed \bp s are due to non-magnetic phenomena \citep[e.g. non-magnetized vortices;][]{Giagkiozis2018}.
As an additional consideration, observations show that, when a \bp\ disappears, a re-appearance of a \bp\ in that location is more likely to occur soon, within $\sim 10$~min, than over the subsequent hour \citep{Bodnarova2014}.
This suggests that the underlying flux tube endures during these disappearances and reappearances, and what is being observed is only a cessation and then resumption of conditions favorable for the producing of a visible-light enhancement.

Despite these limitations, \bp s serve as an extremely valuable and useful proxy for the location of these flux tubes, enabling studies through broad-band observations which tend to feature better resolution or cadence than spectroscopic magnetograms---particularly in pre-DKIST observations.


\section{Measuring Flux-tube Waves, Now and in the Future}

\begin{figure}[t]
	\centering
	\includegraphics[width=3in]{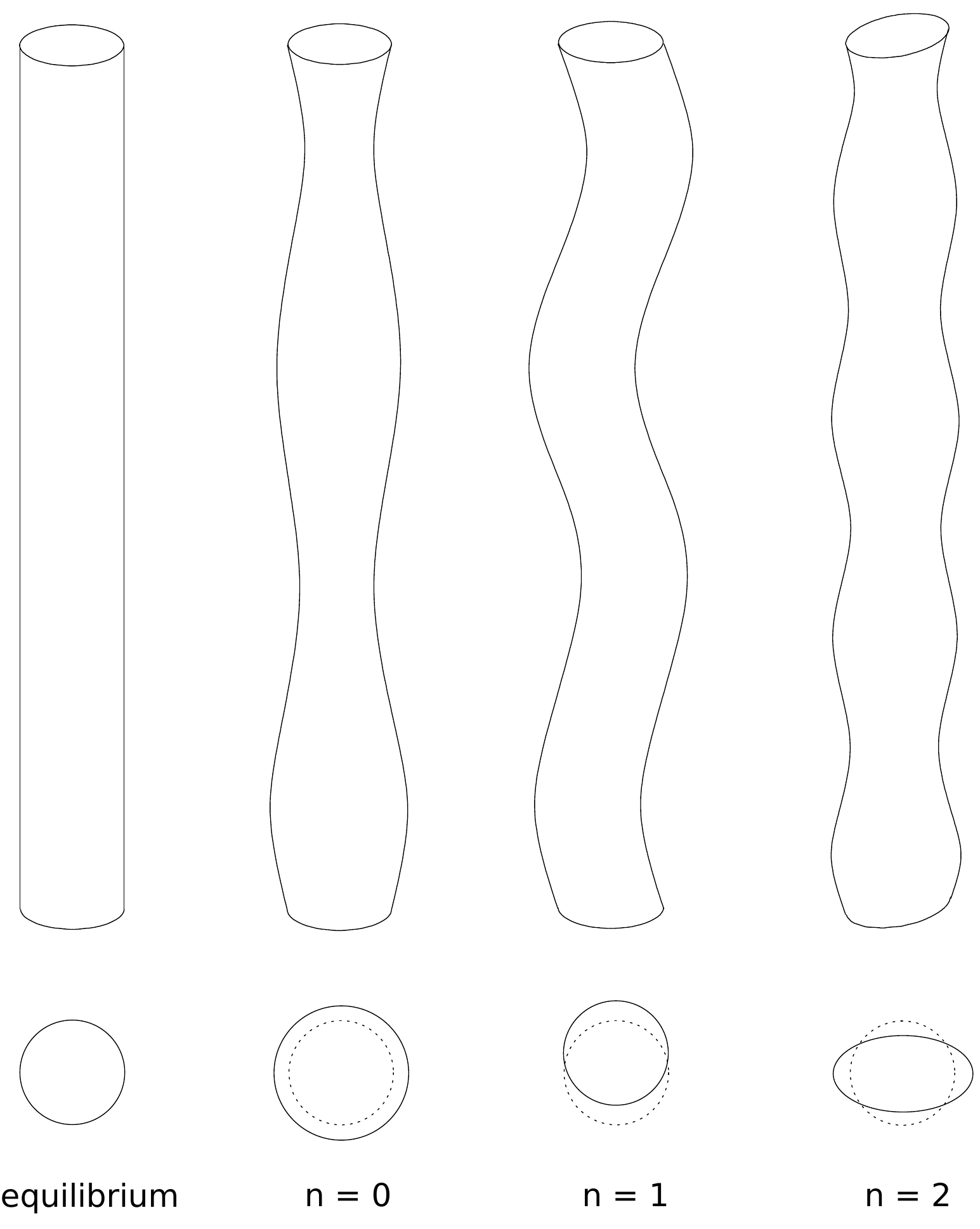}
	\caption[Diagram of $n=0$ through $2$ wave modes]{Diagram of $n=0$ through $2$ wave modes, showing \textit{(top)} the vertical extent of the tube, and \textit{(bottom)} horizontal cross-sections of the tube, with the equilibrium cross-section indicated by the dotted lines. (This figure is adapted from \citealp{Braithwaite2006}).}
	\label{fig:intro-wave-modes}
\end{figure}

Waves are ubiquitous in the solar corona \citep{Tomczyk2007} and chromosphere \citep{Jess2015}.
But directly observing waves in the corona is very difficult, and so \bhp\ motions have long been studied as an observational signature of the driving of waves in the overlying flux tubes.
When a \bp\ is seen to move horizontally, this can be interpreted as transverse motion of a horizontal cross-section of the tube, which is associated with a kink-mode or $n=1$ mode wave that causes uniform translation of any given cross section.
Additional wave modes, illustrated in Figure~\ref{fig:intro-wave-modes}, are possible, including sausage or $n=0$ modes associated with uniform changes in cross-sectional area, $n=2$ modes (sometimes called ``fluting'' modes) associated with elliptical distortions to the cross-sectional area, and many higher-$n$ modes.
These modes are characterized predominantly by radial distortions to the cross-section shape, but torsional modes also exist \citep[e.g.][]{WedemeyerBohm2012,Mumford2015,Mumford2015a} which demonstrate predominantly rotational motion within the flux tube.

There is a long history of measuring the positional offsets of \bp s and of interpreting them as kink-mode wave excitation \citep{Nisenson2003,Cranmer2005,Chitta2012,Yang2014,VanKooten2017,Xiong2017}.
However, the changes in \bhp\ shapes resulting from other modes are much more difficult to observe than the position offsets of kink-modes, especially in current observations, where the typical diffraction limit is comparable to the $\sim100$~km size of \bp s.
The upcoming Daniel K. Inouye Solar Telescope \citep[DKIST;][]{Rimmele2020} will offer unsurpassed resolution in its photospheric observations.
Expected to begin regular observations within a year, the DKIST 4~m aperture promises a diffraction-limited spatial resolution in the G-band of $\sim 15$~km with the Visible Broadband Imager \citep[VBI;][]{Elmore2014}, promising to reveal details of the size and shape of \bp s in ways previously not possible.
This will open many pathways to using \bp s to understand the driving of wave modes other than the kink mode, which can provide new insights on the potential energy budget of wave-based heating of the quiet-Sun corona.
This will be the focus of much of this thesis, to prepare and demonstrate techniques so that I and others are well-positioned to make use of the of high-resolution DKIST observations once they become available.

\section{Convection}

This thesis contains other lines of investigation as well, focused on the properties of surface convection in the Sun and other stars.
Convection dominates the upper third of the Sun by radius, transporting energy from the radiative zone to the solar surface \citep{Miesch2005}.
While convection is an entirely different sub-field from the study of the corona, the two are closely connected in that convective flows are taken to be the source of the energy that eventually heats the corona, with \bhp-forming flux tubes as one possible bridge over which that energy flows.
Convection is also implicated in the dynamo processes that generate solar magnetism and likely drive the Sun's 11-year activity cycle \citep{Charbonneau2020}, making convection a key process to be understood as part of understanding the Sun as a whole.

Convection plays a similarly-important role in the dynamics of stars, with surface convection present on red giants and stars on the main sequence from F dwarfs down to M dwarfs---that is to say, most stars---and core convection is common in other stars.
Understanding this convection is therefore key to interpreting observations of these stars as well as modeling their structure and evolution.
Convection can manifest as low-amplitude, high-frequency variations in stellar light curves, called ``flicker'' \citep{Bastien2013,Bastien2016}, which can become significant when seeking to interpret the details of exoplanetary transits \citep[e.g.,][]{Sulis2020}, meaning understanding stellar convection is even relevant for exoplanetary science.

In recent years, helioseismology and asteroseismology have provided new tools for probing the interiors of the Sun and stars, opening new frontiers for testing models of convection.
One result of these new observations is what is known as the convective conundrum, which is the fact that simulations of stellar convection predict the existence of ``giant cells,'' large convective structures deep in the convective zone, which are not seen in observations.
(For a more detailed overview of this conundrum, see \citealp{Anders2020,Rast2020}.)
I describe this not to suggest that I will help resolve this question, but to demonstrate that there is still much to learn about solar and stellar convection.

In the course of my investigations of \bhp\ motions, I will investigate some of the details of how \bhp\ motions relate to the nature of solar surface convection and the granular pattern.
I will also present a study of the convective signal in \kep\ light curves, testing and improving the ability of a physical model to predict the dependence of this signal on stellar parameters, and demonstrating specific aspects of convection which must be better understood to improve our ability to model stellar convection.

\section{Contents of this Dissertation}

Much of this thesis revolves around the use of \muram\footnote{The \textbf{M}ax-Planck-Institute for Aeronomy / \textbf{U}niversity of Chicago \textbf{Ra}diation \textbf{M}agneto-hydrodynamics code \citep{Vogler2005}} simulations of DKIST-like resolution as stand-ins for observations to prepare techniques and make predictions for what DKIST will see.
In Chapter~\ref{chap:bp-centroids} I apply centroid tracking to these simulated high-resolution observations, and I compare the centroid motions to the motions of passive tracers allowed to flow through the \muram\ simulation.
I also present a simplified, toy model of granulation that, when tuned to match observed granulation, allowed me to explore how different granular properties---and in particular, the presence of turbulence---affect passive-tracer motion.

In Chapters \ref{chap:bp_tracking}--\ref{chap:method-comp-and-conclusions}, I develop two techniques for moving beyond centroid tracking and interpreting changes to the shapes of resolved \bp s as the driving of wave modes other than the kink mode, and I show that these modes may make non-negligible contributions to the energy budget of wave heating of the corona.
The first technique (Chapter~\ref{chap:ellipse-fitting}) builds directly upon centroid tracking, and calculates moments across the measured shapes of \bp s, the evolution of which are connected to $n=0$ through $2$ wave modes.
The second technique (Chapter~\ref{chap:emd}) employs algorithmic tools from the field of computer vision related to the earth-mover's distance (or EMD).
I use these tools to infer a pixel-by-pixel velocity field that can advect a \bp\ from one shape to another---a use for which I am not aware of any precedent in solar physics.
From these velocity fields, I estimate a mixed-mode energy flux.
Chapter~\ref{chap:method-comp-and-conclusions} compares these two techniques and the energy fluxes estimated by each.

For all this work, I rely on a \bp\ tracking algorithm which I initially developed in Chapter~\ref{chap:bp-centroids}, and which I refine significantly in Chapter~\ref{chap:bp_tracking}.
In the latter chapter I also discuss some of the decisions which guide my approaches to analyzing shape changes.

Chapter~\ref{chap:flicker} diverges from the preceding chapters.
In it, I consider the flickering, granular signal seen in \kep\ light curves, and I compare it to predictions from a physically-based model.
I make a number of refinements to this model, which improves the model--observation agreement, and I explore possible explanations for the remaining disagreement, including specific aspects of stellar convection which must be better understood to resolve some questions.

Finally, in Chapter~\ref{chap:conclusions} I discuss the implications for my work, across the two themes of \bhp\ wave excitation and convection, and I discuss what the future may hold for these lines of inquiry.

\biblio

\chapter{\BHP\ Centroid Tracking}
\label{chap:bp-centroids}

The following is the full text of my manuscript published in the Astrophysical Journal \citep{VanKooten2017} as it was when accepted for publication, save for light adaptations for the formatting requirements of this thesis and a few added footnotes.
This text is based on the project I defended for the second part of my comprehensive exam.

\section{Introduction}
\label{sec:intro}
The surface of the Sun is the upper boundary of its convective zone, where convective cells called granules appear on spatial scales of 1~Mm and change on timescales of 10~minutes.
In the dark lanes between granules, bright, high-contrast structures are found, ranging from isolated points to extended structures wrapping around the granules.
These features, called ``G-band bright points,'' ``network bright points,'' ``magnetic bright points,'' ``magnetic bright features,'' ``inter-granular bright points,'' and ``filigree'', will be referred to simply as \textit{\bp s} in this chapter.
Small, isolated \bp s are of order 0.1~Mm across and are visible for a few minutes at a time.
\Bp s are the footpoints of kilogauss-strength magnetic flux tubes extending into the corona.
In \bp s, the concentrated magnetic flux reduces gas pressure, lowering the optical-depth-unity surface below that of the surrounding areas by about 350~km \citep{Carlsson2004,Nordlund2009}.
This lower depth corresponds to a higher temperature (an $\sim800$ K difference), producing enhanced brightness relative to the surrounding downflow lanes.
This effect is visible in continuum imaging with sufficient resolution, but the contrast between \bp s and even the brightest granular areas is significantly enhanced in wavelengths such as the 430.5~nm G-band.
The G-band is dominated by lines from CH, a molecule whose abundance is significantly reduced at the optical bottom of \bp s through dissociation \citep{Steiner2001}.
This reduced CH opacity allows more light to shine through from the continuum below, producing significantly enhanced G-band contrast.

The dynamics of \bp s are important to study as a compelling potential source of coronal heating.
\Bp s move due to buffeting from the convective churning of the photosphere \citep{Berger1996}.
This motion transverse to the emerging flux tubes excites transverse magnetohydrodynamic (MHD) waves in the tubes.
These waves propagate up to the corona, where they are thought to turbulently dissipate, depositing heat.
Thus the power spectrum of \bhp\ motion is directly related to the power spectrum of these transverse waves.
This wave-heating power spectrum serves as an important boundary condition for models of transverse MHD wave propagation in the solar environment \citep[e.g.][]{Cranmer2005} and thus informs models of MHD-wave based coronal heating as well as models of the heliosphere.

A number of observations have been made of \bp s.
Approaches to understanding their motion include the use of passive tracers that follow the gradient of magnetic intensity images \citep{VanBallegooijen1998}, an approach which can handle extended structures; and the tracking of features identified by hand in individual frames \citep{Nisenson2003,Chitta2012}, which excels in directly tracking isolated \bp s of limited size.
More recent studies improve this latter approach with automatic identification of \bp s, producing much larger datasets \citep{Utz2010,Abramenko2011,Keys2011,Yang2014}.

Observational determinations of the \bhp\ power spectrum are limited by the resolution and cadence of the observations.
Seeing conditions and pointing jitter can also limit the ability to track \bp s accurately over long durations.
In this chapter, we use high-resolution simulations of the photosphere to produce a power spectrum of \bhp\ motion tracked at higher spatial resolution than has been done before.
This serves as a prediction of the observational \bhp\ power spectrum that the Daniel K. Inouye Solar Telescope \citep[DKIST;][]{Tritschler2016} will obtain with unparalleled spatial and temporal resolution.
We also introduce a simplified model of granulation, which allows investigation of the interplay between \bhp\ dynamics and the complex properties of the granulation pattern.
Section \ref{sec:muram} will describe our analysis of the high-resolution \muram\ simulation.
Section \ref{sec:rough} will introduce our simplified, laminar \RF\ model.
Section \ref{sec:results} will present our results, and Section \ref{sec:conclusion} will summarize our conclusions.

\section{\muram\ Simulations}
\label{sec:muram}
\subsection{Introduction}
\label{sec:muram_intro}

\begin{figure*}[t]
	\centering
	\includegraphics[width=0.32\linewidth]{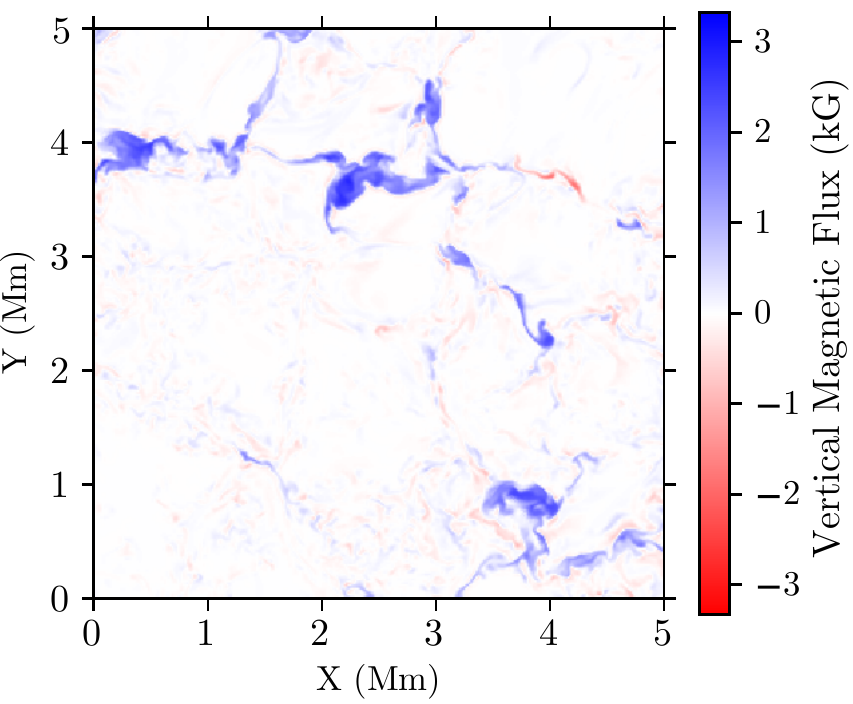}
	\includegraphics[width=0.32\linewidth]{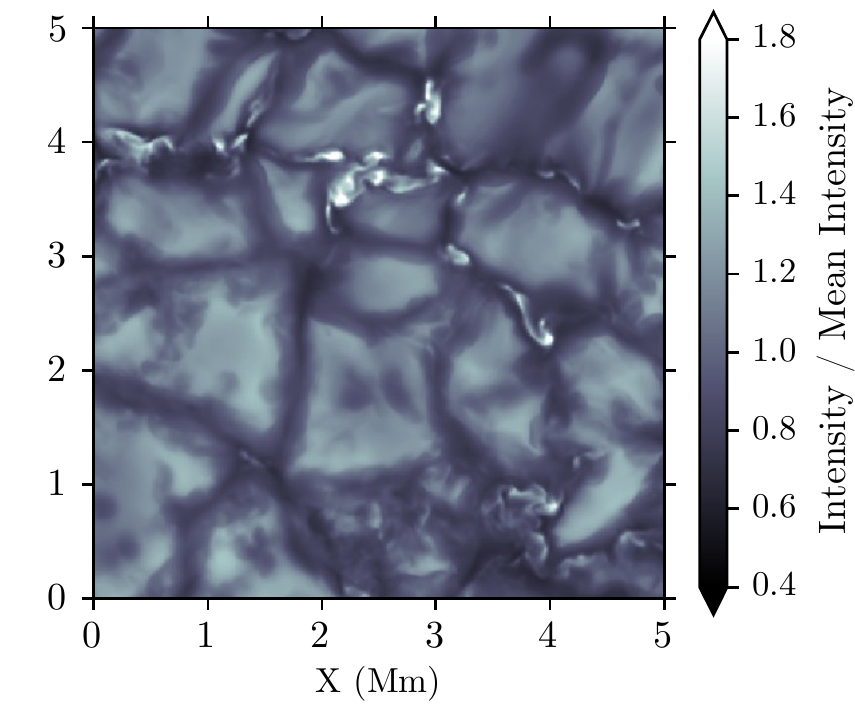}
	\includegraphics[width=0.32\linewidth]{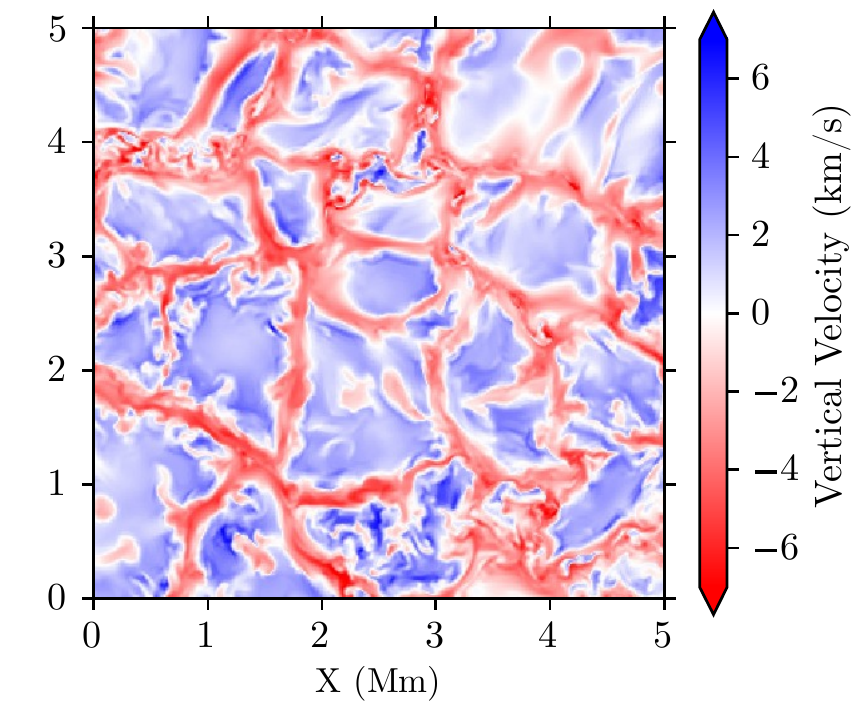}
	\caption[A $5\times 5$~Mm portion of the \citet{Rempel2014} \muram\ simulation]{A common $5\times 5$~Mm portion of the \citet{Rempel2014} \muram\ simulation showing, from left to right, vertical magnetic flux, white-light intensity, and vertical plasma velocity at the beginning of the analyzed time range (time stamp 040000).  Over the full frame, peak values for $I/I_\text{mean}$ are near 2.75, peak values for \bz\ are near 3.3~kG, and peak values for \vz\ are near 9.9~km~s$^{-1}$ and -11.9~km~s$^{-1}$. (The full $24.5 \times 24.5$~Mm simulation is used in this chapter.)}
	\label{fig:rempel_3_panel}
\end{figure*}

\citet{Rempel2014} uses \muram , software for performing radiative MHD simulations, to simulate the upper convective zone and photosphere.
The version of \muram\ used is extensively modified from the original of \citet{Vogler2005}.
In this chapter we analyze the results of a run with a domain size of $24.576 \times 24.576 \times 7.680 \; \text{Mm}^3$, at a horizontal resolution of 16~km per pixel.
The data set covers 8280~s, or 2.3~hours, of time at a cadence of one snapshot every 20.7~seconds (for 401 total snapshots).
The simulation run begins with an imposed mean vertical magnetic field of 30 G (as discussed in Section 4.3 of \citet{Rempel2014}), but is otherwise identical to the ``O16bM'' described in the paper's Table 1.
The imposed flux is quickly swept into the downflow lanes, resulting in an enhancement of kilogauss features.
The top boundary is positioned 1.5~Mm above the $\tau = 1$ surface of the photosphere, allowing a convective region depth of 6.2~Mm.
From this simulation we take the $\tau=1$ surface as our solar-observation analogue.
Within this slice, we use the upward-pointing white light intensity and the three components of both plasma velocity and magnetic field.
A sample patch of the simulation is shown in Figure \ref{fig:rempel_3_panel}.

\subsection{Analysis Algorithms}
\label{sec:muram_algorithms}

From these \muram\ simulations, we extracted granule size distributions, the motion paths of passive tracers, and the motion paths of automatically-identified \bp s.
This section describes the algorithms used for collecting this information, while the data products themselves are analyzed in Section \ref{sec:results}.

For the identification of granules, observers such as \citet{Abramenko2012} use intensity from non-spectrographic data to segment granules.
With the \muram\ data, we segment granules based on vertical velocity instead, using just a positive-negative threshold.
While deviating from observational approaches, this produces very similar results with a simple, more direct technique.

We track passive, point-like test particles, called ``corks,'' as one way to analyze the plasma flows in the \muram\ data.
We begin with a uniformly-spaced grid of 4,900 corks.
(We find the power spectrum of cork motion to be very insensitive to the initial arrangement.)
Once initially placed and released, corks move according the horizontal portion of the velocity vector.
We linearly interpolate in time to a timestep of 4~s when propagating corks, to ensure corks in small, high-plasma velocity regions (e.g. turbulent whirls) do not maintain that high velocity after quickly exiting the region.
However, when computing power spectra, we sample the cork velocity every 20~s in order to avoid exceeding the Nyquist rate of the data.
With a timestep of 4~s, a pixel size of 16~km, and typical plasma velocities of $\sim2$~km~s$^{-1}$, a cork stays near its initial location during a timestep.
Cork locations are stored to floating-point precision, and the velocity field is interpolated spatially from the four nearest grid points.
Corks are allowed to travel through the periodic boundaries of the simulation.
Cork velocities used for computing power spectra are those read out of the interpolated velocity field at cork locations.

For tracking \bp s themselves, we use an algorithm\footnote{A refined version of this algorithm is presented in Chapter \ref{chap:bp_tracking} of this thesis. The improved algorithm is much better at robustly detecting the apparent edges of \bp s.} that first identifies features within a given frame, following \citet{Feng2012}, and then links identified features between frames, following \citet{Yang2014}.
The intra-frame algorithm keys in on the fact that \bp s are small and are characterized by high contrast (they are ``bright'' relative to their immediate surroundings, but not always in absolute terms---especially in continuum images).
In each step, we err toward preventing false positive detections at the cost of more false negatives, since there is such a wealth of \bp s available to be detected.

The single-frame stage begins by identifying ``seed'' pixels in a white-light intensity map.
A discrete analogue of the Laplacian is calculated by subtracting from each pixel's intensity the mean intensity of the eight surrounding pixels.
A local maximum in intensity will produce a significantly positive value, and this indicates a likely \bhp\ center.
We use a threshold of three standard deviations from the mean in the Laplacian to identify a set of likely centers.

\begin{figure}[t!]
	\centering
	\includegraphics[width=\linewidth]{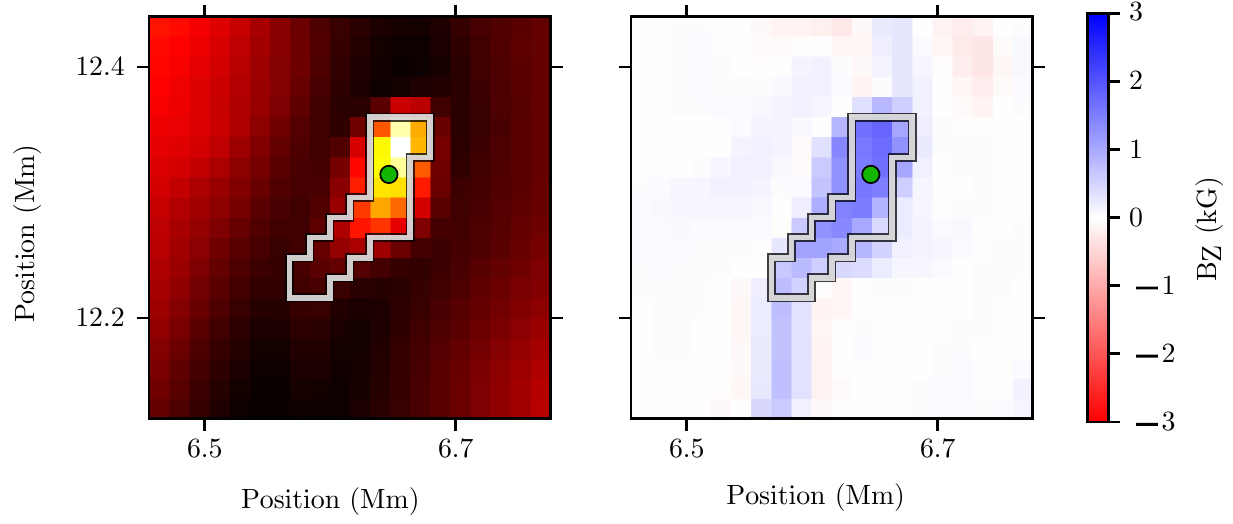}
	\caption[Example of an identified \bp]{Intensity (left) and vertical magnetic field (right) for an example identified \bp. The gray line marks the borders of the identified region, while the green dot marks the intensity-weighted centroid. Identification is on the basis of intensity, and the vertical magnetic field shows the expected concentration.}
	\label{fig:bp_border}
\end{figure}

Next, the set of candidate pixels is dilated to include pixels that both exceed a lower threshold and neighbor an already-selected pixel (including neighbors along a diagonal).
We repeat this dilation process three times, and the threshold for inclusion in this stage is a positive Laplacian value.
This process expands the set of selected candidate pixels from the ``seed'' pixels, which are quite likely to be in \bp s, to a nearby cloud of pixels whose likelihood of being in a \bp\ is enhanced by virtue of being near a pixel already considered a candidate.
Since the structure of a \bp\ tends to be a decrease in intensity from a local maximum to low values at the edge, the use of the Laplacian $>$ 0 threshold aims to ensure that the dilation includes pixels within the \bp\ but stops expanding where it reaches the edge of the \bp.
At this point, each contiguous block of selected pixels can be identified and labeled as a unique candidate \bp.
An example of such a \bp\ is shown in Figure \ref{fig:bp_border}.

The next step is to remove as false positives local intensity maxima inside granules.
After the previous three rounds of dilation, these local maxima will produce identified features surrounded by further bright pixels.
True \bp s, however, tend to be mostly or entirely included within the identified set of pixels at this stage, so the identified feature will be surrounded by dark intergranular lane pixels.
(A \bhp\ diameter is of order 100~km, and a pixel in this simulation is 16~km across.
Three rounds of dilation allow a maximum expansion of $16\times3=48$~km or one \bhp\ radius, meaning a seed pixel at the center can be expanded to cover a typical \bp.
Often more than one seed pixel will be identified within a feature, allowing features on the large end of the spectrum to easily be captured.)
We compute the fraction of a feature's neighboring pixels which would be added to the feature if a fourth round of dilation occurred (still using the Laplacian $>$ 0 threshold).
We reject as false positives any features for which more than $20\%$ of these new neighboring pixels have positive Laplacians.

We apply two more criteria to ensure a high-quality selection of \bp s.
First, we reject features within four pixels of another feature.
This ensures the analysis is of \bp s that are cleanly and unambiguously separated from other \bp s.
We reject as short-lived transients features below 4 total pixels in area (36~km equivalent diameter).
We also reject features above 110 total pixels in area (189~km equivalent diameter) or above 20 pixels (320~km) across (as determined by finding the diagonal of the smallest rectangle to contain the feature), to ensure identified features are not so large or extended that their motion cannot be represented by the motion of their centroid.
(This necessarily eliminates from consideration the long chains of connected \bp s that can occur in regions of strong background magnetic flux; see Section \ref{sec:conclusion}.)

This completes the intra-frame identification step.
Each frame now has a set of uniquely-identified features, but these features must be linked in the temporal dimension for any analysis.
This linking step, following \citet{Yang2014}, is short: if a feature in frame $n$ overlaps, by at least one pixel, the location of a feature in frame $n-1$, the two features are considered to be the same.
This is supported by observations \citep[e.g.][]{Chitta2012,Nisenson2003}, which find typical \bhp\ sizes of 100-150~km and typical velocity distributions with maxima around 7~km~s$^{-1}$. 
In the simulation's 20-second time steps, a \bp\ would move a maximum of 140~km (with most \bp s moving much less), ensuring that most \bp s will overlap with their prior position.
In situations where a feature in frame $n$ overlaps multiple features in frame $n-1$ (e.g. two \bp s merge), or where multiple \bp s in frame $n$ overlap one feature in frame $n-1$ (e.g. a \bp\ splits in two), linkage is ambiguous and is not attempted---instead, the feature(s) in frame $n$ are considered newly-found features.
Once linking is completed across all frames, we reject \bp s with lifetimes under 5 frames (1.6~minutes) as a further guard against transient features.

The result of this algorithm is a set of \bhp\ identification numbers, with each number corresponding to a set of coordinates within each frame of the \bp's lifespan.

\section{\RF\ Model}
\label{sec:rough}

In an attempt to accurately and efficiently capture the laminar portion of the solar granulation pattern, we developed a new Monte Carlo-based model called \RF\ (Random, Observationally-motivated, Unphysical, Granulation-based Heliophysics).
\RF\ simulations are highly phenomenological, and the properties of the simulation are drawn from observations of solar granulation.
The model aims to produce a reasonable representation of granulation at a single slice of constant height while offering very direct control over the properties of that granulation pattern.

In a \RF\ simulation, a set number of granule centers are randomly distributed over a region of space and throughout a given time span.
Granule centers are assigned a horizontal drift speed, drawn from a Gaussian distribution of zero mean and width 0.25~km~s$^{-1}$ \citep{Roudier2012}, as well as a drift direction drawn from a uniform distribution.
Each center is also assigned a lifetime randomly drawn from the lifetime distribution of \citet{DelMoro2004}.
We use the exponential fit to lifetimes $>$ 5~minutes ($e^{-t/2.7 \; \text{min}}$, $t>5$~min), as this produces excellent agreement between simulated and observed granulation movies.
This produces a mean lifetime of 7.7~min.
At the end of a granule's lifetime, a granule can fade away or it can fragment.
\citet{Lemmerer2017} show that the majority of granules fragment into two or more child granules.
We give each simulated granule a 70\% chance of fragmenting into exactly two children.
The child granules each have new lifetimes drawn again from the \citet{DelMoro2004} distribution.

Each granule center is taken as a point source of radially-diverging, horizontal flow with a time-dependent amplitude of
\begin{equation}
	\vr (t) = V\max \; e^{- ( (t-t_\text{mid})/\delta)^4},
	\label{eq:V_r(t)}
\end{equation}
where $t_\text{mid}$ is the midpoint of the granule center's lifetime and $\delta$, set to a third of the granule's lifetime, determines the falloff rate with time.
$V\max$ is set to 5.1~km~s$^{-1}$ to match the integrated \RF\ power spectrum to the integrated \muram\ cork-tracking power spectrum.
The radial flow is given a spatial dependence of
\begin{equation}
	\vr (r, t) = \vr (t) \; \frac{r}{r_0} \; \frac{1}{(1 + (r/r_0)^a)^b},
	\label{eq:V_r(r, t)}
\end{equation}
where $r_0=500\;\text{km}$, $a=4$ and $b=1/a$ determine the shape of the function, and $r$ is the horizontal distance from a location to the center of the granule.
This functional form is chosen to match the observed granulation pattern \citep[see also][]{Simon1989}.
At each pixel in the simulation, the horizontal velocity is taken to be the velocity corresponding to the largest value of $\vr (r, t) / r$ produced at that pixel by any one granule center.
In other words, each granule center has a ``sphere of influence'' in which it is both the sole producer and the strongest possible producer of horizontal flow.
$\vr /r$ is used for this step rather than \vr\ itself to flatten the initial ramp-up from $\vr (r=0)=0$ to the function's maximum value.

\begin{figure}[t!]
	\centering
	\includegraphics[width=0.49\linewidth]{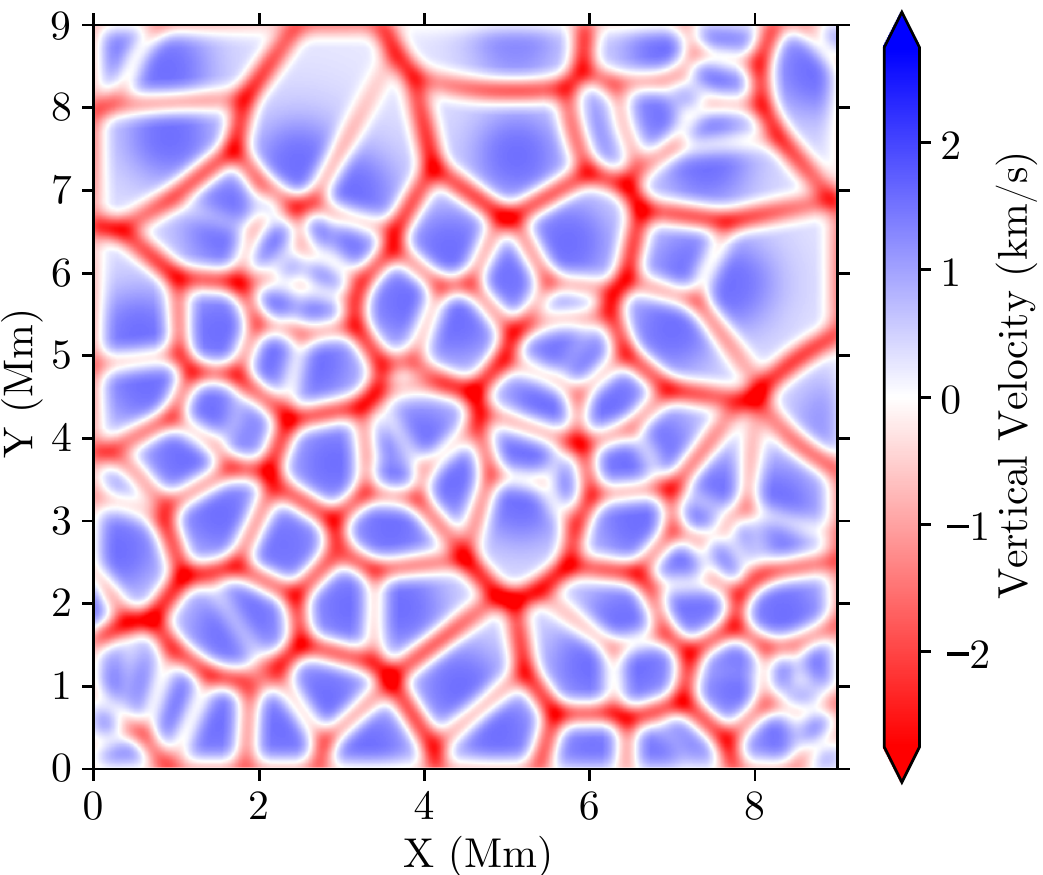}
	\includegraphics[width=0.49\linewidth]{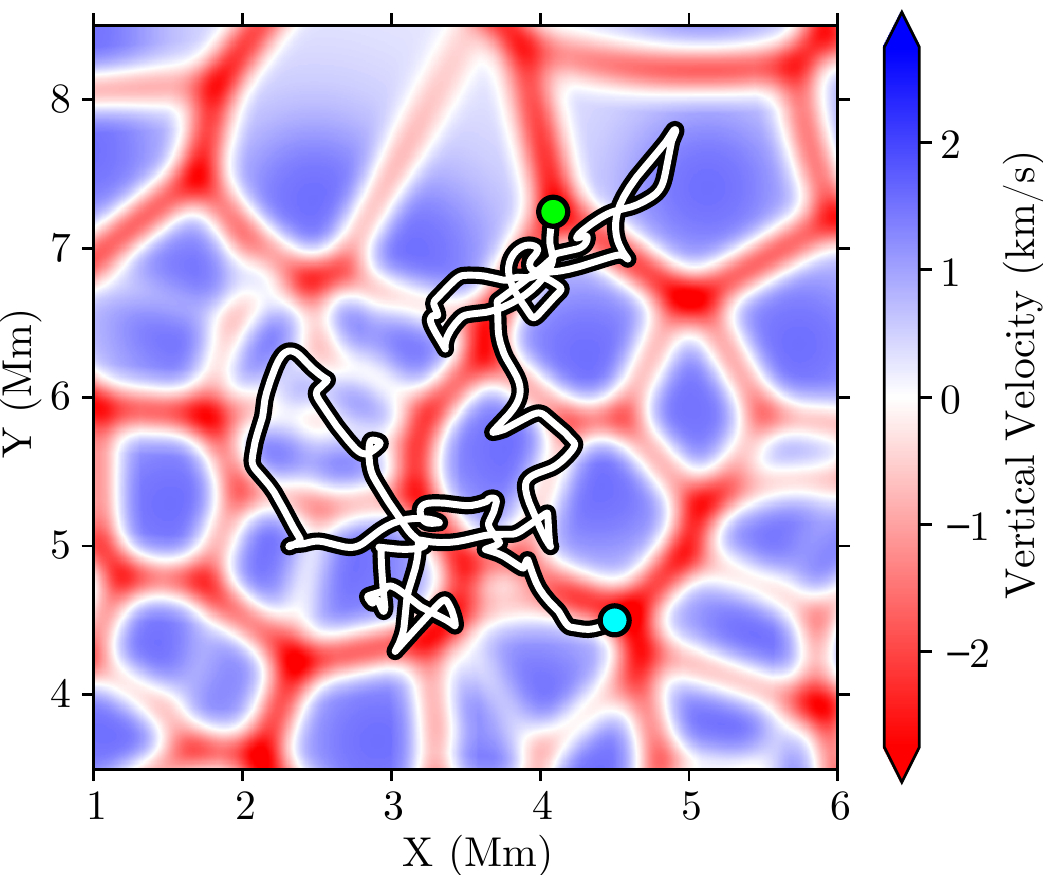}
	\caption[Sample granulation pattern produced by our \RF\ model]{Sample granulation pattern produced by our \RF\ model. On the left, the full simulation domain. On the right, a magnified portion showing the motion track of a single cork over 166~minutes. The cork moved from the cyan dot on the bottom to the green dot on the top. The granulation pattern shown in both plots is that of the final frame.}
	\label{fig:Granulation_Pattern}
\end{figure}

At this point, the modeled granulation pattern will have discontinuous transitions at granule boundaries.
Gaussian smoothing with a $1/e$ full-width of 317~km is applied to the horizontal velocities to produce realistic inter-granular lanes of finite size.
An example of the resulting granulation pattern is shown in Figure \ref{fig:Granulation_Pattern}.

As a final step, vertical velocities are calculated by mass conservation in an assumed hydrostatic density profile, using
\begin{equation}
	\vz = H_\text{eff} \left(\frac{d\vx}{dx} + \frac{d\vy}{dy}\right)
	\label{eq:vertical velocity}
\end{equation}
\citep[see][]{Simon1989}, where the effective scale height $H_\text{eff}$, defined in terms of the pressure and vertical velocity scale heights by $1/H_\text{eff} = 1/H_P + 1/H_{\vz}$, is set to 75~km computed from values of $H_{\vz}$ $\approx$ 150~km \citep{Oba2017} and $H_P$ $\approx$ 150~km.

\RF\ cannot contain explicit \bp s, so instead we analyze the horizontal flow fields using passive tracers (``corks'').
A cork is placed initially at the center of the simulation domain.
At each time step, the horizontal velocity at the cork's location is calculated by a two-dimensional interpolation from the four nearest pixels, and the cork moves with that velocity until the next timestep.
The cork's location is not constrained to discrete pixel locations, but is calculated and saved to full numeric precision.
When calculating power spectra, we remove the initial portion of each cork's path, up until the cork first reaches a pixel with a downward vertical velocity.
This is more representative of actual \bp s, though we find that doing so produces little change in the final power spectrum.
An example cork motion track is shown in Figure \ref{fig:Granulation_Pattern}.

The \RF\ model contains a number of adjustable parameters that control the granulation pattern, and varying those parameters yields physical insight.
We find that the cork power spectra are insensitive to the granule drift velocity distribution, consistent with the fact that granule drift is slow compared to granular flow velocities and to the timescale of change in the granulation pattern.
We find increased power in cork motion when the horizontal plasma velocities are increased, with the integrated power spectrum following a $\propto V_\text{max}^2$ scaling for moderate changes in velocity (reflecting the nature of kinetic energy).
We also find that total power is decreased when the granule emergence rate (per unit area, per unit time) is increased or when the granule scale size $r_0$ (Equation \ref{eq:V_r(r, t)}) is increased.
Increasing the emergence rate increases the number of granules and thus the fraction of area occupied by stationary inter-granular lanes.
Lanes are areas of converging horizontal flow, so a cork in a lane is in a relatively stable position and will experience less motion.
Further, having more granules causes each individual granule to be smaller.
When a new granule emerges and re-arranges the local intergranular lanes, then, the lanes will be shifted by a smaller distance and any displaced corks will be dragged along for a shorter distance before being in a stable location again.
An increase in the granule scale size pushes the maximum of the radial velocity function into the region affected by the Gaussian smoothing, thus extending the low-horizontal-velocity region from the center of granules to the edges.
This means that when the locations of the lanes changes, the granule-edge, horizontal flows that push corks along with the moving lane are weaker, and so corks move more slowly.

The analysis in this chapter is of 3072 independent \RF\ simulation runs, with one cork per run.
We use a spatial resolution of 50~km/pixel, a time step of 2~seconds, a simulation duration of 167~minutes, and a granule emergence rate of $5.9 \times 10^{-9}$~km$^{-2}$~s$^{-1}$.

\section{Results}
\label{sec:results}

\begin{figure*}[t!]
	\centering
	\begin{tabular}{ccc}
	\includegraphics[trim={0 0 .9cm .8cm},clip,width=0.31\textwidth]{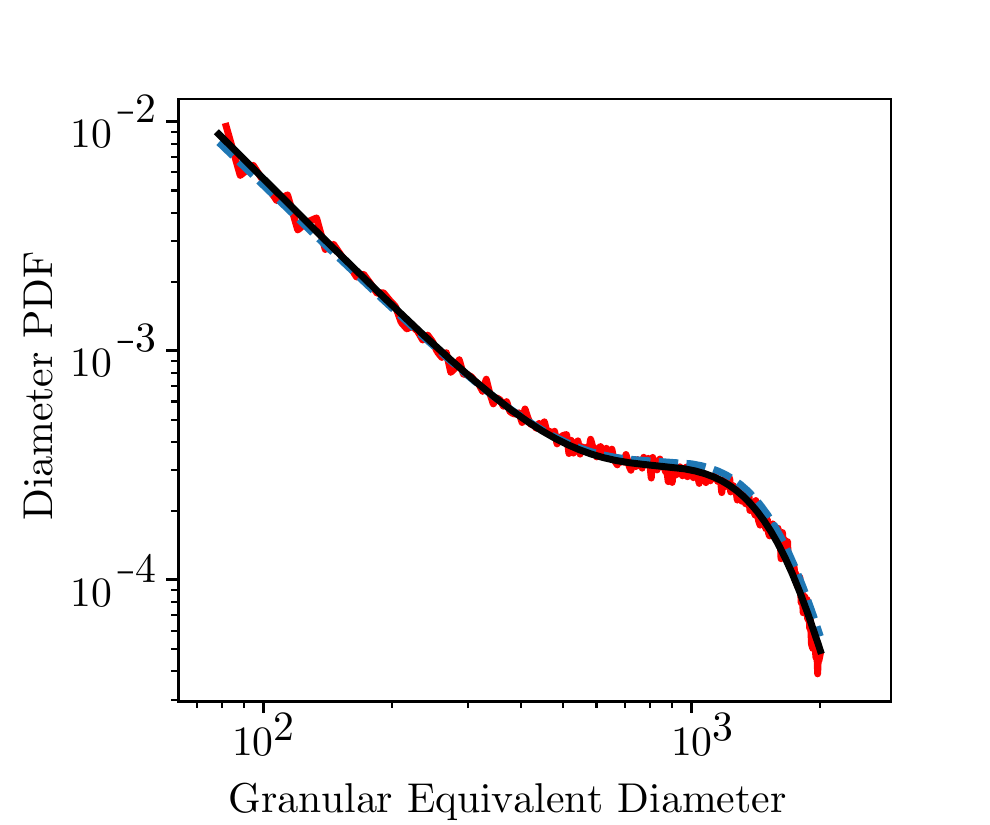}
	&
	\includegraphics[width=0.31\textwidth]{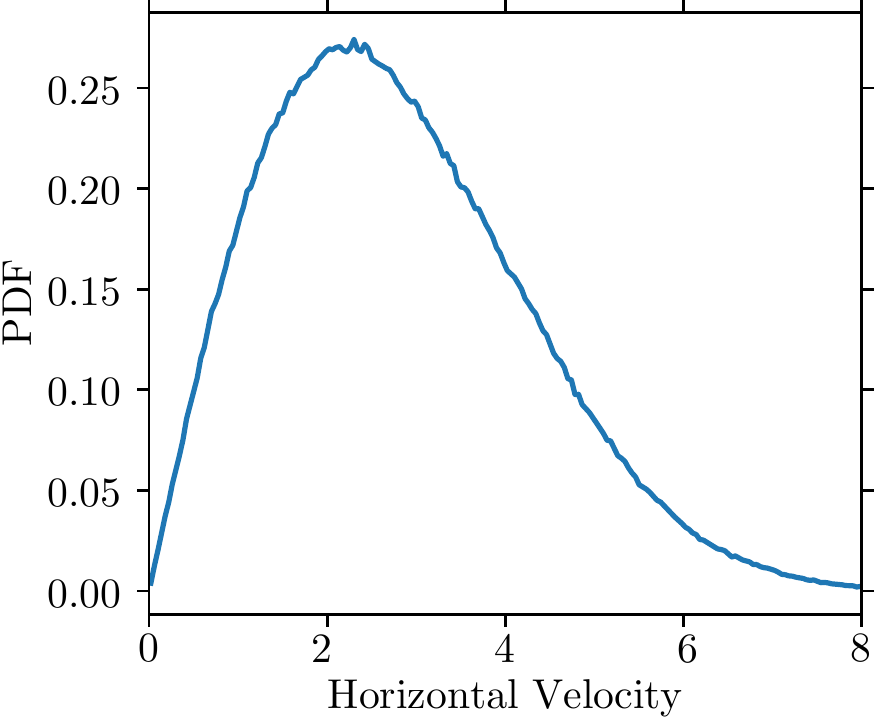}
	&
	\includegraphics[trim={0 0 0 .11cm},clip,width=0.31\textwidth]{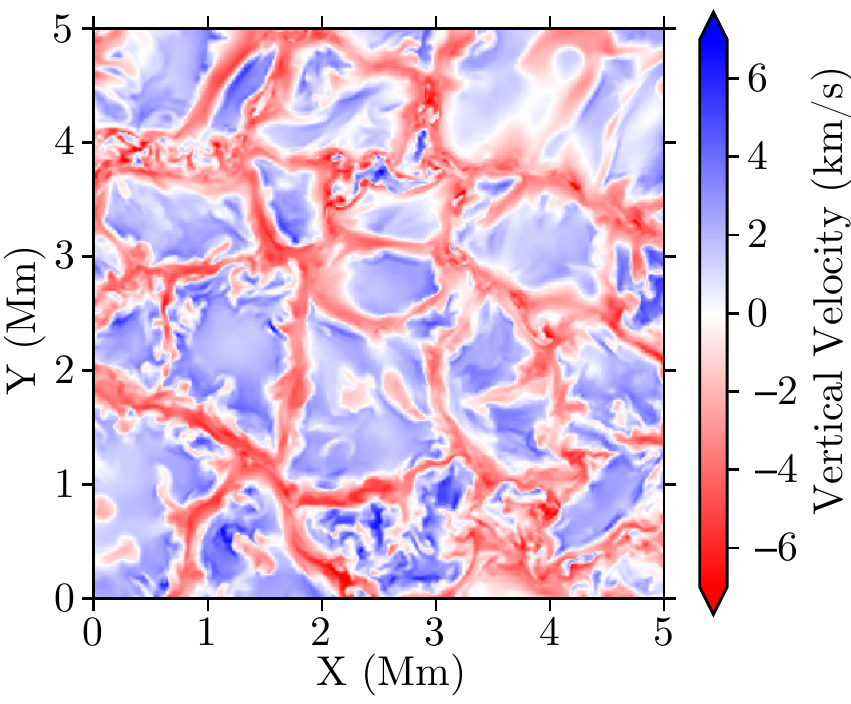}
	\\
	\includegraphics[trim={0 0 .9cm .8cm},clip,width=0.31\textwidth]{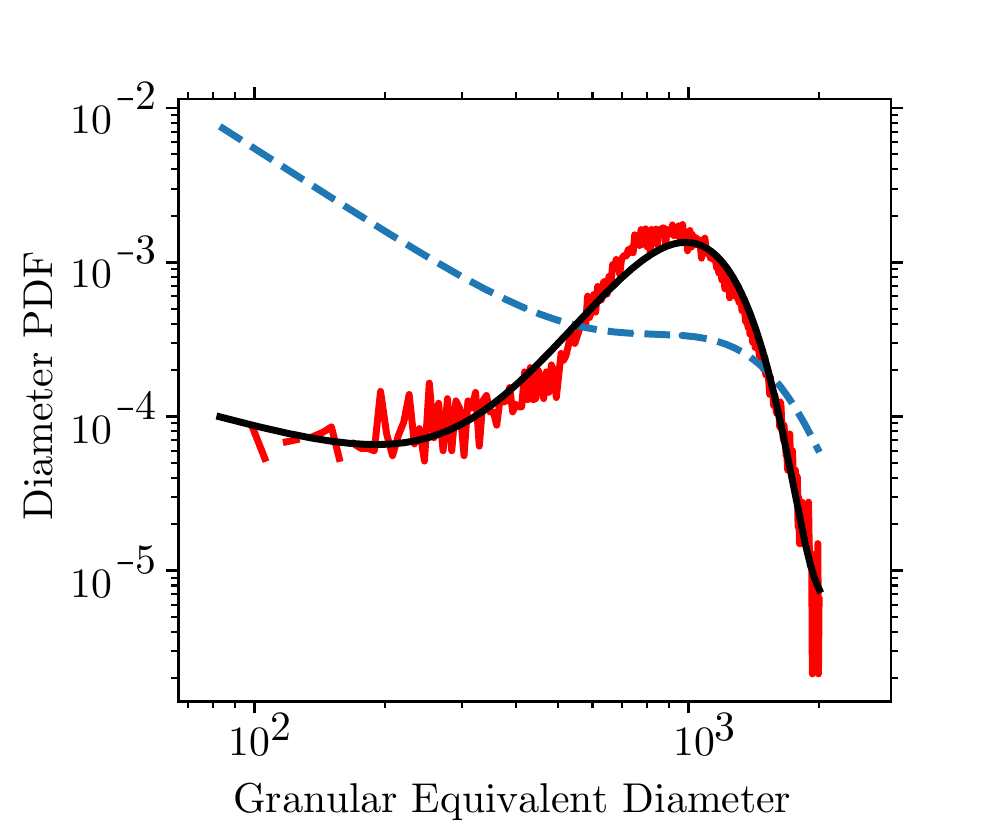}
	&
	\includegraphics[width=0.31\textwidth]{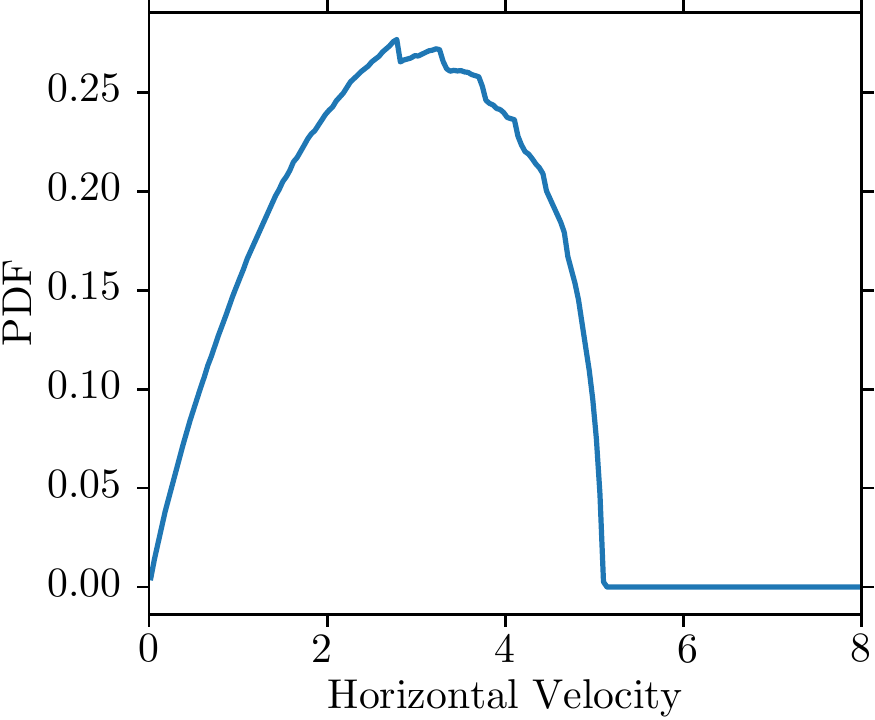}
	&
	\includegraphics[trim={0 0 .6cm 1.0cm},clip,width=0.31\textwidth]{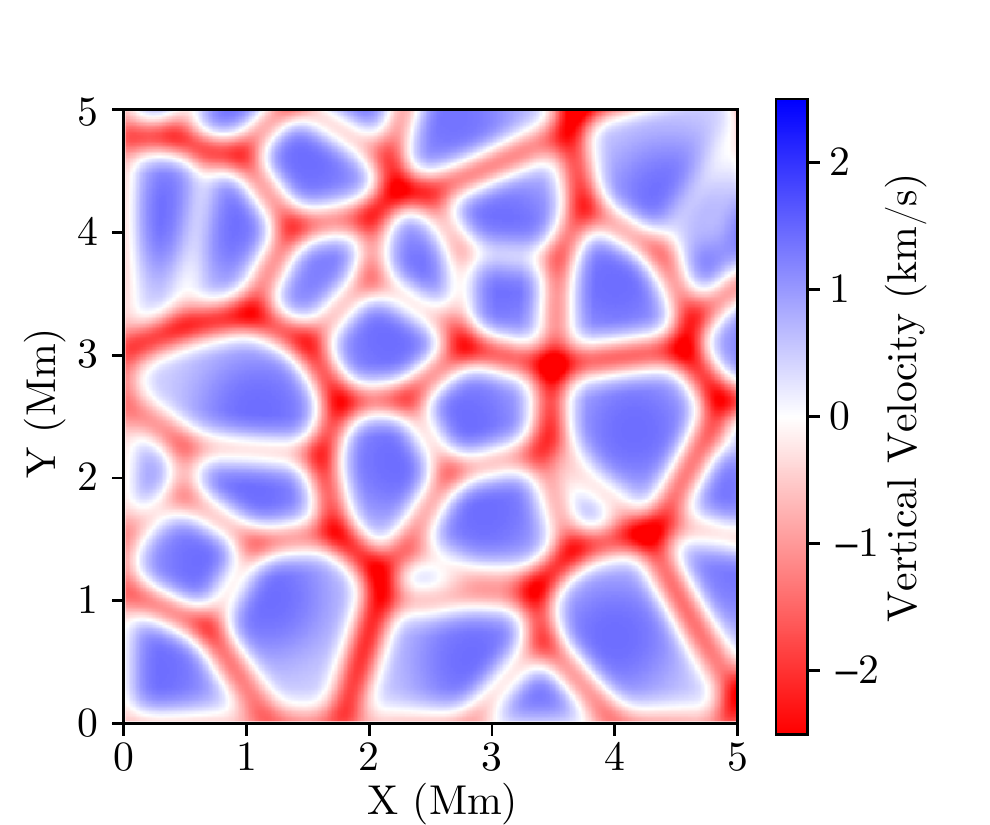}
	\end{tabular}
	\caption[Comparison of \muram\ and \RF\ granulation]{Comparison of \muram\ (top) and \RF\ (bottom) granulation. On the left is the distribution of granule sizes. In red is the PDF of granule sizes. The dashed line is the fitted distribution of \citet{Abramenko2012} (re-normalized to this plotting range, and obscured by the other curves in the \muram\ panel), while the solid curve is the same model fitted to our histogram. In the center is the histogram of horizontal velocity magnitudes at all pixels in the simulation. On the right is a sample vertical velocity map. Sizes and horizontal velocities are drawn from the full 401 frames (2.3 hours) of \muram\ data, and from 1500 frames (2.5 hours) of \RF\ simulation.}
	\label{fig:six-pan}
\end{figure*}

\subsection{Granulation}
\label{sec:results_granulation}
Here we present a comparison of the granulation patterns in \RF\ and \muram\ as a validation of the \RF\ approach.
In Figure \ref{fig:six-pan}, we compare the distribution of granule sizes between the two simulations and the observational values.
The observational values, from \citet{Abramenko2012}, draw on a field of view comparable to the \muram\ simulation domain and a diffraction-limited resolution of 77~km.
\citet{Abramenko2012} use intensity thresholding to separate granules from intergranular lanes.
As described in Section \ref{sec:muram_algorithms}, we instead use a $\vz=0$ threshold, which provides more straightforward granule segmentation.

Observed granule sizes are fit by the sum of a Gaussian distribution of large granules and a power law distribution of small granules \citep{Abramenko2012}, thought to reveal the presence of two distinct populations of granules \citep[as in][]{Hirzberger1997}.
We find that the \muram\ distribution reproduces very well the observational distribution.
However, the \RF\ distribution contains only the Gaussian portion.
This suggests that the small granules with a power-law size distribution have their origins in the turbulent phenomena not included in the \RF\ model, whereas the Gaussian-distributed large granules are produced by the large-scale, convective structures of the photosphere.

Figure \ref{fig:six-pan} also compares the distribution of horizontal velocities in the two simulations.
While the shape of the \RF\ distribution is set by the radial velocity functions (Equations \ref{eq:V_r(t)} \& \ref{eq:V_r(r, t)}), the horizontal scaling is set by the matching of integrated power spectra.
This produces good agreement between the two horizontal velocity distributions.
Notable, however, is the hard cutoff at $V\max$ in the \RF\ plot, whereas the \muram\ plot has a high-velocity tail due to turbulent regions.
Figure \ref{fig:six-pan} finally compares two granulation images, mapped in terms of vertical velocity, in which the qualitative agreement is good (aside from the presence of turbulence).

\subsection{\BP\ Statistics}
\label{sec:results_stats}
\begin{figure*}[t!]
	\centering
	\includegraphics[width=0.32\linewidth]{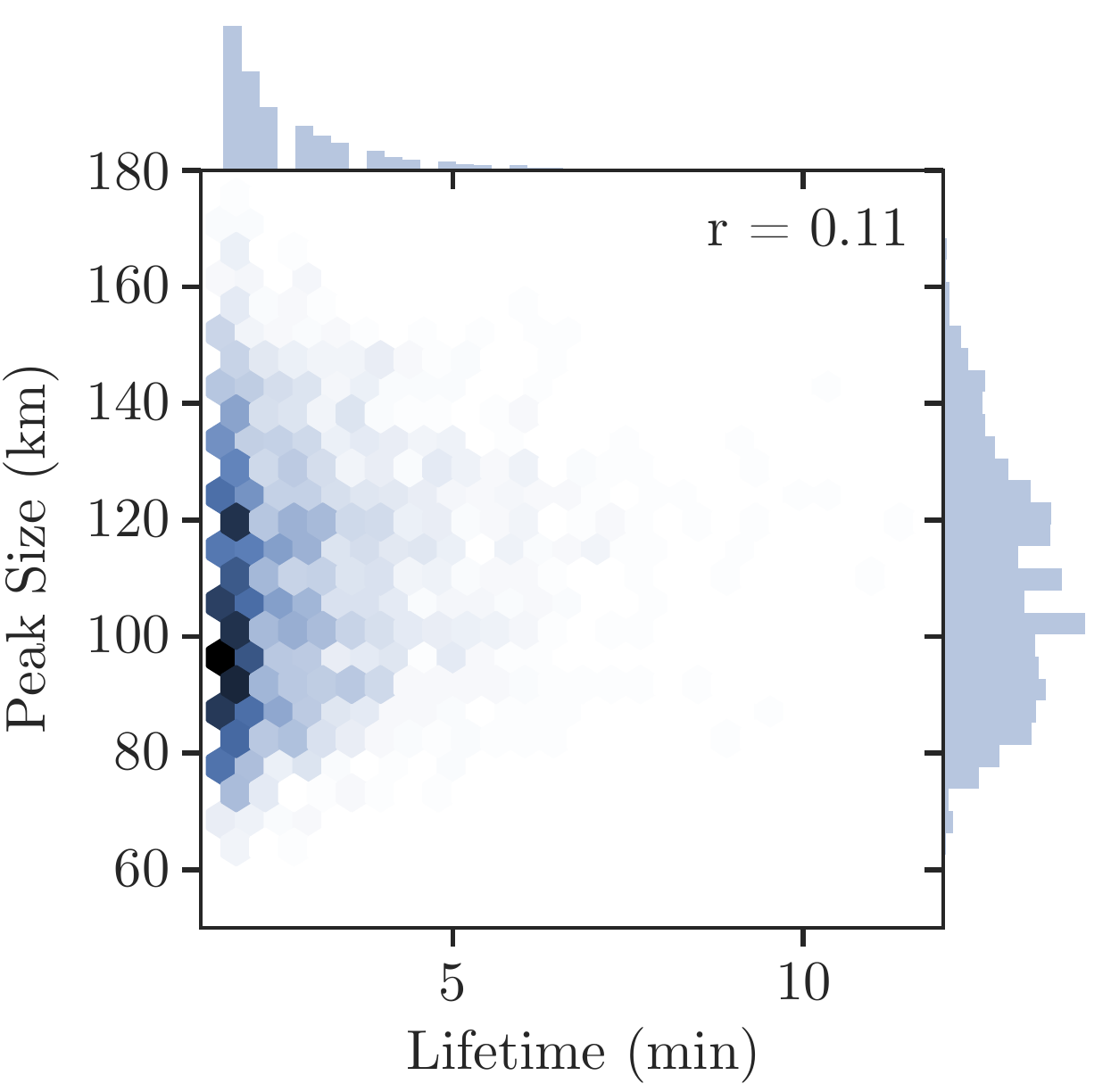}
	\includegraphics[width=0.32\linewidth]{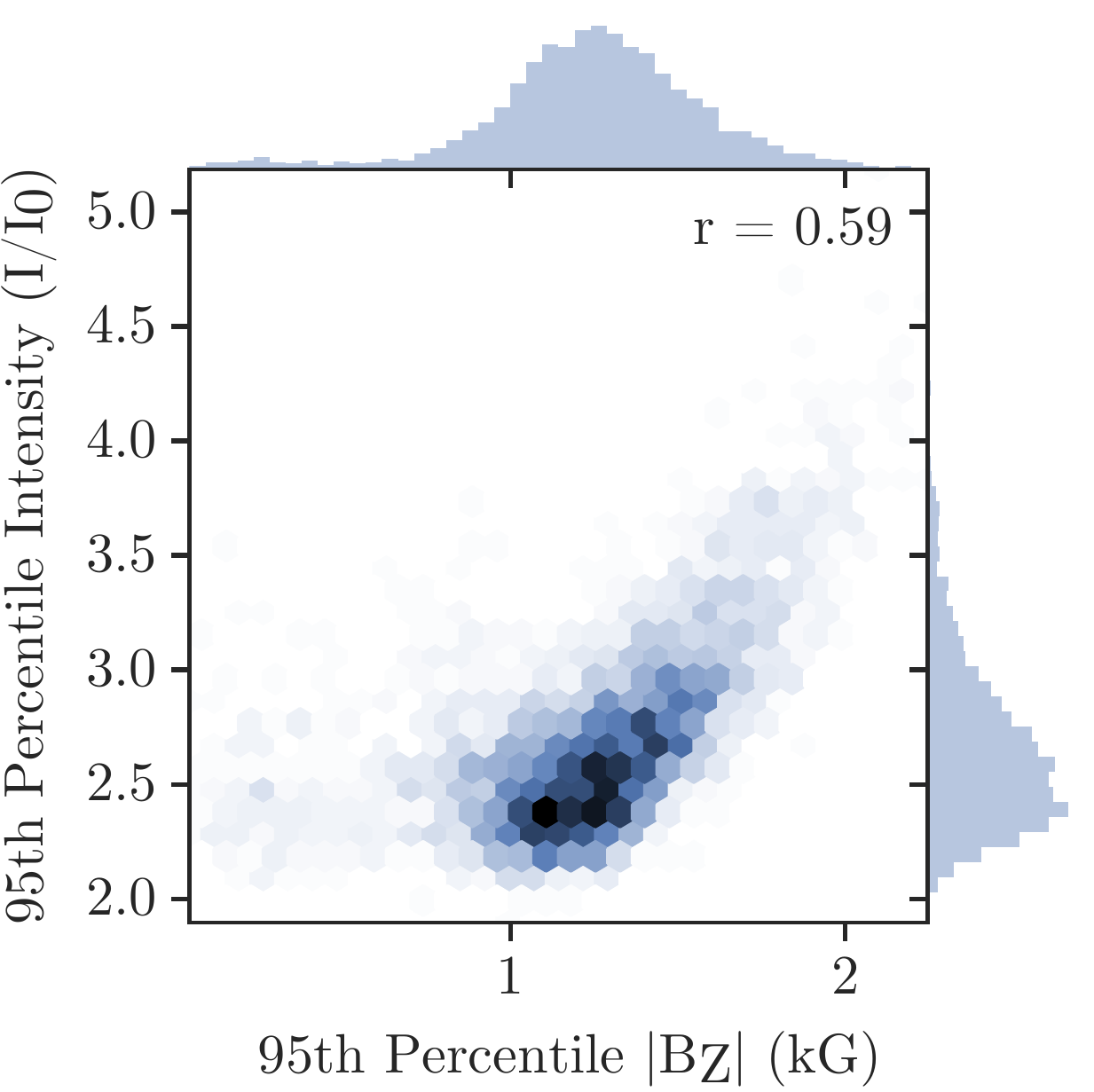}
	\includegraphics[width=0.32\linewidth]{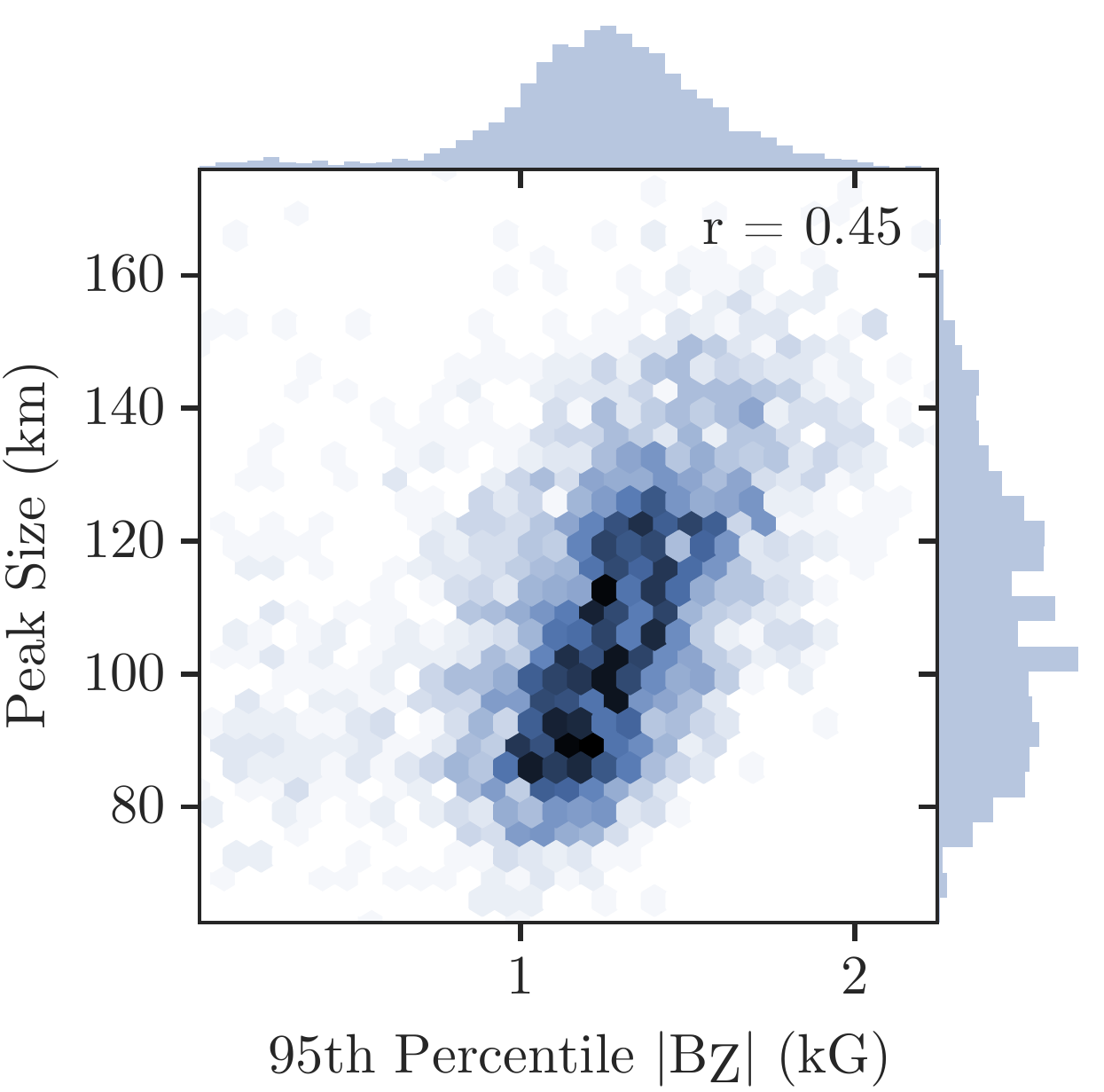}
	\caption[Distributions of \bhp\ statistics in the \muram\ data]{Distributions of \bhp\ statistics in the \muram\ data. ``Size'' is an equivalent diameter. The $95^\text{th}$ percentiles are computed from the values at all pixels identified as within a \bp . Marginalized distributions are shown along the edges of the plots. Shown for each distribution is the correlation coefficient; each has a two-sided p-value $<10^{-5}$.}
	\label{fig:bp_stats}
\end{figure*}

Applying the \bhp\ identification algorithm described in Section \ref{sec:muram_algorithms}, we find a total of 3,185 \bp s meeting all our size and lifetime thresholds in the \muram\ simulation data over the simulation time of 2.3 hours (401 snapshots).
Of these thresholds, the required lifetime of at least 1.6~minutes has the largest effect.
35 out of every 36 identified \bp s are removed by this criterion.
(See the steep lifetime distribution in Figure \ref{fig:bp_stats}.)
Of the remaining 3,185 \bp s, 9 were the result of mergers, 34 ended in a merger, 37 were the product of a fragmenting \bp, and 56 ended by fragmenting.

\begin{figure}[t!]
	\centering
	\includegraphics[width=0.6\textwidth]{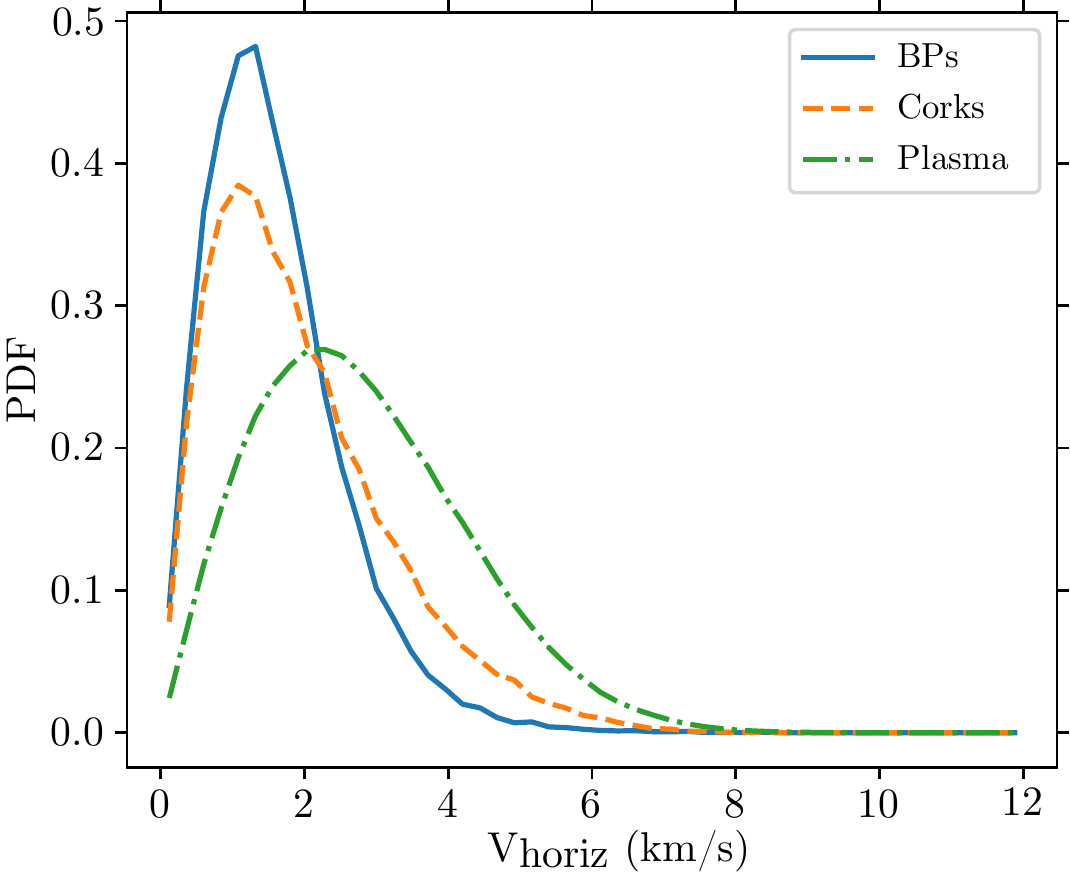}
	\caption[Horizontal velocity histogram for \bp s, corks, and plasma in the \muram\ data]{Horizontal velocity histogram for \bp s, corks, and plasma in the \muram\ data. Plasma velocity is sampled at every pixel in one frame, whereas the \bp and cork distributions use the full-length simulation. The largest value in each histogram is 8.97~km~s$^{-1}$ for \bp s, 10.8~km~s$^{-1}$ for corks, and 11.7~km~s$^{-1}$ for the plasma.}
	\label{fig:v_horiz_hist}
\end{figure}

In Figure \ref{fig:v_horiz_hist}, we show the distribution of observed horizontal velocities for \bp s, corks, and the plasma flow.
Cork velocities are lower than plasma velocities.
This is consistent with the fact that corks in a high-velocity region will travel at high velocity and quickly leave, while corks in a low-velocity region will linger; low-velocity regions are therefore preferentially sampled by corks.
\Bhp\ velocities may be susceptible to contributions from jitter in the centroid location caused by, e.g., \bhp\ shape changes or variation in the identified edge of the \bp .
We note here that \bhp\ velocities are typically lower than plasma or cork velocities, suggesting this is not a strong effect.
In the Appendix, we show more evidence that this effect appears to be insignificant in our data.

In Figure \ref{fig:bp_stats}, we show selected distributions of \bhp\ lifetimes, intensities, sizes, and vertical magnetic field strengths.
The distributions of \bhp\ size, vertical flux, and intensity all cluster near a central value (respectively, 109~km, 1.25~kG, and 2.7~$I_0$).
The \bhp\ lifetime, however, shows a sharp decay from the minimum lifetime threshold (1.6~minutes), with nearly all \bp s falling below a 5-minute lifetime.
The mean \bhp\ lifetime (for \bp s living longer than 1.6~min) is 2.7~min, while the maximum is 14.5~min.
This mean is comparable to the mean lifetime of $\sim$3~min for the longer-lived granule population in \citet{DelMoro2004}, indicating that \bp s evolve on the same timescale as the granulation pattern.
Our mean \bhp\ lifetime is notably shorter than the 9.33~minutes of \citet{Berger1998}.
However, they take steps to track \bp s through merging and splitting as well as through short ($\sim 40$ s) disappearances of \bp s.
We do not attempt to match these capabilities, as doing so would complicate the calculation of power spectra.

Neither \bhp\ lifetime nor \bhp\ velocity show a strong correlation with any other measured value.
This may be expected for lifetime due to our inability to track \bp s through splitting and merging events, meaning that our reported lifetimes do not all correspond to the true lifetimes of \bp s, and due to the fact that \bp s themselves are only one stage in the constant evolution of magnetic flux.
The non-correlation may also be expected for velocity, since \bhp\ motion is controlled by external forces.
However, strong correlations emerge between \bhp\ intensity, size, and magnetic field strength.
We use for each \bp\ the maximum size and the 95$^\text{th}$ percentile pixel value value for intensity and field strength in order to capture the value when the \bp\ is most fully emerged and vertical but to avoid outlier values.
We find a correlation coefficient of 0.59 between intensity and $|\bz|$, 0.44 between size and $|\bz|$, and 0.62 between size and intensity.
These correlations are all statistically significant and reasonably suggest that a stronger flux concentration will produce a larger, brighter \bp\ \citep[a conclusion supported observationally, e.g.][]{Ji2016}.

The correlations we find between \bhp\ lifetime and both diameter and velocity differ from those found in \citet{Yang2014}.
These Hinode/SOT observations yielded a correlation coefficient between lifetime and size to be +0.83 (contrasted with our value of +0.01 for mean values and +0.1 for peak values) and between lifetime and velocity to be $-$0.49 (contrasted with our $-$0.09 for mean values and +0.14 for peak values).
The large differences here are likely due to a difference in approach.
\Bhp\ lifetimes are discrete multiples of the imaging cadence, and for each lifetime value, \citet{Yang2014} calculate a mean value of diameter and velocity among all \bp s of that lifetime and use these mean values to perform a linear regression.
In contrast, we use our full dataset with all its variation.
Using average values collapses the data to a mean trend line, which risks artificially enhancing the resulting correlation coefficient.
We believe that using the full set of data more accurately represents the level of variation in the data (and, therefore, how well one measurable can predict another).

\subsection{Power Spectra}
\label{sec:results_spectra}

Following the literature \citep{VanBallegooijen1998,Cranmer2005,Chitta2012}, we compute power spectra from velocity sequences via the Wiener-Khinchin theorem, which provides a power spectrum as the Fourier transform of the autocorrelation of a data sequence.
While perhaps a roundabout approach,\footnote{In fact, this is the reverse of the usual application of the Wiener-Khinchin theorem, which is often used as an efficient way to calculate an autocorrelation by way of a Fourier transform, an absolute-square, and an inverse Fourier transform.} this allows individual \bhp\ velocity sequences to be more straightforwardly averaged together.
Since velocity sequences of different lengths are produced, their direct Fourier transforms (and those transforms' absolute-square) have different ranges and spacings in frequency, making a more direct averaging non-trivial.
But as long as different velocity sequences have the same temporal spacing, their autocorrelations will have the same spacing.
This allows averaging to be performed on the autocorrelations (analytically equivalent to averaging the final power spectra), which is computationally simpler.

For the \muram\ \bp s, we calculate power spectra of the finite-difference velocities of the intensity-weighted centroids\footnote{Post-publication addition: in calculating these weights, the intensity values within each \bp\ were normalized, such that the brightest pixel in the feature had a weight of 1, and the dimmest pixel in the feature had a weight of 0.} of \bp s.
Our observational comparison, \citet{Chitta2012}, present their power spectrum and autocorrelation function as fitted, analytical functions.
In this section, we present the power spectrum produced by discretely sampling the \citet{Chitta2012} autocorrelation function at its observational cadence of 5 s out to $\pm$ 105 s (the range used for their fit to the autocorrelation) and Fourier transforming, to provide a more direct comparison to our own power spectra.

\begin{figure}[t!]
	\centering
	\includegraphics[width=0.6\linewidth]{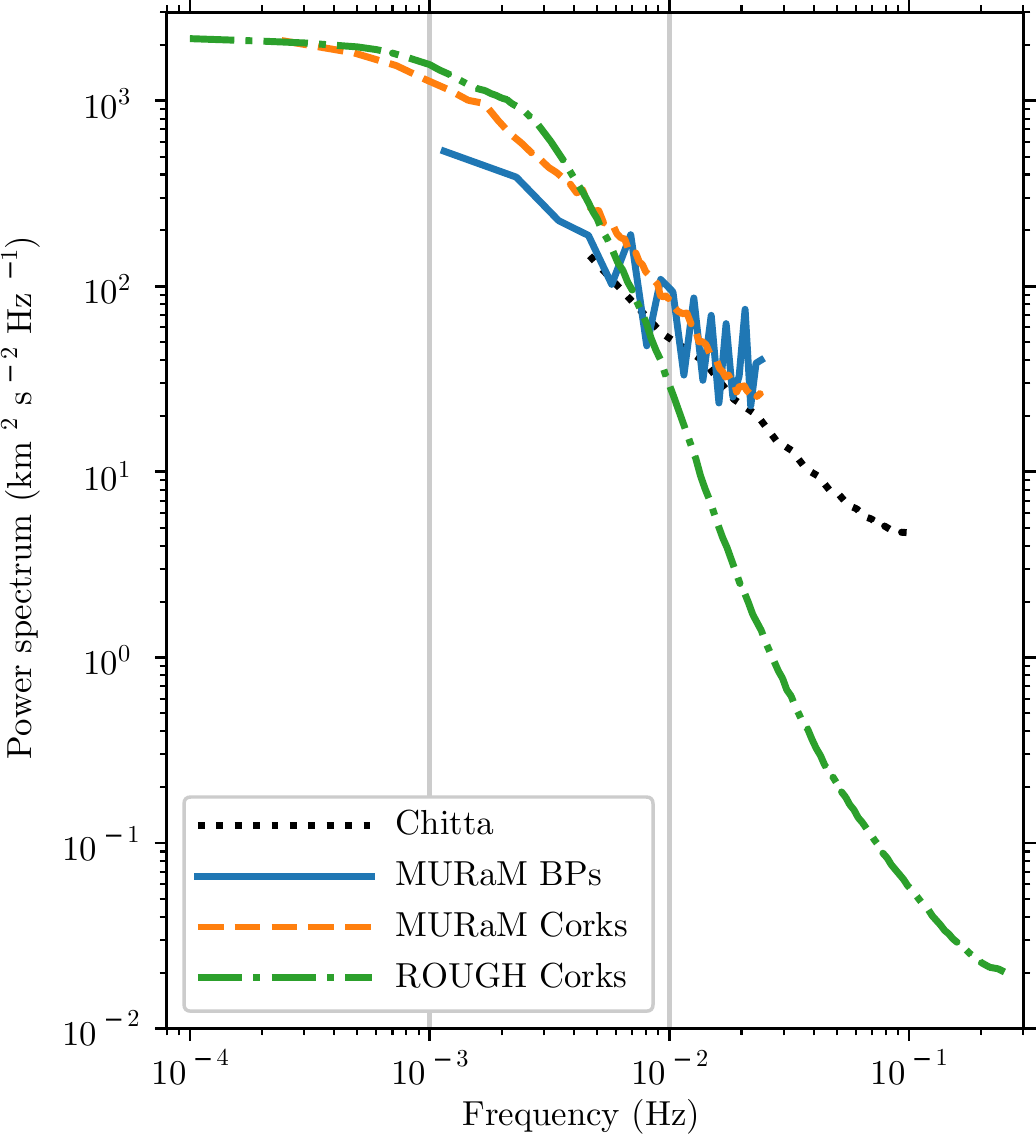}
	\caption[All power spectra mentioned in this chapter]{All power spectra mentioned in this chapter. The \muram\ cork curves are those of Section \ref{sec:muram_algorithms}. The \muram\ curve is from our \bp\ tracking approach described in Section \ref{sec:muram_algorithms}. The Chitta curve is from the observations of \citet{Chitta2012}. 
	The \RF\ curve comes from our simulations, described in Section \ref{sec:rough}. Gray vertical lines mark the frequencies of the slope measurements in Table \ref{tab:integrated_power}.}
	\label{fig:all_spectra}
\end{figure}

Power spectra for observations, \muram\ corks, \muram\ \bp s, and \RF\ corks are shown in Figure \ref{fig:all_spectra}.
\RF\ is configured so that the integrated power matches the integrated \muram\ cork power.

At low frequencies, the \RF\ spectrum is similar to the \muram\ cork spectrum, whereas at high frequencies it drops off precipitously.
This reflects the similarity between the two simulations of the large-scale granulation patterns, which cause low-frequency movement, and the absence in \RF\ of small-scale turbulence \citep[e.g. vortex flows, see][]{Giagkiozis2018}, which drive high-frequency motion.
This comparison also suggests that these two phenomena are the dominant drivers of motion in their respective frequency regimes.

The \muram\ cork spectrum shows more low-frequency power than the \muram\ \bp s, suggesting that perhaps corks are more free to drift in steady flows than are \bp s, but the two spectra converge at high frequencies, suggesting that they respond similarly to rapid variation in plasma flow.
Finally, both \muram\ analyses show a consistent enhancement in power over the Chitta data for the frequencies of overlap, quantified later in this section.

The integral of the power spectrum is equal to the variance of the velocity data (via Parseval's theorem).
These integrated values are shown in Table \ref{tab:integrated_power}.
We report both the integral of each spectrum over its full extent and the integral over the frequency range covered by all four spectra (0.005 Hz $\le$ f $\le$ 0.02 Hz).
The former represents a spectrum's lower bound on the ``true'' total power, unconstrained by limited frequency coverage, while the latter allows spectra to be compared over a comparable domain.
These numbers reveal more motion and variation for \muram\ corks than for \bp s, and as noted above, we find a 91\% increase in total power measured in the frequencies of overlap for \muram\ \bp s versus observed \bp s\footnote{The comparison of energy fluxes computed from \muram\ and the Chitta observations is explored further in Section \ref{sec:blurring-data} of this dissertation}.
This increase in motion, in both the power spectra and the integrated spectra, might be explained in part by the calibration offset applied by \citet{Chitta2012}, a step which could tend to damp out some real motion in addition to spurious motion and which is not needed in simulation analysis.
It might also suggest there is, in fact, more power to be detected with higher-quality observations.
Improvements in spatial resolution and sensitivity will make \bhp\ detection and centroid identification more reliable and make smaller \bp s visible, both enabling more robust statistics.
Verification of the presence of unmeasured power must wait until DKIST comes online, currently scheduled for 2019.
The Visible Broadband Imager on DKIST will have a diffraction-limited, G-band resolution of $\sim$16~km (comparable to these \muram\ simulations) and a 3~second cadence \citep{Elmore2014}, improving on both the $\sim$160~km resolution of the Hinode SOT observations of \citet{Yang2014,Yang2015} and the $\sim$100~km resolution and 5~second cadence of the SST observations of \citet{Chitta2012}.  

Both the \muram\ and \citet{Chitta2012} power spectra shown in Figure \ref{fig:all_spectra} indicate power-law behavior with slopes of order $f^{-1}$ between $10^{-3}$ and $10^{-2}$~Hz (see Table \ref{tab:integrated_power}).
This is a much flatter spectrum than the presumed exponential drop-off associated with random-walk-like motions like those found from earlier measurements of bright points \citep[e.g.,][]{VanBallegooijen1998,Cranmer2005}.
However, whether this points to the presence of an active turbulent cascade \citep{Petrovay2001} in the photosphere is still not known.
Reduced MHD turbulence simulations for solar flux tubes \citep{VanBallegooijen2011,Woolsey2015} showed that low-frequency random-walk motions rapidly produce a steep power-law spectrum ($P \sim f^{-4.5}$) in the photosphere.
This may be modified for the flatter bright-point driving spectrum found in this work.
There is a long history of interplanetary $f^{-1}$ fluctuations being interpreted as fossil remnants of oscillations that map down to the solar surface \citep{Matthaeus1986,Velli1989,Dmitruk2007,Verdini2012}.
However, those fluctuations occur at even lower frequencies than those studied in this chapter (e.g., $f < 10^{-4}$~Hz) and their physical origin is still being debated.

\begin{table*}[t]
	\centering
	\begin{tabular}{l|cc|cc}
		& \multicolumn{2}{c}{Integrated Power (km$^2$ s$^{-2}$)} & \multicolumn{2}{c}{Slope at} \\
		Spectrum  & over all frequencies &  0.005 Hz $\le$ f $\le$ 0.02 Hz & 0.001 Hz & 0.01 Hz \\
		\hline
		Chitta & 1.87 & 0.56 & \dots & -1.25 \\
		\muram\ BPs & 3.46  & 1.07 & \dots & -1.27 \\
		\muram\ Corks & 5.42  & 1.21 & -0.37 & -1.59 \\
		\RF\ Corks & 5.42 & 0.57 & -0.45 & -3.45\\
	\end{tabular}
	
	\vspace{.1in}
	
	\caption[Properties of the spectra in Figure \ref{fig:all_spectra}]{Properties of the spectra in Figure \ref{fig:all_spectra}. Note that \RF 's $V\max$ value is set so its corks' integrated spectrum matches that of the \muram\ corks.}
	\label{tab:integrated_power}
\end{table*}

\section{Discussion and Conclusions}
\label{sec:conclusion}

The goal of this chapter was to better understand the horizontal motions of photospheric intergranular \bp s.
We characterized this motion using simulations that exceed present observational capabilities, and we investigated which aspects of the granular flows drive this motion.

Using high-resolution simulations, we found an increase of 91\% over observations in the power spectrum of \bhp\ motion at all frequencies of overlap.
This suggests the possibility that the limitations of current observations under-represent the motion of \bp s---an idea to be tested by DKIST observations.
The power spectrum of \muram\ \bp s is only negligibly affected when blurring the simulation output to an observational resolution\footnote{Blurring of the \muram\ data is further explored in Section \ref{sec:blurring-data} of this dissertation.}.
However, the expected increase in signal-to-noise ratio for DKIST observations also plays a significant role in improving \bhp\ tracking reliability.

We compared the power spectra of passive tracers in \muram\ flows and \RF 's simplified, turbulence-free flows.
This comparison suggested that low-frequency motion of passive tracers (and thus \bp s) is driven by long-term granular evolution, while high-frequency motion is due to turbulent effects.

We found that the power spectra of \bhp\ motion and passive-tracer motion are very similar at high frequencies, but passive tracers display about 2.5 times more power at low frequencies.
One possible explanation for this is that the passive tracers move according to the plasma flow field at the $\tau = 1$ surface only, whereas \bp s are pushed at multiple depths along the flux tube.
Any depth dependence to the horizontal plasma velocity will result in competing forces on the flux tube, producing muted \bhp\ motion.
That this difference is seen only at low frequencies may suggest that rapid changes to the flow field (producing high-frequency motion) are either coherent across the relevant depths or are sufficiently strong at one depth to dominate over the flows at other depths, whereas long-term flows have more depth dependence and so produce net smaller effects on \bp s.

We found that \muram\ effectively reproduces the granule size distribution of the observations in \citet{Abramenko2012}.
However, of this two-component distribution, \RF\ reproduces only one component---the Gaussian-distributed, large granules.
We proposed that the two components to the granule size distribution represent the large convective cells and the small, turbulent features.
Separately, \citet{DelMoro2004} finds that granule lifetimes are best fit by a two-exponential model, finding a different fit for the $> 5$~minute and $< 5$~minute lifetimes.
We speculate that these two fits match the two classes in the granule size distribution---the small, turbulent features correspond to the short-lived features, and the large-scale convective features correspond to the long-lived features.
Indeed, \citet{DelMoro2004} finds a correlation between granule lifetime and size that is positive for short-lived granules, but which flattens for lifetimes over $\sim 8$~minutes.

We found that \bhp\ size, brightness, and magnetic flux are all positively correlated (with $r \sim 0.5$).
This can be made sense of, remembering that \bp s are bright because a reduction in gas density lowers the $\tau =1$ surface to hotter, brighter regions.
A larger-diameter flux tube will tend to allow observations deeper into the tube before bending cuts off the line of sight, while stronger magnetic fields will produce lower gas pressure and density in the flux tube that allows a clearer view deeper into the tube.

It is worthwhile to use the results found above to investigate some properties of \Al\ fluctuations that propagate up into the corona and give rise to dissipative heating.
Following \citet{Cranmer2005}, it is possible to estimate the net upward energy flux $F$ of \Al\ waves given the properties at the photosphere,
\begin{equation}
  F \, = \, \rho \langle v_{\perp}^2 \rangle V_{\rm A}
  \left( \frac{1 - {\cal R}}{1 + {\cal R}} \right) \,\,\, ,
  \label{eq:flux}
\end{equation}
where we assume a photospheric density $\rho = 2 \times 10^{-7}$ g~cm$^{-3}$ and typical bright point field strength $B = 1500$ G in order to estimate the \Al\ speed $V_{\rm A} = B/\sqrt{4\pi\rho}$, and the velocity variance $\langle v_{\perp}^2 \rangle$ is the integrated power given in Table \ref{tab:integrated_power}.
The above expression also depends on the reflection coefficient for wave energy ${\cal R}$, which \citet{Cranmer2005} found to be approximately 0.90 below the transition region.

Note that Equation (\ref{eq:flux}) gives the energy flux {\em inside} a vertically oriented bright-point flux tube.
However, in the corona, these flux tubes are expected to broaden out to eventually fill the entire volume.
Thus, a more relevant energy flux for coronal heating is the surface-averaged value $\langle F \rangle \approx f_{\ast} F$, where $f_{\ast}$ is the photospheric filling factor of bright points, which we treat as a simple, unweighted area fraction.
In the \muram\ simulation analyzed above, $f_{\ast} = 0.00474$ for all identified bright points, including those with lifetimes below 1.6~minutes that did not yield useful velocity information.
Combining the above quantities, we find $\langle F \rangle \approx 860$ W~m$^{-2}$ for the \citet{Chitta2012} integrated power, 1590 W~m$^{-2}$ for the power in \muram\ bright points, and 2490 W~m$^{-2}$ for the power in \muram\ cork motions.
These values are well above the classical \citet{Withbroe1977} empirical requirements for coronal heating in the quiet Sun (300 W~m$^{-2}$) and coronal holes (800 W~m$^{-2}$), indicating that only a fraction of the wave energy flux needs to be dissipated as heat in these regions.

One aspect not considered in this work is the impact of long chains of connected \bp s seen in regions of strong background magnetic flux \citep{Dunn1973,SanchezAlmeida2004}.
These features were considered by \citet{VanBallegooijen1998}, but they generally require a different approach than that used in this chapter for small, isolated \bp s.
Differences may be expected between the dynamics of the isolated and the extended types of \bp s, due to both interactions between flux elements in extended structures, and the observed trend for \bp s to move more slowly in regions of increased magnetic flux \citep{Ji2016,Yang2016}.

While the transverse \bhp\ motion studied in this chapter is useful, it does not tell the complete story.
MHD waves can also be excited both inside and outside the \bhp\ flux tubes by compressive and longitudinal fluctuations.
This fact will become more prominent when DKIST turns traditionally-point-like \bp s into resolved structures in which the motions necessary to excite these additional modes will become observable.
In fact, as discussed in the Appendix, the ability to track higher-order motion may become necessary.
We plan to study these wave modes in future work.
Beyond MHD waves, \bhp\ motion can also be analyzed in bulk, looking at the diffusive statistics of the \bp s \citep[e.g.][]{Abramenko2011,Ji2016,Jafarzadeh2017,Agrawal2018}.

Future efforts at \bp\ tracking may include algorithm refinements.
\citet{Berger1998} differ from our approach in their ability to sensibly track \bp s through merging and splitting events, whereas we simply consider the products of such events (which are fortunately rare in our \muram\ data set) to be newly-found \bp s.
Their algorithm is also resilient to \bp s which disappear for up to a few frames before reappearing.
However, this improvement may become unnecessary with the high cadence and sensitivity of DKIST data.
Algorithmic improvements to improve robustness or detection quality might also be adapted from the recommendations of \citet{DeForest2007}, benefiting from existing efforts in large-scale feature detection.

\paragraph{Acknowledgements}

The authors warmly thank Matthias Rempel for sharing the results of his \muram\ simulations, and Mark Rast and Piyush Agrawal for discussing with us the results of their own work and inspiring the contents of this chapter's appendix.
SJV thanks Craig DeForest and Derek Lamb for a productive summer spent learning many coding techniques used in this analysis.
This work was supported by NSF grant 1613207 (AAG).
SRC's work was also supported by NASA grants {NNX\-15\-AW33G} and {NNX\-16\-AG87G}, NSF grant 1540094 (SHINE), and start-up funds from the Department of Astrophysical and Planetary Sciences at the University of Colorado Boulder.
Portions of this work utilized the RMACC Summit supercomputer, which is supported by the National Science Foundation (awards ACI-1532235 and ACI-1532236), the University of Colorado Boulder, and Colorado State University. The Summit supercomputer is a joint effort of the University of Colorado Boulder and Colorado State University.
This work made extensive use of the Scientific Python ecosystem and NASA's Astrophysics Data System.

\section{Appendix: Centroid Jitter}
\label{sec:jitter}

The tracking of centroids in identified features is susceptible to jitter in the centroid motion due to the effect of feature shape changes on the computed centroid location.
Whether the shape change is due to twisting or compression of the \bp , the merging of sub-resolution \bp s with the tracked \bp , or is just an artifact of the tracking routine due to changes in the identified edge of the \bp , a change in shape can produce additional centroid motion beyond idealized, rigid-body advection of the \bp .
This additional motion may represent real motion of magnetic field lines and a source of higher-order MHD waves.
However, in the context of tracking \bhp\ centroids as representative of the flux tube's bulk motion (the context of this chapter and most papers tracking \bp s as MHD wave sources), this additional motion is spurious.
Centroid tracking becomes vulnerable to this effect only when increased spatial resolution changes \bp s from points into well-defined features---i.e. high-resolution simulations and DKIST observations.
This effect is explored more fully in \citet{Agrawal2018}; here we attempt to constrain the influence of this effect on power spectra such as those presented in this chapter.

Between two consecutive Frames 1 and 2, a \bp\ will have measured areas $A_1$ and $A_2$, with a size change $\Delta A \equiv A_2-A_1$.
In this section we use a simple model of the Frame 1 \bp\ as a square of side length $\sqrt{A_1}$ in Frame 1, and the \bp\ in Frame 2 as a rectangle of side lengths $\sqrt{A_1}$ and $\sqrt{A_1} + \ell$, where $\ell = \Delta A / \sqrt{A_1}$ so that the rectangle's area is $A_1 + \Delta A = A_2$.
This is an intermediate case for the effect of a size change $\Delta A$ on the centroid location, where the extremes are uniform, radial expansion having no effect on the centroid, and the addition of a long and thin filament to the shape of Frame 1 having an extreme effect on the centroid.
In this model, and assuming no advection of the \bp , the unweighted centroid will move a distance of $\Delta x = \Delta A / (2 \sqrt{A_1})$ along the major axis of the rectangle.

We calculated $\Delta x$ from measured areas for each \bp\ and for each timestep in our 20-second-cadence \muram\ simulation.
We did the same after applying our tracking routine to a 2-second-cadence \muram\ run (without the 30~G net injected flux of the 20-second-cadence run; see \citet{Rempel2014,Agrawal2018}).
While this 2-second-cadence run has a smaller spatial domain (6~$\times$~6~Mm, versus the 24.5~$\times$~24.5~km of the 20-second-cadence simulation used in this chapter) and thus provides fewer \bp s and poorer statistics, it allows us to investigate how DKIST measurements might be affected by this centroid jitter. 

\begin{figure}[t!]
	\centering
	\includegraphics[width=0.65\textwidth]{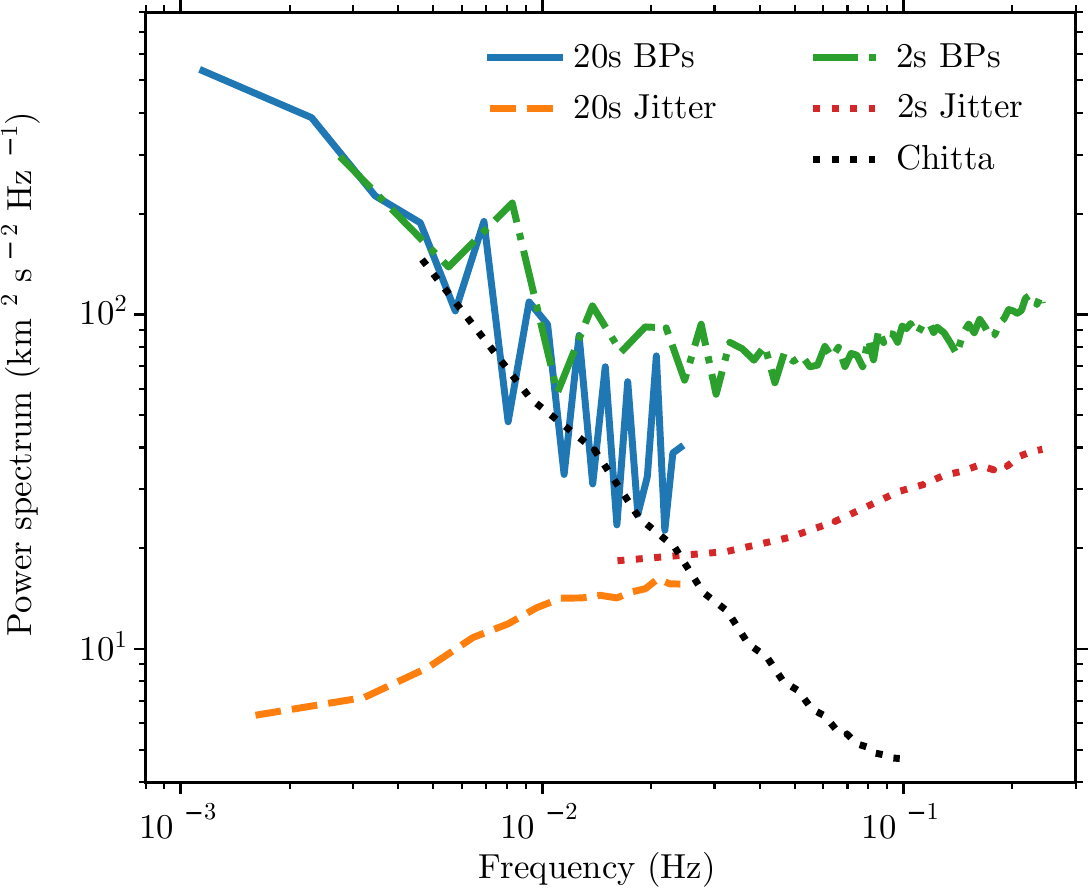}
	\caption[Power spectra of total \bp\ (BP) motion and the modeled contribution of centroid jitter]{Power spectra of the total \bp\ (BP) motion and the modeled contribution to that motion of centroid jitter for \bp s in 20-second- and 2-second-cadence \muram\ runs. The 20-second-cadence \bp\ spectrum is the same as is shown in Figure \ref{fig:all_spectra}, and the observational \citet{Chitta2012} spectrum of that figure is also reproduced to facilitate comparison.}
	\label{fig:jitter}
\end{figure}

To simulate the contribution of centroid jitter to the motion of a \bp , we then average the power spectra of every 30-sample sequence of $\Delta x$ values.
These spectra are shown in Figure \ref{fig:jitter}.
The jitter spectra for the two different cadences produce a nearly-continuous line.
The spectrum for the 20-second-cadence data exceeds the jitter spectrum while showing a consistent slope of opposite sign.
The spectrum for the 2-second-cadence data, however, coincides with the 20-second-cadence spectrum for low frequencies, but levels off and even takes on a positive slope for the highest frequencies.
In the mid- to high-frequency regime, this spectrum suggestively takes the shape of the jitter spectrum.

We interpret this to mean that, at a 20-second-cadence, the jitter in a \bp 's centroid is small compared to the advective motion between frames, and the analysis in this chapter is safe from misinterpreting jitter as bulk advection.
However, the additional frequencies probed by the 2-second-cadence data appear to be dominated by jitter rather than advection.
We base this on the similarity in slope between the \bhp\ and jitter spectra and attribute the factor of $\sim 2$ difference in magnitude to the simplicity of our jitter model.
This effect may not have been visible in the high-cadence \citet{Chitta2012} observations due to their spatial resolution rendering unresolvable the shape changes that give rise to this jitter.
However, we anticipate both high cadence and high spatial resolution in DKIST data.
Therefore, analysis of MHD wave excitation by \bp s observed by DKIST must take a more advanced approach\footnote{This will be partially addressed in the next chapter of this thesis by refinements to the automatic tracking algorithm.} or it will risk assigning the energy in higher-order modes to simple, bulk advection.

\biblio

\chapter{Beyond Centroid Tracking}
\label{chap:bp_tracking}

As discussed in previous chapters, the upcoming DKIST solar telescope will reveal the shapes and sizes of \bp s.
This will provide a new source of information and new ways to estimate the energy flux from waves excited by \bp\ motions.
In this and the following chapters, I present my work to develop exploratory techniques for making use of this new information.
This chapter discusses requisite improvements in my \bp\ tracking code and the motivation for new analysis techniques.
Chapters \ref{chap:ellipse-fitting} and \ref{chap:emd} present two novel techniques for analyzing \bhp\ shape- and size-changes, and Chapter~\ref{chap:method-comp-and-conclusions} compares the two techniques and presents concluding thoughts.
After the defense of this thesis, these chapters will be adapted and submitted for publication, with Steve Cranmer as co-author.
At that time, the code used in this analysis will be made public.

\section{Improved Identification of \BHP\ Boundaries}
\label{sec:improved-boundaries}

\subsection{\muram\ Simulation}
\label{sec:2s-muram}
To demonstrate and analyze the techniques presented in the following chapters, we use a \muram\ simulation run of DKIST-like resolution.
\citet{Rempel2014} used an extensively-modified version of \muram\ \citep{Vogler2005,Rempel2009}, a radiative MHD simulation code.
We use the outputs of a \muram\ run similar to the O16b run of \citet{Rempel2014} but with an expanded vertical domain above the photosphere and with radiative transfer computed in four opacity bins.
This run was also analyzed by \citet{Agrawal2018}.
The simulation domain is $6\times6\times4$~Mm$^3$ at a 16~km grid spacing, with the photosphere approximately 1.7~Mm below the top of the domain.
Snapshots were saved every two seconds over about one hour of simulated time.
The simulation includes convective flows at well-resolved granular scales, with \bp s arising naturally from a magnetic field produced by a small-scale dynamo.
The upward-directed, white-light intensity is computed through radiative transfer with four opacity bins and produces an analog of observational images which we use in our analysis.
This simulation was used in the Appendix of Chapter~\ref{chap:bp-centroids}.
Compared to the simulation used in the main body of Chapter~\ref{chap:bp-centroids}, this simulation has a faster cadence but a reduced spatial domain.
This simulation also does not feature an imposed net vertical magnetic flux.

\begin{figure}[t]
	\centering
	\includegraphics{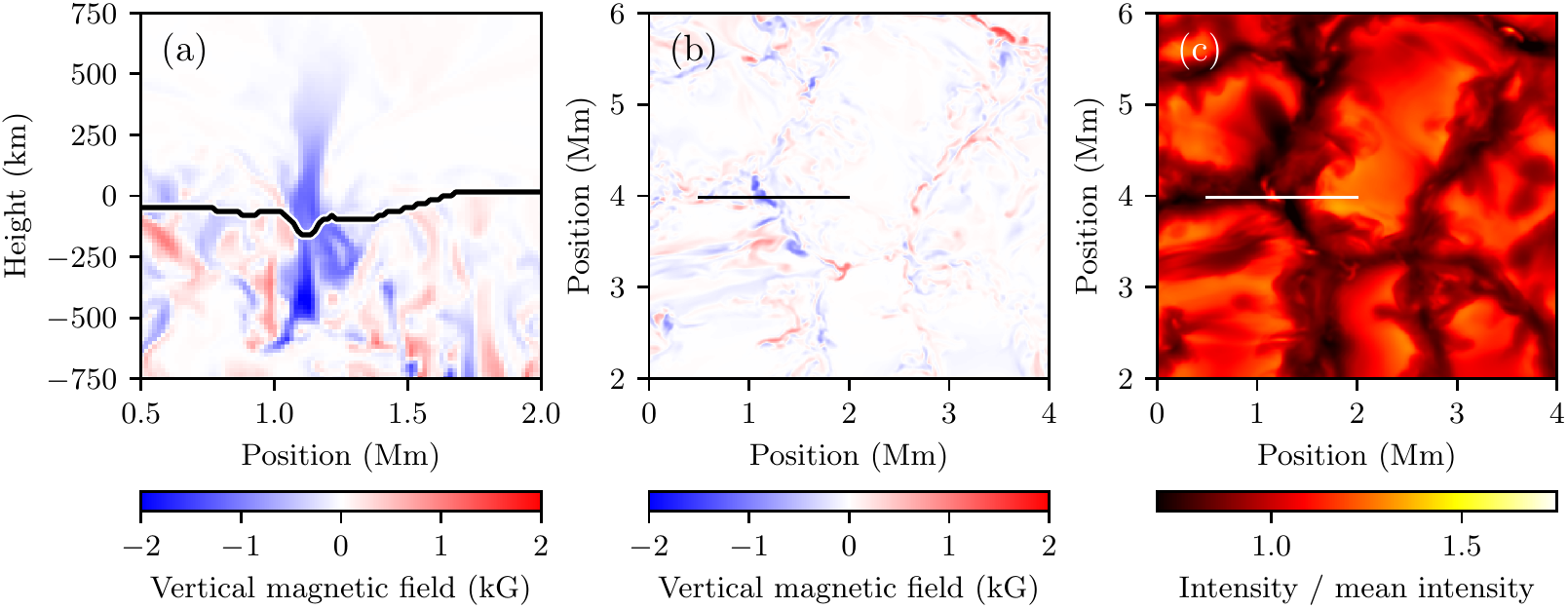}
	\caption[A sample \bp\ seen in the \muram\ simulation]{A sample \bp\ seen in the \muram\ simulation. Panel (a) shows the vertical magnetic field strength on a vertical slice through a \bp, with the $\tau=1$ surface marked by the black line. A strong concentration of vertical flux is seen, coinciding with a lowering of the optical surface by a $\sim100$~km. Panels (b) and (c) show the vertical field strength at the average $\tau=1$ level and the upward, white-light intensity for a subsection of the full domain. The granular pattern is seen in the intensity image, and vertical flux concentrations are confined to the downflow lanes. Horizontal lines in these plots indicate the location of the vertical slice in panel (a).}
	\label{fig:muram-sample}
\end{figure}

We show sample data from the simulation in Figure~\ref{fig:muram-sample}.
It can be seen that small concentrations of vertical flux abound in the granular downflow lanes and are often associated with intensity enhancements.
The flux tube associated with a \bp\ is shown.
The strength and verticality of the flux tube is greatest near the optical surface.
Deeper down, convective forces bend and twist the flux tube out of the plane of the plot.
Above the photosphere, the flux tube rapidly expands as the plasma pressure drops, reducing the field strength.

\subsection{Updated Tracking Algorithm}
\label{sec:new-tracking-algo}

To identify \bp s, we employ a slightly modified version of the tracking code of Chapter~\ref{chap:bp-centroids}, which consists of the following steps. (Values for all parameters will be given in Table \ref{table:tracking_params}.)
\begin{enumerate}
\item The code calculates the intensity of each pixel relative to the mean of its eight neighboring pixels (a discrete analog of the Laplacian), and it identifies those pixels for which this value lies $n_\sigma$ standard deviations above the mean of this value (calculated across that frame).
This provides a collection of ``seed'' pixels which are significantly brighter than their immediate surroundings, which are taken to likely be in the core of a \bp.
The use of this approach, rather than simply identifying the brightest pixels in the image, allows for the identification of \bp s which stand out strongly against the surrounding dark lanes but which are not all necessarily bright in absolute terms.

\item Each contiguous set of seed pixels is expanded to include neighboring pixels that satisfy a set criterion, which in our past work was that the pixel exceed in brightness the average of its neighbors by any amount.
In the present work, for each contiguous feature, we instead identify the maximum brightness $B\max$ of the constituent pixels as well as the minimum brightness $B\min$ of the pixels in a box surrounding those seed pixels (specifically, the minimal rectangle containing the seed pixels, widened in all directions by $n_\text{rect} \cdot n_\text{exp}$ pixels).
In each round of expansion, pixels are eligible to be added if their brightness exceeds a threshold set at $B\min + f_\text{contour} (B\max - B\min)$.
Effectively, expansion is constrained to be within a contour drawn at that threshold value.
Setting this threshold with local brightness values ensures that it is tuned to each \bp 's own contrast with its immediate surroundings.
We also define a parameter $\Delta B\max$: for very bright \bp s, when $B\max-B\min > \Delta B\max$, the contour is instead set at $B\min + f_\text{contour} \Delta B\max$.
Since very bright \bp s would otherwise have their contours drawn at relatively high levels, this step allows $f_\text{contour}$ to be kept appropriately high for dimmer \bp s without constricting the shape of the brightest \bp s by placing their contours too high.
With either the contour or Laplacian criterion, a fixed number $n_\text{exp}$ of expansion rounds are conducted, and the neighboring pixels added to the feature in each round act as seed pixels for the next round of expansion.
The number of rounds of expansion is set based on the typical size of \bp s and the pixel size.

\item Following expansion, a ``false-positive'' rejection step eliminates features that could continue to expand significantly, which are typically features which are actually especially-bright portions of a granule, or \bp s which have begun expanding into granular pixels.
These features are relatively large features existing within or including part of the large, bright area of a granule, while true and properly-identified \bp s are small and exist within dark downflow lanes.
Thus, after the fixed number of expansion rounds is conducted, \bp s should be largely or fully marked and unable to further expand, whereas these intra-granular false positives will be surrounded by a large number of pixels eligible to be included if expansion were allowed to continue.
Our code calculates for each contiguous feature the fraction of the feature's immediately surrounding pixels which would be added to the feature if another round of expansion were conducted.
If that fraction falls above a set threshold $f_\text{fp}$, the feature is rejected.

\item Features are also rejected if they touch the edge of the frame, if they are very large or very small (an area $A$ outside the range $A\min<A<A\max$ or a minimal bounding rectangle with a diagonal length exceeding $d\max$), or they they are extremely close to another \bp\ (within a distance $d_\text{prox}$) and thus possibly form a larger, complex feature that was identified as multiple fragments.

\item Surviving features are connected from frame to frame based on mutual overlap.
Merging and splitting events (that is, instances in which a feature in one frame overlaps multiple features in the preceding or following frame) are treated as the end of the feature(s) going into the event and the birth of the new feature(s) arising from the event.
A final lifetime criterion eliminates features that are tracked for fewer than $t\min$ seconds.
Newly added in the present work are the parameters $\Delta s_\text{max,\%}$ and $\Delta s_\text{max,px}$. 
If, between two frames, a feature's size changes by more than $\Delta s_\text{max,px}$ pixels and that change in area, measured as a percentage relative to the smaller of the two sizes, is greater than $\Delta s_\text{max,\%}$, the feature is not connected across those two frames.
That is, the first feature is regarded as ending in the earlier frame, and the second, very differently-sized feature is regarded as beginning in the second frame.
This avoids very large, impulsive velocities inferred when a sudden, large change in \bhp\ area occurs, usually due to a large number of pixels being very close to a relevant threshold rather than a large change in the underlying intensity distribution.
Manual inspection of a sample of such events supports this characterization.

\end{enumerate}

\begin{table}[t]
	\centering
	\begin{tabular}{lcc}
		Quantity & Value in previous chapter & Value in current chapter \\
		\hline
		$n_\sigma$ & 3 & 7.5 \\
		$n_\text{exp}$ & 3 & 9 \\
		$n_\text{rect}$ & --- & 1 \\
		$f_\text{contour}$ & --- & 0.65 \\
		$\Delta B\max$ & --- & 1.2 \\
		$f_\text{fp}$ & 0.2 & 0.2 \\
		$d_\text{prox}$ & 4~px & 2~px \\
		$A\min$ & 4~px$^2$ & 8~px$^2$ \\
		$A\max$ & 110~px$^2$ & 200~px$^2$ \\
		$d\max$ & 20~px & 30~px \\
		$t\min$ & 5 frames & 5 frames \\
		$\Delta s_\text{max,\%}$ & --- & 50\% \\
		$\Delta s_\text{max,px}$ & --- & 10~px \\
		Diagonal connections? & Yes & No \\
		
	\end{tabular}
	\caption[Tracking parameters used in Chapter~\ref{chap:bp-centroids} and in the following chapters]{Tracking parameters used in Chapter~\ref{chap:bp-centroids} and in the following chapters. The meaning of each quantity is explained in the text. Dashes indicate quantities not used in the previous work.}
	\label{table:tracking_params}
\end{table}

In the present work, we tune many of the thresholds and cutoff values (listed in Table \ref{table:tracking_params}) to better match the tracking to our data set and to improve the subjective quality of the tracking, and when considering ``contiguous'' or ``neighboring'' pixels, we use only the four above/below/left/right pixels and do not include the four diagonal connections.
Our most significant change, however, is the new expansion criterion of step (2).
This change is intended to better identify the edges of each \bp, supporting our analysis of shapes and sizes in the present work.
Our past work, by contrast, focused on intensity-weighted centroids, which are controlled more closely by the bright centers of \bp s than the darker edges, and so precise boundary identification was much less of a concern.
In comparing \bp\ boundaries drawn with the old and new approaches, we find from manual inspection that in the majority of cases the two criteria produce comparable boundaries, with each approach at times performing better in the subjective judgment of the viewer, but we also find a significant minority of cases in which our new approach produces significantly more plausible boundaries, whereas we did not identify any cases in which the new approach performed significantly worse.

Determining values for each of these tracking parameters is an imprecise process, a difficulty compounded by the absence of an objective metric of quality for identified boundaries.
Often, adjusting a parameter will improve the boundaries drawn for some \bp s but worsen the boundaries for others.
As an example, increasing $f_\text{contour}$ from 0.65 to 0.75 will prevent some fainter \bp s from ``spilling out'' into surrounding pixels that are visually unrelated, but it will cause many feature boundaries to be drawn too tightly, removing from those features some pixels along the edge which appear appropriate to include.
And with an eye toward how well our boundaries align with the underling magnetic flux concentrations, at the lower value, 3.8\% of all vertical, unsigned magnetic flux is included in an identified \bp, with a mean absolute vertical field strength of 855~G per \bp.
At the higher value, 3.4\% of all vertical, unsigned flux is included, with a mean field strength of 915~G per feature, suggesting that the higher value causes the identified \bp\ boundaries to exclude some weak-flux pixels, increasing the average field strength, but at the cost of capturing less of the total magnetic flux.
In this example and in many other ways, the values of tracking parameters will necessarily represent a compromise between eliminating instances of very bad boundary identification and slightly harming instances of good boundary identification.

\subsection{The Identified Boundaries}
\label{sec:new-boundaries}

\begin{figure}[p]
	\centering
	\includegraphics[trim=0 .34cm 0 0,clip]{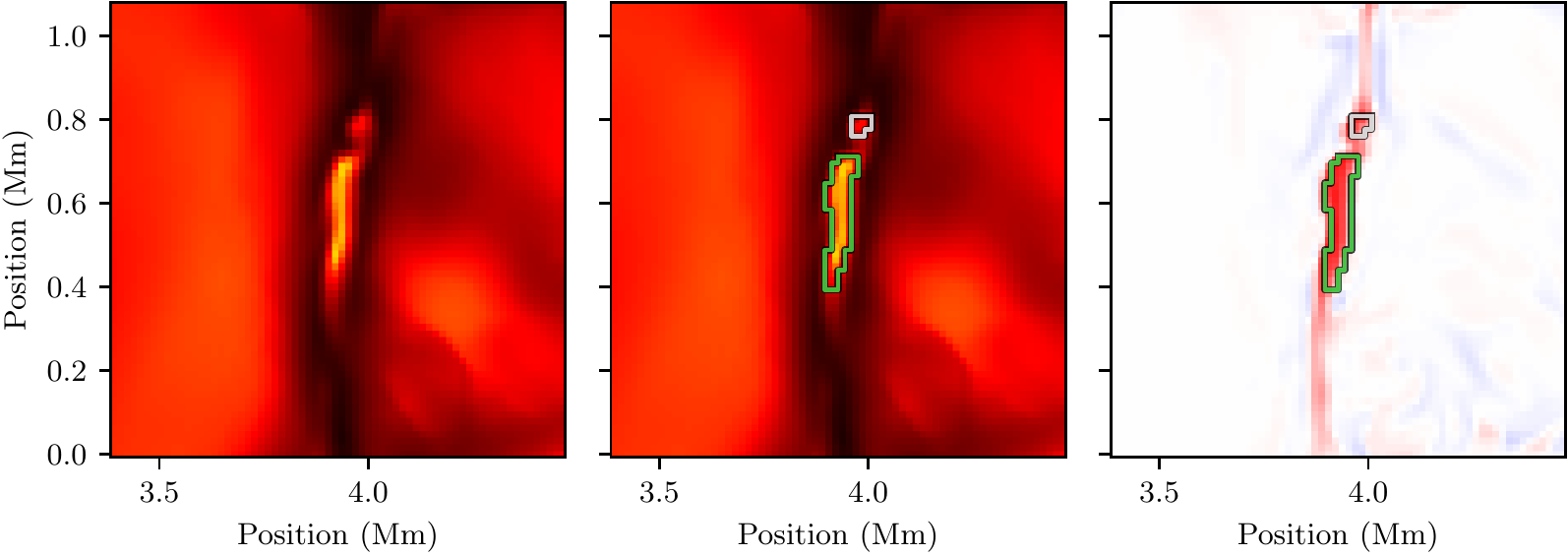}
	\includegraphics[trim=0 .34cm 0 0,clip]{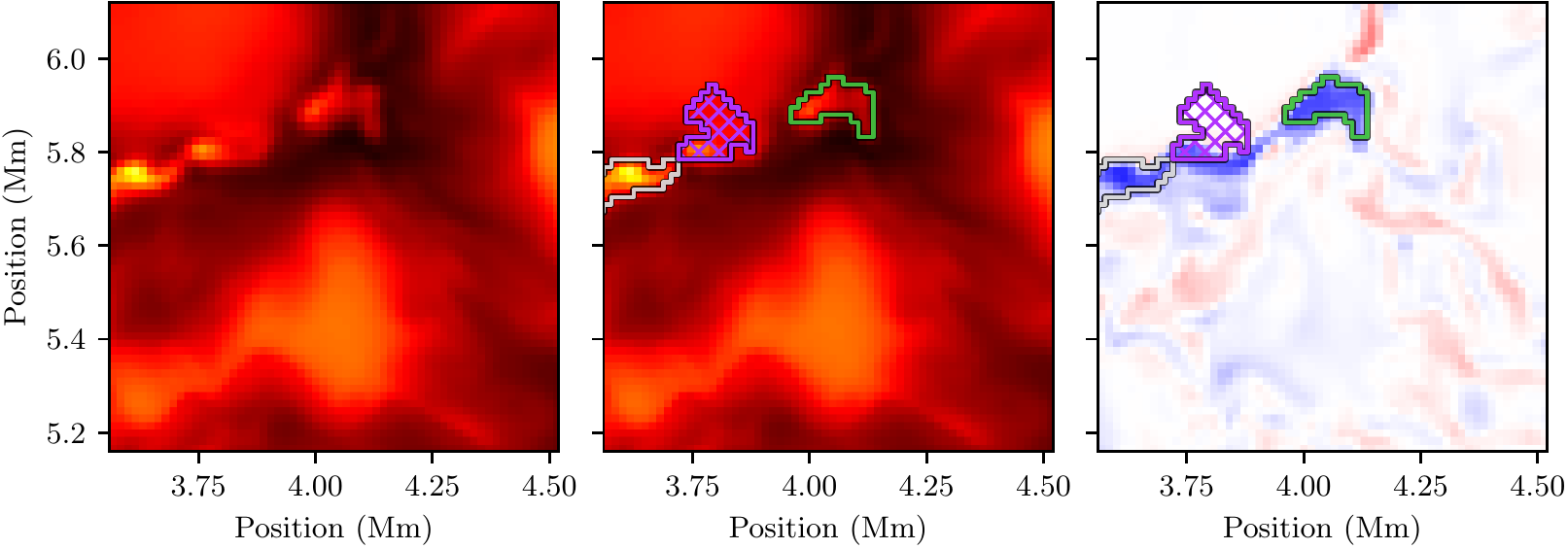}
	\includegraphics{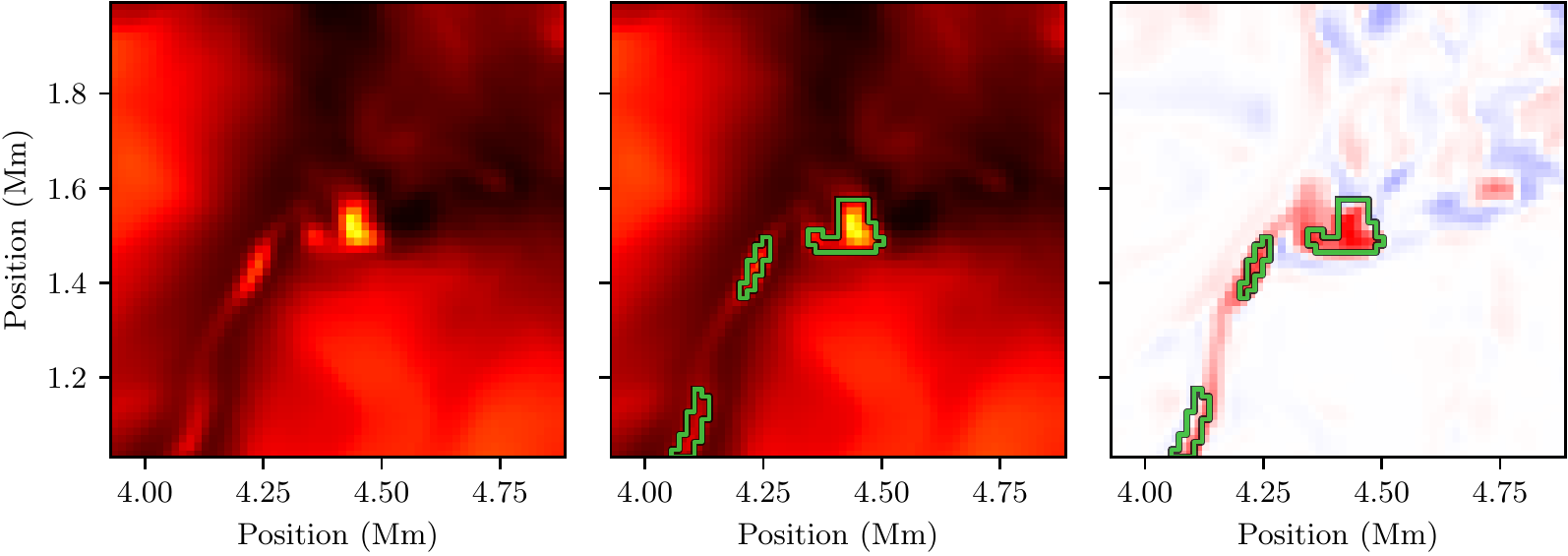}
	\caption[Sample of identified \bp s]{Sample of identified \bp s (continued on the next page). These \bp s were selected from a sample of 30 randomly-chosen \bp s to illustrate different shapes and scenarios. The left column provides an unobstructed view of the intensity pattern, and the center and right columns show \bp\ boundaries overlaid on intensity and vertical magnetic field strength at $\tau=1$, respectively. Green lines mark the boundaries of accepted features. Magenta lines (with crosshatches) mark regions rejected by our false-positive rejection step. White lines mark regions rejected due to the minimum-lifetime constraint. Features rejected for other reasons are not present in these samples. Color maps range from 0.7~(black) to 1.75~(white) times the mean intensity for the intensity maps, and -2~(red) to 2 (blue)~kG for magnetic field strength. Animations of these \bp s are available on the author's website at \href{http://samvankooten.net/thesis}{samvankooten.net/thesis} and are archived at \href{http://doi.org/10.5281/zenodo.5238943}{doi.org/10.5281/zenodo.5238943}.}
	\label{fig:tracking-bp-sample}
\end{figure}

\begin{figure}[p]
	\ContinuedFloat
	\centering
	\includegraphics[trim=0 .45cm 0 0,clip]{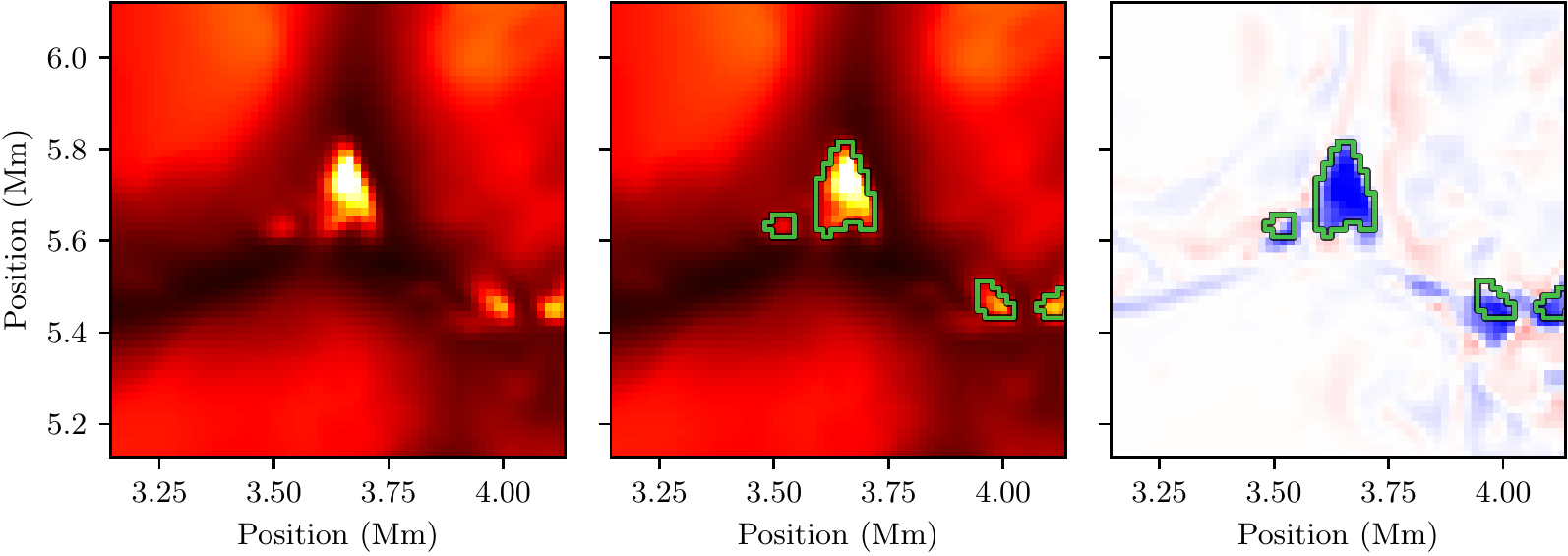}
	\includegraphics[trim=0 .45cm 0 0,clip]{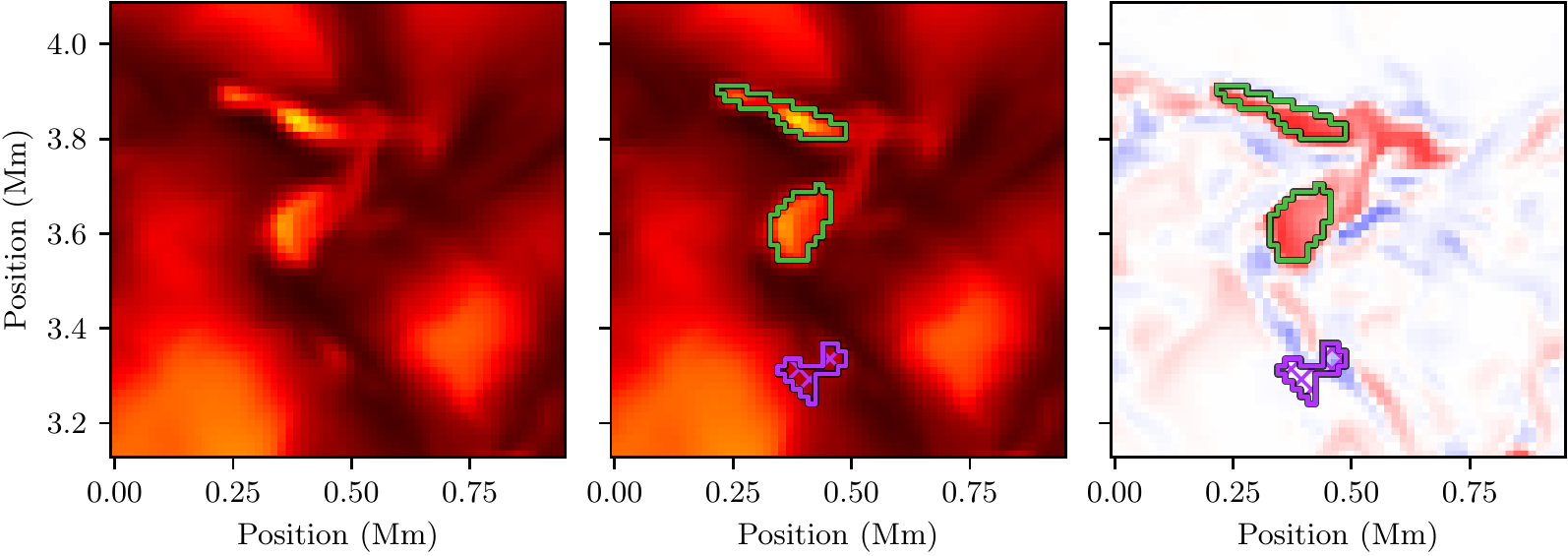}
	\includegraphics[trim=0 .45cm 0 0,clip]{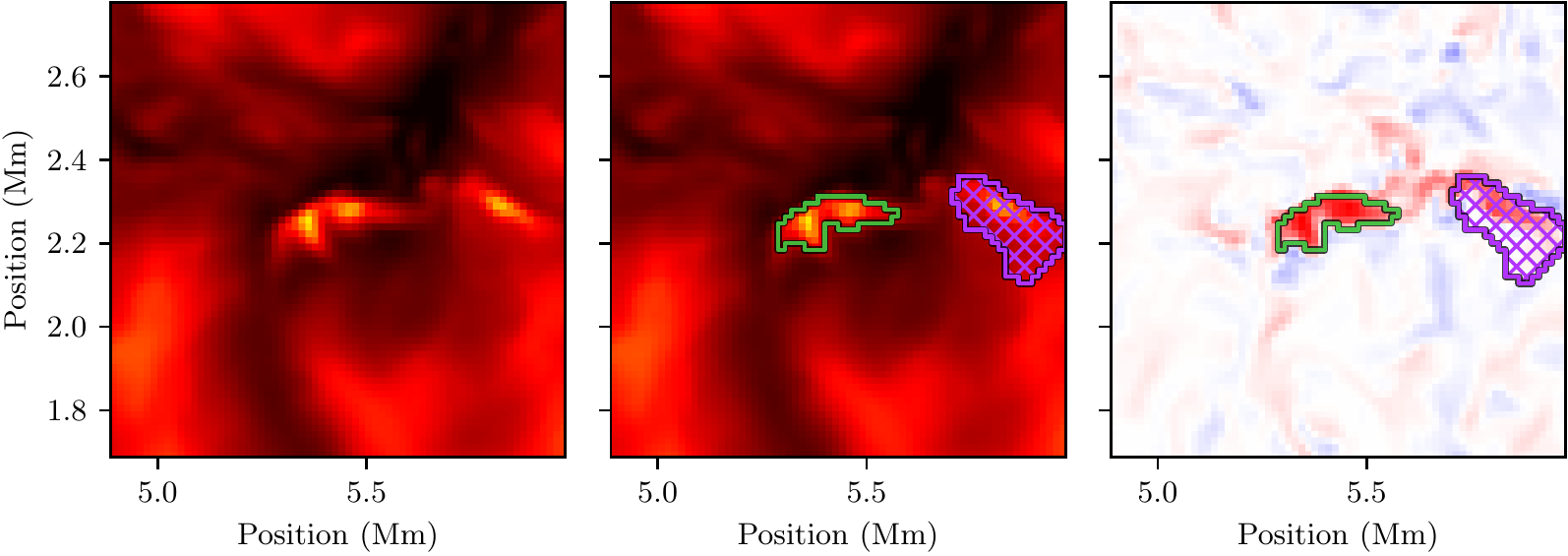}
	\includegraphics{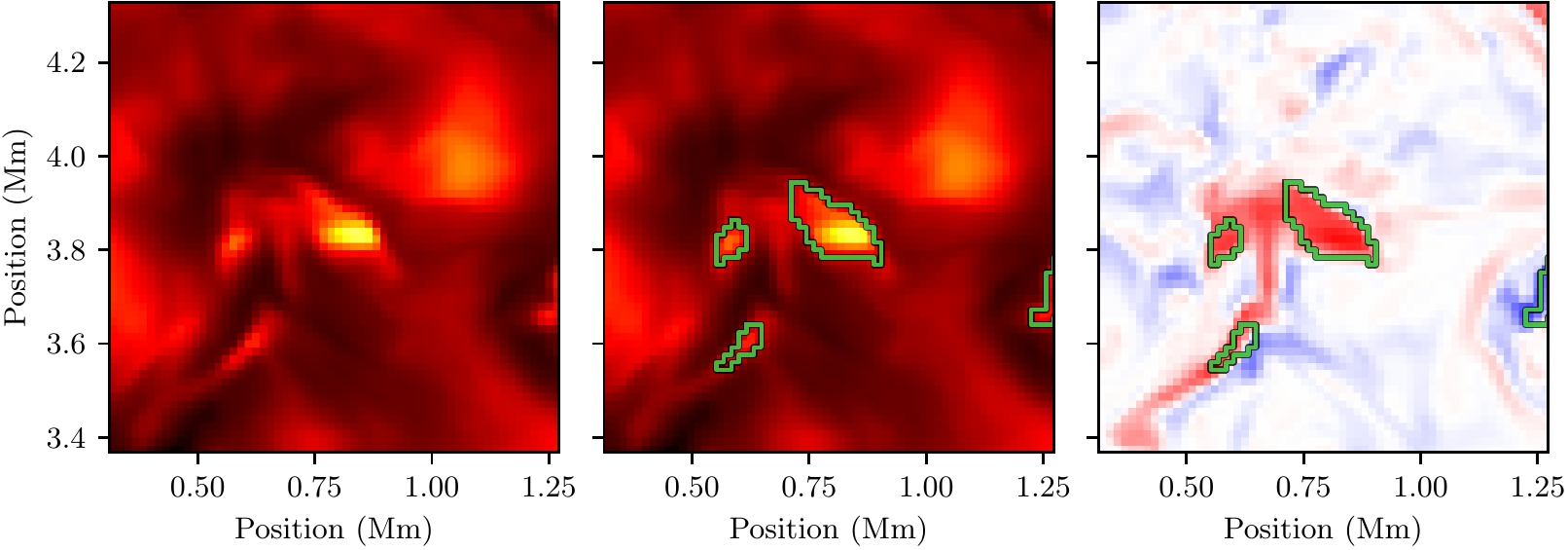}
	\caption[]{Sample of identified \bp s (cont.)}
\end{figure}

In Figure~\ref{fig:tracking-bp-sample} we show a sample of identified \bhp\ boundaries.
Across these samples it can be seen that the boundaries typically align quite well with the boundary that one would draw by-eye.
Additionally, comparison with the maps of vertical magnetic field strength shows that \bhp\ boundaries are typically, but not always, well-aligned with the boundaries of the magnetic enhancement, though it can also be seen that the flux concentration sometimes extends, in a weakened form, to other pixels not identified as part of the \bp.
It can also be seen that weak-field regions not associated with a white-light enhancements are abundant.
(These relations between intensity and magnetic flux at the smallest scales are relations that should be verified observationally with DKIST, and not necessarily accepted as fact based on these simulations.)

\subsection{\BP\ Statistics}
\label{sec:new-boundary-statistics}

\begin{figure}[t!]
	\centering
	\includegraphics{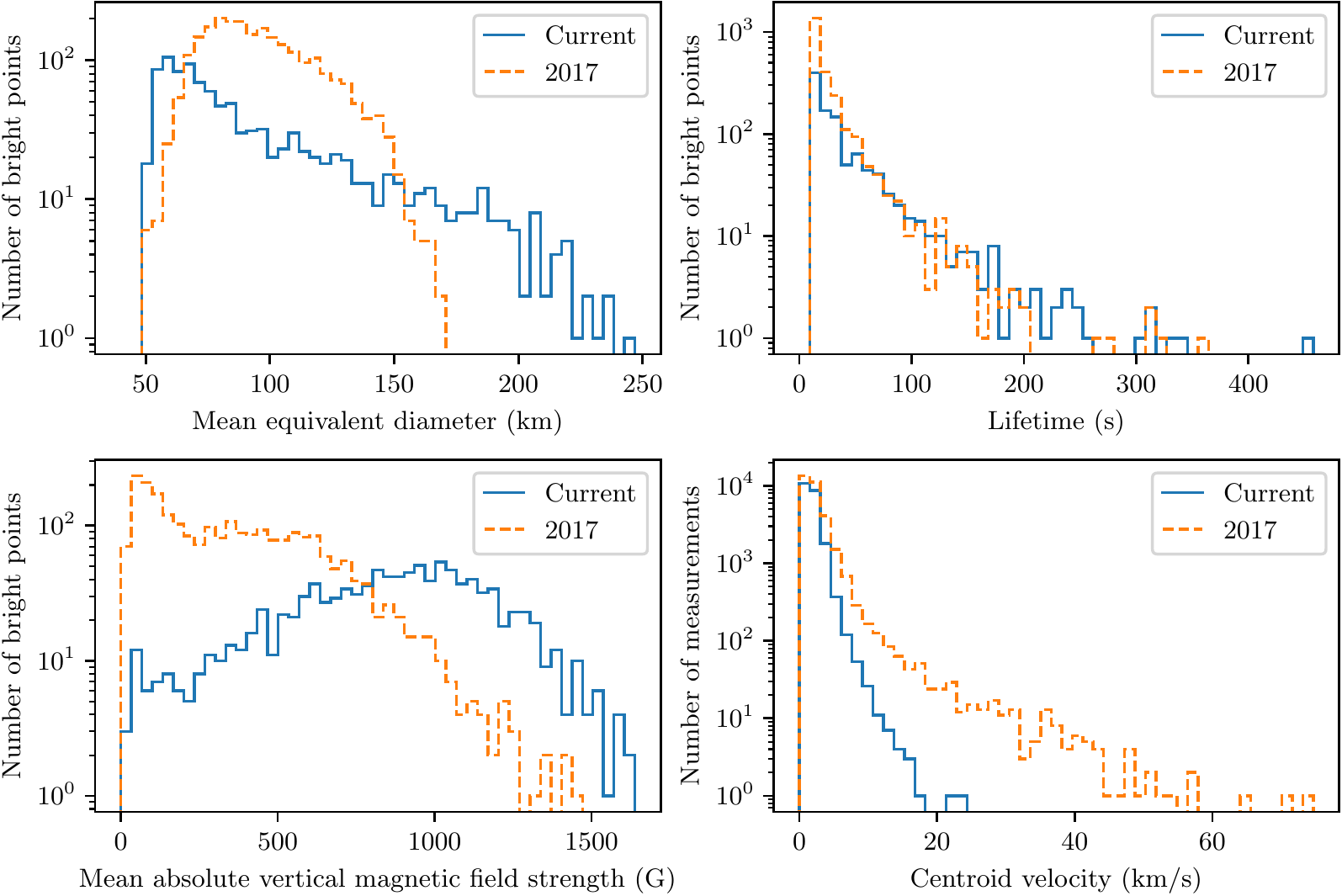}
	\caption[Distributions of \bp\ properties using our 2017 tracking and our current tracking]{Distributions of \bp\ properties using our 2017 tracking and our current tracking. Equivalent diameter is the diameter of a circle of the same area as an identified \bp, and we plot the mean equivalent diameter across the lifetime of each \bp. Mean absolute vertical magnetic field strength is the mean of the absolute value of each pixel across the lifetime of the \bp\ and is measured at the $\tau=1$ surface. Centroid velocities are measured and plotted in between each pair of subsequent time steps.}
	\label{fig:tracking-distributions}
\end{figure}

We identify 1,064 \bp s in the 2~s cadence \muram\ run.
(This is reduced from 2,425 \bp s using the tracking approach of Chapter~\ref{chap:bp-centroids}.)
In Figure~\ref{fig:tracking-distributions} we show the distribution of \bhp\ sizes, lifetimes, and centroid velocities.
We include the distributions using our current tracking approach and that of Chapter~\ref{chap:bp-centroids} for comparison.
It can be seen that larger \bp s are produced by our new approach (driven directly by the increase in $A\max$, $d\max$, and $n_\text{exp}$, though the effect of increasing $n_\text{exp}$ is mitigated by disallowing diagonal pixel connectivity).
Lifetimes with the new tracking are very slightly skewed toward longer lifetimes, suggesting that the updated feature identification makes the inter-frame linking more robust.
Centroid velocities are significantly reduced by the new tracking, which we interpret as indicating that the identified \bhp\ boundaries are more consistent from frame to frame---this was a strong motivation for the contour-based expansion step.
This reduction in centroid velocities is likely linked to the strong reduction in centroid jitter noted earlier in the centroid-velocity power spectra.
The distribution of per-\bhp\ mean vertical magnetic field strengths is shifted toward higher values with our updated tracking.
This may be because our new approach is more selective in the features it identifies, and the more marginal features that are now rejected may be less likely to be true or strong-field \bp s.

\begin{figure}[t!]
	\centering
	\includegraphics{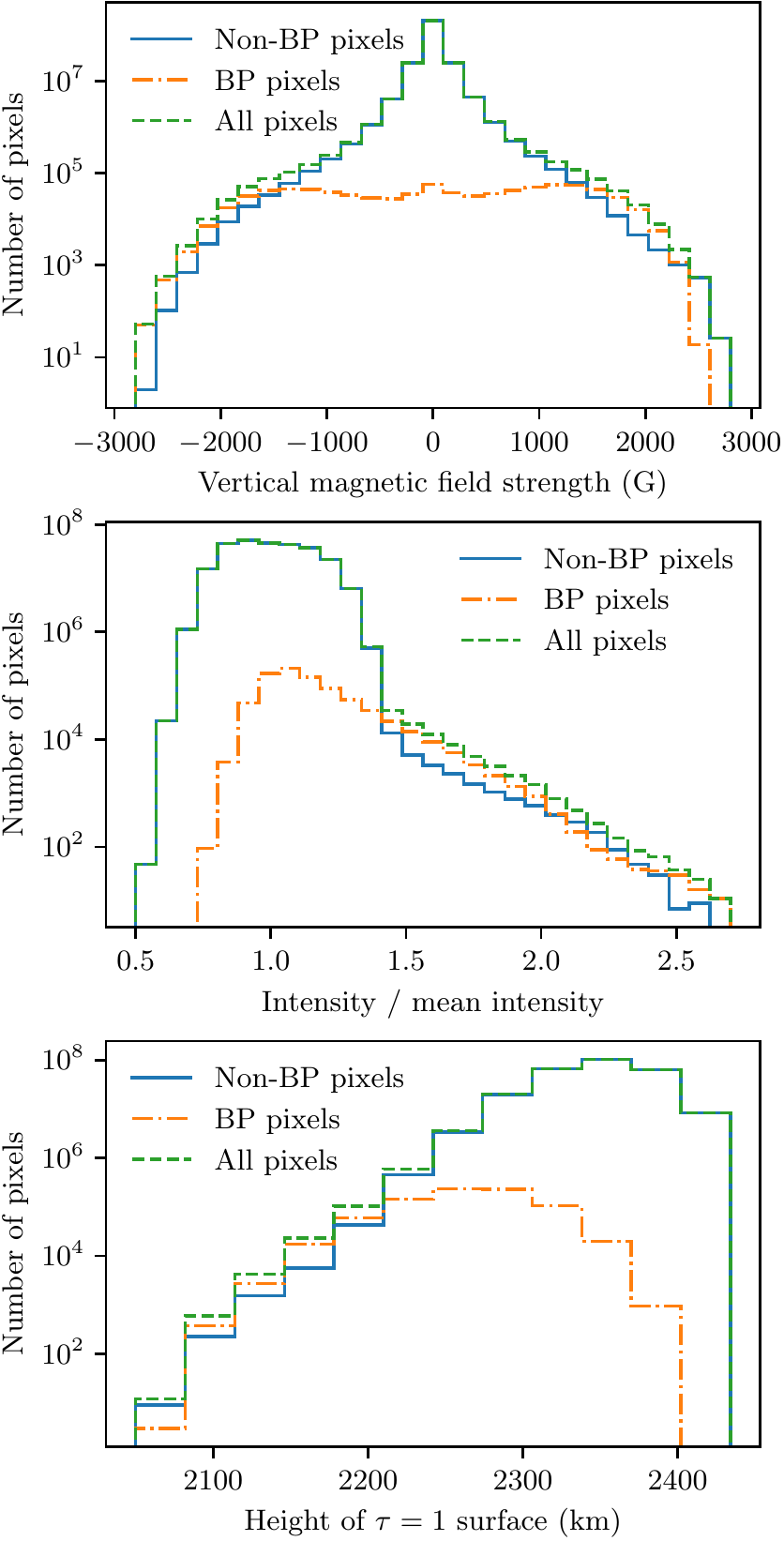}
	\caption[Distributions of pixel values for \bp s and full images]{Distributions of pixel values for \bp s and full images. For each quantity, we show the distributions of the values of each pixel included in a \bp\ across all \bp s and all snapshots, the values of each pixel not included in a \bp, and the values of all pixels. Vertical magnetic field strength is measured at the $\tau=1$ surface. The height of the $\tau=1$ surface, computed independently at each pixel for a vertical line of sight, is measured relative to the bottom of the simulation box (which has a total height of $\sim 4$~Mm).}
	\label{fig:tracking-pixel-distributions}
\end{figure}

In Figure~\ref{fig:tracking-pixel-distributions} we show the distribution of per-pixel values within our identified \bp s.
It can be seen that these values skew heavily toward strong vertical magnetic field strengths, high intensities, and low depths of continuum formation.
(Recall that \bp s appear bright because the high magnetic pressure offsets gas pressure and reduces gas density, lowering the $\tau=1$ surface to deeper, hotter plasma.)
Additionally, the clear majority of strong-field pixels are included in identified \bp s (though not the majority of flux: only 3.8\% of unsigned vertical magnetic flux is within a \bp).
Of note, however, these distributions do still include significant numbers of pixels with values that would not be expected for the ``ideal'' \bp, namely weak fields, low intensity, and high continuum formation.
While some of these values are no doubt due to imperfect tracking, some are due to the fact that, while the intensity enhancement and the magnetic flux concentration are related and coupled phenomena, there is no exact relation between the two quantities.
With this caveat in mind, the strong-field nature of most detected \bhp\ pixels suggests that the intensity enhancements do serve as a useful proxy of the flux concentration, even at these small scales.
(Observational validation of this claim must await adequate DKIST data.)

\begin{figure}[t!]
	\centering
	\includegraphics{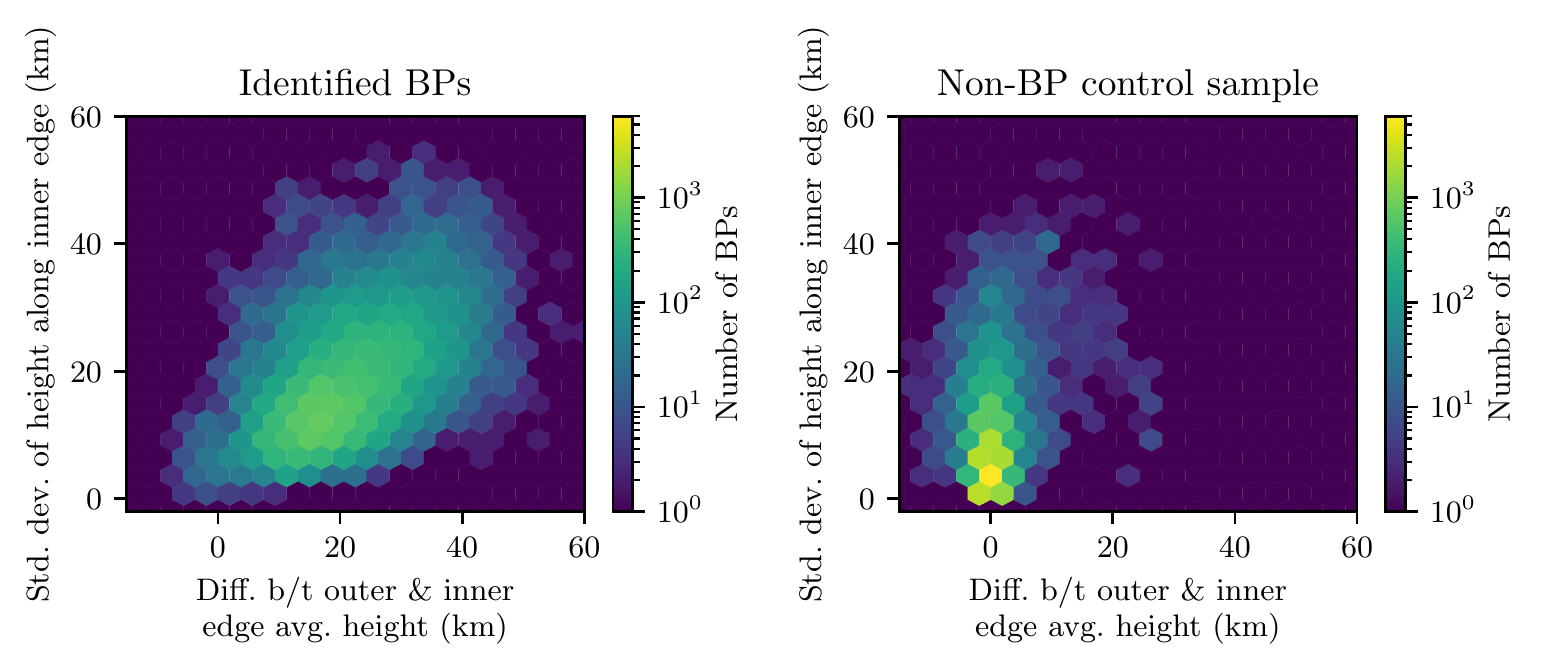}
	\includegraphics{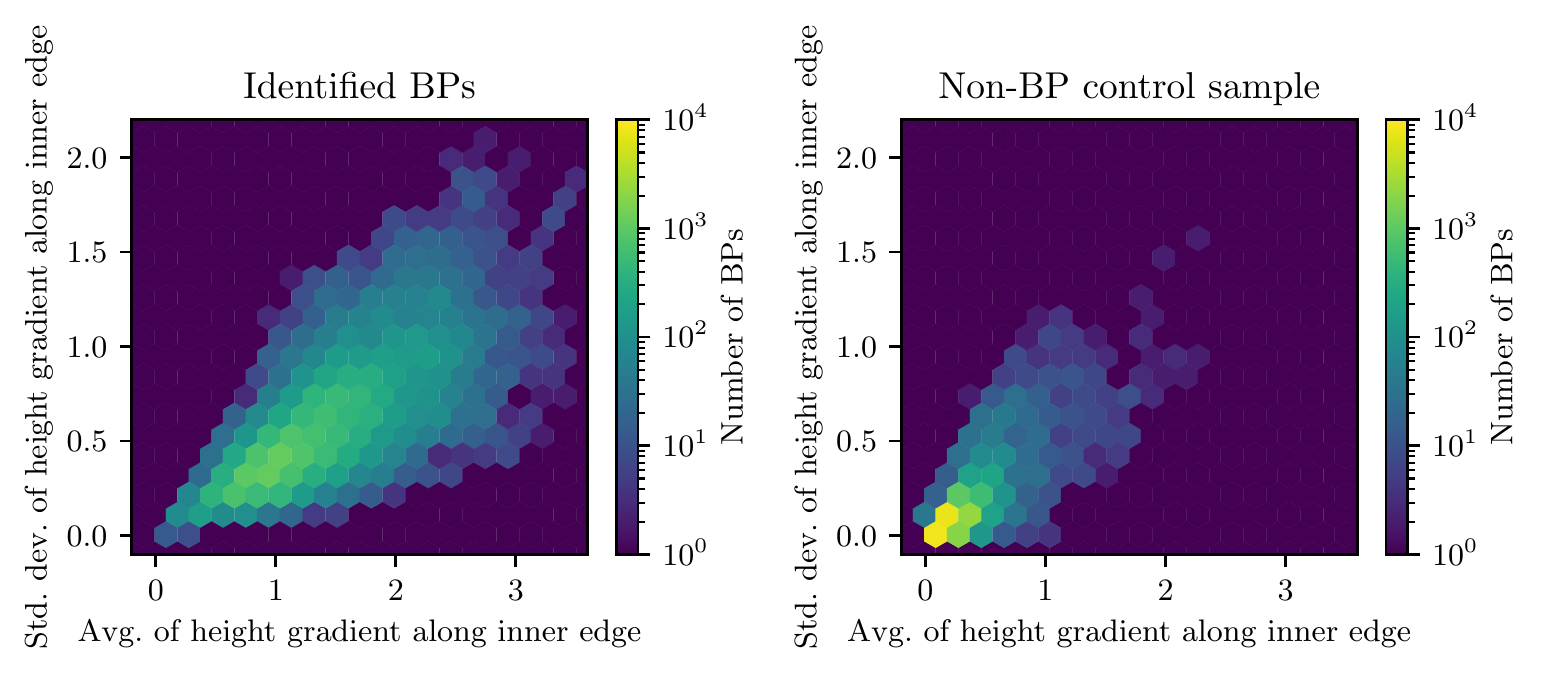}
	\caption[Variation in $\tau=1$ height at \bp\ edges]{Variation in $\tau=1$ height at \bp\ edges. For each \bp, we show in the upper panels the difference between the average $\tau=1$ height along the outer edge and the average along the inner edge. The lower panels show the average value of the magnitude of the gradient of the $\tau=1$ height along the inner edge. (The gradient is unitless, as it is height divided by horizontal distance.) The ``control sample'' is discussed in the text.}
	\label{fig:tracking-bp-GOF-heights}
\end{figure}

\begin{figure}[t!]
	\centering
	\includegraphics{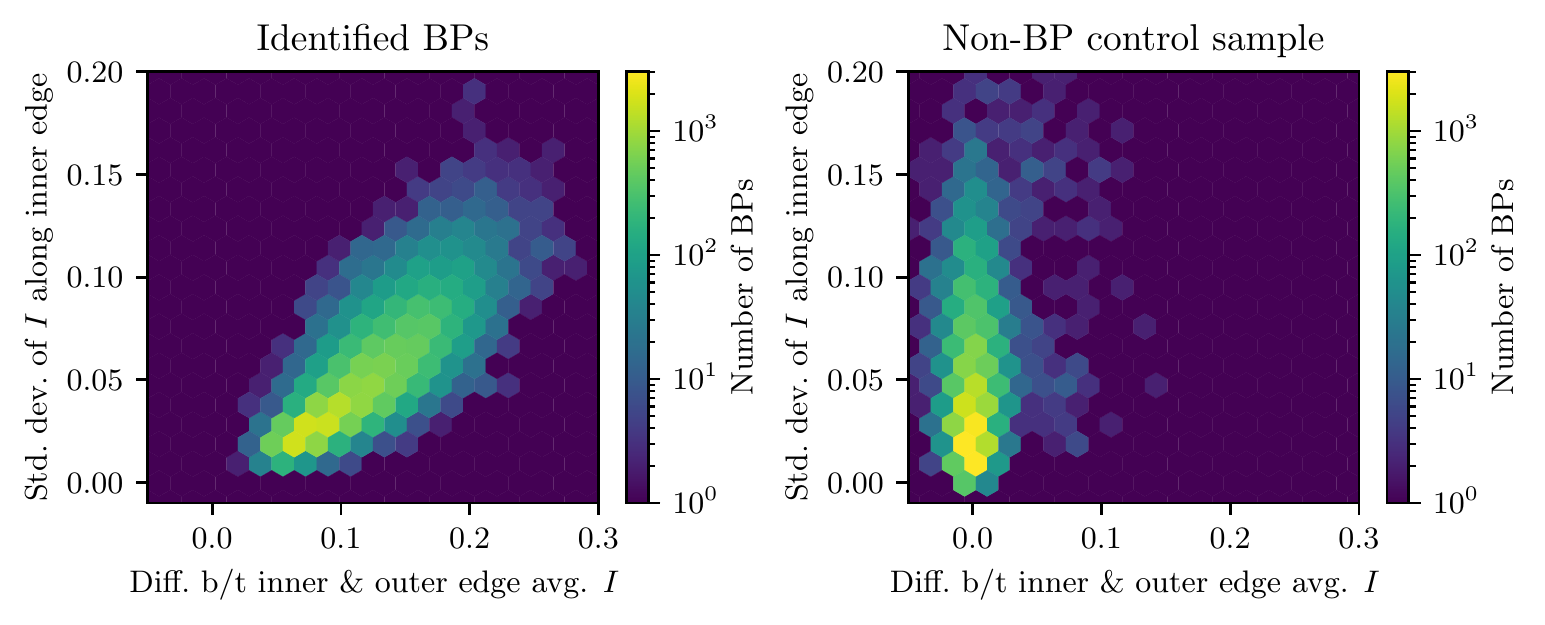}
	\includegraphics{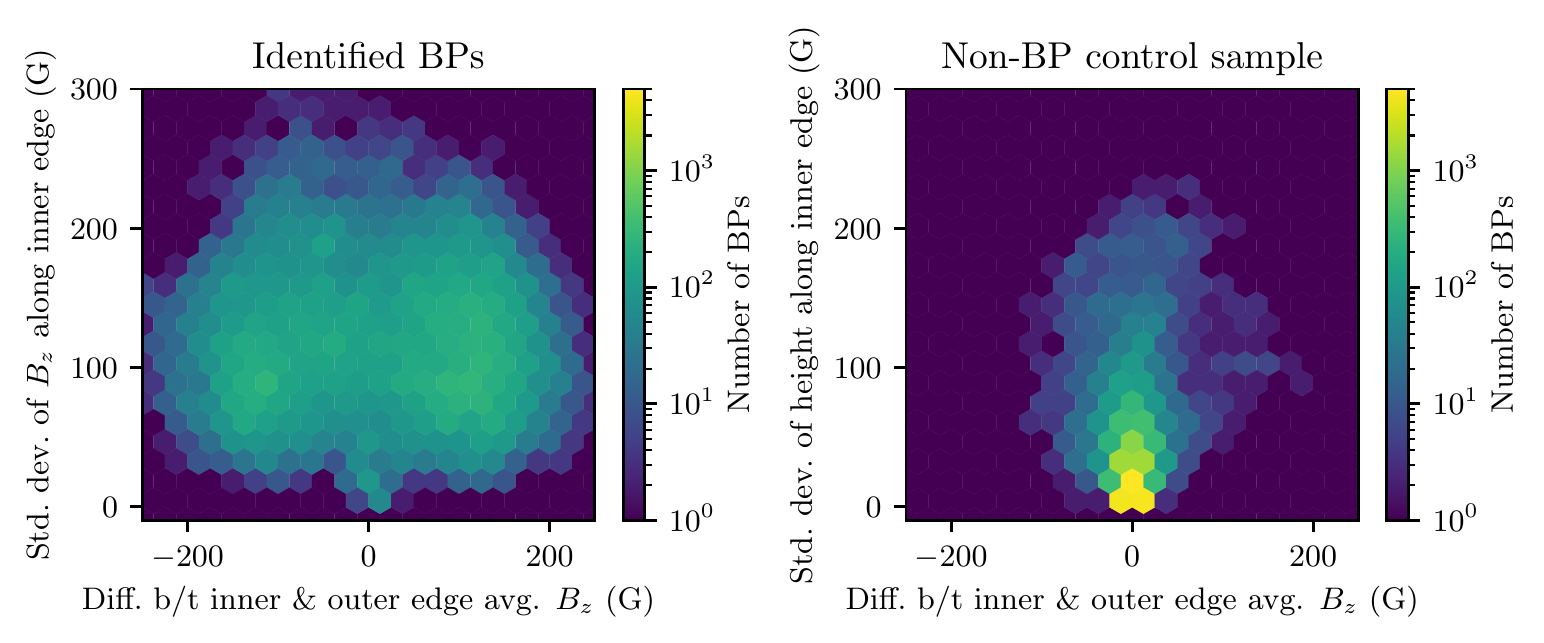}
	\caption[Variation in $I$ and $B_z$ at \bp\ edges]{Variation in intensity $I$ and vertical magnetic field strength $B_z$ at \bp\ edges. For each \bp, we show in the upper panels the difference between the average value of $I$ along the inner edge and the average along the outer edge (in units of the image's average intensity). The lower panels show the difference between the average value of $B_z$ along the inner edge and the average along the outer edge. The ``control sample'' is discussed in the text.}
	\label{fig:tracking-bp-GOF-BI}
\end{figure}

To probe these trends at the level of individual \bp s, we identify for each \bhp\ outline the pixels that are touching the identified boundary and either inside or outside the feature---the ``inner'' and ``outer'' edges, respectively.
In Figures~\ref{fig:tracking-bp-GOF-heights} through \ref{fig:tracking-bp-GOF-BI}, we compare various quantities between the inner and outer edge of each \bp.
We also show ``control samples'' produced by transposing the coordinates of each \bp's pixels ($x \rightleftharpoons y$), to show what values are typical for non-\bhp\ areas with \bhp-like shapes.
The first quantity we show is the height of the $\tau=1$ surface, where there is typically a difference of 10--30~km (or 1--2 vertical pixels) between the average value of this height along the outer edge and the average value along the inner edge.
While only a few pixels, this level of variation corresponds to slopes in the $\tau=1$ surface steeper than $45^\circ$, and it is much larger than the variation by a small fraction of a pixel typical of the control sample.
This indicates that almost all of our identified boundaries mark an area where the $\tau=1$ surface tends to be decreasing toward the \bhp\ center.
Also shown is the standard deviation of this height along the inner edge, showing that the level of variability in this height along our \bhp\ edges is only somewhat higher than is seen elsewhere on the Sun.
This shows that not only is the $\tau=1$ height decreasing across the boundary, but that the boundaries tend to form a reasonable contour line for the $\tau=1$ surface.
(A similar analysis of the standard deviation values holds for the following quantities as well.)
The second quantity we show is the average of the magnitude of the gradient of the $\tau=1$ height along the inner edge, where the difference between the \bp s and the control sample bolsters the interpretation that our \bhp\ boundaries mark regions where the $\tau=1$ height is changing significantly.
The third quantity is the ratio of the average white-light intensity measured along the inner and outer edges, showing that all of our boundaries mark locations where the intensity is increasing across the boundary.
The fourth quantity is the difference between the average $B_z$ along the inside and outside edges, showing that the field strength is increasing across the boundary of most \bp s.
This last trend is the weakest of these four quantities, consistent with the fact that enhancements in the magnetic field and the white-light intensity are not always perfectly aligned at the pixel level (as seen in Figure~\ref{fig:tracking-bp-sample}).
Nevertheless, the clear majority of our \bhp\ outlines coincide with a transition in the magnetic field strength, suggesting they correlate with the edge of a field strength concentration (as opposed to falling along a more homogeneous region within a concentration or lying in a weak-field region).

In future work, metrics such as these might be able to serve as a figure of merit for optimizing the parameters of our (or any other) tracking algorithm.
However, such a use must include a more careful exploration of these metrics to explore and justify whether they are the best metrics to be optimized.
For instance, one must explore whether the ``best'' \bhp\ outline occurs where the gradient of the $\tau=1$ height is steepest, or if it occurs where the gradient first begins to steepen relative to the flatter gradient outside the \bp.
Our presentation of these metrics here is not meant to show that our boundary identification is optimized, but rather that our boundaries tend to fall in places where these quantities are varying in ways expected for a \bhp\ edge.

\subsection{Impact of Algorithmic Changes}
\label{sec:algo-change-impacts}

In Section~\ref{sec:jitter}, we noted that the power spectra of centroid velocities followed a power law with a slope of order $f^{-1}$ over frequencies of order $10^{-3}$ and $10^{-2}$~Hz, but the spectra became flat and even began a slight increase at frequencies of order $10^{-1}$~Hz (extending the analysis to these higher frequencies by using the same 2~s cadence \muram\ run used in the present work).
In light of the work of \citet{Agrawal2018}, this change in the power-law slope was interpreted as the effect of centroid jitter: movement in the centroid location not corresponding to true, bulk motion of the feature
This jitter can be due to algorithmic effects (such as incorrect identifications or pixels with values near relevant thresholds suddenly jumping into or out of the feature) or physical effects (such as emerging flux or the merging into the feature of smaller flux elements that were unresolved or untracked).
Many of these effects can be broadly described as centroid motion due to the expansion (or contraction) of the feature boundary that is asymmetric relative to the centroid, and a simple model of this general phenomenon and its effect on the centroid was shown to reproduce the eventual, upward slope of the centroid-velocity power spectra at high frequencies.
Here we revisit this analysis in light of our new tracking algorithm.

\begin{figure}[t]
	\centering
	\includegraphics{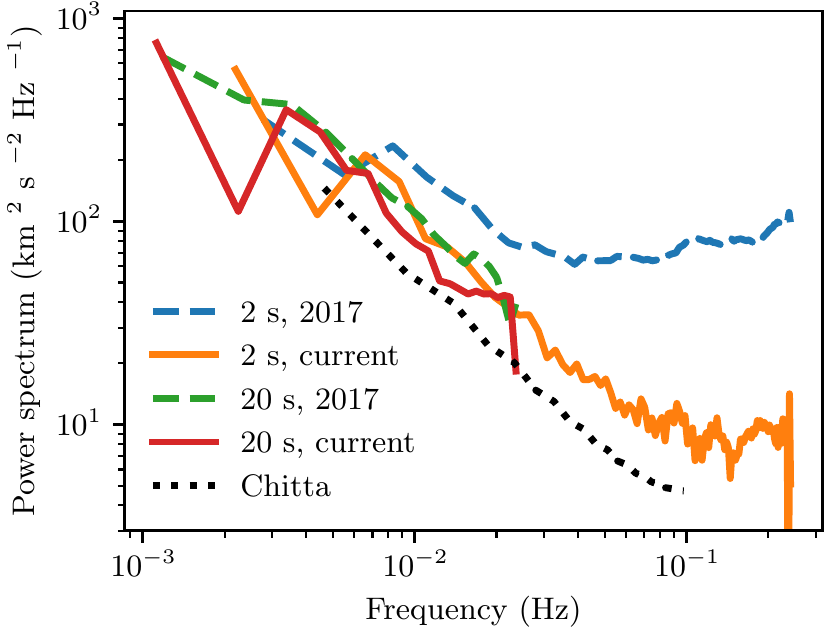}
	\caption[Comparison of power spectra from our previous and current tracking algorithm]{Comparison of power spectra from our previous and current tracking algorithm. Shown are power spectra of the velocities of the intensity-weighted centroids for the \bp s in the 20~s and 2~s cadence \muram\ runs. The spectra labeled ``2017'' are those of Figure~\ref{fig:jitter}, while the spectra labeled ``current'' use our updated tracking. Also shown is the observational spectrum of \citet{Chitta2012} for reference. Boxcar smoothing with a width of 7 frequency bins has been applied to all spectra to produce a clearer plot.}
	\label{fig:tracking_jitter_old_new_cf}
\end{figure}

In Figure~\ref{fig:tracking_jitter_old_new_cf} we show power spectra of centroid velocities computed using the \bhp\ tracking of Chapter~\ref{chap:bp-centroids} as well as using the tracking described in this chapter.
Included are both the 20~s cadence \muram\ run primarily analyzed in Chapter~\ref{chap:bp-centroids} and the 2~s cadence run which is the primary focus of our subsequent work.
It can be seen that the flattening of the 2~s spectrum at high frequencies is significantly reduced, with that spectrum more closely following the observational spectrum over a wider range of frequencies, and eventually flattening at a value of the power an order of magnitude lower than before.
The 20~s spectrum, however, sees little difference from the updated algorithm, likely because it is not very susceptible to centroid jitter.

\begin{figure}[t!]
	\centering
	\includegraphics{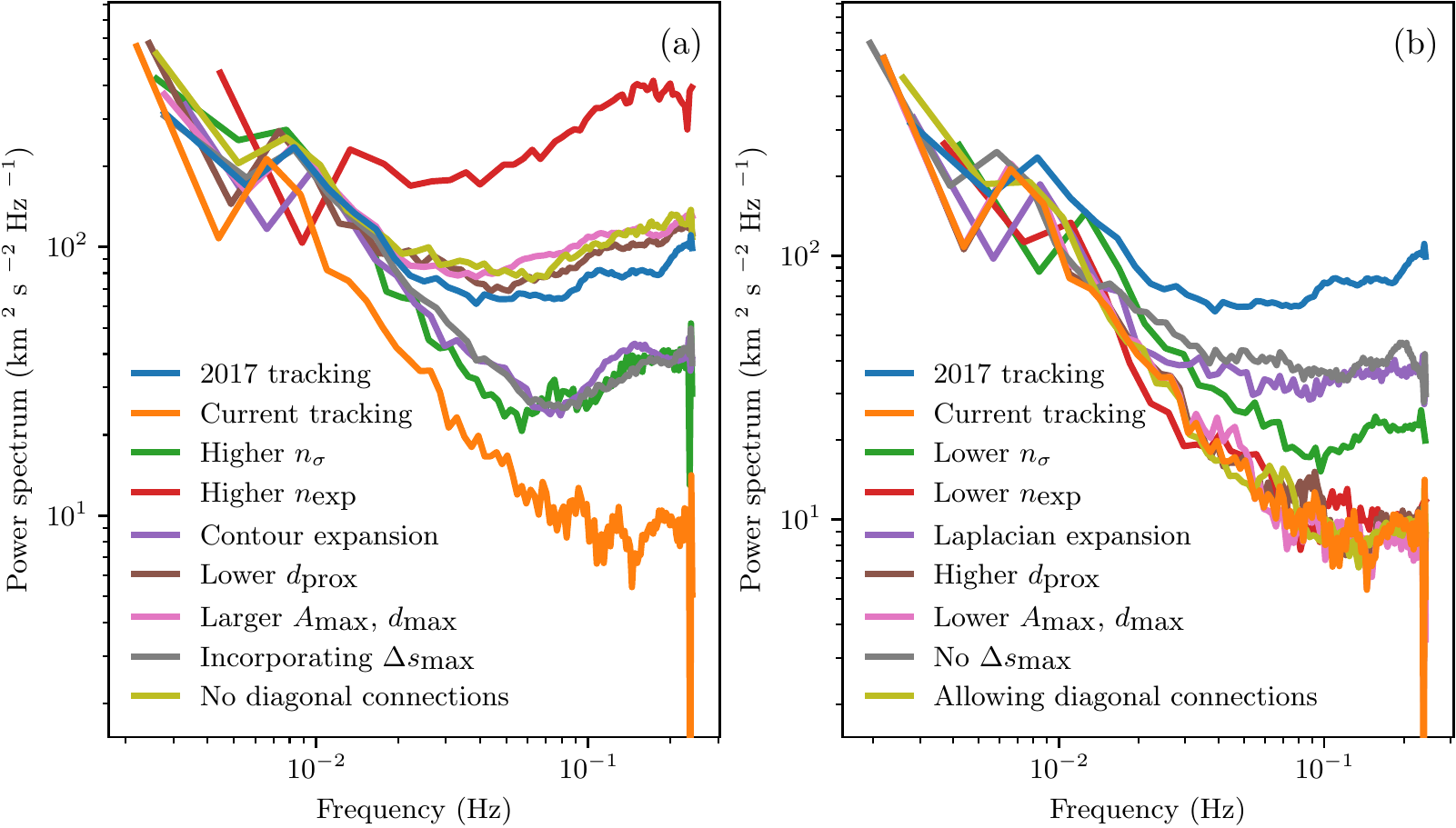}
	\caption[Effects on power spectra of each tracking algorithm change]{Effects on power spectra of each tracking algorithm change. Shown are spectra produced with our old and new tracking algorithm (curves labeled ``2017 tracking'' and ``current tracking,'' as well as spectra produced after changing one of the tracking parameters. In panel (a), each curve uses the old tracking but changes a single parameter to its value in the new tracking, and vice-versa in panel (b). All spectra use the 2~s cadence \muram\ run, and boxcar smoothing with a width of 7 frequency bins has been applied to each spectrum. When diagonal connections are disallowed in (a), the number of pixels that may be expanded to in each round is halved, so $n_\text{exp}$ is doubled to isolate the effect of diagonal connectivity from the change in growth potential; likewise, $n_\text{exp}$ is halved in (b) when diagonal connections are allowed.}
	\label{fig:tracking_jitter_algo_changes}
\end{figure}

To further investigate this improvement, we show in Figure~\ref{fig:tracking_jitter_algo_changes}a the effect of taking our 2017 tracking algorithm and changing any one of the tracking parameters to its current value, and in Figure~\ref{fig:tracking_jitter_algo_changes}b the effect of changing any one of the current parameter values to its 2017 value.
It can be seen that the rejection of size changes greater than $\Delta s_\text{min,\%}$ or $\Delta s_\text{min,px}$ (that is, size changes which a very likely due to bad tracking rather than true \bhp\ evolution), the increase in $n_\sigma$ (causing the tracking to be more selective in which features it identifies and to ignore more subtle features) and the replacement of the Laplacian-based expansion step with contour-based expansion (causing identified \bhp\ boundaries to better align with the edges of the intensity enhancements) all significantly reduce the flattening of the 2017 spectrum, and that removing any of these changes from the new tracking worsens the flattening.
It can also be seen that the other parameter changes, taken in isolation and applied to the old tracking, worsen the spectral flattening; however, reversing any one of these changes from the new tracking does not significantly affect the spectrum.
This resiliency of our updated tracking is due, at least in part, to the interaction between these parameters.
For instance, with the old tracking in Figure~\ref{fig:tracking_jitter_algo_changes}a, increasing $n_\text{exp}$ from its old value of 3 to its current value of 9 dramatically worsens the spectrum because it allows some features to expand spuriously into nearby pixels that have only a tenuous connection to the feature (where the Laplacian is only just above zero), and these weakly-associated pixels are more likely to jump into and out of the feature over time, resulting in very high centroid velocities changing at high frequencies.
Under the new tracking, the increase in $n_\text{exp}$ is balanced in part by the removal of diagonal connectivity, reducing the number of pixels eligible to be expanded to in each round.
Additionally, the contour-based criterion for expansion provides a more robust boundary to limit expansion, and the rejection of large frame-to-frame size changes eliminates much of the effect of any remaining, tenuously-connected pixels that jump into and out of the feature.
This allows $n_\text{exp}$ to be safely increased to allow the largest \bp s to be more fully identified without an adverse effect on the tracking of other \bp s.

Ultimately, we are pleased to find that our improved tracking algorithm dramatically reduces the amount of centroid jitter seen in our centroid velocity power spectra, which mitigates much of the concern over centroid tracking applied at high resolution which was raised in Chapter~\ref{chap:bp-centroids}.
The onset of spectral flattening is now near a frequency of 0.1~Hz, 5--10 times higher that the previous onset frequency.
This new value suggests that, in interpreting \bhp\ motions by any means, signals on time-scales up to 10~s are largely reliable, while signals at higher frequencies may be largely spurious and ought to be removed.
This constraint might be able to be relaxed through further refinements to the tracking algorithm, but there may be diminishing returns in such efforts as one approaches the shortest timescales on which resolvable physical processes occur within \bp s.
At these scales, centroid velocities may become noisy as the discretization of \bhp\ boundaries turns the underlying slow, smooth shape changes into a series of impulsive changes.

\subsection{Flux Element Tracking}
\label{sec:fe-tracking}

Since the \muram\ simulation includes magnetic field information, we can also perform tracking on the strong-field features (or \textit{flux elements}) as an alternative metric for the motions of flux-tube bases.
While the bulk of this work is focused on \bp s (as discussed in Section~\ref{sec:bps-vs-magneograms}), we also perform flux element tracking so that we can provide a sense of how our results vary between the two approaches.

We implement flux element tracking through a set of light modifications to our \bp\ tracking code.
All pixels with an absolute value of the vertical magnetic field strength exceeding 1000~G are marked as seed pixels (rather than identifying seed pixels via the Laplacian).
The parameter $f_\text{contour}$, controlling the location of the contours that limit seed pixel expansion, is changed from 0.65 to 0.6, and the parameter $\Delta B_\text{max}$ is not used.
Feature identification and tracking is run independently for positive and negative polarities, so that each feature contains only pixels of a single polarity across its entire life time.
With these modifications, our algorithm produces flux element outlines that tend to agree with a by-eye assessment of the magnetic flux concentrations, using the same system of manual spot checks we used to tune our \bp\ tracking.
This development and tuning was performed without reference to the agreement between flux-element and \bhp\ outlines, so as to not bias the comparison between the two, but we used the \bhp\ tracking parameters as a starting point in an attempt to minimize the influence of algorithmic changes on the eventual comparison.

We apply this tracking to the \muram\ data through $\tau=1$ maps of the vertical magnetic field strength---that is, each pixel contains the field strength from that column at the height where $\tau=1$.
This is not a complete replacement for synthesized, line-of-sight integrated polarimetric images (analogous to the line-of-sight intensity maps we use for \bp\ tracking), but it is a better stand-in for observational magnetograms than simple horizontal slices through the simulation.

Our tracking produces 1.61 times as many flux element sequences as \bp s (1,718 flux elements versus 1,064 \bp s).
The average lifetime of an identified flux element is 45.1~s (versus 42.4~s for \bp s), and the average area is 26.2~px (versus 35.5~px for \bp s).
There are 1.26 times as many pixels included in all flux elements as there are pixels in a \bp s.
66\% of flux-element pixels are not part of an identified \bp, while 57\% of \bhp\ pixels are not part of a flux element.
(The former number can be at most 79\%, due to the larger number of flux-element pixels.)

\begin{figure}[p]
	\centering
	\includegraphics[trim=0 .34cm 0 0,clip]{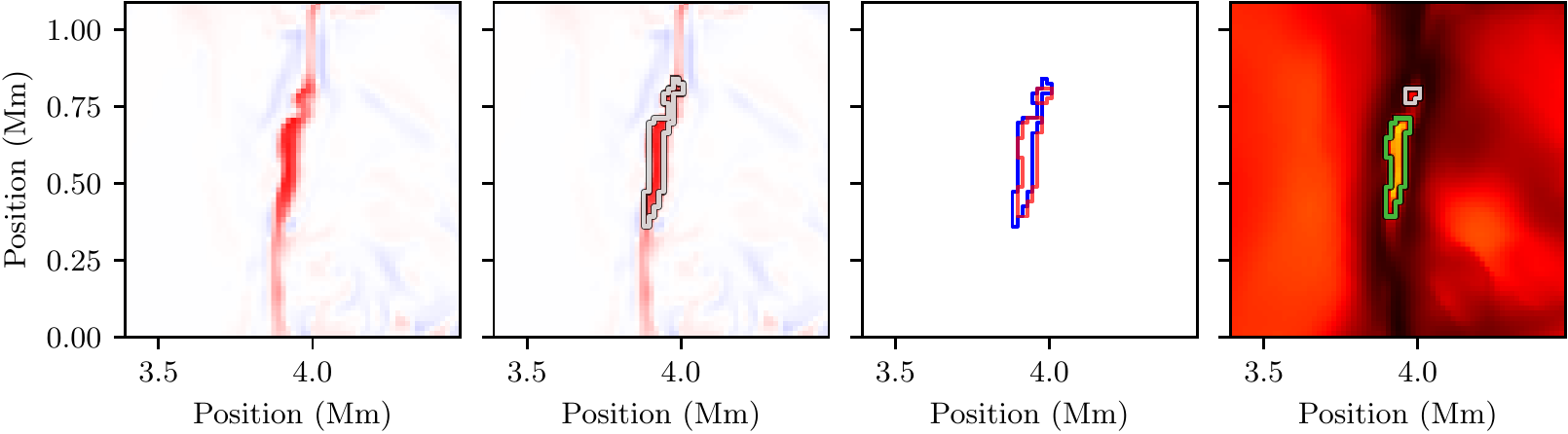}
	\includegraphics[trim=0 .34cm 0 0,clip]{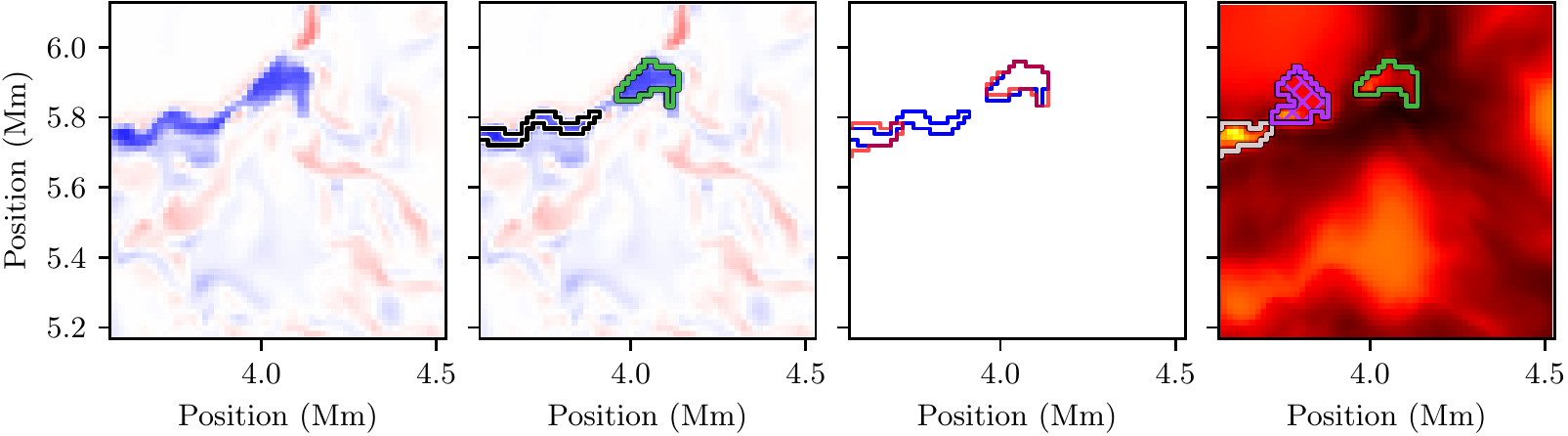}
	\includegraphics{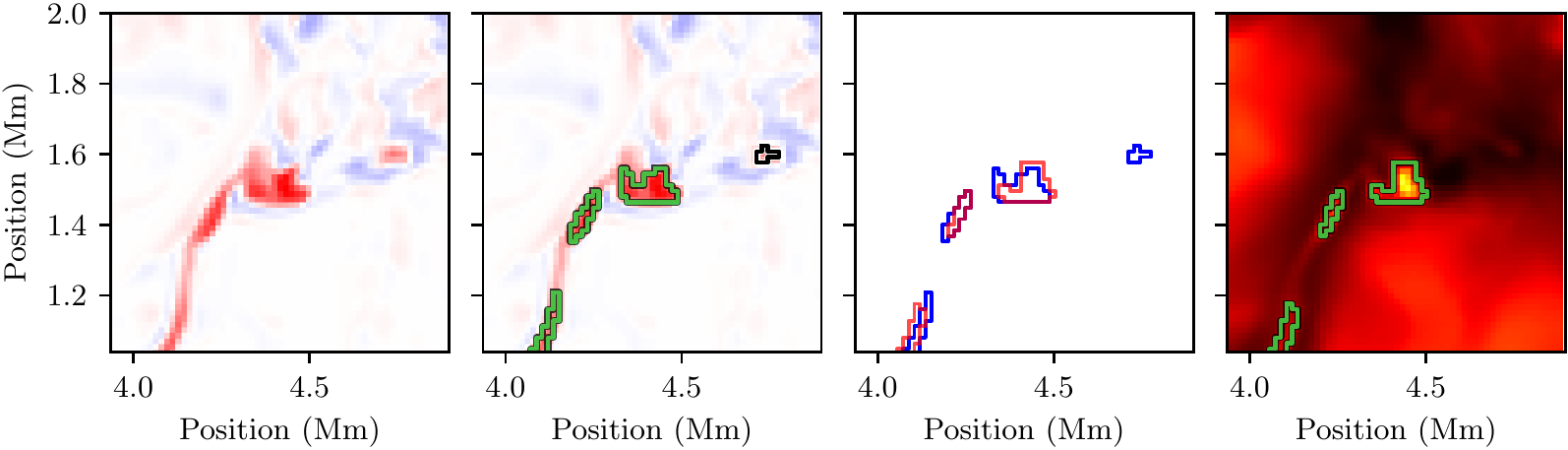}
	\caption[Sample of identified flux elements]{Sample of identified flux elements (continued on the next page). These are the same cut-out regions as in Figure~\ref{fig:tracking-bp-sample}. The left column provides an unobstructed view of the vertical magnetic field strength at the $\tau=1$ surface, and the second column shows the identified flux-element boundaries. The fourth column shows the identified \bp s (and is a repeat of the central column of Figure~\ref{fig:tracking-bp-sample}), while the third allows direct comparison of the flux-element and \bhp\ boundaries (in blue and red, respectively). Unobstructed views of the intensity patterns can be found in Figure~\ref{fig:tracking-bp-sample}. In columns 1, 2, and 4, green lines mark the boundaries of accepted features, magenta lines (with crosshatches) mark regions rejected by our false-positive rejection step, white lines mark regions rejected due to the minimum-lifetime constraint, and black marks features rejected for any other reason (e.g., the maximum-size criterion). Color maps range from 0.7~(black) to 1.75~(white) times the mean intensity for the intensity maps, and -2~(red) to 2 (blue)~kG for magnetic field strength. Animations of these \bp s are available on the author's website at \href{http://samvankooten.net/thesis}{samvankooten.net/thesis} and are archived at \href{http://doi.org/10.5281/zenodo.5238943}{doi.org/10.5281/zenodo.5238943}.}
	\label{fig:tracking-fe-sample}
\end{figure}

\begin{figure}[p]
	\ContinuedFloat
	\centering
	\includegraphics[trim=0 .34cm 0 0,clip]{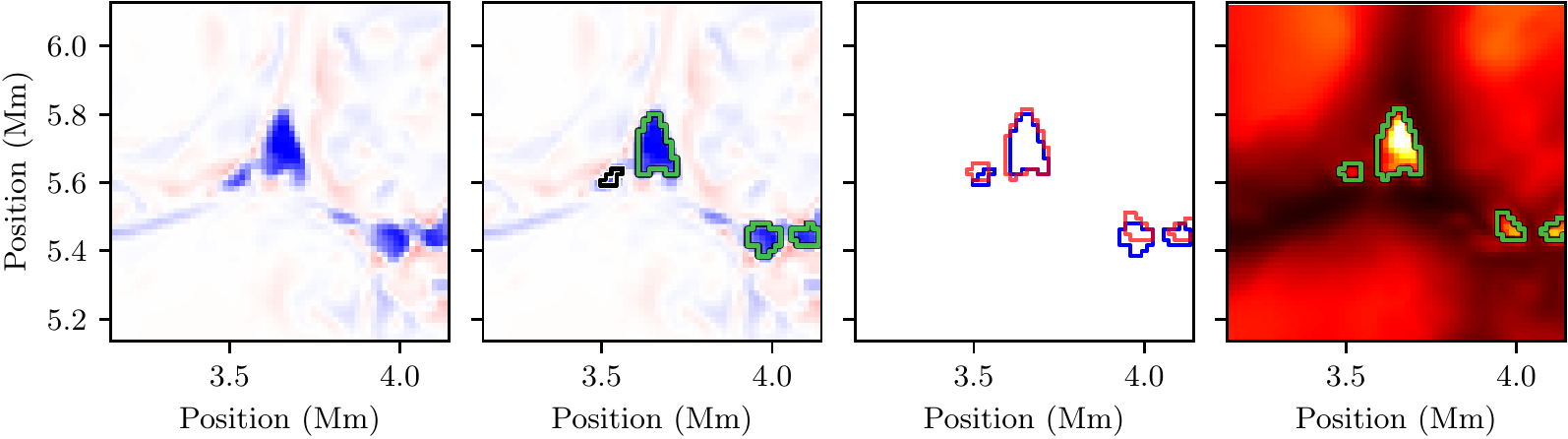}
	\includegraphics[trim=0 .34cm 0 0,clip]{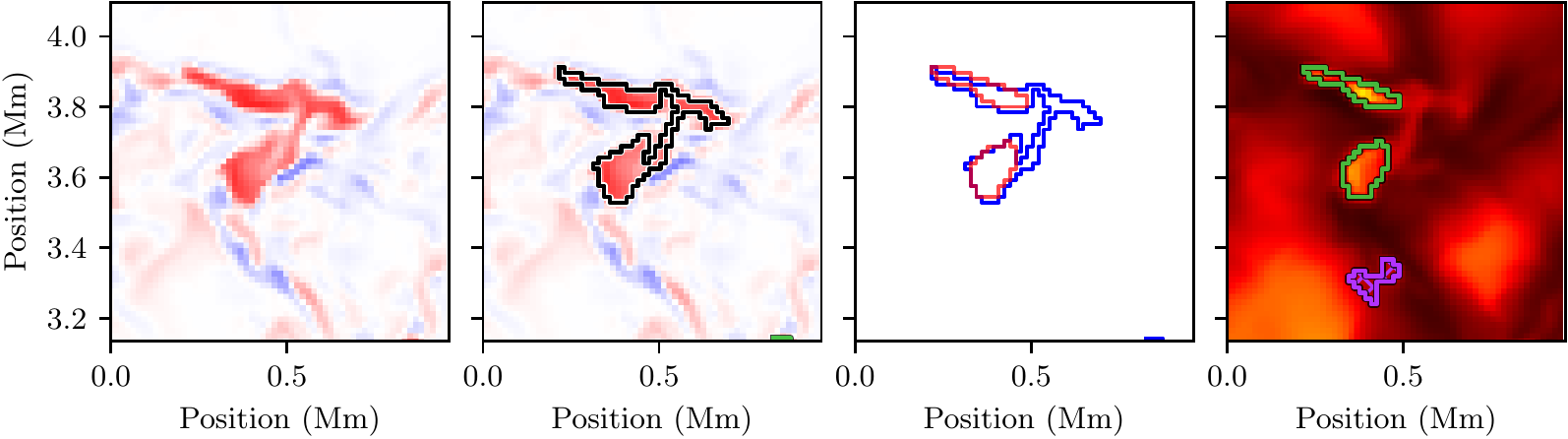}
	\includegraphics[trim=0 .34cm 0 0,clip]{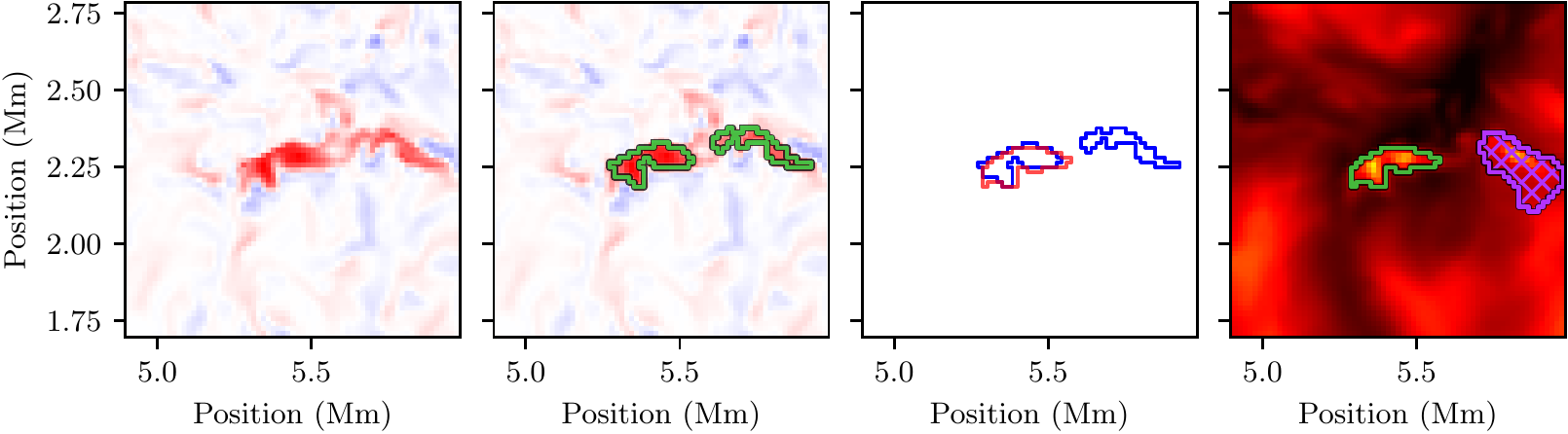}
	\includegraphics{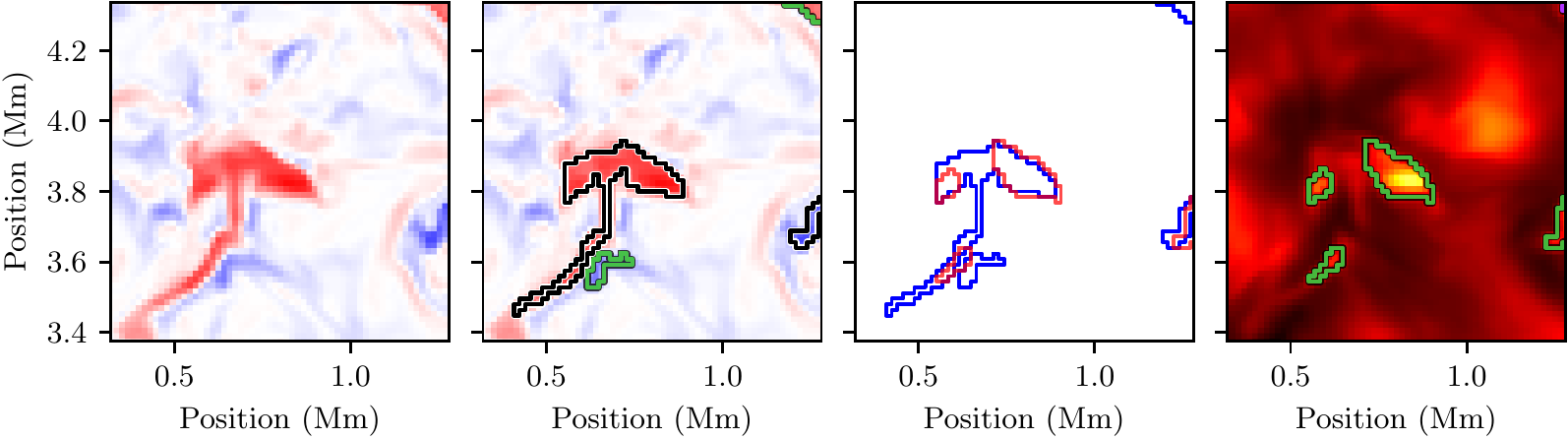}
	\caption[]{Sample of identified flux elements (cont.)}
\end{figure}

We show a sample of identified flux elements in Figure~\ref{fig:tracking-fe-sample}.
A key feature of this plot is a comparison of the outlines of identified flux elements and \bp s.
It can be seen that in many cases, the two types of features align quite well as identified.
In many other cases, however, only a subset of the flux element is included as a \bp.
In these cases, the ``missing'' portion of the \bp\ is usually much less bright than pixels included in the \bp, indicating that the incongruity is due to the lack of an intensity enhancement in those pixels, rather than being a failing of our tracking.

These identified flux elements will not form a cornerstone of this work, but in Chapter~\ref{chap:method-comp-and-conclusions} we will compare the energy fluxes computed from the our identified \bp s and flux elements.
Further exploration of that flux comparison, as well as further comparison of the identified feature boundaries, may be a very compelling direction for future work.
Such work would help to understand the advantages, disadvantages, and degree of interchangeability of these two ways of estimating flux-tube boundaries.

\section{On the Need for New Analysis Techniques}
\label{sec:need-new-techniques}

\subsection{Overview of Existing Techniques}

When estimating wave energy fluxes from \bhp\ motions, the key quantity required is the magnitude of the plasma velocities associated with changes in the \bhp\ shape.
A number of existing algorithms for inferring plasma flows from sequences of images have been used throughout the field.
However, we believe these algorithms have various limitations when applied to the problem of interpreting \bhp\ shape and size changes.
One such limitation of all of these techniques, when applied to simple images of \bp s (i.e., not to magnetograms of the underlying flux element) is that the production of a visible-light enhancement by a flux concentration is a relatively complex phenomenon, and so there is not a direct correlation between visible-light enhancement and magnetic field strength.
Therefore each of these techniques risks over-interpreting changes in the appearance of a \bp.

Local correlation tracking (LCT) was first used by \citet{Leese1970} to analyze the motion of clouds in satellite observations of the Earth, and one of its earliest applications to solar observations was by \citet{November1988}.
For each location in an image, LCT finds the motion vector such that, when a small region of the image centered on the location in question is offset by that vector, the cross-correlation (or a similar measure of similarity) is maximized between that offset piece of the image and the contents of the following image frame at the offset location.
In other words, LCT takes the image pattern at a given location and looks for that pattern in the surrounding region in the following image, and it produces a motion vector connecting the pattern's original location to the location of the best-match pattern in the following image.
As this technique has long been used in the field, a variety of implementations have been employed.
One illustrative difference in implementation is tile-based techniques \citep[e.g.,][]{Berger1998} versus Fourier-based techniques \citep[e.g.,][]{Welsch2004}.
The limitation of this approach when applied to \bhp\ shape changes is that, at DKIST resolutions, \bp s as a whole will be resolved, but details of their internal structure might not be resolved, and so there may not be much resolved structure that can be looked for in following images.
Additionally, LCT requires some level of rigidity on scales comparable to the windowing function of the algorithm (i.e., the smallest scales in the image) so that the structure in any small patch moves coherently, whereas \bp s exist at or near those scales, and their evolution is expected to occur at those scales.
Put simply, LCT works best when small-scale regions of the features in question are distinct and quasi-rigid, which is not expected to be the case for \bp s.

Optical flow techniques \citep{Horn1981} estimate local velocities based on each pixel's value for the spatial gradient and time derivative of the pixel value.
(These might be distinguished from LCT as \textit{differential} optical flow techniques.)
As a simple example, if a dark pixel is next to a bright pixel (such as at the edge of a feature in the image) and the dark pixel grows brighter in the following frame, it is inferred that motion is occurring from the bright pixel to the dark pixel.
To overcome an inherent under-determination of the technique, implementations may include constraints to prefer, e.g., a globally smooth velocity field \citep[e.g.,][]{Colaninno2006} or local uniformity \citep[e.g.,][]{Gissot2007}.
Since optical flow techniques depend on the local derivatives in intensity, these techniques are also difficult to apply in \bp s, where significant structure within the \bp\ may not occur.
In particular, if a \bp 's intensity profile is flat within the body of the \bp\ and the \bp\ is advecting uniformly, optical flow may produce a doughnut-like velocity map, with strong velocities near the edge of the feature and zero velocity in the center, where the derivatives vanish.

Another class of motion-inference algorithms, often used with magnetograms, attempts to use the ideal-MHD magnetic induction equation to impose constraints on the inferred flow field in an attempt to derive more physically-realistic flows.
One such algorithm is the Induction LCT (or ILCT) approach of \citet{Welsch2004}, which uses flow fields generated by LCT as well as measured values of the vertical magnetic field strength and its time derivative as inputs to constraints derived from the induction equation, and from this ILCT generates a uniquely-specified flow field which is self-consistent with respect to the induction equation.
Another approach is the differential affine velocity estimator \citep[DAVE;][]{Schuck2006} and the related nonlinear affine velocity estimator \citep[NAVE;][]{Chae2008}, which apply the induction equation as the constraint to fully determine the optical flow equations.
While valuable approaches in the study of magnetograms, these techniques do not assist the analysis of \bp s, as the intensity enhancement does not obey any equivalent of the induction equation.

A novel technique called balltracking \citep{Potts2004,Attie2009} has been employed to estimate flows at the granular and supergranular scales.
In this approach, image intensity is imagined as the geometric height of a fluid surface (i.e., brighter pixels are higher than darker pixels), with the evolving granular pattern imagined as ripples in that fluid, and a number of passive spherical tracers of a given volume are allowed to float in the fluid and to be pushed about by the evolving pattern of the intensity ripples, tending to slide toward and follow intensity minima.
As the balls carry momentum between frames, this approach can deal well with noise and missing frames.
Additionally, the approach is much more computationally efficient than correlation-tracking, enabling easy application to larger data sets.
This approach has been shown to be effective at granular and supergranular scales but is believed to be less effective at smaller scales.
Additionally, when applied to \bp s, and with the intensity-to-height relation reversed to that balls flow toward, rather than away from, \bp s, these balls may attempt to closely follow the brightest part of the \bp\ (typically the center), meaning the results of balltracking may end up closely related to the traditional tracking of the intensity-weighted centroids of \bp s.

Neural networks have also been applied in reconstructing horizontal velocities \citep[e.g., DeepVel,][]{AsensioRamos2017,Tremblay2021}.
Neural networks can be powerful tools, but they require large quantities of appropriate training data.
Since \bp s constitute only a small fraction of the pixel in any given simulation, extremely large quantities of simulated data are required to build up a proper training set for intra-\bp\ velocity inference, making this application very difficult.

\subsection{Motivation and Goals for a New Technique}
\label{sec:motivation-for-new-technique}

Given the limitations of existing techniques when applied to \bhp\ evolution, we believe that a new technique is required for the particular problem of inferring velocities within \bp s, with the goal of using these velocities to understand the driving of waves in the flux tube, and we propose two such techniques in the following chapters.
The primary differences between \bhp\ evolution and other regimes where horizontal velocities are desired (e.g., supergranular flows) are the need to extract velocities at the smallest scales in the image, and the fact that the observed intensity pattern within a \bp\ is only indirectly related to the underlying flux tube and its evolution.
The boundary of a \bp\ may serve as a reasonable proxy for the location of the flux tube (though observational verification of this at the highest resolutions with DKIST will be a valuable study), but the intensity pattern within the boundary of the \bp\ and the evolution of this pattern do not necessarily have any relation to the magnetic field configuration or the internal motions of the flux tube.
To address these matters, we believe an effective strategy is to rely solely on the identified border of the \bp\ and not on the intensity pattern within the \bp.
Following the evolution of this boundary serves as a probe of the plasma velocities driving the evolution of the flux-tube shape and therefore the observed \bhp\ boundary, while ignoring the intensity pattern within the \bp\ avoids the risk of over-interpreting patterns and changes which may be more indirectly related to the plasma flows we seek.
The end goal of our techniques is to estimate velocity vectors within the flux tube that can be interpreted as the characteristic velocities of flux-tube waves caused by the convective buffeting of the tube.
Such vectors can be used to estimate the vertical energy flux of those flux-tube waves.
It is very difficult in such an analysis, if not impossible, to fully incorporate the complex dynamics of \bp s and the mechanisms by which the visible-light enhancement connected to plasma flows, and we do not claim to do so in the present work.
Instead, we emphasize that our approach will be a proof-of-concept, to be developed further after DKIST data becomes available.
Our overarching goals are to produce an initial estimate of the energy flux which could not be measured before DKIST, in order to determine whether further study is warranted, and to propose ideas which may inspire further analyses.
We thus will not fully solve the problem of analyzing resolved \bp s, but we hope to take the first steps into a post-centroid-tracking world.

\section{\BP s Versus Magnetograms}
\label{sec:bps-vs-magneograms}

While \bhp\ motions are commonly studied through the observed intensity enhancements, one could also study the motions of the corresponding flux concentrations at the photosphere via magnetograms, and this may be a more direct way to probe the location and shape of a flux tube base.
In this dissertation, we continue to favor the former approach, however.
Our primary reason is that the \muram\ data we use is accompanied by synthesized white-light images, providing a clear observational analogue for \bp s.
While we can easily extract the locations of strong-field regions, horizontal slices through the \muram\ data cube are not a strong analogue for the spectropolarimetric observations required to detect magnetic field concentrations in observations, as they do not incorporate any line-of-sight effects.
We can extract the magnetic field strength at the $\tau=1$ height for each column of pixels, and while this may better approximate the magnetic field pattern that would be seen observationally, it is not clear how well it approximates the line-of-sight integrated Stokes vector.
Synthesizing spectropolarimetric observations from the \muram\ cubes would provide a true observational analogue and would allow interesting comparisons between the locations of \bp s and magnetic flux concentrations as they would be observed, but such a synthesis is outside the scope of the present work.

Additionally, tracking flux elements in magnetograms with DKIST poses certain observational trade-offs relative to white-light images of \bp s with the Visible Broadband Imager (VBI).
For example, the Visible Spectro-Polarimeter (ViSP) achieves a design spatial resolution of approximately twice the telescope's diffraction limit \citep{DeWijn2012}.
Magnetogram images from ViSP will therefore provide a reduced spatial resolution compared to white-light images from VBI.
Additionally, ViSP must scan across its field of view to build up the second spatial dimension.
Using ViSP's minimum allowable integration time of 0.1~s and its smallest slit size of 0.0284\arcsec, the magnetically-sensitive, photospheric Fe~I line at 620.2~nm will achieve signal-to-noise values of $\sim200-300$ in intensity and $\sim100$ in the other Stokes components\footnote{These numbers were obtained with the ViSP Instrument Performance Calculator (IPC).}.
However, to achieve a 3-second cadence (the quoted cadence for speckle-reconstructed VBI images; \citealp{Elmore2014}), the instrument can observe at only approximately 10 slit positions.
With a step size matched to the slit width, this corresponds to output maps of approximately $0.28\arcsec$ or 200~km in width.
While the large height of the ViSP field of view will still provide an area within an order of magnitude of that of the \muram\ simulations we analyze, the very narrow shape of the maps means it is very likely that many or most \bp s will straddle or cross the edge of the field of view, limiting the ability to produce long time series of observed \bp s.
To avoid this problem, a larger number of slit steps must be used, at the cost of observational cadence relative to what VBI can provide.
Alternatively, the Visible Tunable Filter (VTF) requires approximately 13~s to produce a full Stokes map with a 60\arcsec\ field of view and a resolution at the diffraction limit \citep{Elmore2014}.
This is a much more favorable field-of-view geometry, but it still comes at the cost of observational cadence relative to VBI.
While a three-second cadence is not necessarily required for resolving \bhp\ motions, a longer cadence may limit the ability to apply some of the temporal smoothing techniques we will discuss in the following chapters, and this might result in some high-frequency ``jumpiness'' in the inferred \bhp\ boundaries and velocities.

While using magnetograms sidesteps the complication of the indirect linkage between the evolution of the flux tube and the evolution of a visible \bp, it must be remembered that magnetograms are derived through an inversion technique, rather than being a ground-truth measurement.
\citep[For a review of inversions, see][]{DelaCruzRodriguez2017}.
Our work is focused on the location of the outer boundary of the observed \bp, and at this edge the magnetic field and other plasma properties may vary rapidly over the vertical column responsible for the observed spectrum in each pixel---particularly if the flux tube is not strictly vertical---making the required inversions non-trivial.
It is not immediately clear whether the boundaries of the inverted magnetic field concentration will be better or worse than the boundary of the white-light feature at approximating the flux tube's cross section.
(Indeed, investigating the degree of co-occurrence of white light and magnetic field features observationally will be an important project once DKIST begins operations.)

We thus continue to focus our efforts on white-light observations of \bp s.
However, our proposed techniques could just as well be applied to magnetograms (whether observational or synthesized from \muram), and we are hopeful that such analyses are performed once DKIST observations become available.

\chapter{Velocity Fields via Moments and Wave Modes}
\label{chap:ellipse-fitting}

\section{Overview}
\label{sec:ellipse-overview}

In this chapter, we present the first of our two proposed methods of analyzing \bhp\ shape changes.
In this method, we calculate moments over the area of the \bp, which describe the large-scale shape of the \bp.
Treating changes in these moments as describing changes in the cross-sectional shape of a flux tube, we connect these changes changes to MHD flux-tube waves of various modes.
We focus on $n=0,1,2$ modes, which are connected to the zeroth through second moments, but the approach can be extended to higher modes.
The zeroth through second moments can be used to produce an ellipse fit to the \bhp\ shape, so this approach can be thought of as analogous to approximating the \bp\ as an ellipse.
(However, it is important to note that the $n=2$ mode we use only appears elliptical for small-amplitude waves.
For larger-amplitude waves, where some of the approximations we will use break down, it takes on a two-lobed, peanut-like shape\footnote{This distinction between low- and high-amplitude waves also appears for the more commonly-studied $n=1$ mode, where low-amplitude waves appear as a simple translation of the circular flux-tube cross section, whereas high-amplitude waves also affect the shape of the cross section, turning a circular cross section into something almost heart-shaped.}.)

\Bp s have long been studied by measuring the motion of their centroids or barycenters \citep[e.g.,][for a small sampling, as well as Chapter~\ref{chap:bp-centroids} of this thesis]{Muller1994,Nisenson2003,Utz2010}.
The motion of the centroid is often taken as a proxy for the bulk motion of the flux tube associated with the \bp, and so this motion is assumed to correspond to transverse perturbations (or shaking) of the flux tube, which is then expected to excite kink-mode ($n=1$) waves that propagate upward, carrying energy to the corona.
The centroid of a \bp\ can be calculated from the first moments computed across the shape.
The moment-based approach we present in this chapter can therefore be thought of as building upon this traditional centroid-tracking by considering the zeroth and second moments of the \bp 's shape (and laying the framework for considering additional moments, as well).
As with the kink-mode ($n=1$) wave fluxes estimated by centroid tracking, the properties of the sausage-mode ($n=0$) waves, associated with changes in the zeroth moment, and the $n=2$ (sometimes called ``fluting mode'') waves, associated with changes in the second moments, can be treated analytically (in the regime near the photosphere where the flux tube can be approximated as a cylinder), and distinct energy fluxes can be calculated for each wave mode.
The expected degree of upward propagation and reflection of the $n=0$ and $n=2$ wave modes must be separately modeled for appropriate conditions \citep[this has been done for the $n=1$ mode, see e.g.,][]{Cranmer2005}, as these wave modes cannot be expected to propagate through the varying plasma conditions above the photosphere in exactly the same way as the $n=1$ waves which are more commonly studied in the context of \bp s.
We do not perform this modeling, and so the energy fluxes we will estimate cannot be directly connected to the coronal energy budget.
Instead, the results of this chapter will represent a sort of lower boundary condition, and they will help motivate such modeling in the future.

In our approach, we assume that changes in the \bhp\ boundary are strictly caused by, and directly connectable to, motions of the flux-tube wall, and that these changes occur uniformly at one distinct height along the strictly-vertical flux tube.
These assumptions are not guaranteed to hold (and, in fact, it would be astonishing if they were universally true), but the assumptions we require in this approach are reasonable extensions of the assumptions in traditional centroid tracking (i.e., that motion of the observed \bhp\ centroid corresponds directly to horizontal motion of the tube cross section).
In Section~\ref{sec:emd-comp-but-should-we}, after presenting our second proposed method for analyzing shape changes, we will discuss in detail ways that these assumptions might break down.

We are not the first to consider incorporating the shapes of \bp s to more accurately treat their behavior and evolution (see, e.g., \citealp{Keys2020}, who proposed modeling \bp s as ellipses), but we believe that our approach of connecting changes in moments (which is analogous to changes in elliptical properties for the zeroth through second moments in the low-amplitude limit) to estimated flux-tube wave energy fluxes is a novel contribution.

In Section~\ref{sec:ellipse-derivation} we present a summary of the properties of flux-tube waves in the thin-tube approximation, we define the moments used in this chapter, and we connect changes in the value of those moments to key quantities for calculating the energy flux of the associated waves.
In Section~\ref{sec:ellipse-results} we apply the technique to \bp s in a \muram\ simulation, producing estimated energy fluxes and associated power spectra.
We summarize our work in Section~\ref{sec:ellipse-summary}, and we discuss our results further in Chapter~\ref{chap:method-comp-and-conclusions}.

\section{Method}
\label{sec:ellipse-derivation}

\subsection{Summary of Linear Oscillation Properties}

The properties of small first-order wave-like perturbations have been studied extensively, and we will use much of the notation from \citet{Spruit1982} and \citet{Edwin1983}.
To begin, we will summarize key quantities from these works.
We assume the zeroth-order (unperturbed) flux tube is cylindrical, with radius $r_0$, and oriented vertically in the solar photosphere.
We will use cylindrical coordinates $(r, \phi, z)$ to describe positional variations inside ($r < r_0$) and outside ($r > r_0$) the tube.
We assume the background magnetic field is always pointed along the $z$-axis of the cylinder.
The pressure, mass density, and field strength are $P_0$, $\rho_0$, and $B_0$ inside the tube, and $P_e$, $\rho_e$, and $B_e$ outside.
We also define the interior and exterior (adiabatic) sound speeds
\begin{equation}
  c_0 \, = \, \sqrt{\frac{\gamma P_0}{\rho_0}}
  \qquad
  \mbox{and}
  \qquad
  c_e \, = \, \sqrt{\frac{\gamma P_e}{\rho_e}}
\end{equation}
and \Al\ speeds
\begin{equation}
  V_{{\rm A}0} \, = \, \frac{B_0}{\sqrt{4\pi\rho_0}}
  \qquad
  \mbox{and}
  \qquad
  V_{{\rm A}e} \, = \, \frac{B_e}{\sqrt{4\pi\rho_e}},
\end{equation}
and we will often assume that the exterior region is field-free, such that $B_e \approx V_{{\rm A}e} \approx 0$.

The derivation for compressive MHD waves begins by assuming that the divergence of the velocity fluctuations behaves as
\begin{equation}
  \nabla \cdot \vec{v} \, = \, \Delta \, = \,
  R(r) \, \exp \left( i \omega t \, + \, i n \phi \, + \, ikz \right),
  \label{eq:Delta}
\end{equation}
where $R(r)$ is a yet-undetermined function, and the azimuthal mode number $n$ is the primary identifier of the mode of the oscillation.
We will solve a dispersion relation for the frequency $\omega$ as a function of the axial wavenumber $k$.
Two other key quantities derived by \citet{Spruit1982} and \citet{Edwin1983} are the radial wavenumbers for the interior and exterior of the tube:
\begin{equation}
  m_0^2 \, = \,
  \frac{(k^2 c_0^2 - \omega^2)(k^2 V_{{\rm A}0}^2 - \omega^2)}
  {(k^2 c_{{\rm T}0}^2 - \omega^2)(c_0^2 + V_{{\rm A}0}^2)}
  \qquad
  \mbox{and}
  \qquad
  m_e^2 \, = \,
  \frac{(k^2 c_e^2 - \omega^2)(k^2 V_{{\rm A}e}^2 - \omega^2)}
  {(k^2 c_{{\rm T}e}^2 - \omega^2)(c_e^2 + V_{{\rm A}e}^2)},
\end{equation}
where the tube speed $c_{\rm T}$ is defined inside and outside the tube as
\begin{equation}
  c_{{\rm T}0} \, = \, \frac{c_0 V_{{\rm A}0}}
  {\sqrt{c_0^2 + V_{{\rm A}0}^2}}
  \qquad
  \mbox{and}
  \qquad
  c_{{\rm T}e} \, = \, \frac{c_e V_{{\rm A}e}}
  {\sqrt{c_e^2 + V_{{\rm A}e}^2}}
  \, .
\end{equation}
Note that $m_e^2$ is assumed to always be positive, while $m_0^2$ can be positive or negative. When $m_0^2 < 0$, we define $\mu_0^2 = -m_0^2 > 0$.

The solutions for $R(r)$ take the form of Bessel functions.
Both the sign of $m_0^2$ and the boundary conditions determine which type of Bessel functions to use:
\begin{equation}
  R(r) \, = \, A \left\{
  \begin{array}{ll}
     I_n (m_0 r) \, , &
     r < r_0 \,\,\, \mbox{and} \,\,\, m_0^2 > 0 \\
     J_n (\mu_0 r) \, , &
     r < r_0 \,\,\, \mbox{and} \,\,\, \mu_0^2 = -m_0^2 > 0 \\
     K_n (m_e r) \, , &
     r > r_0 
  \end{array}
  \right.
  \label{eq:Rbess}
\end{equation}
where $A$ is a normalization constant with units of s$^{-1}$.
\citet{Spruit1982} derived the expressions for all three components of the first-order velocity perturbation.
Leaving off the exponential term in Equation~\eqref{eq:Delta}, these are
\begin{equation}
  v_z \, = \, -A \, \frac{c^2}{\omega^2} \, ik {\cal B}_n
  \, , \qquad
  v_r \, = \, A \, \frac{\omega^2 - k^2 c^2}{m^2 \omega^2}
  \frac{d}{dr} {\cal B}_n
  \, , \qquad
  v_{\phi} \, = \, iA \, \frac{\omega^2 - k^2 c^2}{m^2 \omega^2} \,
  \frac{n}{r} {\cal B}_n
  \, ,
  \label{eq:vdef}
\end{equation}
where the general Bessel function ${\cal B}_n$ must be replaced with a function and argument as in Equation~\eqref{eq:Rbess} above.
Note also that the interior or exterior versions of $m$ and $c$ must be substituted as well.

The last key quantity is the dispersion relation for $\omega(k)$.
It is given by ensuring that both $v_r$ and the total pressure are continuous across the boundary at $r_0$.
\citet{Edwin1983} specifies two versions for two distinct cases.
For ``surface waves'' ($m_0^2 > 0$),
\begin{equation}
  \rho_0 (k^2 V_{{\rm A}0}^2 - \omega^2) \, m_e \,
  \frac{K'_n (m_e r_0)}{K_n (m_e r_0)}
  \,\, = \,\,
  \rho_e (k^2 V_{{\rm A}e}^2 - \omega^2) \, m_0 \,
  \frac{I'_n (m_0 r_0)}{I_n (m_0 r_0)}
\end{equation}
and for ``body waves'' ($m_0^2 < 0$),
\begin{equation}
  \rho_0 (k^2 V_{{\rm A}0}^2 - \omega^2) \, m_e \,
  \frac{K'_n (m_e r_0)}{K_n (m_e r_0)}
  \,\, = \,\,
  \rho_e (k^2 V_{{\rm A}e}^2 - \omega^2) \, \mu_0 \,
  \frac{J'_n (\mu_0 r_0)}{J_n (\mu_0 r_0)}
\end{equation}
where ${\cal B}'_n (m r_0)$ is $(d/dx) {\cal B}_n (x)$ evaluated at $x = m r_0$.

\subsection{The Thin Tube Approximation}

We will assume that the arguments of the Bessel functions are all $\ll 1$ (the thin-tube limit), and we will evaluate the appropriateness of this assumption shortly.
Table~\ref{table:bessel-functions} shows small-argument expansions for the required Bessel functions and ratios in the dispersion relation.

\begin{table*}[!t]
\begin{center}
\begin{tabular}{r|ccccc}
 & $J_n(x) \approx I_n(x)$ & $K_n(x)$ & $J'_n/J_n$ & $I'_n/I_n$ & $K'_n/K_n$\\
\hline
$n=0$ &  1  & $-\ln x$ &
  $-x/2$  &  $+x/2$  &  $1/(x \ln x)$ \\
$n=1$ &  $x/2$  &  $1/x$  &
  $1/x$  &  $1/x$  &  $-1/x$  \\
$n=2$ &  $x^2/8$  &  $2/x^2$  &
  $2/x$  &  $2/x$  &  $-2/x$  \\
$n=3$ &  $x^3/48$  &  $8/x^3$  &
  $3/x$  &  $3/x$  &  $-3/x$  \\
$n=4$ &  $x^4/384$  &  $48/x^4$  &
  $4/x$  &  $4/x$  &  $-4/x$  \\
$n \geq 1$ & $x^n / (2^n n!)$ & $2^{n-1} (n-1)! / x^n$ &
  $n/x$  &  $n/x$  &  $-n/x$
\end{tabular}
\end{center}
\caption{Bessel Functions for Thin Tubes ($x \ll 1$)}
\label{table:bessel-functions}
\end{table*}

In this limit, the surface-wave and body-wave dispersion relations produce the same expressions for $\omega^2/k^2$, and these expressions depend on the value of $n$.
For $n=0$ (``sausage-mode'' waves),
\begin{equation}
  V_{\rm ph}^2 \, = \, \frac{\omega^2}{k^2} \, = \, c_{{\rm T}0}^2
  \, .
\end{equation}
For $n \geq 1$ (general ``kink-mode'' waves),
\begin{equation}
  V_{\rm ph}^2 \, = \, \frac{\omega^2}{k^2} \, = \, c_{\rm k}^2
  \, ,
  \qquad
  \mbox{where}
  \qquad
  c_{\rm k}^2 \, = \,
  \frac{\rho_0 V_{{\rm A}0}^2 + \rho_e V_{{\rm A}e}^2}
  {\rho_0 + \rho_e}
\end{equation}
defines the kink speed $c_{\rm k}$.
The phase velocity is constant, indicating these are dispersionless waves with $V_{\rm gr} = V_{\rm ph}$.

With these phase velocities in hand, we now calculate values for many of these quantities and evaluate the appropriateness of the thin-tube approximation.
Drawing from the similar model of \citet{Cranmer2005}, at the photosphere of a typical intergranular \bp\ surrounded by a field-free granule, we have $B_0 = 1430$~G, $\rho_e = 3 \times 10^{-7}$~g~cm$^{-3}$, and a density contrast of $\rho_0 / \rho_e = 0.316$.
If the temperature is the same inside and outside the tube, then
\begin{equation}
  V_{{\rm A}0} = \mbox{13.1 km~s$^{-1}$}
  \, , \qquad
  V_{{\rm A}e} = 0
  \, , \qquad
  c_0 = c_e = \mbox{8.13 km~s$^{-1}$} \, ,
\end{equation}
where we also assume $T = 5770$~K and the mean atomic weight $\mu = 1.2$ for the mostly neutral photosphere.
The gas pressures ($P = \rho k T / \mu m_{\rm H}$) inside and outside
the tube are roughly $3.765 \times 10^4$ and $1.190 \times 10^5$
dyne~cm$^{-2}$, respectively.
Thus,
\begin{equation}
  c_{{\rm T}0} = \mbox{6.91 km~s$^{-1}$}
  \, , \qquad
  c_{{\rm T}e} = 0
  \, , \qquad
  c_{\rm k} = \mbox{6.42 km~s$^{-1}$} \, .
\end{equation}
The field-free limit for the external medium simplifies the evaluation of $m_e$, since in this case
\begin{equation}
  m_e^2 \, \approx \, k^2 \left( 1 - \frac{V_{\rm ph}^2}{c_e^2} \right)
\end{equation}
which is indeed always positive for the above values.
For $n=0$, $m_e \approx 0.528 k$, and for $n \geq 1$, $m_e \approx 0.614 k$.

The interior radial wavenumber $m_0$ can be written as
\begin{equation}
  m_0^2 \, = \, k^2 \,
  \frac{(c_0^2 - V_{\rm ph}^2)(V_{{\rm A}0}^2 - V_{\rm ph}^2)}
  {(c_{{\rm T}0}^2 - V_{\rm ph}^2)(c_0^2 + V_{{\rm A}0}^2)} \, ,
  \label{eq:m0vph}
\end{equation}
and we examine two cases:
\begin{itemize}
\item
For $n=0$, with $V_\text{ph} = c_\text{T0}$, the denominator of Equation~\eqref{eq:m0vph} is zero.
However, \citet{Edwin1983} show numerical solutions where $c_{{\rm T}0}^2 < V_{\rm ph}^2 < c_0^2$ (with $V_{\rm ph}$ very close to $c_\text{T0}$ for small values of the argument to the Bessel functions) producing $m_0^2 < 0$ and requiring $J_n$ Bessel functions.
Later, we will show that $m_0$ (or rather, $\mu_0$) is not present in the final equations for the $n=0$ modes, so that the resulting very large magnitude for $m_0$ is not physically relevant.
\item
For $n \geq 1$, $V_{\rm ph} = c_{\rm k}$, and Equation~\eqref{eq:m0vph} produces $m_0 \approx 1.448 k$.
Since $m_0^2 > 0$, the $I_n$ Bessel functions are used for kink modes.
\end{itemize}
To evaluate the suitability of the thin-tube limit, we take $r_0 = 50$~km (for a flux tube of order 100~km across) and $V_{\rm ph} = 6.5$~km~s$^{-1}$ for the $n\ge1$ modes.
Taking $m \approx k$, the Bessel-function argument $x \approx k r_0 = \omega r_0 / V_{\rm ph}$, which depends on the frequency.
Representative timescales for \bhp\ evolution range from about 2 to 20 minutes.
This produces a span from $x \approx 0.04$ (for the longer period) to $x \approx 0.4$ (for the shorter period).
For the $n=0$ mode, taking $V_{\rm ph} = 1.1 c_{\rm T0}$, those same representative timescales produce values from $x \approx 0.02$ (for the longer period) to $x \approx 0.2$ (for the shorter period).
We see that the thin-tube approximation is not suitable for any faster flux-tube motions, but it is certainly appropriate for representative motions over the bulk of a \bp 's life.

\subsection{Kink Modes: Velocity Amplitudes and Energy Fluxes}

Here we analyze the $n \geq 1$ (kink) waves, with $n=0$ waves deferred for later.
First, we derive the velocity field both inside and outside the tube.
In the thin-tube limit, $I'_n (x) \approx n I_n (x) / x$, so that Equation~\eqref{eq:vdef} gives $v_\phi = i v_r$.
Additionally, the axial (i.e., vertical) velocity amplitude is negligible for a kink-mode wave, since
\begin{equation}
  \frac{v_z}{v_r} \, = \, -i N_0^2
  \left( \frac{c_0^2}{V_{\rm ph}^2 - c_0^2} \right)
  \frac{kr}{n} \, ,
\end{equation}
where $N_0 = m_0 / k$ is a dimensionless number.
For the example numbers given above, $N_0 \approx 1.448$ and $|v_z / v_r| \approx 5.57 kr/n$.
We are treating $n$ values of order one, and the thin-tube limit sets $kr \ll 1$.
Thus, we can ignore the contributions of $v_z$ motions to the total energy fluxes of modes with $n \geq 1$.

\begin{figure}[t]
	\centering
	\includegraphics{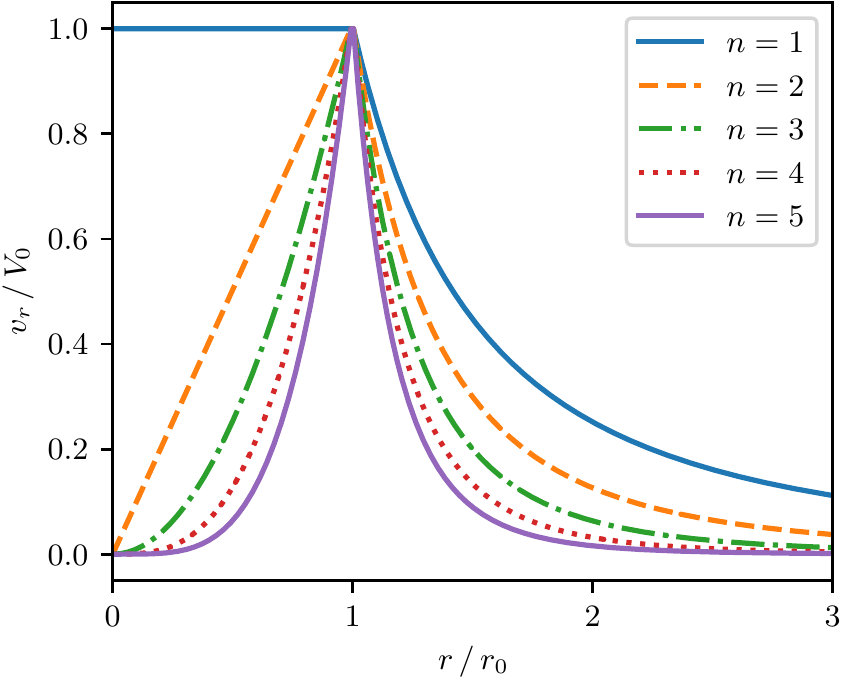}
	\caption[Velocity profiles given by Equation~\eqref{eq:vrsimple} for $n=1$, 2, 3, 4, and 5]{Velocity profiles given by Equation~\eqref{eq:vrsimple} for $n=1$, 2, 3, 4, and 5.}
	\label{fig:moments-kink-modes}
\end{figure}

We aim to derive the full eigenfunctions for $v_r(r)$ and $v_{\phi}(r)$ for different values of $n$.
The constants $A$ given in Equation~\eqref{eq:vdef} are not continuous across the boundary of the tube, but $v_r$ itself must be continuous.
Thus, we define $A_0$ inside the tube and $A_e$ outside the tube, with
\begin{equation}
  v_r (r < r_0) \, = \, \frac{A_0}{m_0}
  \left( 1 - \frac{c_0^2}{V_{\rm ph}^2} \right)
  \frac{n \, (m_0 r)^{n-1}}{2^n \, n!}
\end{equation}
\begin{equation}
  v_r (r > r_0) \, = \, -\frac{A_e}{m_e}
  \left( 1 - \frac{c_e^2}{V_{\rm ph}^2} \right)
  \frac{n \, 2^{n-1} \, (n-1)!}{(m_e r)^{n+1}}
\end{equation}
where, outside the tube, we use $K'_n (x) \approx -n K_n (x) / x$.
Equating the two velocities at $r_0$, and assuming for now that $c_0 = c_e$, we find
\begin{equation}
  \frac{A_e}{A_0} \,\, = \,\, -\frac{m_e}{m_0} \,
  \frac{(m_0 r_0)^{n-1} (m_e r_0)^{n+1}}{2^{2n-1} \,
  (n-1)! \, n!} \,\, .
\end{equation}
However, if we choose to normalize everything by the value of $v_r$ at the tube boundary (which we label $V_0$), the expression simplifies to
\begin{equation}
  v_r(r) \,\, = \,\, V_0 \left\{
  \begin{array}{ll}
    (r/r_0)^{n-1} \, , & \mbox{if $r < r_0$} \\
    (r_0/r)^{n+1} \, , & \mbox{if $r > r_0$}
  \end{array}  \right.
  \label{eq:vrsimple}
\end{equation}
(note that the complete expression for $v_r(r,\phi,z,t)$ is the above function multiplied by $\exp ( i \omega t + i n \phi + ikz )$).
Figure~\ref{fig:moments-kink-modes} shows example radial profiles of $v_r$ for various modes.
It can be seen that the $n=1$ mode displays uniform radial motion throughout the body of the flux tube (consistent with its nature of offsetting the entire tube cross section).
Higher-$n$ modes are increasingly concentrated at the tube boundary.
Observationally, a uniform shift in the cross-section position is more easily detectable than a more finely-detailed perturbation along the tube edge, particularly in low-resolution data, meaning these radial profiles are consistent with the observational focus thus far on the $n=1$ mode.

As an aside, we showed earlier that inside the flux tube, $v_{\phi} = i v_r$.
However, outside the flux tube, $v_{\phi} = -i v_r$.
Thus, if $v_r$ is forced to be continuous at the boundary, then $v_{\phi}$ must be discontinuous (although it only flips sign).
This was discussed by \citet{Goossens2009}.

The instantaneous kinetic energy density carried by a wave with complex amplitudes is defined as
\begin{equation}
  \varepsilon \, = \, \frac{1}{2} \rho \langle
  {\cal R}(\vec{v}) \cdot {\cal R}(\vec{v}) \rangle \, ,
  \label{eqn:epsilon}
\end{equation}
where $\rho$ is the zero-order density ($\rho_0$ inside the tube and $\rho_e$ outside the tube), ${\cal R}({\bf v})$ is the real part of vector {\bf v}, and angle brackets denote an average over one wave period.
For sinusoidal time dependence, this is equivalent to
\begin{equation}
  \varepsilon \, = \, \frac{1}{4} \rho
  ( \vec{v} \cdot \vec{v}^{\ast} )
\end{equation}
\citep[see, e.g.,][]{Mihalas1984,Walker2005}.
Note that the above expressions for $\varepsilon$ are fully spatially dependent, but since we have used Equation~\eqref{eq:Delta}, the only remaining spatial variation is in the $r$ direction.
Thus, if
\begin{equation}
  v_r (r,\phi,z,t) \,\, = \,\, V_0 \, f(r) \,
  \exp \left( i \omega t \, + \, i n \phi \, + \, ikz \right) \, ,
\end{equation}
where we define $f(r)$ as the dimensionless quantity to the right of the braces in Equation~\eqref{eq:vrsimple}, then
\begin{equation}
  {\bf v} \cdot {\bf v}^{\ast}
  \,\, = \,\,
  v_r v_r^{\ast} \, + \, v_{\phi} v_{\phi}^{\ast} 
  \,\, = \,\,
  2 v_r v_r^{\ast}
  \,\, = \,\,
  2 | V_0 \, f(r) |^2 \,\, .
  \label{eq:v-dot-v}
\end{equation}

Traditionally, one multiplies $\varepsilon$ by the group velocity to obtain a kinetic energy flux (with units W~m$^{-2}$) of
\begin{equation}
  F \,\, = \,\, \frac{1}{2} \rho | V_0 \, f(r) |^2 \, V_{\rm gr} \,\, .
\end{equation}
However, because each isolated flux tube contains both internal and external fluctuations (as shown in Figure~\ref{fig:moments-kink-modes}), we integrate over the horizontal plane to obtain the power (in W) associated with that one tube.
Later, we can sum this power over all detected \bp s and divide by the total area under study to produce a spatially-averaged energy flux across a patch of the solar surface.
We write the power (associated with the kinetic energy) as
\begin{equation}
  \dot{E}_{\rm K} \,\, = \,\, \int dA \,\, F
  \,\, = \,\, 2\pi \int_0^{\infty} dr \,\, r \,\, F
  \,\, .
\end{equation}
Note that the two relevant integrals over $f^2$ are identical:
\begin{equation}
  \int_0^{r_0} dr \,\, r \, \left( \frac{r}{r_0} \right)^{2n-2}
  \,\, = \,\,
  \int_{r_0}^{\infty} dr \,\, r \, \left( \frac{r_0}{r} \right)^{2n+2}
  \,\, = \,\,
  \frac{r_0^2}{2n} \, ,
\end{equation}
so the power is given by
\begin{equation}
  \dot{E}_{\rm K} \,\, = \,\, \frac{\pi r_0^2}{2n} \,
  \left( \rho_0 + \rho_e \right) \, V_0^2 \, V_{\rm gr}
  \,\, .
  \label{eqn:Edot}
\end{equation}
Usually, for linear MHD waves, the kinetic energy is exactly half of the total energy carried by all forms of variability (i.e., magnetic and thermal energy; see Section~2.5 of \citealp{Kulsrud2005}), so $\dot{E}_{\rm tot} = 2 \dot{E}_{\rm K}$.
In any case, if we can measure $V_0$ for each kink-mode ($n \geq 1$), we can compute each mode's contribution to the total power
\citep[see also][]{Goossens2013,VanDoorsselaere2014}.

\subsection{Moments: Determining Velocities from Shape Changes}
\label{sec:moments-n12-velocities}

We can define general moments of an observed flux-tube cross-section (which is assumed to correspond to an observed \bhp\ outline) as
\begin{equation}
	M_{i,j} \,\, = \,\, \int dA \,\, W(x, y) \, x^i \, y^j \, ,
	\label{eqn:moments}
\end{equation}
where $x$ and $y$ are the coordinates of a pixel, $W(x, y)$ is a weight assigned to each pixel, and the integral is evaluated over all pixels in an identified \bp\ at one instant in time.
If $W=1$ uniformly, $M_{0,0}$ is the area of the \bp, and $M_{1,0} / M_{0,0}$ and $M_{0,1} / M_{0,0}$ are the coordinates of the centroid of the \bp.
$M_{2,0}$, $M_{1,1}$ and $M_{0,2}$ are connected to the ellipticity or oblateness of the shape of the \bp, and they can be used (with the lower moments) to produce the parameters of an ellipse fitted to the \bhp\ shape \citep{Teague1980}.

Since these moments are readily measurable from \bhp\ observations, we would like to connect changes in these moments to the driving of waves (considering only the $n\ge1$ modes in this section).
For these modes, the radial velocity at the tube edge ($r=r_0$) and at one fixed height is
\begin{equation}
	\label{eqn:velocity-perturbation-wave-form}
	v_r(r=r_0, \phi, t) \, = \, V_0 \exp(i\omega t + in\phi + \psi) \, ,
\end{equation}
where $\psi$ is a general phase offset.
As shown earlier, $v_z$ is negligible, and $v_\phi = i v_r$.
($v_\phi$ is not considered in this section. Instead we find the amplitude of the radial velocity perturbation, $V_0$, and the equivalent amplitude of $v_\phi$ is accounted for in Equation~\eqref{eq:v-dot-v}.)
Integrating the real part of $v_r$ gives the radial position of the tube edge as
\begin{align}
	\label{eqn:radial-displacement-wave-form}
	r(\phi, t) \, &= \, r_0 + \frac{V_0}{\omega} \cos(\omega t + n\phi + \psi) \\
	&= \, r_0 \left[ 1 + \alpha \cos(\omega t + n\phi + \psi) \right] \, ,
\end{align}
where $\alpha \equiv V_0/\omega r_0$, which is assumed to be small, and the origin is taken to be the center of the unperturbed tube (for which the perturbed centroid is substituted in practice).
This describes a locus of radial positions outlining a cross-section of the cylindrical tube with a radial perturbation due to a wave.

We can evaluate a moment $M_{i,j}$ of this expression (assuming, for now, uniform values for the weights~$W$):
\begin{align}
	\label{eqn:calc_a_moment}
	M_{i,j} \, &= \, \int dA \; W x^i \, y^j
	\, = \, W \int_0^{2\pi} d\phi \int_0^r dr' \, r' \, (r'\cos\phi)^i \, (r'\sin\phi)^j \\
	&= \, W \int_0^{2\pi} d\phi \, \cos^i\phi \, \sin^j\phi \int_0^r dr' \, (r')^{i+j+1} \\
	&= \, W \int_0^{2\pi} d\phi \, \cos^i\phi \, \sin^j\phi \, \frac{r^{i+j+2}}{i+j+2} \\
	&= \, W \frac{r_0^{i+j+2}}{i+j+2} \int_0^{2\pi} d\phi \, \cos^i\phi \, \sin^j\phi \; [1 + \alpha \cos(n\phi + \omega t + \psi)]^{i+j+2} \, .
\end{align}
Assuming small distortions ($\alpha \ll 1$), we can expand the term in brackets with the binomial expansion as
\begin{equation}
	M_{i,j} \, \approx \, W \frac{r_0^{i+j+2}}{i+j+2} \int_0^{2\pi} d\phi \, \cos^i\phi \, \sin^j\phi \; [1 + (i+j+2) \,\alpha \cos(n\phi + \omega t + \psi)] \, ,
	\label{eqn:moments-by-wave}
\end{equation}
which is solvable analytically.
In the small-distortion limit, the zeroth-moment is the (weighted) area of the unperturbed \bp, $M_{00} = W \pi r_0^2$.

For the $n=1$ mode, the first moments are the quantities affected by the wave.
Evaluating Equation~\eqref{eqn:moments-by-wave} produces
\begin{equation}
	\frac{M_{1,0}}{M_{0,0}} \, = \, \frac{V_0}{\omega} \, \cos(\omega t + \psi) \, ,
	\qquad
	\frac{M_{0,1}}{M_{0,0}} \, = \, -\frac{V_0}{\omega} \, \sin(\omega t + \psi) \, .
	\label{eqn:n1_moments}
\end{equation}
In Equation~\eqref{eqn:calc_a_moment}, we assumed that the coordinates $x$ and $y$ are measured relative to the unperturbed center of the flux tube.
For the $n\ne1$ modes and in the $\alpha \ll 1$ limit, the perturbed centroid of the \bp\ can be used as a substitute for the unperturbed center, and so when calculating values of $M_{i,j}$ directly from identified \bhp\ boundaries, the centroid's coordinates can first be subtracted off each pixel's coordinates.
However, this cannot be done when calculating moments for the $n=1$ mode, as this mode is the \textit{cause} of the offset between the unperturbed and perturbed centers of the flux tube (and this mode is unique in being the only mode to perturb the centroid location).
Using coordinates relative to the perturbed centroid location would therefor subtract out the very oscillation we seek to measure, and so this moment must use coordinates relative to some fixed location.
Therefore, the expressions of Equation~\eqref{eqn:n1_moments} should include additional (constant) terms $x_0$ and $y_0$ on their right-hand sides, representing the offset of the unperturbed center relative to the origin.
However, these constants will be immediately lost, since we next take the time derivative of these expressions:
\begin{equation}
	v_{1,0} \, \equiv \, \frac{d}{dt} \left( \frac{M_{1,0}}{M_{0,0}} \right) \, = \, - V_0 \, \sin(\omega t + \psi) \, ,
	\qquad
	v_{0,1} \, \equiv \, \frac{d}{dt} \left( \frac{M_{0,1}}{M_{0,0}} \right) \, = \, -V_0 \, \cos(\omega t + \psi) \, .
\end{equation}
We then add these terms in quadrature,
\begin{equation}
	\sqrt{v_{1,0}^2(t) + v_{0,1}^2(t)} \, = \, V_0(t) \, ,
	\label{eqn:V0eff_n1}
\end{equation}
and obtain an expression for $V_0$ synthesizing the measurements of the two first moments.
We have indicated here that $V_0$ is a function of time, despite its use as a constant in Equation~\eqref{eqn:velocity-perturbation-wave-form}.
For an ideal instance of an $n=1$ wave, with constant amplitude $V_0$ and with a single frequency being driven, the $V_0$ value produced by Equation~\eqref{eqn:V0eff_n1} will indeed be constant with time.
However, if waves are being driven with mixed frequencies or time-varying amplitudes, Equation~\eqref{eqn:V0eff_n1} can produce a time series of distinct $V_0$ values from a time series of $M_{1,0}$ and $M_{0,1}$ measurements.
In this case, an RMS of that $V_0$ time series can provide a mean effective amplitude, from which a mean energy flux can be produced.

It is also worth noting that we assumed a functional form of the radius (and velocity) perturbation that corresponds to a circularly-polarized $n=1$ wave.
A linearly-polarized wave (akin to the flux tube being shaken back and forth, rather than being shaken around in a circle), even if ideally expressed, will see time-variability in the observed amplitude of the velocity perturbation, as the fluctuations will pass through a null point between each extreme.
Such a wave can be written as the sum of two circularly-polarized waves of equal amplitudes (each half the total amplitude of the wave) and equal but oppositely-signed frequencies.
In this case, with a linearly-polarized amplitude of $\mathcal{V}$, each of the two constituent waves will have amplitude $\mathcal{V}/2$ and will contribute a time-averaged power of $\mathcal{V}^2/4$, for a total power of $\mathcal{V}^2/2$.
The RMS of Equation~\eqref{eqn:V0eff_n1}, meanwhile, will appear as, for example,
\begin{alignat}{2}
	\rms(V_0) \, &= \, \sqrt{0 + \frac{1}{2} \mathcal{V}^2} \, = \, \frac{\mathcal{V}}{\sqrt{2}}
	&& \qquad \mbox{for vertical polarization,} \\
	\rms(V_0) \, &= \, \sqrt{\frac{1}{2} \frac{\mathcal{V}^2}{2} + \frac{1}{2} \frac{\mathcal{V}^2}{2}} \, = \, \frac{\mathcal{V}}{\sqrt{2}}
	&& \qquad \mbox{for $45^\circ$ polarization,}
\end{alignat}
where we have used the fact that the variance of a sinusoid is half of its amplitude.
This shows that the effective $V_0$, from taking the RMS of the time series of $V_0(t)$ values, produces the correct time-averaged power of $\mathcal{V}^2/2$ in this mixed-frequency case.

Equation~\eqref{eqn:V0eff_n1} allows for the quantification, whether as a power spectrum or as a time-averaged energy flux, of the of $n=1$ wave power driven by an observed time series of \bhp\ shapes.
This is not novel, as $n=1$ modes have been analyzed through changes to a \bp 's centroid location by many authors in many studies using expressions of the same form (including Chapter~\ref{chap:bp-centroids} of this thesis).
However, it is valuable to fit this wave mode into our treatments of moments, which we next extend to $n=2$ modes.

For the $n=2$ mode, the second moments are the moments which are varied by the wave, and their values, from Equation~\eqref{eqn:moments-by-wave}, are
\begin{gather}
	\label{eqn:second-moments-start}
	\frac{M_{2,0}}{M_{0,0}}
	\, = \, \frac{r_0^2}{4} + \frac{r_0V_0 \, \cos(\omega t + \psi)}{2 \omega} \, ,
	\qquad
	\frac{M_{0,2}}{M_{0,0}}
	\, = \, \frac{r_0^2}{4} - \frac{r_0V_0 \, \cos(\omega t + \psi)}{2 \omega} \, ,
	\\
	\frac{M_{1,1}}{M_{0,0}}
	\, = \, - \frac{r_0V_0 \, \sin(\omega t + \psi)}{2 \omega} \, .
	\label{eqn:second-moments-end}
\end{gather}
We again take the time derivatives of these quantities.
\begin{gather}
	v_{2,0} \, \equiv \, \frac{d}{dt} \left( \frac{M_{2,0}}{M_{0,0}} \right)
	\, = \, - \frac{r_0V_0 \, \sin(\omega t + \psi)}{2} \, ,
	\qquad
	v_{0,2} \, \equiv \, \frac{d}{dt} \left( \frac{M_{0,2}}{M_{0,0}} \right)
	\, = \, \frac{r_0V_0 \, \sin(\omega t + \psi)}{2} \, ,
	\\
	v_{1,1} \, \equiv \, \frac{d}{dt} \left( \frac{M_{1,1}}{M_{0,0}} \right)
	\, = \, - \frac{r_0V_0 \, \cos(\omega t + \psi)}{2} \, .
\end{gather}
We combine these terms as follows:
\begin{align}
	\sqrt{v_{2,0}^2 + 2v_{1,1}^2 + v_{0,2}^2} \label{eqn:second-moments-combine}
	\, &= \, \frac{r_0V_0}{2}
	\sqrt{ \sin^2(\omega t + \psi) + 2\cos^2(\omega t + \psi) + \sin^2(\omega t + \psi)} \\
	&= \, \frac{r_0V_0}{\sqrt{2}} \\
	\therefore \, V_0(t) \, &= \, \frac{\sqrt{2}}{r_0} \sqrt{v_{2,0}^2 + 2v_{1,1}^2 + v_{0,2}^2} 
	\label{eqn:V0eff_n2}
\end{align}

As before, this equation provides a way to use moments calculated across a time series of \bhp\ shapes to produce an effective value of $V_0$ for the $n=2$ wave mode.
The expression for $V_0(t)$ now includes $r_0$, which is the radius of the unperturbed flux tube.
It can be written as $r_0 = \sqrt{M_{0,0} / \pi}$, and we use the average value of $r_0$ across all frames to determine a \bp 's effective value of $r_0$.
(Changes in the value of $r_0$ are accounted for by $n=0$ waves, which we treat in the following sections.)

Equations \eqref{eqn:V0eff_n1} and \eqref{eqn:V0eff_n2} allow time-averaged values of $V_0$ (that is, the RMS of $V_0(t)$) to be calculated directly from a time series of observed \bhp\ shapes, and these values of $V_0$ can be inserted into Equation~\eqref{eqn:Edot} to calculate an estimated rate of energy transfer, integrated spatially and averaged temporally, associated with each of the $n=1,2$ wave modes.

This technique can also be extended to higher-order wave modes, though we do not do so currently.
In order to do this extension, the $n=2$ mode must be ``subtracted out'' of the \bhp\ shape in some way, similar to how the $n=1$ centroid motion is subtracted out by using coordinates centered on the centroid when conducting the $n=2$ analysis.
This requirement is due to the non-orthogonality of the moment calculations,\footnote{To be clear, the wave modes themselves are an orthogonal basis---it is only the moment calculations that lack this property.} which means an outline containing only an $n=2$ perturbation will have non-zero fourth moments, for instance.
It is not immediately apparent how this subtraction should be done in our moment-based approach.
An alternative approach may be to take the polar coordinates $(r, \theta)$ of the boundary pixels of a \bp, and then take the Fourier transform of $r(\theta)$---in effect, ``unrolling'' the feature boundary into a one-dimensional function, and then directly extracting the perturbation amplitudes in an orthogonal way.
However, this faces at least two difficulties.
First, the $r(\theta)$ points will not be sampled at a uniform spacing.
This may be treatable via the nonuniform discrete Fourier transform; however, one must be sure to understand both this technique and its outputs before applying it.
Second, if a \bp\ outline has concavity, its outline in $r(\theta)$ space might ``double back'' on itself, resulting in multiple $r$ values for a given $\theta$.
This may be treatable by using something similar to the convex hull to produce a convex approximation of the \bhp\ shape.
However, only concavity that produces a multiply-valued $r(\theta)$ outline requires this treatment, while other concave regions are perfectly acceptable.
Therefore, this possible alternative approach must be thought through carefully before it can be employed.

\subsection{Sausage Modes: Velocity Amplitudes and Energy Fluxes}
\label{sec:sausage-modes}

We now consider the separate case of $n=0$ modes.
The velocity amplitudes are given by Equation~\eqref{eq:vdef} with appropriate choices of the Bessel functions.
Note that $v_{\phi} = 0$ for these circularly symmetric oscillations.
In the thin-tube limit, the vertical velocity amplitudes are
\begin{equation}
  v_z (r < r_0) \,\, = \,\,
  -A_0 \, \frac{c_0^2}{\omega^2} \, ik \, J_0 (\mu_0 r)
  \,\, \approx \,\,
  -i A_0 \, \frac{c_0^2}{c_{{\rm T}0}^2} \, \frac{1}{k}
  \label{eq:sausageVZIN}
\end{equation}
\begin{equation}
  v_z (r > r_0) \,\, = \,\,
  -A_e \, \frac{c_e^2}{\omega^2} \, ik \, K_0 (m_e r)
  \,\, \approx \,\,
  i A_e \, \frac{c_e^2}{c_{{\rm T}0}^2} \, \frac{1}{k}
  \, \ln (m_e r) \, ,
  \label{eq:sausageVZOUT}
\end{equation}
and the radial velocity amplitudes are
\begin{equation}
  v_r (r < r_0) \,\, = \,\,
  A_0 \, \left( \frac{\omega^2 - k^2 c_0^2}{-\mu_0^2 \, \omega^2} \right) \, 
  \frac{d}{dr} J_0 (\mu_0 r)
  \,\, \approx \,\,
  A_0 \, \left( \frac{c_{{\rm T}0}^2 - c_0^2}{c_{{\rm T}0}^2} \right)
  \, \frac{r}{2}
  \label{eq:sausageVRIN}
\end{equation}
\begin{equation}
  v_r (r > r_0) \,\, = \,\,
  A_e \, \left( \frac{\omega^2 - k^2 c_e^2}{m_e^2 \, \omega^2} \right) \, 
  \frac{d}{dr} K_0 (m_e r)
  \,\, \approx \,\,
  -A_e \, \left( \frac{c_{{\rm T}0}^2 - c_e^2}{c_{{\rm T}0}^2} \right)
  \, \frac{1}{m_e^2 \, r} \, .
\end{equation}
Note also that because
\begin{equation}
  \frac{d}{dr} J_0 (\mu_0 r) \,\, \approx \,\,
  -\frac{\mu_0^2 \, r}{2}
\end{equation}
the explicit dependence of $v_r$ on $\mu_0$ (in the $r < r_0$ case) cancels out in the thin-tube limit, so its value need not be computed.

\begin{figure}[!t]
	\centering
	\includegraphics{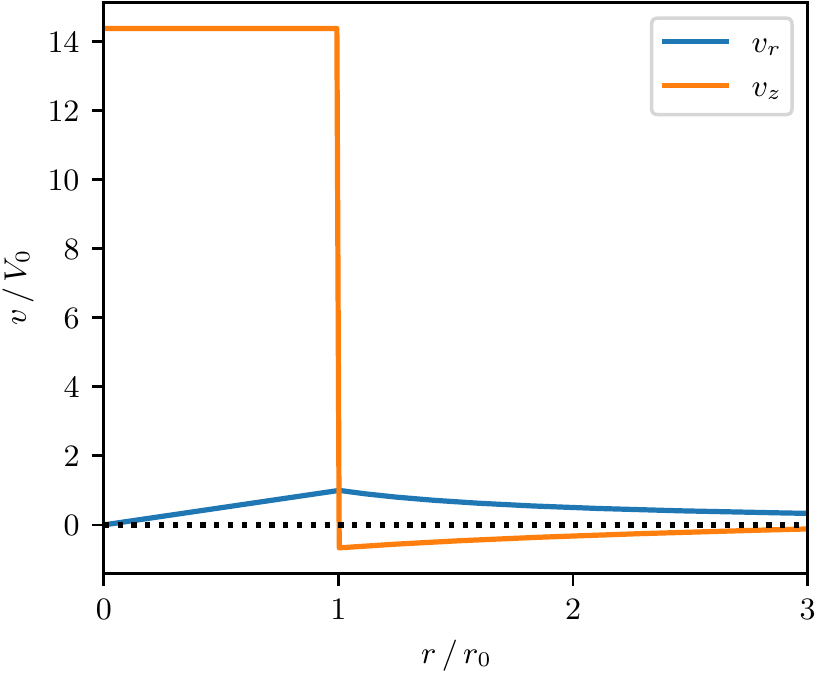}
	\caption[Radial dependence of sausage-mode ($n=0$) velocity amplitudes for standard photospheric values]{Radial dependence of sausage-mode ($n=0$) velocity amplitudes for the standard photospheric values of $c_0 = c_e$, $c_{{\rm T}0}$, and $m_e$. For this plot, we also assume $k r_0 = 0.5$. A smaller $k r_0$ would further enhance the relative dominance of the interior $v_z$ amplitude over the other components.
	\label{fig:moments-sausage-components}}
\end{figure}

As with the $n \ge 1$ case, we can write the radial velocity in terms of its value at the tube boundary:
\begin{equation}
  v_r(r) \,\, = \,\, V_0 \left\{
  \begin{array}{ll}
    r/r_0 \, , & \mbox{if $r < r_0$} \\
    r_0/r \, , & \mbox{if $r > r_0$}
  \end{array}  \right.
\end{equation}
By requiring $v_r$ be continuous at $r=r_0$, we can find expressions for $A_0$ and $A_e$ (assuming $c_0 = c_e$ as before):
\begin{equation}
  A_0 \, = \, \frac{2 V_0}{r_0} \left(
  \frac{c_{{\rm T}0}^2}{c_{{\rm T}0}^2 - c_0^2} \right)
  \, , \qquad
  A_e \, = \, - m_e^2 \, r_0 V_0  \left(
  \frac{c_{{\rm T}0}^2}{c_{{\rm T}0}^2 - c_0^2} \right) \, .
\end{equation}
Figure~\ref{fig:moments-sausage-components} shows an example set of radial profiles for both $v_z$ and $v_r$ after substituting these expressions for $A_0$ and $A_e$ into Equations (\ref{eq:sausageVZIN})--(\ref{eq:sausageVZOUT}) and normalizing by $V_0$.
Note that in the thin-tube limit, the dominant velocity is the vertical $v_z$ inside the flux tube, and this quantity is also spatially constant inside the tube.

To compute the upward kinetic energy flux, we assume that $\vec{v} \cdot \vec{v}^{\ast} \approx |v_z^2|$ in the interior of the tube, and $\vec{v} \cdot \vec{v}^{\ast} \approx 0$ outside, so that
\begin{equation}
  F \,\, \approx \,\, \left\{
  \begin{array}{ll}
    \frac{1}{4} \rho_0 \, | v_z^2 |  \, V_{\rm gr} \, ,
      & \mbox{if $r < r_0$} \\
    0 \, , & \mbox{if $r > r_0$}
  \end{array}  \right.
\end{equation}
and
\begin{equation}
  \dot{E}_{\rm K} 
  \,\, = \,\, 2\pi \int_0^{\infty} dr \, r \, F
  \,\, = \,\,
  \frac{\pi r_0^2}{4} \, \rho_0 \, | v_z^2 |  \, V_{\rm gr} \, ,
  \label{eq:EKn0}
\end{equation}
which agrees with Equation~(24) of \citet{Moreels2015}.
(Recall that $\dot{E}_{\rm K}$ must be doubled to account for all forms of energy.)

\subsection{Sausage Modes: Determining Velocities from Shape Changes}
\label{sec:moments-n0-velocities}

Despite $v_z$ being the dominant velocity in an $n=0$ mode, only $v_r$ can be directly inferred from changes in the circular area of the flux tube.
At the tube boundary (where $|v_r| = V_0$), we can use Equations (\ref{eq:sausageVZIN}) and (\ref{eq:sausageVRIN}) above to write
\begin{equation}
  \frac{v_z}{v_r}
  \,\, = \,\,
  - i \frac{2}{k r_0} 
  \left( \frac{c_0^2}{c_{{\rm T}0}^2 - c_0^2} \right)
  \,\, = \,\,
  i \frac{2}{k r_0} 
  \left( \frac{c_0^2 + V_{{\rm A}0}^2}{c_0^2} \right) \, .
  \label{eq:defouw}
\end{equation}
The above expression reproduces Equation~(12) of \citet{Defouw1976}.

The relevant moment for the $n=0$ wave mode is $M_{0,0}$.
Evaluating Equation~\eqref{eqn:moments-by-wave}, we find
\begin{equation}
	M_{0,0} \, = \, \pi r_0^2 \, + \, 2 \pi r_0 \frac{V_0}{\omega} \, \cos(\omega t + \psi) \, .
	\label{eqn:M00}
\end{equation}
Contrary to the $n\ge1$ modes, we do not take the time derivative of this equation, as the factor of $1/\omega$ will be useful later.
The constant term $\pi r_0^2$ is the mean value of $M_{0,0}$.
It represents a zero-frequency component to what will become the plasma velocity, and this component does not contribute to wave motion, so we subtract it out.
Taking the RMS after subtracting the mean is equivalent to taking the standard deviation, which we denote as $\sd(\;\cdot\;)$, and we take the standard deviation of the zeroth moment,
\begin{equation}
	\sd(M_{0,0}) \, = \, \sqrt{2} \, \pi r_0 \frac{V_0}{\omega}
	\label{eqn:sd-M00}
\end{equation}
(using the fact that the standard deviation of a sinusoid is $1/\sqrt{2}$ times its amplitude).
With this expression, an effective value of $V_0$ can be computed from a time series of $M_{0,0}$ measurements.
$V_0$ is the magnitude of the radial velocity oscillations at the flux-tube boundary, but we require a value for the dominant vertical velocities.
With Equation~\ref{eq:defouw}, we find
\begin{equation}
	\left| v_z \right|
	\, = \, \frac{\sqrt{2} \; \sd(M_{0,0}) }{\pi r_0^2} \, \frac{\omega}{k} \left( \frac{c_0^2 + V_{A0}^2}{c_0^2} \right)
	\, = \, \frac{\sqrt{2} \; \sd(M_{0,0}) }{\pi r_0^2} \left( \frac{V_{A0}^2}{c_{\rm T0}} \right),
	\label{eqn:vmag-n0}
\end{equation}
where we have used $\omega / k \, = \, c_{\rm T0}$, a constant.
This expression serves as an estimate of the amplitude of $v_z$.
An observed sequence of $M_{0,0}$ and $v_r$ values will, in fact, be a sum of these sinusoidal oscillations at varying frequencies.
Taking advantage of this sinusoidal nature, $|v_z| / \sqrt{2}$ can be taken to be an effective RMS of $v_z$, and it can be used directly in Equation~\eqref{eq:EKn0} to compute a mean upward energy transfer rate over time for the sum of the individual wave components.

If we desire a power spectrum of the energy flux, however, a different approach must be taken.
Each observed $v_r$ value will be due to the contributions of those multiple sinusoids of varying frequencies, and it derives not from the amplitudes of those sinusoids but from individual values taken from any point along those sinusoidal curves.
Additionally, because of the presence of multiple frequency components, the frequency-dependent Equation~\eqref{eq:defouw} cannot be used to convert these instantaneous $v_r$ values into $v_z$ values.
(It is this requirement to convert the measurable $v_r$ values into the dominant $v_z$ values that sets the $n=0$ mode apart from the $n\ge1$ modes.)
Instead, we re-write Equation~\eqref{eqn:M00} as
\begin{equation}
	M_{0,0} \, = \, \pi r_0^2 \, + \, 2 \pi r_0 r_1(t) \, ,
\end{equation}
which expresses the area of the flux-tube cross section as the area of the unperturbed cross section and an area perturbation due to a small perturbation $r_1(t)$ in radius.
Taking the time derivative yields
\begin{equation}
	\frac{d}{dt} \, M_{0,0} \, = \, 2 \pi r_0 \frac{d}{dt} \, r_1(t) \, = \, 2 \pi r_0 v_r(t)  \, ,
	\label{eqn:vr_time_series}
\end{equation}
where we have used $v_r(r=r_0) = dr_1(t)/dt$, which follows from the radial position of the tube edge, $r(t) = r_0 + r_1(t)$.
Equation~\eqref{eqn:vr_time_series} directly connects the derivative of a time series of $M_{0,0}$ measurements to a time series of $v_r$ measurements (assuming a constant value of $r_0$, which we take to be the average of $r(t) = \sqrt{M_{0,0}/\pi}$).
We then take the Fourier transform of the $v_r$ time series, which then allows us to re-scale each frequency bin from $v_r$ to $v_z$ using Equation~\eqref{eq:defouw}, with $k=\omega/c_{\rm T0}$.
We then take an inverse-Fourier transform to recover a time series of estimated $v_z$ values.
Repeating for each \bp\ then allows us to produce an average $v_z$ (or $\dot{E}$) power spectrum for the $n=0$ mode.
(It may seem more straight-forward to produce power spectra directly from our re-scaled $v_r$ spectrum, rather than returning to the time domain with $v_z$ values.
However, having time series of $v_z$ values allows us to follow our procedure of producing average power spectra via the Wiener-Khinchin theorem, which much more easily allows the production of average spectra from time series of unequal duration---and therefore uneven frequency spacing of their spectra.)

It is notable that Equation~\eqref{eqn:vmag-n0}, the RMS amplitude of $v_z$ for a single-frequency wave, depends only on the range of \bhp\ sizes observed, and not on the speed at which the \bhp\ size changes (that is, it depends on the amplitude of the radial size perturbation, not on the radial velocity perturbation).
This follows directly from the presence of the vertical wave number $k$ in Equations \eqref{eq:sausageVZIN} and \eqref{eq:defouw}, which produces a constant when combined with the $\omega$ of Equation~\eqref{eqn:sd-M00}.
(The $v_r$ amplitudes that were dominant for the $n\ge1$ modes had no such $k$ term.)
This independence from wave frequency can also be justified through a heavily-simplified conceptual picture of a sausage-mode wave.
As the wave propagates, any one horizontal cross-section of the tube will alternate between expanding and contracting.
Since we assume $\rho_0$ is constant throughout the tube, any volume where the tube is expanding must have a net in-flowing plasma to maintain its density, and any contracting volume must have net out-flowing plasma.
These flows will be vertical, carrying plasma from contracting regions to expanding regions.
Any given volume will expand or contract for half of a wave period before beginning the reverse process, and so the vertical plasma flows have $\pi/\omega$~seconds to carry in or out an appropriate amount of plasma.
Doubling $\omega$ halves this amount of time.
However, doubling $\omega$ also doubles $k$ and halves $\lambda$ (since $w/k \, = \, c_{\rm T0}$ is constant), meaning that each expanding or contracting region of the tube has half the volume, and so half the amount of plasma must flow in or out during the reduced amount of time.
Thus, for a given oscillation amplitude, the vertical plasma velocities necessary to maintain a constant $\rho_0$ do not depend on the oscillation frequency.

\subsection{Pixel Weighting}
\label{sec:moments-pixel-weighting}

Our definition of the moments in Equation~\eqref{eqn:moments} includes a weight $W(x,y)$ assigned to each pixel.
We use these weights to implement a temporal smoothing of our \bhp\ shapes.
We choose a smoothing window $\Delta t$ and a normalized smoothing kernel $F(t')$, defined over the range $-\Delta t < t' < \Delta t$, and we define a shape function $S(x, y, t)$, which is defined for each pixel in our sequence of \bhp\ shapes.
With one particular \bp\ chosen for smoothing, $S=1$ if the pixel at $(x, y)$ is part of that \bp\ at time $t$, and it is zero otherwise.
We then calculate a weight $W$ for each pixel in that \bp\ as
\begin{equation}
	W(x, y, t) \, = \,
	\sum _{t' = t - \Delta t / 2} ^ {t + \Delta t / 2} S(x, y, t') \, F(t' - t) \, .
	\label{eqn:ellipse-smoothing}
\end{equation}
This produces a weight $W$ for each pixel that is the weighted mean of the binary status of that pixel as a member of the feature, calculated over the time steps within the smoothing window.
For the smoothing kernel $F$, we use a Gaussian, truncated at two standard deviations, with a total width of 5~time steps (or 10~s).
The choice of this width is discussed further in Section~\ref{sec:moments-effect-of-smoothing}.

With the smoothing kernel $F$ normalized, values of the weight $W$ are constrained to the range $[0, 1]$ and can be thought of as describing partial membership in the identified shape, allowing pixels to gradually join or leave the feature over time.
This avoids problems of over-discretization of the \bp 's shape changes by a very rapid cadence.
Consider, for example, the case of uniform rotation of an ellipsoidal \bp.
While the underlying motion is smooth, pixels must enter or leave the identified \bp\ at discrete points in time, and this causes a degree of ``jumpiness'' in the values of the moments calculated across this artificial \bp.
By applying this smoothing, pixels gradually enter and leave the feature, ensuring the orientation angle of the calculated $n=2$ perturbation varies smoothly with time and suppressing high-frequency noise in the time series of calculated moments.
(An alternative approach to avoiding this over-discretization is discussed in Section~\ref{sec:emd-cadence-reduction}.)

\section{Results}
\label{sec:ellipse-results}

We present here results from applying this moment-based technique to \bp s identified in simulated observations from the \muram\ simulation described in Section~\ref{sec:2s-muram}.

\begin{figure}[p]
	\centering
	\includegraphics[trim=0 .34cm 0 0,clip]{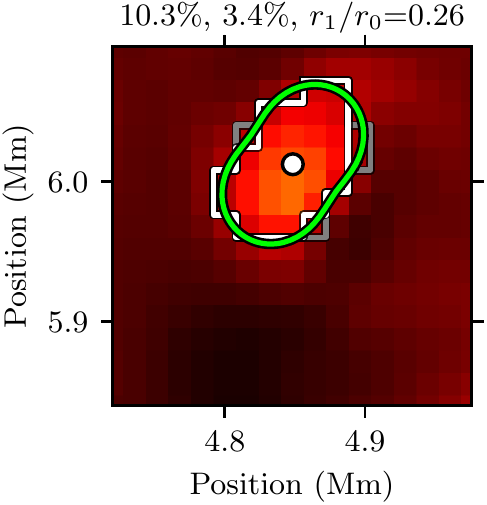}
	\includegraphics[trim=.34cm .34cm 0 0,clip]{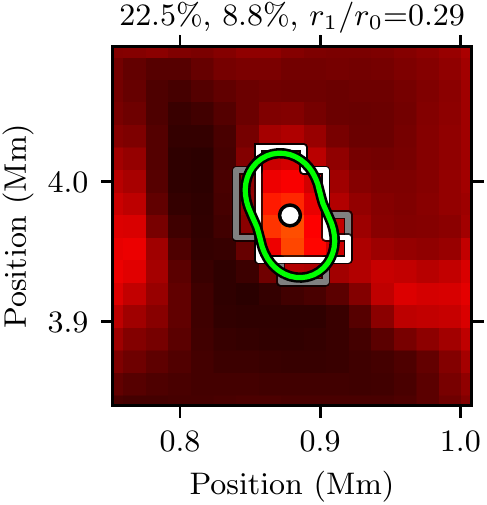}
	\includegraphics[trim=.34cm .34cm 0 0,clip]{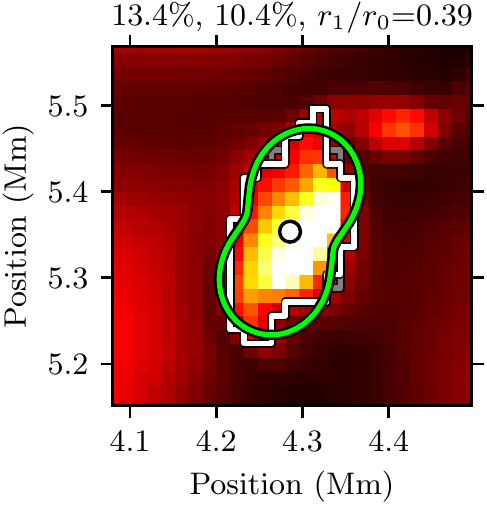}
	\includegraphics[trim=0 .34cm 0 0,clip]{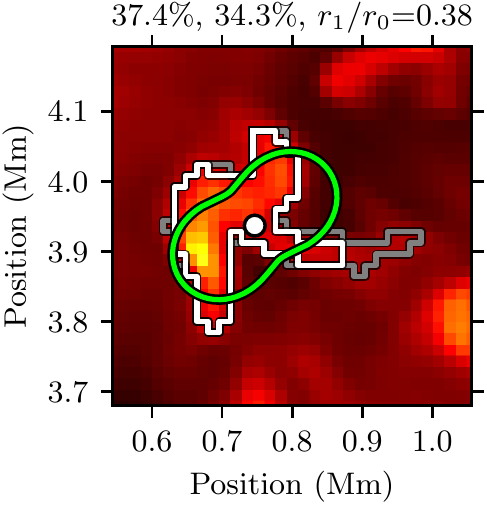}
	\includegraphics[trim=.34cm .34cm 0 0,clip]{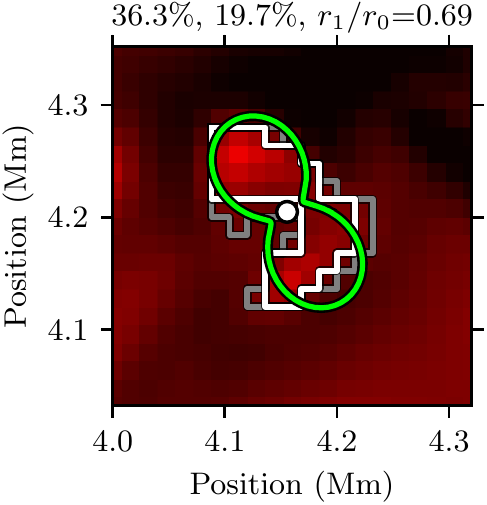}
	\includegraphics[trim=.34cm .34cm 0 0,clip]{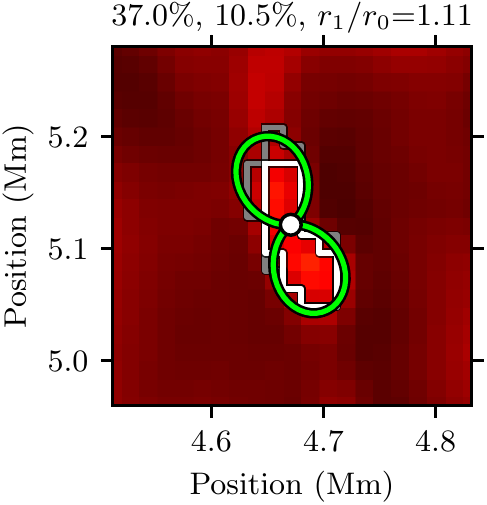}
	\includegraphics[trim=0 0 0 0,clip]{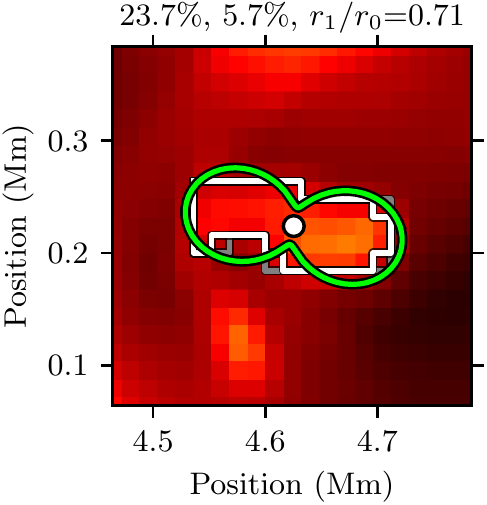}
	\includegraphics[trim=.34cm 0 0 0,clip]{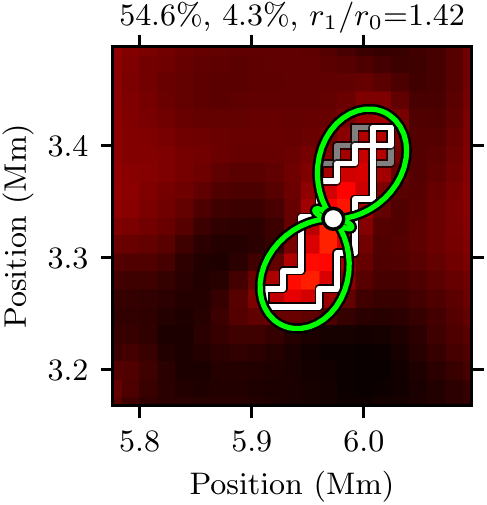}
	\includegraphics[trim=.34cm 0 0 0,clip]{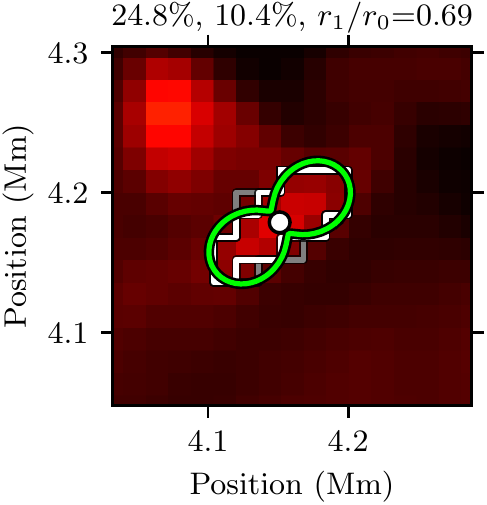}
	\caption[A selection of \bp s with fitted $n=2$ wave outlines]{A selection of \bp s with fitted $n=2$ wave outlines computed from the second moments. Marked are the wave outline (green) and the centroid of the \bp\ (white dot), as well as the pixel-by-pixel outline of the identified shape of the \bp. Since temporal smoothing of \bhp\ outlines is applied as described in the text, the pixelated outline drawn in white surrounds all pixels with weights $\ge 0.5$, while a gray outline surrounds all pixels with non-zero weight. Above each plot, two measures of fit quality are indicated: the wave outline's over-coverage and under-coverage, which are defined in the text. (Since not every pixel within the white boundaries has a weight of one, these fractions are sometimes different from what might be expected from these visualizations.) Also indicated is the value of $r_1/r_0$. Animations of these \bp s are available on the author's website at \href{http://samvankooten.net/thesis}{samvankooten.net/thesis} and are archived at \href{http://doi.org/10.5281/zenodo.5238943}{doi.org/10.5281/zenodo.5238943}.}
	\label{fig:moments-example-fits}
\end{figure}

\subsection{Fitted Wave Outlines}

In Figure~\ref{fig:moments-example-fits} we show a sample of $n=2$ wave perturbations fitted to \bp s by way of the second moments.
(The $n=2$ amplitudes in spatial units---rather than velocity units---are produced by solving Equations \eqref{eqn:second-moments-start} through \eqref{eqn:second-moments-end} for $r_1 \equiv V_0/\omega$ and combining these three expressions as in Equation~\eqref{eqn:second-moments-combine}.
The amplitudes are then plotted as perturbations to a circular background state using the form of Equation~\eqref{eqn:radial-displacement-wave-form}.
The phase angle is determined by taking the orientation angle of an ellipse fitted to the \bp.)
When calculating energy fluxes for the $n=2$ mode, it is the evolution of the amplitudes of these $n=2$ perturbations---relative to the assumed-circular background state---that governs the waves being modeled (while evolution of the size and location of the outline correspond to $n=0,1$ waves).
It can be seen than many of these \bp s have shapes that are well-approximated by the long, thin shape of an $n=2$ perturbation.
This might be expected, since \bp s are confined within downflow lanes where converging horizontal velocities can restrict most \bp s to having thin, elongated shapes.
However, some \bp s (typically larger ones, or \bp s which could also be considered as two nearby \bp s with a bridge of relatively weaker intensity enhancement connecting them) have more complex shapes which are not well-treated by this approach.

\begin{figure}[t]
	\centering
	\includegraphics{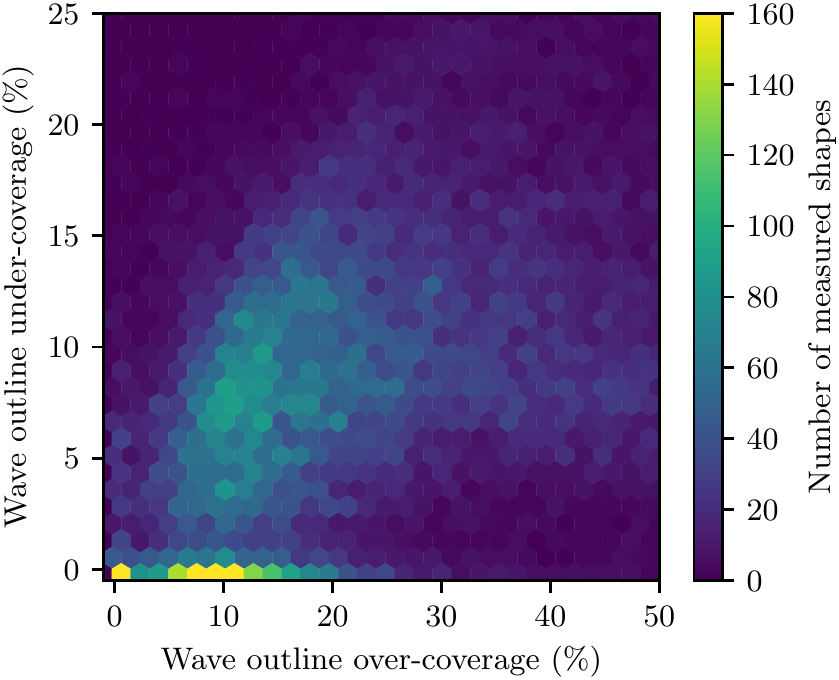}
	\caption[Quality metrics for $n=2$ perturbation fitting]{Quality metrics for $n=2$ perturbation fitting. For each measured shape of each \bp, we compute the two metrics described in the text.}
	\label{fig:ellipse-fit-quality}
\end{figure}

To quantify the goodness-of-fit of these perturbed outlines over the entire \bhp\ population, we plot two quality metrics in Figure~\ref{fig:ellipse-fit-quality}.
To measure the extent to which the perturbed outline extends beyond the identified boundary of the \bp, which we term ``over-coverage,'' we compute the fraction
\begin{equation}
	\frac{N \, - \, \sum_{i=1}^{N} W(i)}{N} \, ,
\end{equation}
where $N$ is the number of pixels whose center is within the perturbed outline, the sum is over all $N$ pixels, and $W(i)$ is the weight assigned to an individual pixel (0 if it is not within the identified \bhp\ boundary; otherwise a value within the range $(0, 1]$).
This effectively assigns a weight of 1 to every pixel in the outline and measures the fraction of that outline weight not counted as part of the \bp.
For the second metric, we compute the fraction of the \bp 's total weight that belongs to pixels whose centers do not fall inside the perturbed outline, which we term ``under-coverage''.
These two metrics measure how much the outline extends beyond the \bp, and how much the \bp\ extends beyond the outline.
It can be seen that typical values fall within 0--15\% for the former metric, and 0--10\% for the latter, indicating that most \bp s are well-approximated by this $n=2$ shape.

\begin{figure}[t]
	\centering
	\includegraphics{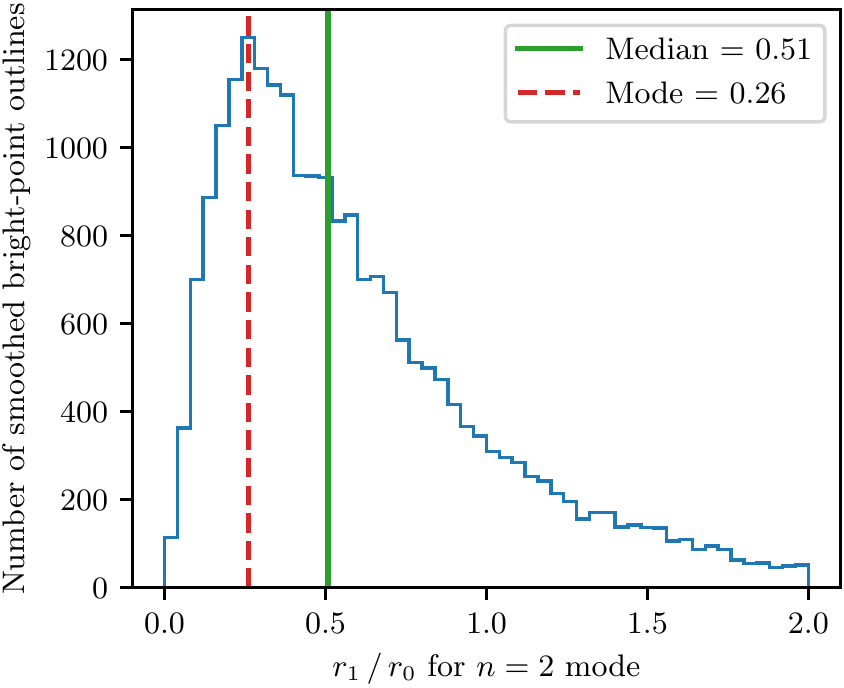}
	\caption[Amplitudes of fitted $n=2$ perturbations]{Amplitudes of fitted $n=2$ perturbations across all smoothed \bhp\ outlines.}
	\label{fig:ellipse-fit-n2-amplitudes}
\end{figure}

As an additional probe into the computed perturbation amplitudes, we plot in Figure~\ref{fig:ellipse-fit-n2-amplitudes} the distribution of the relative perturbation amplitude $r_1/r_0$ over all smoothed \bhp\ outlines.
We find that a value of $r_1/r_0=0.2$ is a threshold for which lower values correspond to perturbed outlines that appear elliptical, whereas higher values display concavity along the minor axis.
At the resolution of our histogram bins, the most common value is $r_1/r_0=0.26$---a marginally-elliptical shape.
The median value is $r_1/r_0=0.51$, indicating that half of the \bp\ outlines are at least quasi-consistent with the small-amplitude approximation used in our derivation.
Further exploration of the implications of these larger amplitudes and the effects of relaxing the small-amplitude approximation may be excellent topics for future work.

\subsection{Energy Fluxes}

\begin{table}[t]
	\centering
	\begin{tabular}{l|ccc}
		Quantity & $n=0$ & $n=1$ & $n=2$ \\
		\hline
		\rule{0pt}{.9\normalbaselineskip}
		Mean flux (W~m$^{-2}$) & $1.01 \times 10^4$ & $3.03 \times 10^4 $ & $2.28 \times 10^4$ \\
		\rule{0pt}{0cm}
		Mean $\dot{E}$ (W) & $3.13 \times 10^{16}$ & $9.93 \times 10^{16} $ & $7.09 \times 10^{16}$ \\
	\end{tabular}
	\caption[Mean vertical flux and $\dot{E}$ values for the three wave modes]{Mean vertical flux and $\dot{E}$ values for the three wave modes, $n=0$ through 2. The fluxes are averaged over the full simulation domain, and the mean $\dot{E}$ values are across all \bp s and time steps. The large difference in the magnitudes of these quantities is due to the very small filling factor of \bp s.}
	\label{table:mean-moment-values}
\end{table}

We described methods for inferring energy fluxes in Section~\ref{sec:moments-n12-velocities} for $n\ge1$ modes and Section~\ref{sec:moments-n0-velocities} for $n=0$ modes.
For each \bp, we use these techniques to produce time series of $\dot{E}$ values, indicating the rate of vertical energy transfer through the flux tube, integrated over the area of the \bp\ (and surrounding areas, for the $n\ge1$ modes where external perturbations are significant).
These values can be integrated across the lifetime of each \bp, and then summed across all \bp s to produce a total amount of energy transferred across the simulation run-time.
We then divide by the simulation duration and simulation area to produce a spatially- and temporally-averaged vertical energy flux across the simulation domain.
The mean flux values we compute, as well as mean values for $\dot{E}$, are shown in Table \ref{table:mean-moment-values} for the three wave modes.
The $n=1$ mode is clearly dominant in both measures, but only by a factor $<3$, indicating that $n\ne1$ modes may make a significant contribution to the energy flux budget---though the upward propagation of these modes through the rapidly-varying plasma properties of the chromosphere and transition region must be modeled before any implications to the coronal energy budget become clear.
The $n=1$ flux is very similar to that found in Chapter~\ref{chap:bp-centroids}.
(After accounting for different assumed values of $\rho_0$ and $B_0$ in this chapter and removing the term accounting for expected reflection of $n=1$ waves, the adjusted $n=1$ flux for that previous work that may be compared to the present flux is 31~kW~m$^{-2}$, and the adjusted flux using the observational $\left< v_\perp^2\right>$ of \citealp{Chitta2012} is 11~kW~m$^{-2}$.)
This similarity indicates that, as in that chapter, these fluxes compare plausibly to the estimated flux required for coronal heating.
In Chapter~\ref{chap:method-comp-and-conclusions}, we will compare and discuss the fluxes further, alongside the fluxes inferred by our second proposed method in Chapter~\ref{chap:emd}.

\begin{figure}[t]
	\centering
	\includegraphics[width=\linewidth]{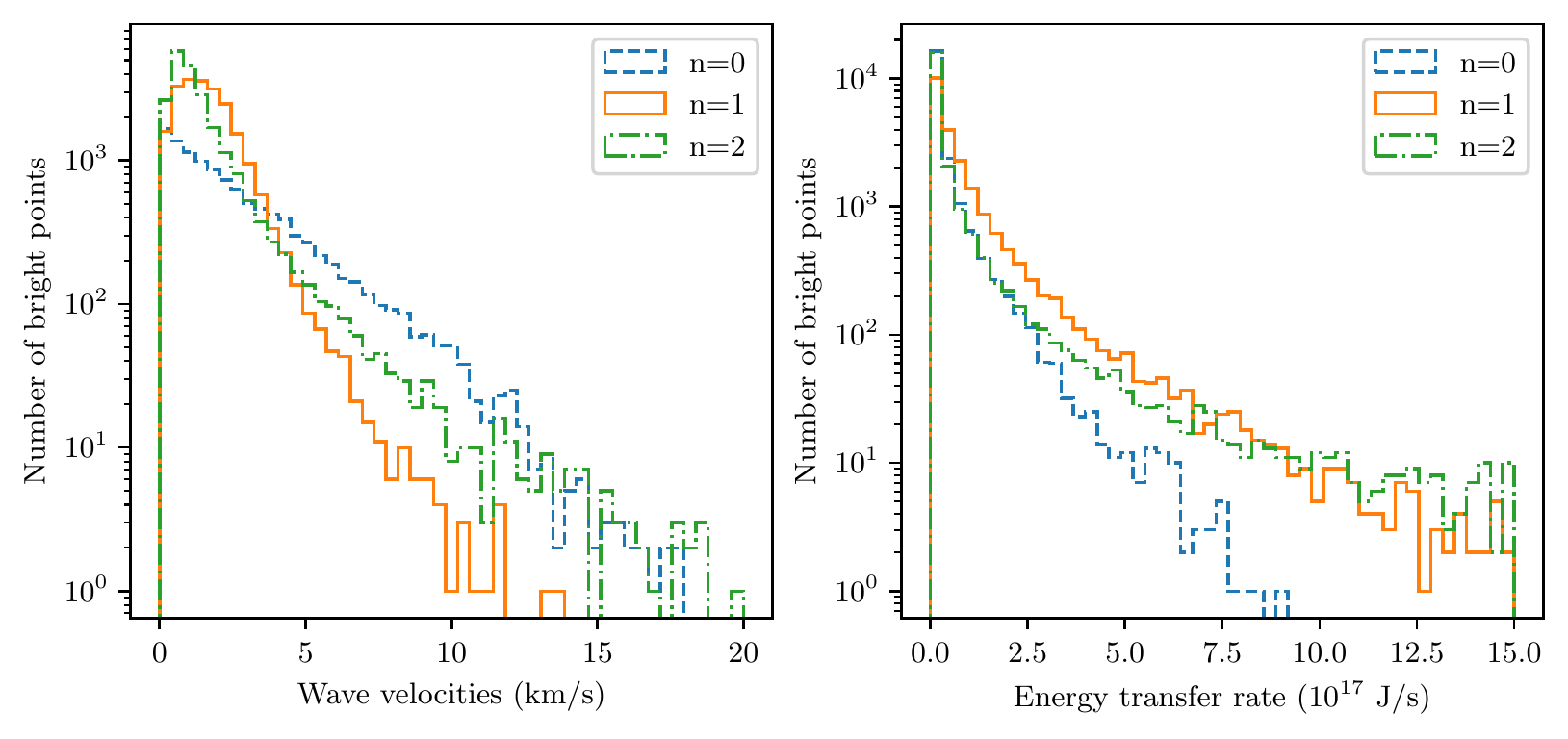}
	\caption[Histograms of wave velocities and energy transfer rates]{Histograms of wave velocities and energy transfer rates for the $n=0$, 1, and 2 modes. Each time step for each \bp\ is represented separately. On the left are the wave perturbation velocities---the instantaneous amplitude of $v_r$ at the tube boundary for $n=1$ and 2 (that is, $V_0$), and of $v_z$ throughout the tube body for $n=0$. On the right are the values of $\dot{E}$ corresponding to each velocity amplitude value. A small number of outliers have been trimmed from each plot.}
	\label{fig:moment-histograms}
\end{figure}

To further explore these fluxes, we plot histograms of velocity amplitudes of instantaneous values of $\dot{E}$ in Figure~\ref{fig:moment-histograms}.
A few observations can be made.
Despite being relatively unremarkable in terms of velocity magnitudes, the $n=1$ mode is clearly dominant in the energy transfer distributions (as with the summary numbers noted earlier).
This is due, in part, to the $1/n$ term in Equation~\ref{eqn:Edot}, which reduces the $n=2$ values of $\dot{E}$ relative to the $n=1$ values.
The $n=0$ mode, meanwhile, uses the form for $\dot{E}$ given by Equation~\ref{eq:EKn0}, and the constants in this equation evaluate to about half of the value of the constants for determining $\dot{E}$ for the $n=1$ mode.
(The conversion from velocity to $\dot{E}$ also depends on the flux tube radius, which inhibits direct comparisons between the two plots.)
Additionally, the $n=1$ velocity distribution has its peak further from 0 than the other modes, which may suggest it is the mode most susceptible to jitter (or simply that \bp s are in nearly-constant motion but that shape changes are less ubiquitous).

\begin{figure}[t]
	\centering
	\includegraphics{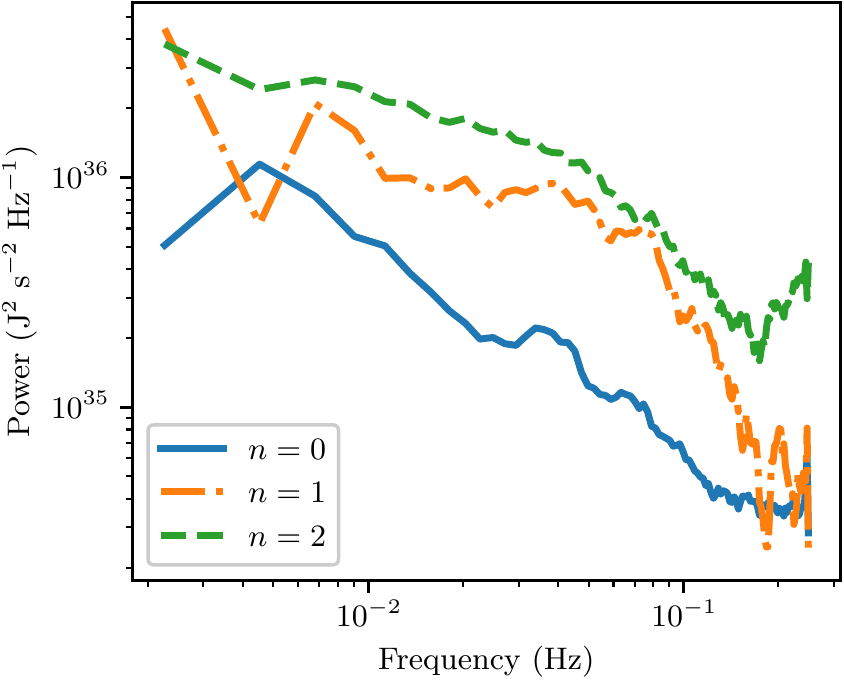}
	\caption[Power spectra of $\dot{E}$ for the $n=0,1,2$ modes]{Power spectra of the $\dot{E}$ time series for the $n=0,1,2$ modes. The spectra have been smoothed with a boxcar of 7~frequency bins prior to plotting, to aid in distinguishing individual curves.}
	\label{fig:moments-power-spectrum}
\end{figure}

In Figure~\ref{fig:moments-power-spectrum} we plot averaged power spectra of the $\dot{E}$ time series for each \bp\ (choosing to show spectra for $\dot{E}$ rather than velocities to account for the varying relation between the two quantities).
Average spectra across these time series of different lengths (and therefore with varying spacings in the frequency domain) are produced by way of the Wiener-Khinchin theorem, as described in Section~\ref{sec:results_spectra}.
The three spectra generally have largely similar shapes, aside from the $n=1$ mode steepening more sharply at the highest frequencies.
(This difference may derive from the different wave modes being affected differently by our temporal smoothing of \bhp\ boundaries.)
The relative magnitudes of these spectra cannot be directly compared with the values listed in Table \ref{table:mean-moment-values}, as the Wiener-Khinchin approach to averaging these spectra (which is also used by, e.g., \citealp{VanBallegooijen1998}) does not ensure that the integral of the averaged spectrum is the mean of the input spectra's integrals.

\subsection{Effect of Smoothing}
\label{sec:moments-effect-of-smoothing}

\begin{figure}[t]
	\centering
	\includegraphics{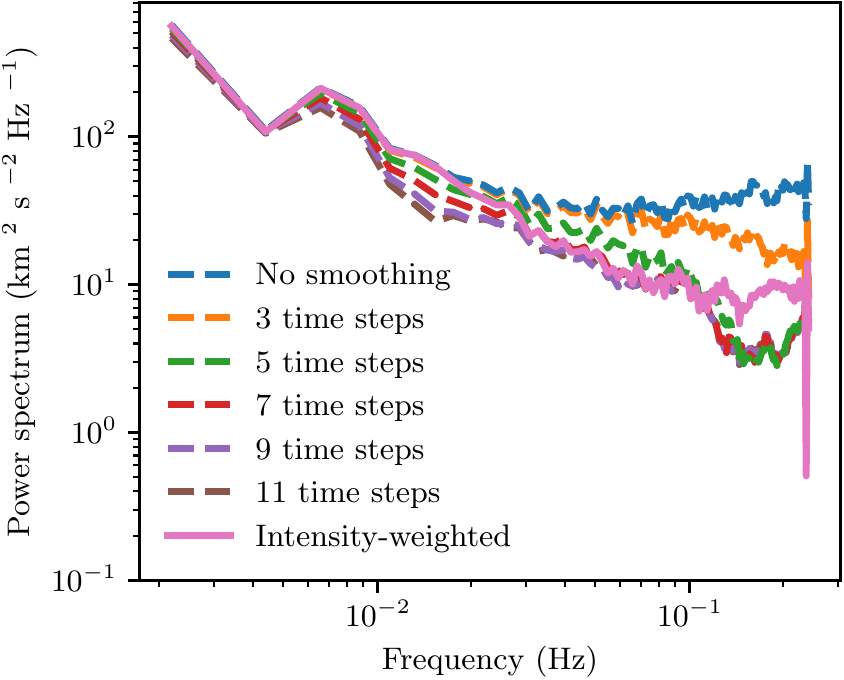}
	\caption[Power spectra of $n=1$ waves for different smoothing windows.]{Power spectra of $n=1$ waves for different temporal smoothing windows. The temporal smoothing window width is given for each spectrum, with a $2 \sigma$ Gaussian kernel fitting in the specified window size. One time step is 2~s. The intensity-weighted spectrum weights each pixel according to its intensity, without any temporal smoothing employed. Boxcar smoothing with a width of 7 frequency bins is applied to each spectrum before plotting to ensure each curve is clearly visible.}
	\label{fig:moments-smoothing-spectra}
\end{figure}

We consider here the effect of the temporal smoothing we employ in this chapter, implemented via the pixel weights described in Section~\ref{sec:moments-pixel-weighting}.
In Figure~\ref{fig:moments-smoothing-spectra} we show power spectra for the $n=1$ mode with various sizes of the temporal smoothing window.
We also plot a spectrum using the intensity-weighted centroids of our past work in Chapter~\ref{chap:bp-centroids}.
In our intensity weighting, intensity values were normalized so that the brightest pixel in the feature had a weight of 1, and the dimmest pixel in the feature a weight of 0.
Since the dimmest pixels in a \bp\ tend to be at the edge, this weighting scheme de-emphasizes the edge pixels, insulating the centroid somewhat from the possibility of jitter as pixels enter and leave the feature, similar to the effect of the temporally-smoothed weighting we employ in this chapter.
This similarity can be seen in the way that the un-smoothed and un-weighted spectrum in Figure~\ref{fig:moments-smoothing-spectra} shows a much higher degree of flattening at high frequencies, suggesting a much stronger contribution from centroid jitter.
A smoothing window of 3~time steps (6~s) can be seen to reduce this white noise somewhat, while a window of 5~time steps (10~s) tends to produce a spectrum very similar to that using intensity weighting over most frequencies, with a notable reduction in jitter (or the flatness of the spectrum) at the highest frequencies.
It is hard to say, though, whether the highest frequencies are over-smoothed, since the spectral shape changes significantly at these frequencies.
Above this 5~time-step window size, spectral power begins to decrease at the middle of the spectrum, suggesting that any larger window may over-smooth the \bp s at all frequencies.
We therefore select a temporal smoothing window of 5~time steps, which seems to be the least amount of smoothing that handles high-frequency jitter well.

\section{Summary}
\label{sec:ellipse-summary}

In this chapter, we introduced what we believe to be a novel method for analyzing the shape changes of \bp s and connecting them to the modeled excitation of flux-tube waves of different modes.
We compute moments of the \bhp\ shape, and we connect variations in these moments to wave-driven perturbations in the cross-section of a thin flux tube.
We considered $n=0$, 1, and 2 modes (the first two are also known as sausage and kink modes), and our framework can be extended to higher-$n$ modes.
Our estimated energy fluxes, when applying this technique to simulated observations from a \muram\ simulation of DKIST-like resolution, show that the $n=1$ mode is dominant, but the $n=0$ and 2 modes make significant contributions to the wave energy budget, together approximately equaling the $n=1$ flux.
This suggests that these wave modes may be significant contributors to wave heating of the coronal in quiet-sun regions, beyond the $n=1$ energy flux that has long been the focus of \bhp\ studies.
These estimated fluxes, though, are to be understood as existing at and immediately above the photosphere.
Further modeling must be done to understand if and how these wave modes propagate upward, and whether they make significant contributions to the heating budget in the corona.

We also explored the effect of temporal smoothing of the shapes of \bp s, and compared it to the intensity weighting we used in the past to compute \bhp\ centroids.
When attempting to reduce white noise due to centroid jitter when computing centroid velocities, we found Gaussian temporal smoothing with a window size 10~s to perform most similarly to intensity weighting, though with a stronger reduction in power at the highest frequencies.

\chapter{Velocity Fields via the Earth-Mover's Distance}
\label{chap:emd}

\section{Overview}
\label{sec:emd-overview}

\begin{figure}[t]
	\centering
	\includegraphics{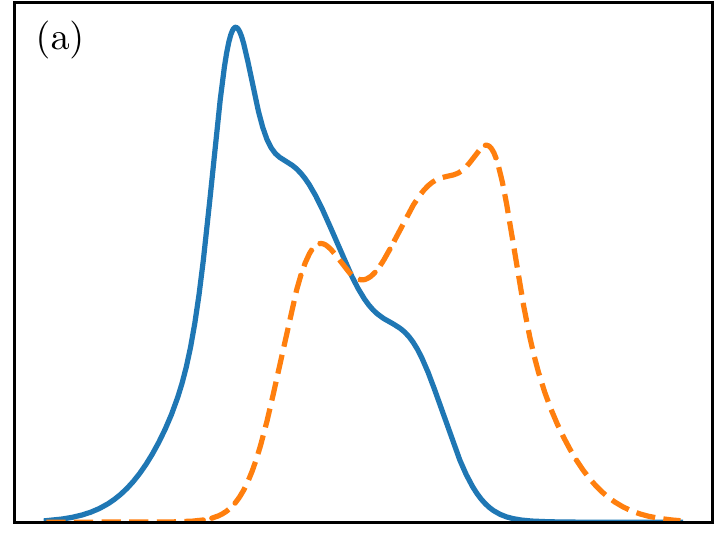}
	\includegraphics{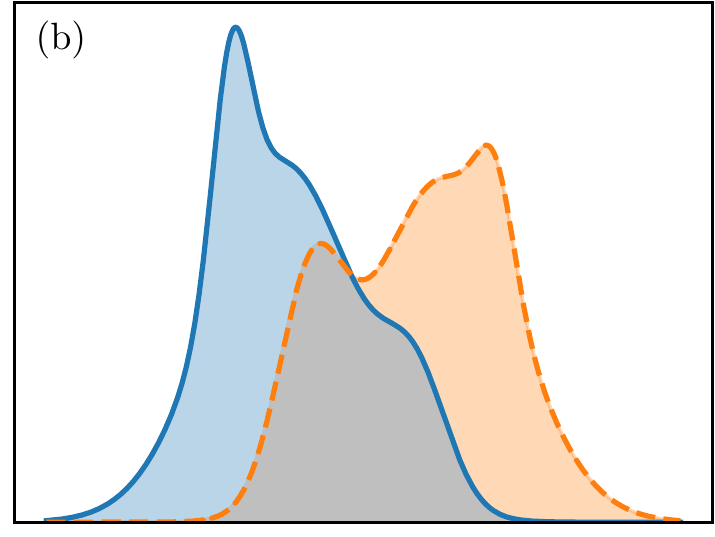}
	\caption[One-dimensional illustration of the earth mover's distance]{One-dimensional illustration of the earth mover's distance. In panel (a), two arbitrary distributions are shown, which can be thought of as elevation profiles of (two-dimensional) hills. The earth mover's distance quantifies the effort required in re-arranging earth to transform one hill into the other. In panel (b), the central portion shaded in gray is included as part of both hills and can remain stationary through the rearrangement. However, the portion of the hill shaded in blue must be moved, as if by shovel, to fill the area shaded in orange.}
	\label{fig:emd-concept}
\end{figure}

\begin{figure}[t]
	\centering
	\includegraphics{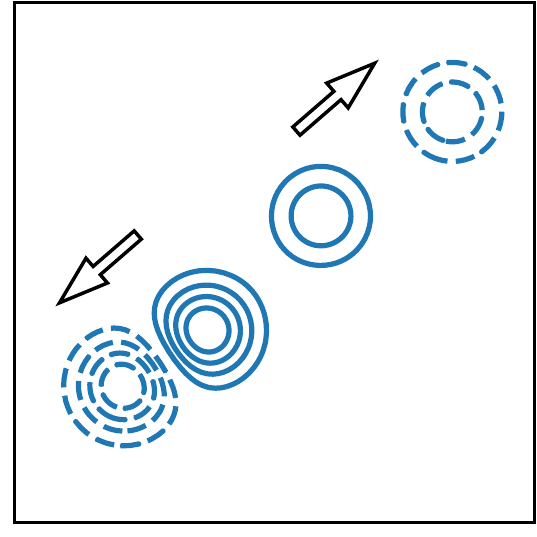}
	\caption[Two-dimensional illustration of the earth mover's distance]{Two-dimensional illustration of the earth mover's distance. Contours are shown of two bimodal distributions, in solid and dashed lines. These can be thought of as elevation maps for two hills. The hills must be moved from the locations marked by solid lines to those marked by dashed lines. The hill in the lower-left is higher and contains more earth, but the hill in the upper-right must be moved further. In this case, the mass and distance ratios are set so that each hill requires equal total earth-moving effort and so contributes equally to the total earth mover's distance between these two distributions.}
	\label{fig:emd-concept-2d}
\end{figure}

The earth mover's distance (or EMD; also known as the Wasserstein distance), is a metric for comparing the similarity between two distributions.
It was first described conceptually by \citet{Monge1781}.
Early applications in the field of image analysis include \citet{Werman1985} and \citet{Peleg1989}, and the name ``earth mover's distance'' was introduced to the literature by \citet{Rubner2000}, who attribute the name to J. Stolfi.
A detailed overview from this image-processing perspective is provided by \citet{Rubner2001}.

The EMD imagines the value of a distribution at any one location as describing an amount of dirt or earth at that location.
The similarity between the distributions is measured as the effort required in transforming one distribution to the other, which is imagined as the process of physically rearranging that earth from one distribution to the other.
(In the usual case, earth is neither created nor destroyed, requiring that the integral of the two distributions over the relevant domain be equal.)
This is illustrated in Figure~\ref{fig:emd-concept} for a one-dimensional distribution, and in Figure~\ref{fig:emd-concept-2d} for a two-dimensional distribution.
The amount of effort required to perform this rearrangement is taken to be $\int r \; dm$, where $dm$ is the mass of each quantum of earth and $r$ is the distance that quantum is moved.
The earth mover's distance between these two distributions is then the minimal value of this integral, attained by finding the optimal mapping between locations that minimizes the total effort required during this rearrangement.
Finding this optimal mapping becomes an optimal transportation problem, for which a rich body of literature exists \citep[for reviews, see][]{Kolouri2017,Peyre2019}.
In our application, this optimal mapping itself is our goal, whereas in many other applications, computing the optimal mapping is just a means to the end of calculating the minimal value of the EMD.

\begin{figure}[t]
	\centering
	\includegraphics{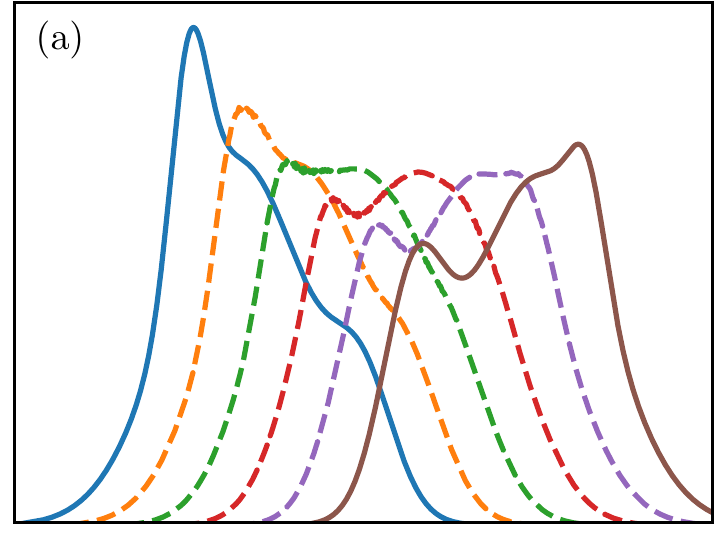}
	\includegraphics{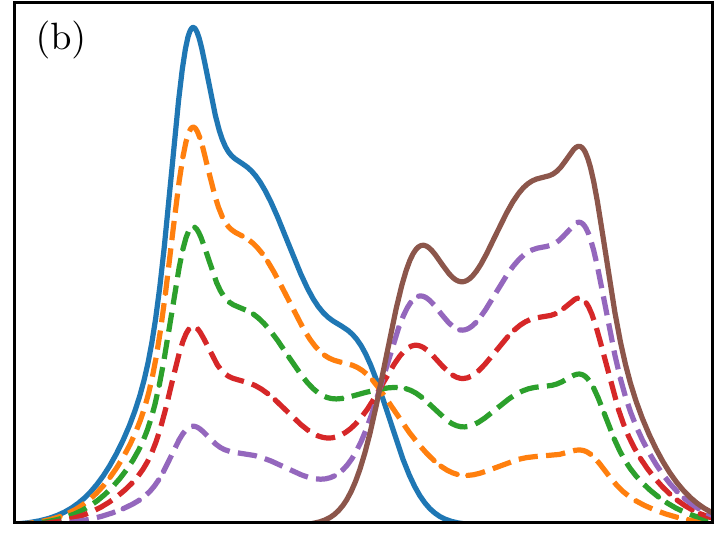}
	\caption[Interpolation between two distributions using an earth mover's solution]{Interpolation between two distributions using an earth mover's solution. In panel (a), the two solid lines mark the starting and ending distributions, and the dashed lines are interpolated distributions produced by positioning each unit of earth a fixed fraction of the way between its starting and ending position. In panel (b), the same two solid curves are interpolated as a linear combination of the two distributions with varying relative amplitudes.}
	\label{fig:emd-concept-interpolation}
\end{figure}

Computed mappings for discrete scenarios are typically formatted as a two-dimensional matrix, in which each row represents a source location, each column a destination location, and each entry represents the amount of earth to be moved from the source to the destination location.
This can be used to produce a displacement vector for each unit of earth.
One interesting use of these vectors is to interpolate between the two distributions (in the sense of producing a distribution ``in between'' the two given distributions---for example, supposing that the second distribution is the first distribution after evolving over time, interpolation would find a version of the distribution at some intermediate point in time).
This is done by positioning each unit of earth a fixed fraction of the way along its displacement vector.
(For more detailed treatments, see, e.g., \citealp{Bonneel2011} or \citealp{Solomon2015}).
This is illustrated in Figure~\ref{fig:emd-concept-interpolation}.
This interpolation between distributions causes the first distribution to smoothly morph into the second, as if it is made of clay and being pressed into a new shape.
This is as opposed to the more typical way of linearly combining the distributions, in which the interpolation between distributions $f$ and $g$ would be $(1-A)f + Ag$---that is, a simple sum of the two distributions with varying amplitudes.

In this chapter, we outline a use of the earth mover's distance to infer horizontal velocities within the evolving boundary of a photospheric \bp.
We do this by treating each \bhp\ boundary as a binary histogram, with values of one within the identified boundary of the \bp\ and zero outside.
For each pair of successive \bhp\ boundaries over time, this gives two distributions.
The framing of the EMD allows us to consider these shapes as two masses that are morphed from one shape to another, which aligns with a basic model of how the \bp\ is evolving in reality, and the machinery of the EMD allows us to use off-the-shelf algorithms to compute a minimal-effort mapping between locations in these two shapes, which indicates where each parcel of plasma in the first shape should move to in the second shape.
From this mapping, horizontal velocities can be produced.
Our approach does not actually go on to compute the EMD, the metric indicating the total effort required in morphing one shape into the other, and so in truth, our approach might be better described as an \textit{optimal transport} approach rather than an \textit{earth mover's distance} approach; however, the conceptual framing of the EMD is key in translating \bhp\ shapes into an optimal transport problem, and it provides the link between the numerical approach and our physical expectations of what might happen withing \bp s.

As opposed to the approach described in Chapter~\ref{chap:ellipse-fitting}, this approach attempts to capture all available velocity information, rather than just a selection of wave modes.
This approach also does not rely on calculating a centroid, a reliance which can be limiting for, e.g., large or complex \bp s.
However, this approach cannot estimate vertical plasma velocities (the dominant velocities in sausage-mode waves), and interpretation of the resulting horizontal velocities is more challenging (in part because the velocities are not decomposed into wave modes, making modeling of their upward propagation more difficult and \bhp\ specific).

In Section~\ref{sec:emd-method}, we describe our method in detail.
In Section~\ref{sec:emd-results}, we present and discuss inferred velocity fields after applying our technique to data from \muram\ simulations with DKIST-like resolution, with comparisons made to the plasma flows in the \muram\ simulations.
In Section~\ref{sec:emd-comp-but-should-we}, we discuss in detail the limitations of this approach and the ways that plasma flows and \bhp\ evolution could violate the assumptions made by our approach.
We present benchmark energy-flux values in Section~\ref{sec:emd-energy-fluxes}, and we summarize our results in Section~\ref{sec:emd-summary}.

\section{Method Description}
\label{sec:emd-method}

\subsection{Cadence Reduction}
\label{sec:emd-cadence-reduction}

In computing EMD-inferred velocities, our first step is to reduce the cadence of \bhp\ boundary measurements.
We will demonstrate this technique on simulated observations produced by the \muram\ simulations described in Section~\ref{sec:2s-muram}, and at the simulation's native 2~s cadence, changes to \bhp\ boundaries are overly discretized.
(DKIST's VBI instrument will achieve a comparable cadence, and so a similar concern will likely apply to observations.)
For example, if a \bp 's evolution appears to the eye as smooth, uniform, upward expansion of the \bp, which warrants the addition to the feature of the row of pixels immediately above the feature, this short cadence might cause that row of pixels to be added one pixel at a time over several frames rather than all together at once.
This will cause our inferred velocities to be very sporadic and jumpy, likely with spurious horizontal components that oscillate back and forth as pixels on different sides of the expanding \bhp\ edge are added one at a time.
This is in opposition to the uniform, upward velocity field that would be expected from uniform, upward motion of the \bhp\ boundary.
We therefore take five frames at a time, and mark a pixel as included within the \bp\ if that pixel is contained within the identified \bhp\ boundary in a majority of those frames.
This produces a temporally-smoothed \bhp\ boundary every five frames, for a 10~s cadence.
We will be using this reduced-cadence sequence for the rest of this chapter, and we will be computing an inferred velocity map for each pair of successive frames (at this reduced cadence) and for each identified \bp.

\subsection{Positioning of Earth}
\label{sec:emd-positioning-earth}

The (discrete) EMD problem requires we express our \bhp\ shapes as distributions of earth by specifying an amount of earth located at each of a set number of locations in the first frame, and an amount of earth which is to end up at each set location in the second frame.
The most straight-forward approach is to define as one of these earth-containing locations the center of each pixel that is included in one of the \bp s.
This can be extended by marking within each pixel an $n_\ell \times n_\ell$ grid of locations, positioned to ensure even spacing both within each pixel and across pixel boundaries, and one use of this generalization will be shown later.

We position one unit of earth at each marked location in the first frame, and require one unit of earth to arrive at each marked location in the second frame.
In the ``distributions as hills'' framing of the EMD, this imagines the \bp s as more akin to mesas than hills---flat on the top, with abrupt edges.
Alternative choices might include setting amounts of earth proportional to the intensity or magnetic field strength of the corresponding pixel.
However, as discussed in the motivation for our new approaches (Section~\ref{sec:motivation-for-new-technique}), evolution of the finest structure of the intensity pattern is not guaranteed (or necessarily expected) to correlate with horizontal plasma motions, and magnetic field information is difficult to reliably acquire at the necessary resolution and cadence.
We instead use this uniform distribution of earth, meeting our goal of relying only on the identified shapes of \bp s to avoid over-interpreting the finest details of the intensity structure.

One requirement of the EMD problem is that there must be an equal amount of earth in the source and destination distributions.
This will only occur in our approach so far when a \bp\ does not change in area, which is a rare occurrence.
We will present here our chosen solution to this problem, and then discuss alternative solutions for the remainder of this sub-section.
All solutions involve either altering the selection process of marked, earth-containing locations to choose the same number of positions across the two shapes (which requires a way to position a fixed number of points evenly or nearly-evenly throughout an arbitrarily-shaped region, which is not trivial), or altering the amount of earth positioned at each location.

\begin{figure}[t]
	\centering
	\includegraphics[width=\linewidth]{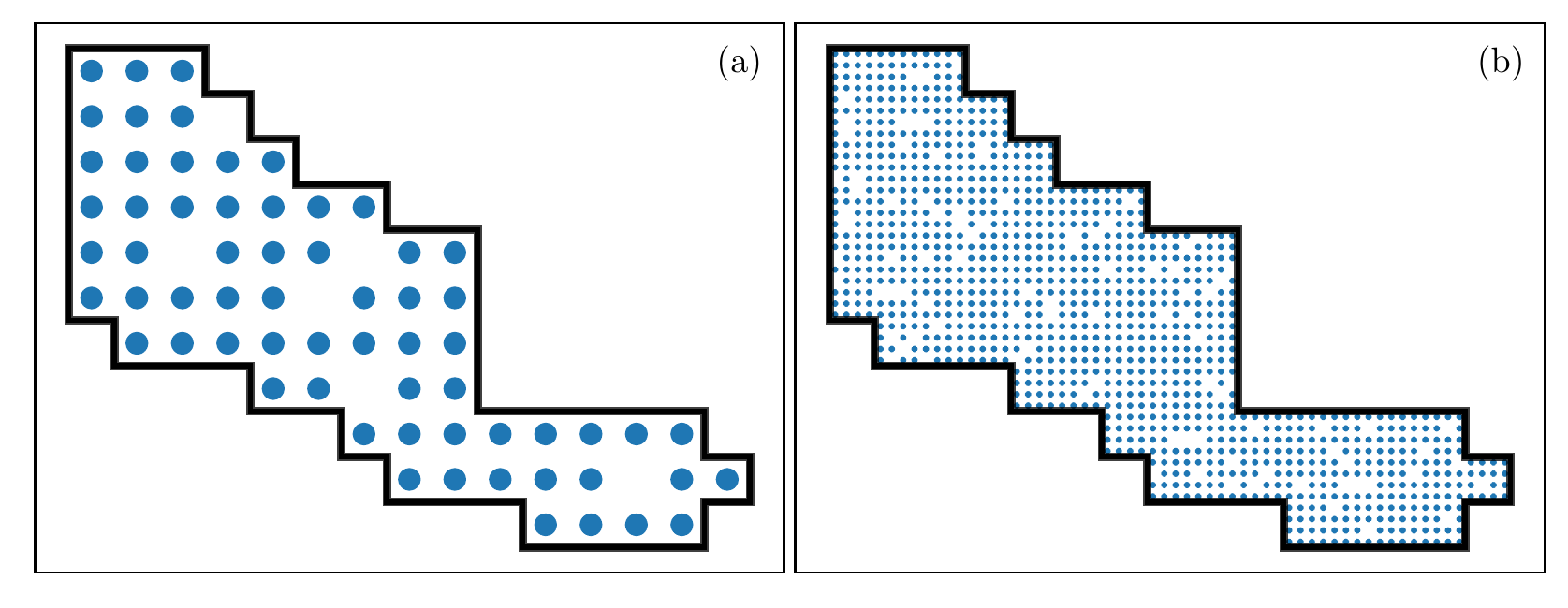}
	\caption[Sample grid of earth-containing locations within a \bp]{Sample grid of earth-containing locations within the outlined shape of a \bp. In panel (a), one position is marked per pixel. In panel (b), a $4 \times 4$ grid of locations is marked within each pixel. In both, positions are randomly un-marked as described in the text so that the total number of locations matches the number of locations marked in the subsequent and smaller shape of this \bp\ (which is not shown).}
	\label{fig:emd-marked-locations}
\end{figure}

Our chosen solution is to mark locations as described above, and then randomly eliminate, or un-mark, some locations from the grid in the larger feature (which may be the first or the second of the two features being considered) until the two features have an equal number of marked, earth-containing locations.
This leaves holes in the grid of marked locations, but they will be smoothed over by the averaging of the eventual motion vectors.
The advantage of this approach over the others we will describe is that the earth-containing locations still fall on a uniform grid for much of the \bhp\ area.
To ensure that the smoothing over these holes is satisfactory, and to mitigate the effect of this randomness on our velocities and improve repeatability, we find it best to initially mark at least a $4 \times 4$ grid of points within each pixel (i.e., the $n_\ell$ of earlier is set to 4).
This increase does, however, come at the cost of increased computation time in finding the EMD solution, though this cost is not burdensome.
A sample grid of locations is shown in Figure~\ref{fig:emd-marked-locations}.

An alternative solution for satisfying the equal-amount-of-earth restriction would be to place points randomly from the outset, by randomly drawing locations within the \bp, as if throwing darts, and rejecting each newly-drawn point if it falls within a certain distance of an already-marked location.
This process would be repeated until a targeted number of accepted locations is achieved in both \bp s.
The threshold distance can be set to ensure the final point distribution has a nearly-uniform density, and it can be slowly decreased during the point-selection process to ensure prompt convergence on the desired number of accepted locations.
This alternative might reduce the need for a high point density by ensuring the gaps in the distribution are more evenly-distributed, though a high density may still be required for approximately-reproducible velocity fields, given the still-random nature of this approach.

Another alternative solution would be to position points using a Sobol sequence \citep{Sobol1967} or any other of what are called low-discrepancy or quasi-random sequences \citep[for overviews, see][]{Bratley1988,Niederreiter1988,LEcuyer2003}.
These are sequences that iteratively fill a region with points in a deterministic manner.
The point clouds appear to the eye as random or arbitrary but avoid the ``clumpiness'' that a truly random cloud of points would display---that is, these sequences produce a distribution of distances between points and their nearest neighbors that is more narrow than for a truly random distribution.
One use for such a sequence is to efficiently sample a domain, such that the numerical integral of a function evaluated at those sample points approximates the true integral of the function, with the estimated integral converging more quickly than a random set of samples as the number of sample points increases, and with the ability to stop sampling once acceptable convergence is seen, which is not an option when integrating with a uniform grid of samples.
Producing points in the plane using such a sequence and positioning earth at points which fall within the boundary of the \bp, and continuing the iterative production of points until the shape reaches a set number of marked locations, would allow the two \bhp\ shapes to be filled with an equal number of points that are close-to-evenly distributed throughout each of the two unequal shapes.
This would avoid both the difficulties of using evenly-spaced grids in two unevenly-shaped and unevenly-sized features, and the randomness of the earlier-mentioned approaches.
However, stopping the point generation once the target number of points is achieved might bias the point cloud toward or away from certain portions of the BP.
The Sobol sequence, for example, only fully realizes its intended properties when the number of points generated is a power of two.
This restriction to a power-of-two number of samples is difficult to reconcile with our need to both select as earth-containing locations only those points that fall within the boundaries of our unevenly-shaped \bp s, and to ensure an equal number of earth-containing locations is selected for the two different \bhp\ shapes of unequal area.
(A possible solution might involve ``unfolding'' each \bp\ by mapping the \bp 's $n$ pixels to the pixels of a $1 \times n$ rectangle.
Points in the two-dimensional Sobol sequence are distributed across a unit square in the plane, but these could be ``stretched'' to fill the $1 \times n$ rectangle, with all points therefore falling within the boundaries of the unfolded \bp, and then re-mapped back to the original \bhp\ shape.
The stretching would cause different horizontal and vertical spacings of points, but this could be mitigated by repeating the process with an $n \times 1$ rectangle, doubling the total number of marked points and ensuring more isotropic spacing.
This would allow a power-of-two number of locations to be selected which are all contained within the \bp, and for the same number of locations to be selected across two unevenly-sized \bp s.)

Another alternative strategy for positioning earth would be to use the same, even grid spacing when marking locations within the two unequally-sized feature boundaries, and rather than adjusting the number of marked locations to be equal, position one unit of earth at each location in the smaller feature, and in the larger feature place at each location an amount of earth equal to the ratio of the two feature's sizes, so that the total amount of earth is equal across the two features.
This approach can be thought of as modeling a change in plasma density between the two unequally-sized features (which may not be appropriate, if plasma moves vertically rather than compressing during this shape change).
This approach would also have the advantage of an even spacing of locations, but it would have the disadvantage that each location in the smaller feature must have multiple mappings to locations in the larger feature, moving a fraction of its earth to or from each of multiple locations.
This proliferation of mappings would increase the challenge of inspecting the resulting mappings, as well as the challenge of reasoning about both the computed and intuitively-expected mappings.

A final alternative for handling the situation of differing \bhp\ sizes would be to allow for the creation or destruction of earth.
Depending on the implementation of the EMD solver one uses, an additional location for earth can be designated for which the cost of moving earth into or out of this location is fixed and independent of distance (and, in fact, concrete coordinates for this location need not be specified), and movement of earth into or out of this abstract location can be thought of as destruction or creation of that earth.
An amount of earth equal to the difference in the amounts of earth constituting the initial and final \bhp\ shapes can be required to begin or end in this extra location, with this earth moving into or out of the \bp\ at locations determined by the EMD solver and the fixed cost assigned to this motion.
This might be thought of as modeling vertical motion of plasma into or out of the plane in response to a changing \bhp\ area.
This approach would have the difficulties of requiring an appropriate cost for this creation or destruction to be identified (that cost may be zero, as used by \citealp{Rubner2001}), and requiring that one ensure, across a range of different types of \bhp\ shape changes, that allowing this creation or destruction does not result in strange or implausible velocity fields.

Each alternative we describe above has strengths and weaknesses.
We use the first approach described, of marking a uniform (and dense) grid of locations within each feature, randomly un-marking locations within the larger feature until an equal number of locations is marked within each of the two features, and requiring exactly one unit of earth to start or end at each marked location.
We feel it is advantageous to ensure an even grid of marked locations to the greatest extent possible, and this approach also avoids the greater complexities of some of the latter alternatives described.

\subsection{Computing the EMD Solution}
\label{sec:emd-computing-solution}

Having determined the initial and final distributions of earth, we now compute the earth-movement mappings that rearrange the initial distribution into the final.
In a traditional use of the EMD, the calculation of these mappings is just a means to an end, borne out of the requirement that one find the minimal mapping in order to know the minimal distance between the two distributions.
However, in our application, these mappings themselves are the goal.

For this calculation, we use the Python Optimal Transport library \citep[POT;][]{Flamary2017}, which implements the algorithm of \citet{Bonneel2011}.
This implementation accepts as inputs a one-dimensional array for each of the starting and ending distributions, with each element representing the amount of earth at one location.
Additionally, a two-dimensional cost matrix provides the cost of moving one unit of earth from any one starting location to any one destination location.
These three inputs specify the earth mover's problem in a very general way, allowing possible extensions such as the creation and destruction of earth discussed earlier.

In computing our cost matrix, we use the square of the distance between any two locations, rather than the location itself.
(This choice is common enough as to be the default in POT's utility function for preparing the cost matrix.)
This ensures that, for example, if a \bp\ expands on one side and contracts on the opposite side, the EMD solution contains a large number of small movements, representing a large-scale flow from the contracting side to the expanding side (conveying the earth from pixel to pixel like a bucket brigade), rather than mapping the pixels in the contracting area directly to those in the expanding area, which would represent earth remaining stationary in the center of the \bp\ while earth from the contracting edge hops over to the expanding edge.
These two types of flows could have equal total cost when using unmodified distances in the cost matrix, but the former is much more desirable in our application.

\subsection{Producing a Uniform Grid of Velocities}
\label{sec:emd-uniform-grid}

The output of the EMD calculation is a two-dimensional matrix, where each row or column represents a source or destination location, respectively, and each matrix element represents a movement of earth between the corresponding source and destination locations (which can include non-movements, mapping a source location to the same location in the destination distribution).
This provides a displacement vector for each unit of earth, and with the time elapsed between the two shape measurements, velocity vectors can be produced in physical units.
These vectors are irregularly and unevenly spaced, however.
A more desirable output is a single velocity vector for each pixel in the feature.
Reducing the velocity field in this way also provides a type of spatial smoothing that can counteract, for example, the effect of unevenly-spaced locations at which earth is positioned.
A similar goal is accomplished in the balltracking approach to velocity inference, with a solution similar to the one we employ \citep{Potts2004}.

To achieve this, we iterate over all pixels in the feature and all velocity vectors produced by the EMD solution.
For each pixel, we identify all velocity vectors for which that vector's closest approach to the pixel center is within one pixel-width of that center.
We average those vectors together, with each vector assigned a weight of $1-d$, where $d$ is the vector's closest-approach distance in pixel widths.
This form for the weight (as opposed to, e.g., a boxcar or a Gaussian) assigns greater weight to vectors passing mostly closely to a pixel center, while ensuring the weighting function drops to zero so that a velocity vector's influence is kept local.

\begin{figure}[t]
	\centering
	\includegraphics[width=\linewidth]{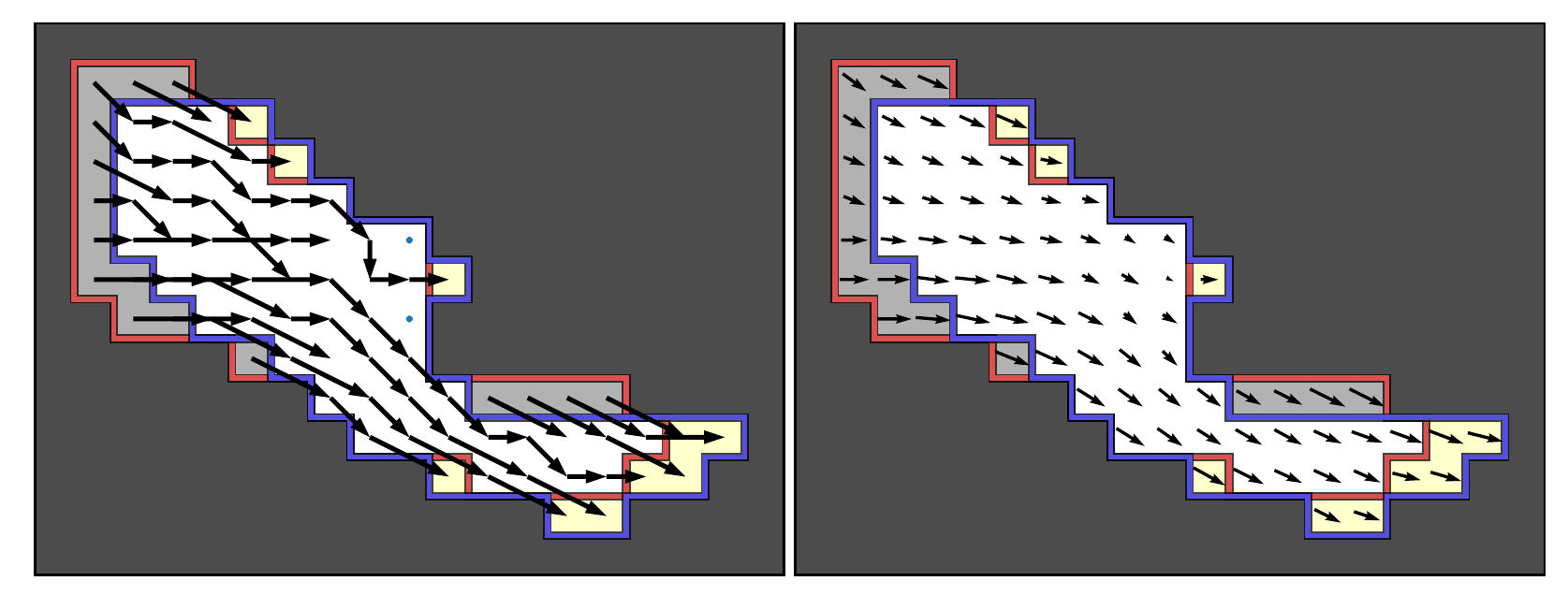}
	\caption[Example EMD velocities]{Example EMD velocities. On the left are the raw displacement vectors computed through the EMD approach, connecting each location in the first distribution to its mapped location in the second. On the right are the velocity vectors smoothed and sampled onto a regular grid. In both, red outlines the ``before'' shape of the \bp, and blue the ``after'' shape. Light gray pixels are being removed from the feature during the time step, and yellow pixels are being added. The \bp\ and time step depicted is arbitrarily selected from those identified in the \muram\ simulation. The initial density of earth-containing locations here is one per pixel, rather than the 16 per pixel we use in the rest of our calculations, to simplify the left-hand panel for visualization.}
	\label{fig:emd-sample-vectors}
\end{figure}

In areas where the \bhp\ shape is concave, the displacement vectors sometimes cross outside the feature and then back in.
This might produce velocity vectors outside the boundaries of the \bp.
In our implementation, pixels fully outside the \bp\ (that is, pixels that are not included in either the first or second shape for a given time step) are excluded from the regularized grid, as our approach is focused on the intra-\bhp\ pixels and can not properly constrain the motions outside the \bp.

This process produces velocity vectors throughout a \bp\ on a regular, pixel-scale grid with physical units, suitable for further analysis.
A sample set of velocity vectors before and after this reduction step is shown in Figure~\ref{fig:emd-sample-vectors}.
It can be seen from the outlined boundaries that the overall motion of the \bp\ is down and to the right, and the EMD-inferred velocity vectors match this overall picture.
Earth is moved out of the pixels that are part of the feature in the first frame but not the second, and likewise moved into pixels that are new to the feature in the second frame.
In between these added and removed pixels, earth is moved toward where it is needed in a ``bucket brigade'' type of motion, with small, collective motion throughout the feature rather than earth from the edges making large jumps over stationary earth in the central pixels.
In the smoothed and regularized grid of pixels, velocities vary smoothly from pixel to pixel, forming a very believable flow of material through the \bp\ that reshapes it from one shape to the next.

\section{Results: Inferred Velocities and Comparisons}
\label{sec:emd-results}

\subsection{Motivation for Comparisons}
\label{sec:emd-comp-motivation}

In this section, much ink will be spilled comparing EMD-inferred velocities to plasma velocities seen in the \muram\ simulation, but we feel it is important to first discuss both the motivation and the limitations of such a comparison.
Doing these comparisons will provide a reasonability check for our inferred velocities, and we will show an appreciable degree of similarity between the velocity fields.
These similarities might be expected for two reasons.
First, in the $\beta \sim 1$ environment of a \bp, external plasma flows can move and bend the flux tube, altering the observed shape of the \bp.
When, for example, a flow normal to the flux-tube edge compresses or advects the tube, those \muram\ plasma velocities just outside the flux tube should match the motion of the flux tube edge.
Second, as the tube shape evolves under this external forcing, the plasma within the tube will rearrange itself to eliminate any horizontal pressure gradients, likely tending toward a uniform plasma density and magnetic field strength along any horizontal slice through the tube.
This can result in internal plasma motion with a horizontal component that is also aligned with the motion of the flux tube edge; for instance, as plasma flows away from an inward-moving edge.
These types of velocity patterns, both within the \bp\ and immediately outside the \bp\ (or more specifically, between the initial and final locations of the \bhp\ edge), are exactly those that our EMD approach is designed to infer, and so under this model of flux-tube evolution, we would expect to see reasonable agreement between our inferred EMD velocities and the corresponding \muram\ plasma velocities within and near the tube.

However, there are many ways that this model oversimplifies both the evolution of flux tubes and the complex process by which a flux tube produces a visible \bp.
We will discuss shortly the requirement to choose only one of the multiple heights within the simulation at which to perform this comparison, and in Section~\ref{sec:emd-comp-but-should-we} we explore several additional ways that the \muram\ velocities might be expected to differ from our EMD velocities, even if the correct height is chosen.
An additional limitation is that the \muram\ velocities must be taken at one instant in time, and there are multiple \muram\ snapshots in between the frames in our reduced-cadence \bhp\ sequences.
The \muram\ velocities do not vary strongly during the 10~s between frames, but there is some variation, and the \bhp\ evolution (and therefore the EMD velocities) is dependent on the sum of the effects of the velocities in each snapshot, for which a single frame can only be representative.
(In our comparisons, we will choose a \muram\ snapshot centered in time between each pair of \bhp\ shapes.)
Despite these limitations, we will show that the agreement between EMD and \muram\ velocities is appreciable and is much better than chance, suggesting that the non-trivial assumptions made by our approach are reasonable, especially as a first attempt at measuring these complex \bhp\ motions.

\begin{figure}[p]
	\centering
	\includegraphics{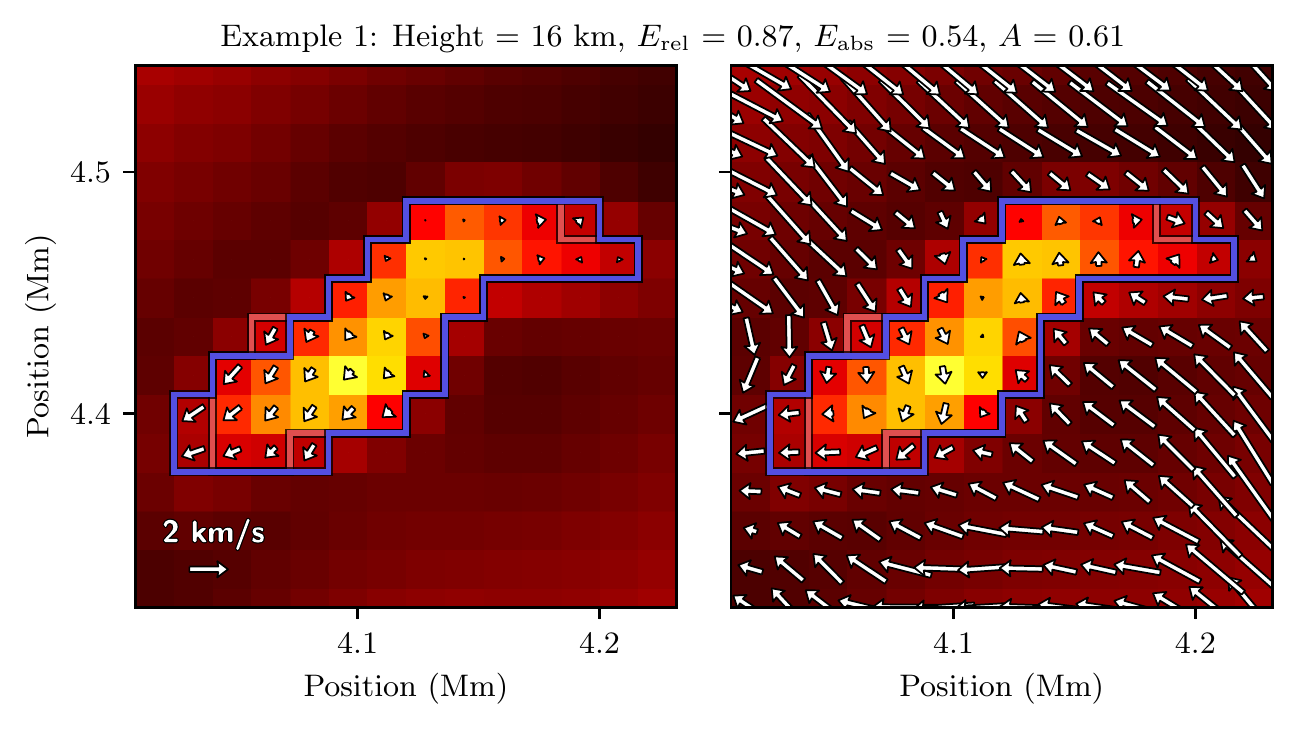}
	\includegraphics{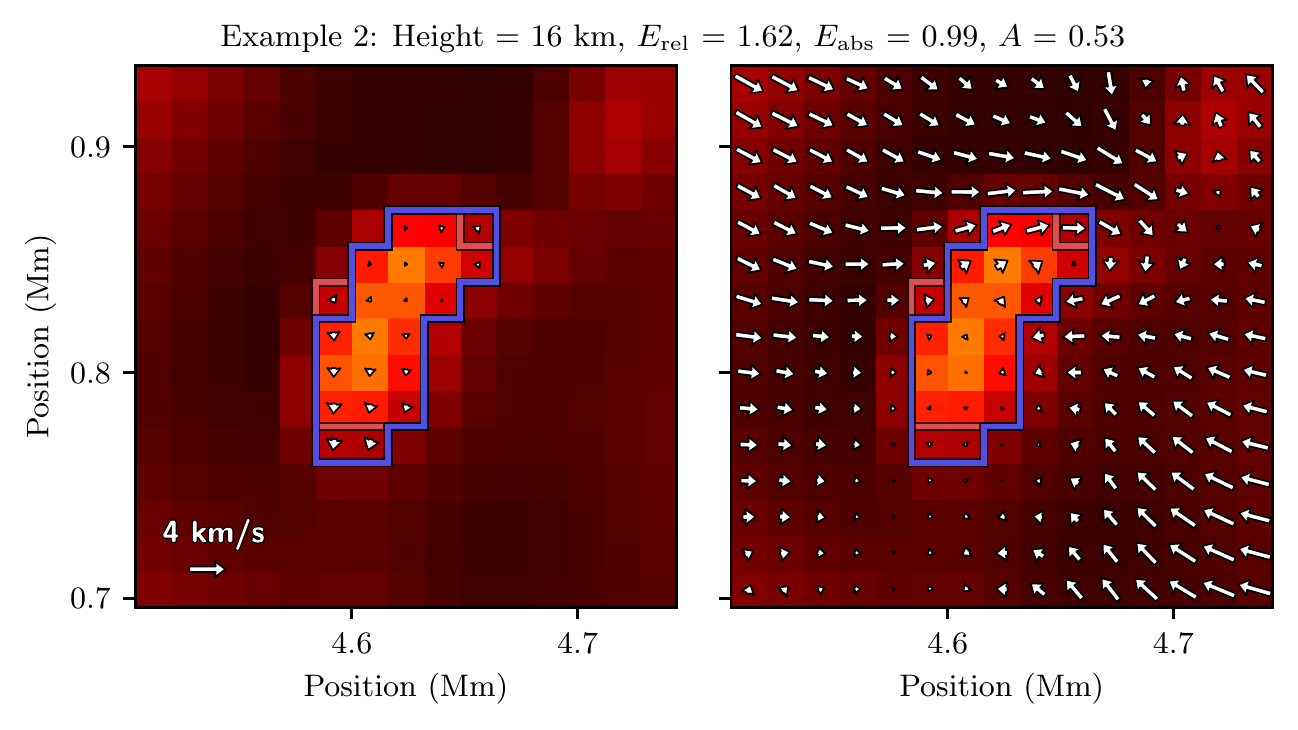}
	\caption[Sample of EMD-inferred velocity fields]{Sample of EMD-inferred velocity fields (continued on the following pages). The left-hand panels are the EMD velocities, and the right-hand panels are horizontal velocities along a slice through the \muram\ simulation at a time step centered between the two frames, provided as a comparison. The red and blue outlines mark the ``before'' and ``after'' shapes of the \bp, and the background color map is the white-light intensity at the ``before'' time step. Indicated above each plot is the height in the \muram\ simulation at which the comparison velocities are taken, and the values of the $E\rel$, $E\abs$ and $A$ metrics comparing the two velocity fields. The first four velocity fields are randomly drawn from the ``plausible'' sub-sample, and the latter four are randomly drawn from the ``implausible'' sub-sample. Note that the velocity arrow scale, shown in the lower-left corner, is adjusted from image to image to ensure the plotted arrows are of appropriate sizes. Animations of these \bp s are available on the author's website at \href{http://samvankooten.net/thesis}{samvankooten.net/thesis} and are archived at \href{http://doi.org/10.5281/zenodo.5238943}{doi.org/10.5281/zenodo.5238943}.}
	\label{fig:emd-real-samples}
\end{figure}

\begin{figure}[p]
	\ContinuedFloat
	\centering
	\includegraphics[trim=0 5pt 0 0, clip]{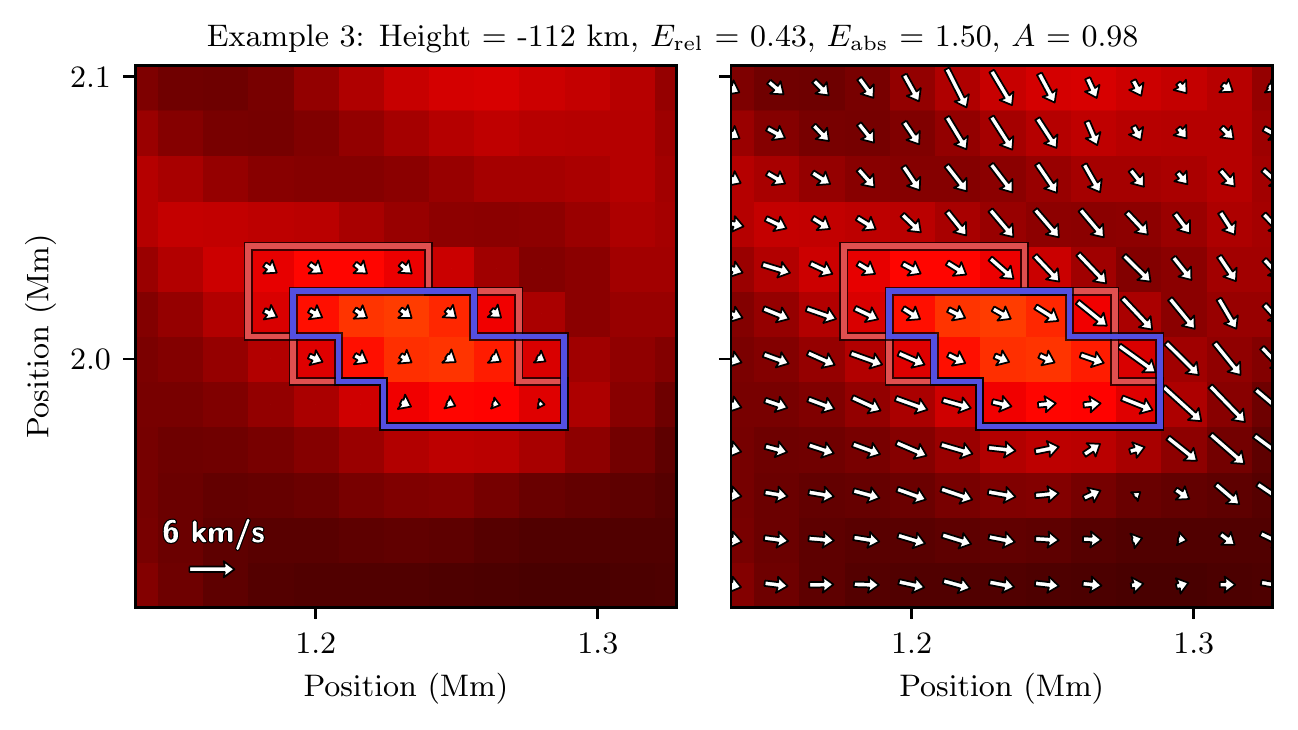}
	\includegraphics[trim=0 5pt 0 0, clip]{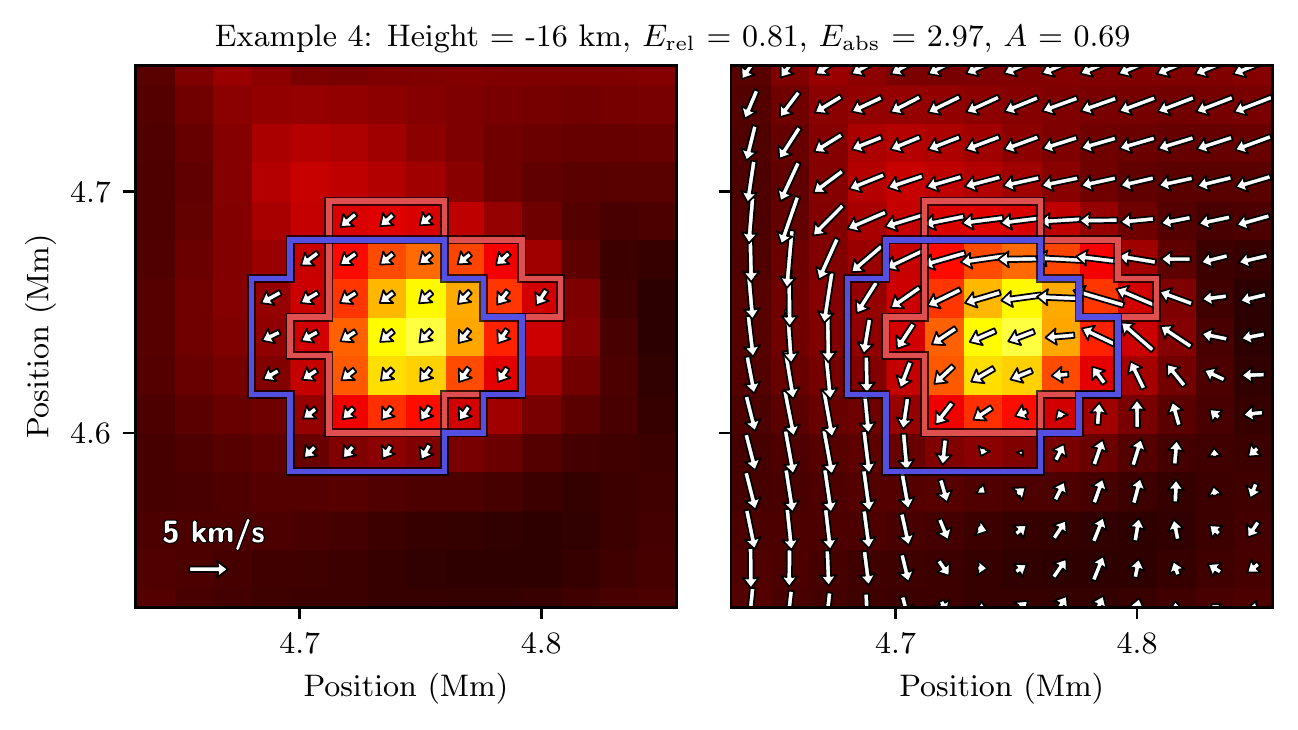}
	\includegraphics[trim=0 6pt 0 0, clip]{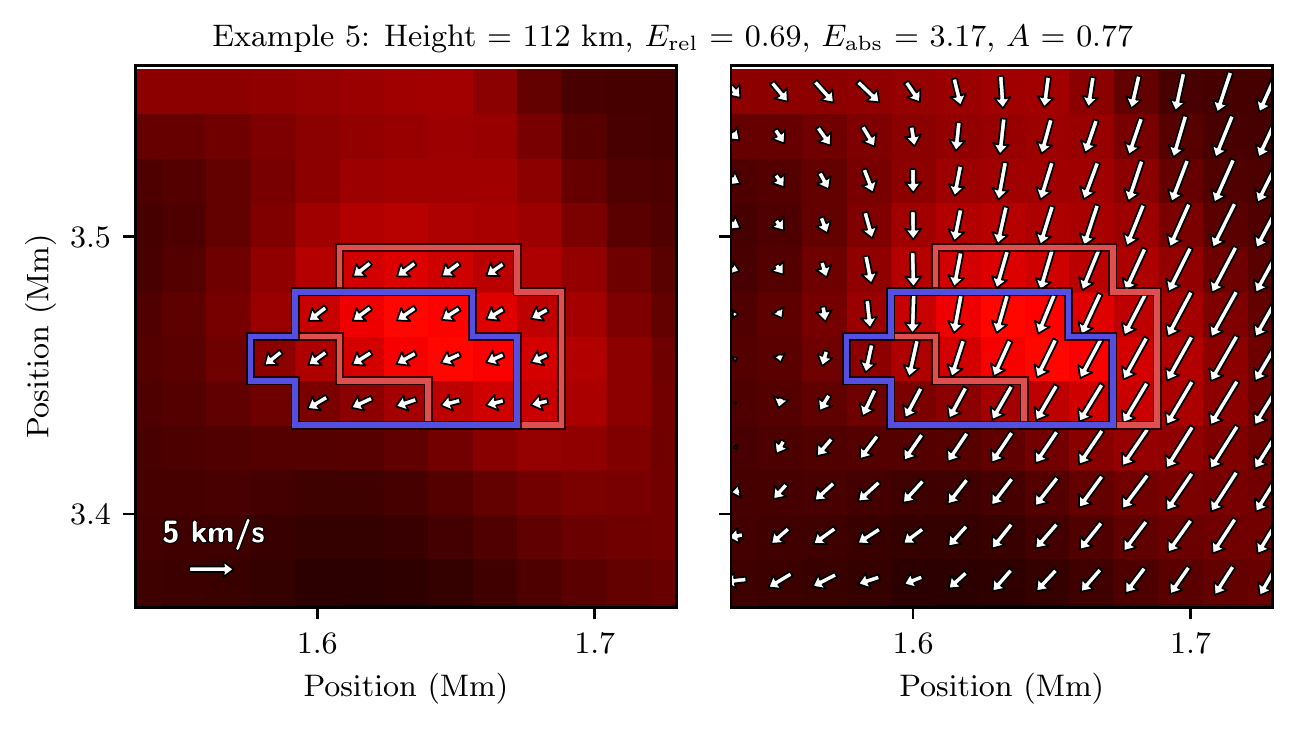}
	\caption[]{Sample of EMD-inferred velocity fields (cont.)}
\end{figure}

\begin{figure}[p]
	\ContinuedFloat
	\centering
	\includegraphics[trim=0 5pt 0 0, clip]{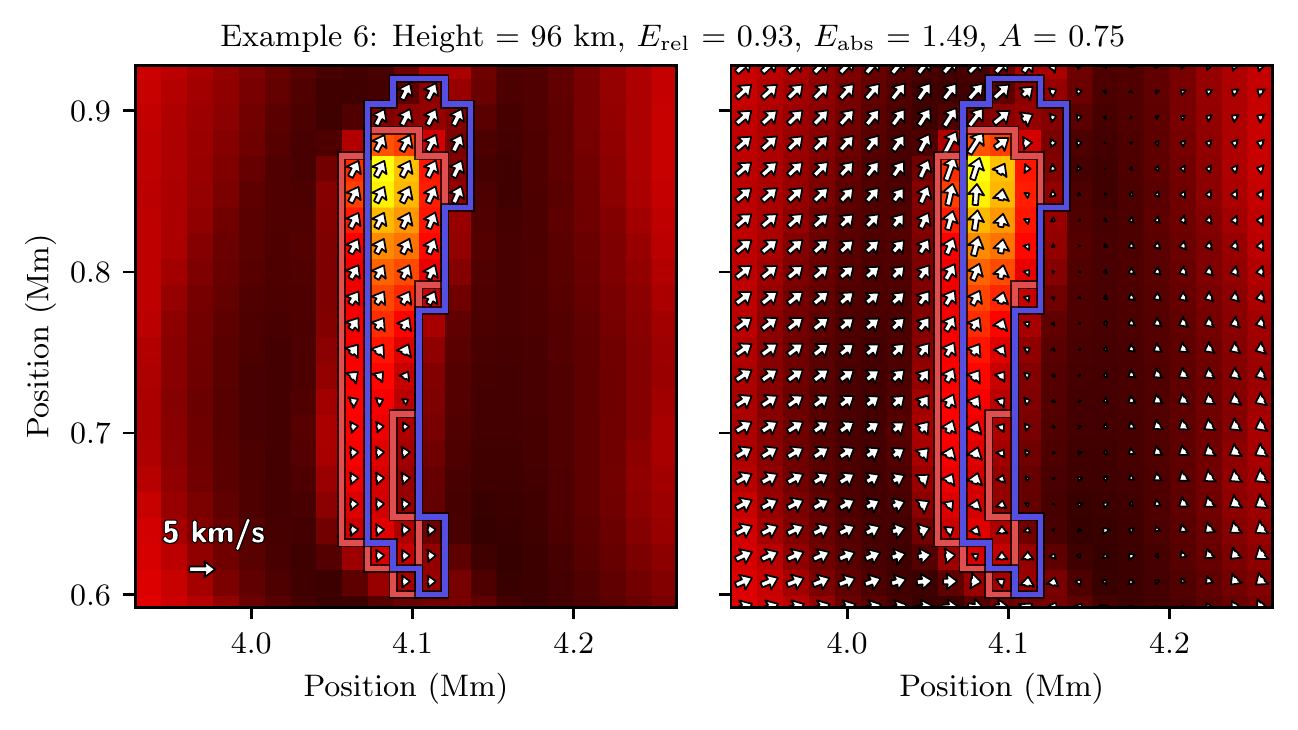}
	\includegraphics[trim=0 5pt 0 0, clip]{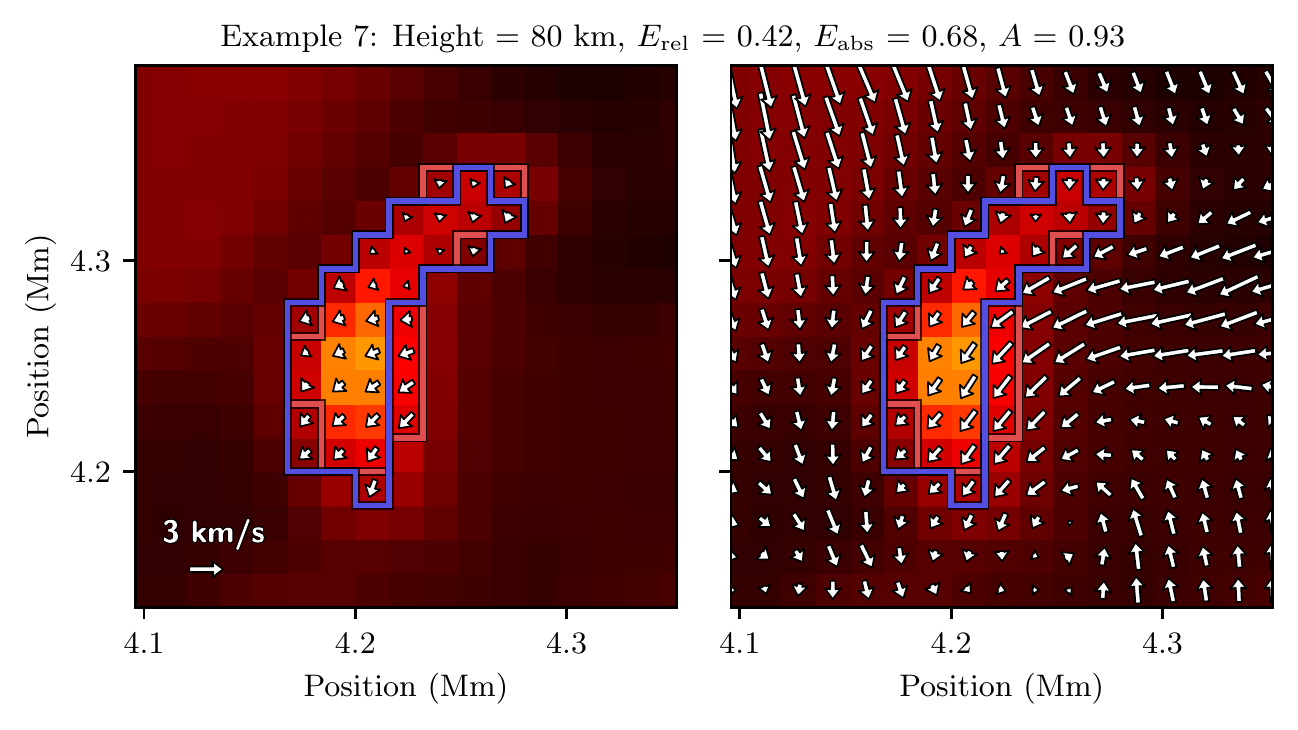}
	\includegraphics[trim=0 6pt 0 0, clip]{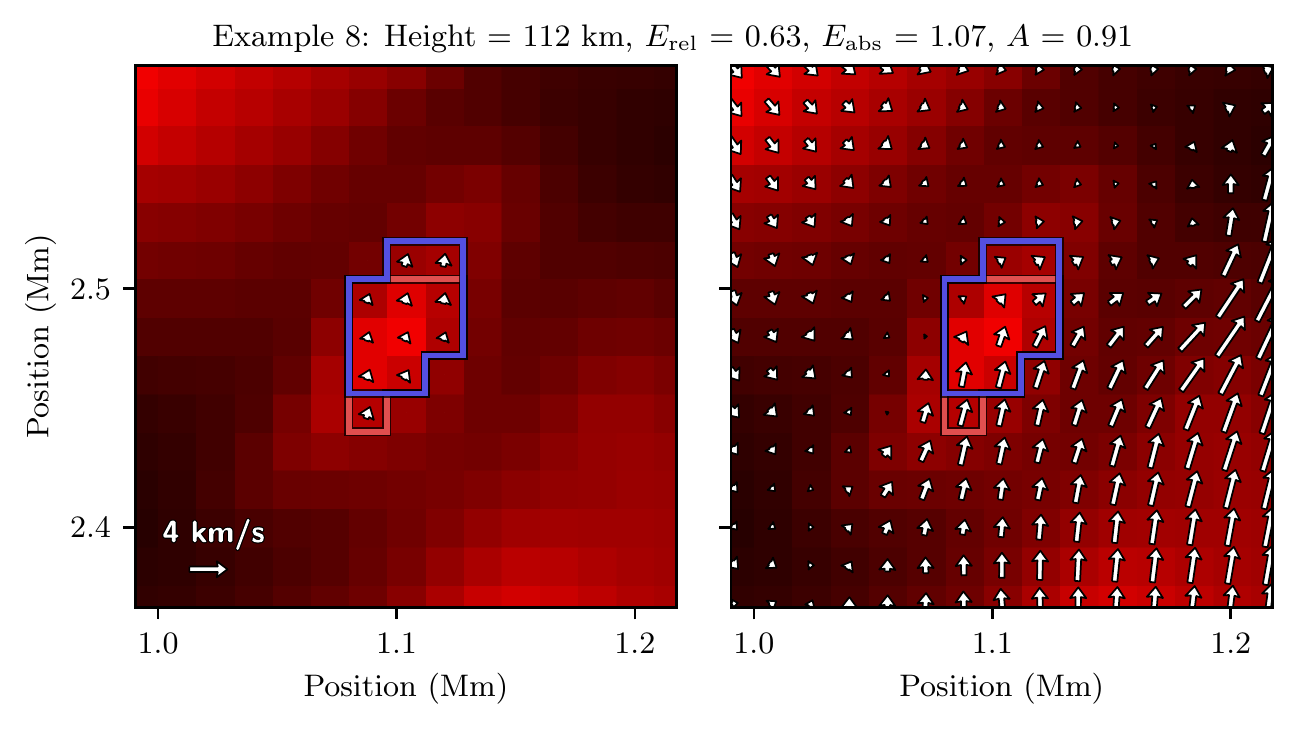}
	\caption[]{Sample of EMD-inferred velocity fields (cont.)}
\end{figure}

\subsection{Initial Comparisons with \muram}
\label{sec:emd-initial-comps}

In Figure~\ref{fig:emd-real-samples}, we show a random sample of EMD-inferred velocity maps, along with corresponding horizontal velocity maps from the \muram\ simulation for comparison.
Later in this section, we will define and discuss the comparison metrics $E\rel$, $E\abs$, and $A$, as well as the restricted range within which the height of the optimal value of $E\abs$ is found---this is the height at which each \muram\ velocity field is shown. 
We will also later discuss dividing the velocity maps into ``plausible'' and ``implausible'' populations.
Better EMD--\muram\ agreement is expected overall in the ``plausible'' population, which is represented by the first four of the eight velocity maps we show.

When viewing these sample EMD velocity fields, a few things can be noted:

First, the amount of change between the two \bhp\ shapes is relatively small, meaning it is unlikely that large, complex changes have been missed during the time step.
The shape changes are also consistent, by which we mean that an entire side of the \bp\ usually changes together in a similar way.
This suggests that our reduced cadence succeeds in ensuring we view motions as a whole and that we avoid the risk of an overly-short cadence resulting in overly-discretized motions, with different parts of the identified \bhp\ boundary changing at slightly different times, resulting in very jumpy and uneven border changes across the very short time steps.

Second, the EMD-inferred velocities appear believable, with amplitudes typical of photospheric flows (as directly evidenced here by the comparable \muram\ velocity magnitudes) and with only gradual variation from pixel to pixel.

Third, it can be seen that the \muram\ flows in and around each \bp\ have significant spatial variability, meaning that matching these flows with our EMD approach is a decidedly non-trivial task---that is, lucky guesses are unlikely.
However, the EMD flows do usually show a reasonable similarity to the \muram\ flows, though sometimes more so in direction than magnitude.
While points of clear disagreement are easy to find, points of strong agreement are also readily found, and it usually feels quite reasonable to describe the EMD and \muram\ velocity fields as distinct but similar or related flow patterns.

\begin{figure}[t]
	\centering
	\includegraphics{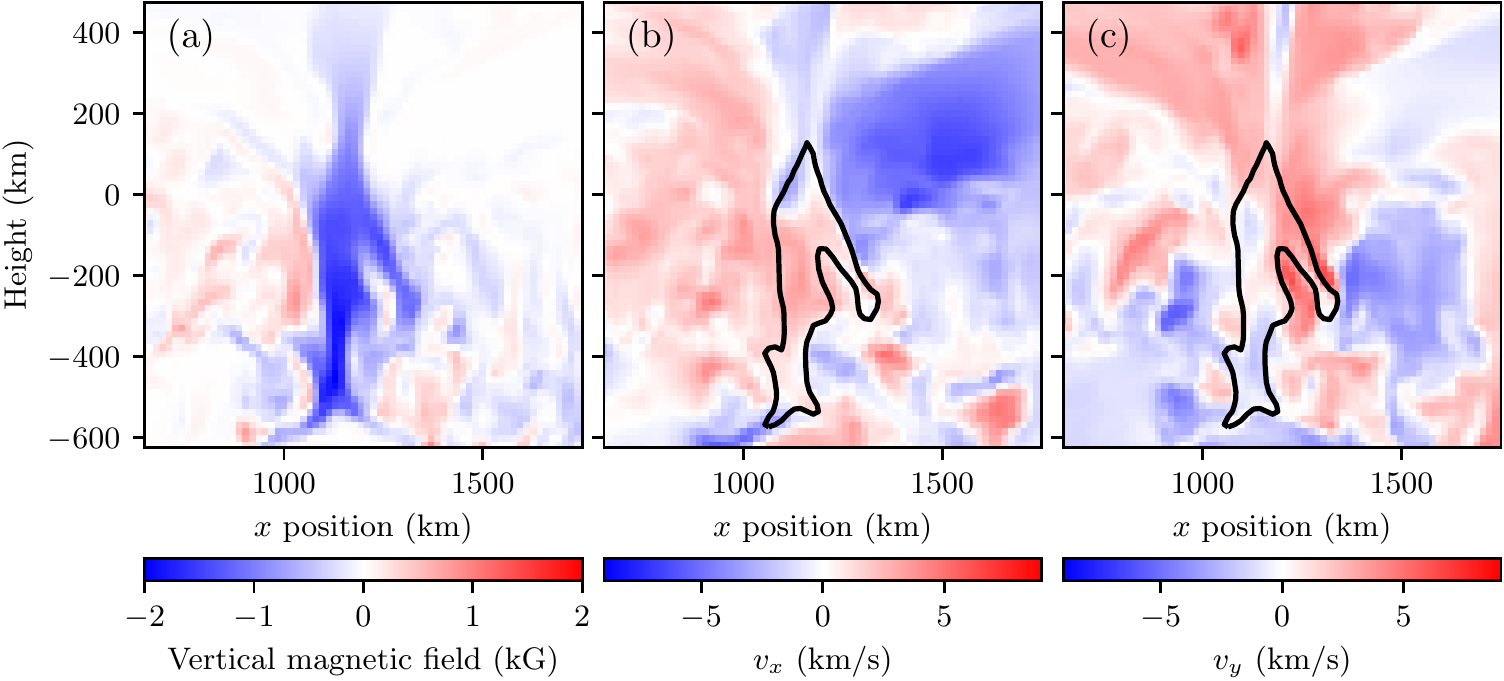}
	\caption[Magnetic field and plasma velocity components on a vertical slice through a \bp]{Magnetic field and velocity components on a vertical slice through the \bp\ shown in Figure~\ref{fig:muram-sample}. In panel (a), the flux tube can be seen by its strong vertical magnetic field, which weakens rapidly above the photosphere as the tube expands and which is bent out of this vertical plane deep below the photosphere. In panels (b) and (c), the two components of the horizontal plasma velocity are shown, and a contour is drawn at $B_z=-900$~G to outline the location of the flux tube.}
	\label{fig:emd-muram-coherence-example}
\end{figure}

\begin{figure}[t]
	\centering
	\includegraphics{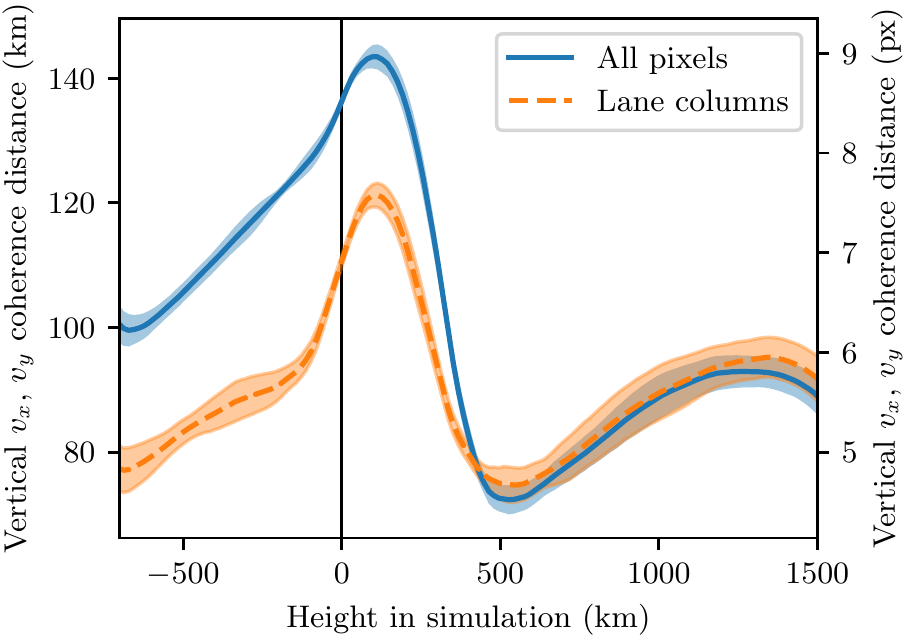}
	\caption[Vertical coherence of horizontal plasma velocities in the \muram\ simulation]{Vertical coherence of horizontal plasma velocities in the \muram\ simulation. For each pixel we test, we determine how far up or down one must look for a horizontal component of the velocity to change by over 50\%. We average this distance over direction (i.e., looking above versus below the pixel), over the values obtained from the $x$ and $y$ velocity components, over each horizontal slice through the cube, and over a sample of 36 cubes evenly-spaced through the \muram\ run. This produces an estimate of the average vertical coherence of the horizontal velocity as a function of height. We show this value across all pixels and also computed using only those columns of pixels for which the vertical plasma velocity is negative at the height of the average optical surface, so as to focus only on the downflow lanes, the regions where \bp s exists. The shaded regions mark $\pm 1$ standard deviation from the mean. Height is measured relative to the average $\tau=1$ level.}
	\label{fig:emd-muram-coherence}
\end{figure}

\subsection{Comparisons Across Height}
\label{sec:emd-comp-across-height}

In making these and other comparisons with \muram, we must choose a single height in the simulation at which to extract horizontal velocities and make the comparison, as our EMD approach produces horizontal velocities along only a single plane.
Intuition suggests that the ``correct'' height---the height at which plasma flows are most directly responsible for evolution of the \bhp\ shape---will be near the average $\tau=1$ surface.
The plasma velocities high above this surface do not have any bearing on the \bhp\ boundaries, which are formed at or below this surface.
Additionally, the flux tube rapidly expands above the photosphere, rendering the coordinates of our identified \bhp\ boundary meaningless at those heights.
Too deep below the $\tau=1$ surface, however, the plasma is not reached by any lines of sight and so has no direct effect on any observations.
Figure~\ref{fig:tracking-pixel-distributions} showed that, in this \muram\ simulation, the $\tau=1$ surface within a \bp\ is lowered by $\sim 100$~km from the average, suggesting that the range of plausible comparison heights is about this size.
Within that range, however, the horizontal velocity field is significantly variable with height.
We illustrate this in Figure~\ref{fig:emd-muram-coherence-example}, where the horizontal plasma velocity components are shown along a vertical slice through a single \bp.
It can be seen that these velocity components vary strongly on scales of $\sim100$~km.
This variability can also be seen in Figure~\ref{fig:emd-muram-coherence}, where we show a measure of the variability of the \muram\ velocities across the whole simulation.
This metric also demonstrates a $\sim 100$~km vertical scale for strong variations.
This variability suggests that multiple, distinct patterns of plasma flows could drive the motion of a flux tube within the range of heights relevant to \bhp\ formation, and that there are multiple distinct comparisons that could be made between our EMD velocities and a set of \muram\ velocities.

We address these facts by comparing the \muram\ and EMD velocities across a range of heights.
We determine for each \bp\ and timestep the height of best agreement between the two velocity fields, which we take as the height most directly responsible for formation of the \bhp\ boundary for that particular \bp\ and at that time, and we look for trends in where agreement is strongest.
As goodness-of-fit metrics, we follow \citet{Tremblay2021}, among others, in using multiple metrics to compare these two vector fields.
We use the mean absolute error,
\begin{equation}
	E\abs \; = \; \left< \sqrt{\left(\vec{v}_\text{\muram} - \vec{v}_\text{EMD} \right)
			      \cdot
		             \left(\vec{v}_\text{\muram} - \vec{v}_\text{EMD} \right)} \right>,
	\label{eqn:emd-Eabs}
\end{equation}
the mean relative error,
\begin{equation}
	E\rel \; = \; \left< \sqrt{\frac{\left(\vec{v}_\text{\muram} - \vec{v}_\text{EMD} \right)
			           \cdot
	                           \left(\vec{v}_\text{\muram} - \vec{v}_\text{EMD} \right)}
			          {\vec{v}_\text{\muram} \cdot \vec{v}_\text{\muram}}} \right>,
	\label{eqn:emd-Erel}
\end{equation}
and the relative orientation of the velocity vectors,
\begin{equation}
	A \; = \; \left< \frac{\vec{v}_\text{\muram} \cdot \vec{v}_\text{EMD}}{||\vec{v}_\text{\muram}|| \;\; ||\vec{v}_\text{\muram}||} \right>.
	\label{eqn:emd-A}
\end{equation}
In these expressions, $\vec{v}_\text{\muram}$ and $\vec{v}_\text{EMD}$ denote the horizontal velocity vector at one pixel, angle brackets denote an average over all the pixels in a feature at one time step, and $||\vec{v}||$ denotes the magnitude of the vector $\vec{v}$.

\begin{figure}[t]
	\centering
	\includegraphics[width=\linewidth]{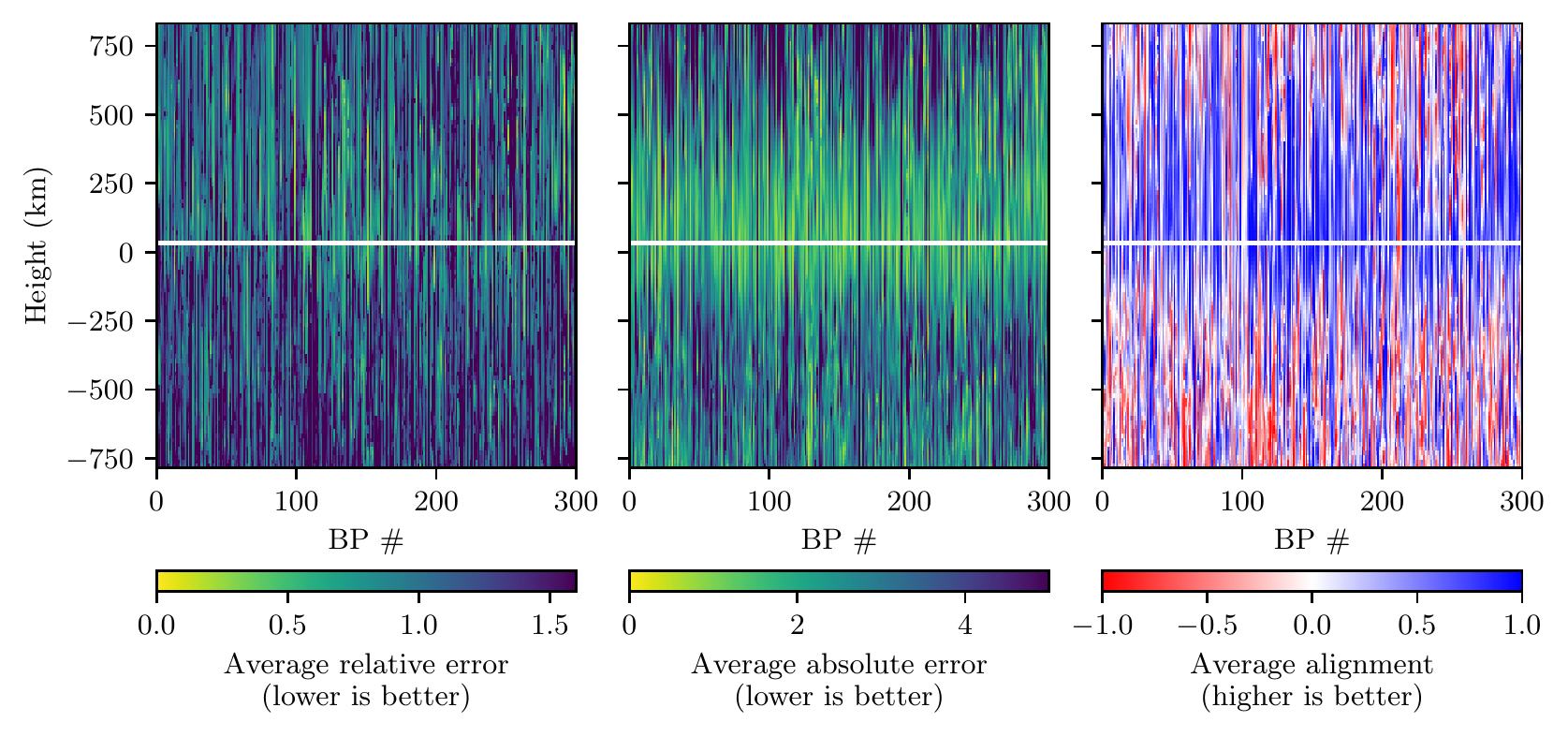}
	\caption[Error metrics vs. height for a sample of \bp s]{Error metrics vs. height for a sample of \bp s. Each vertical strip represents one of 300 randomly-selected \bp s and time steps (that is, one inferred velocity field), and shows the variation of an error metric as a function of the height in the simulation at which the comparison is made between EMD velocities and the simulation's horizontal plasma velocities. The white lines mark the height at which the sum of the error metric across a horizontal row of the plot is optimized. That height is 32~km for all three metrics.}
	\label{fig:emd-comp-strips}
\end{figure}

\begin{figure}[p]
	\centering
	\includegraphics[trim=0 2.5pt 0 0,clip,width=\linewidth]{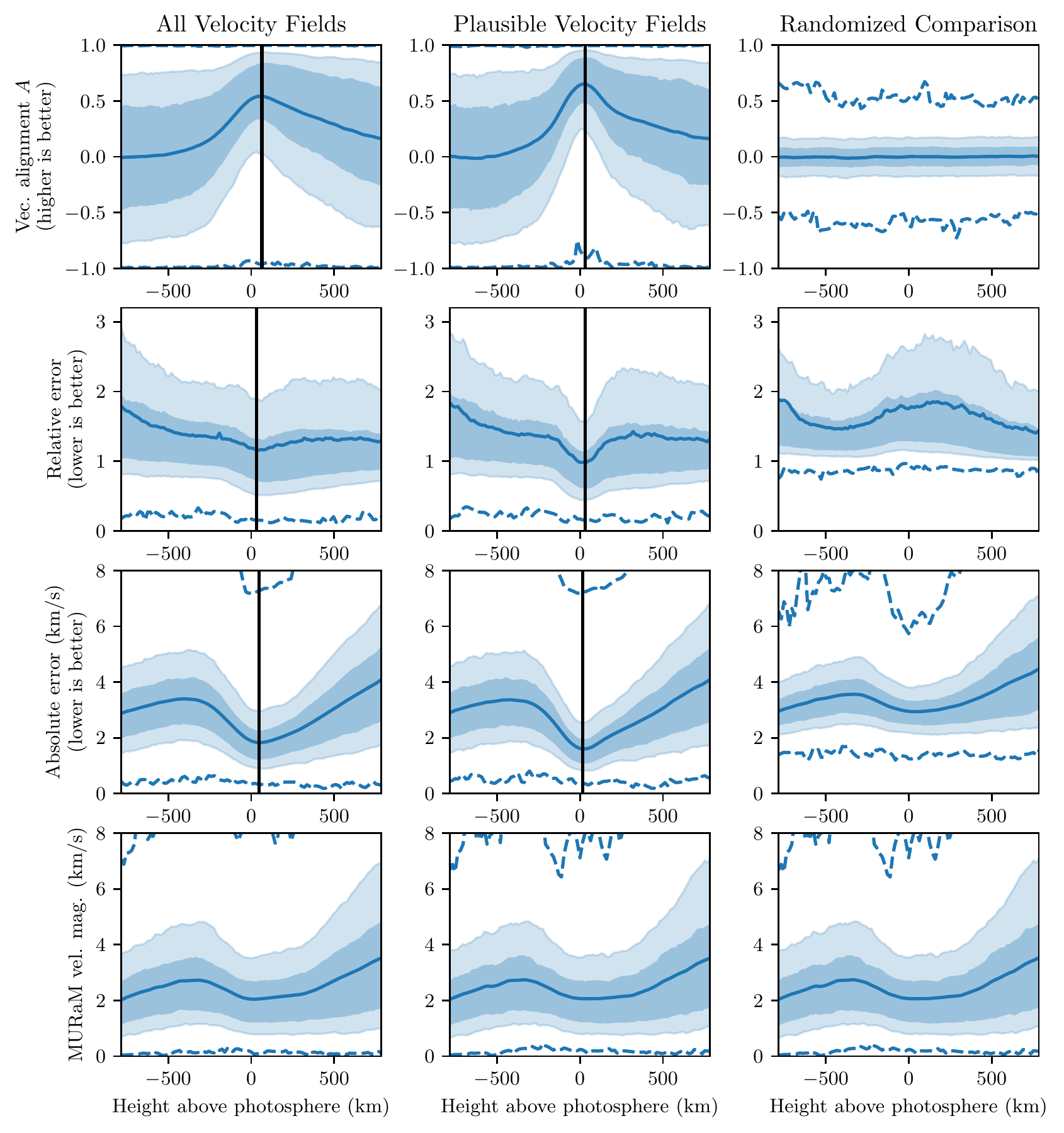}
	\caption[Velocity comparison metrics as a function of height for all \bp s]{Velocity comparison metrics as a function of height for all \bp s and time steps. The top row shows the vector alignment $A$, the second row relative error, and the third row absolute error. The fourth row shows the magnitude of the \muram\ velocities used in the comparison, as a point of reference for the absolute error values. All values are computed separately for each time step in each \bhp\ sequence, and are spatially averaged across all pixels in that velocity field. In each plot, the central line marks the mean value across \bp s and time steps at that height, the dark shading marks the $25^\text{th}$ through $75^\text{th}$ percentiles, the light shading marks the $10^\text{th}$ through $90^\text{th}$ percentiles, and dashed lines mark the minimum and maximum values (which do not always fall within the plotting range). Vertical lines, where present, mark the height at which the mean value of the metric is optimized. The difference between the three columns of plots is described in the text. The photosphere is defined as the height at which the average optical depth $\tau$ is closest to 1.}
	\label{fig:emd-comp-line-plots}
\end{figure}

Our comparisons, using the above-mentioned metrics, are shown on an individual basis for a sample of \bp s in Figure~\ref{fig:emd-comp-strips} and as an ensemble in the left column of Figure~\ref{fig:emd-comp-line-plots}.
(We will return to Figure~\ref{fig:emd-comp-line-plots} later in this section.)
For each comparison between velocity fields, we compare the EMD velocity at each pixel to the \muram\ velocity at each pixel in the feature and at one height, taken from a \muram\ snapshot centered in time between the two identified \bhp\ shapes.
We average across all the pixels in the feature for that timestep to measure the agreement as a whole between that pair of velocity fields.
It can be seen that the best agreement between the EMD and \muram\ velocities occurs near the average $\tau=1$ surface, with EMD velocity vectors typically aligned with 45$^\circ$ of the corresponding \muram\ velocity vector, and with the error in the velocity magnitudes near a factor of two.
Away from the $\tau=1$ surface, the EMD--\muram\ agreement is poor, as expected given that those heights have little or no relation to the observed \bp.
The range of heights at which the agreement is best, within $\sim 100$~km of the $\tau=1$ surface, also coincides with the expected range discussed earlier.
This range of best agreement fades away more gradually above the photosphere than below, which is notable given that \bp s are formed and evolved by processes occurring in the few hundred kilometers below the photosphere.
This upward bias may be expected, however, as plasma velocities are one of the oscillating quantities associated with the flux-tube waves that are expected to be excited by \bhp\ shape changes and to propagate upward, and so it may be expected to see these same velocities over some range above the photosphere.
This upward extent is also consistent with the larger vertical scale for velocity coherence above the photosphere than below, seen in Figure~\ref{fig:emd-muram-coherence}, which may be due, in part, to the drop-off above the photosphere of the convective motions that are responsible for the varied velocities below the photosphere, causing a reduced amount of evolution of the flow field.
(This larger coherence height may also be amplified directly above \bp s by propagating waves influencing the velocities directly above \bp s).
Many \bp s show their best EMD--\muram\ agreement slightly above the photosphere, and one possible explanation for this is that, while the plasma motions driven by changes to the \bhp\ shape do exist at heights below the photosphere, they may be superimposed on plasma velocities driven by other mechanisms (for which possible examples might be horizontally-propagating sound waves, or turbulence driven by convective motions), but those other velocities patterns do not propagate upward and have a reduced level of occurrence above the photosphere, meaning that the wave-associated, shape-change-driven velocities have their cleanest expression after they have propagated slightly upward.

\begin{figure}[t]
	\centering
	\includegraphics{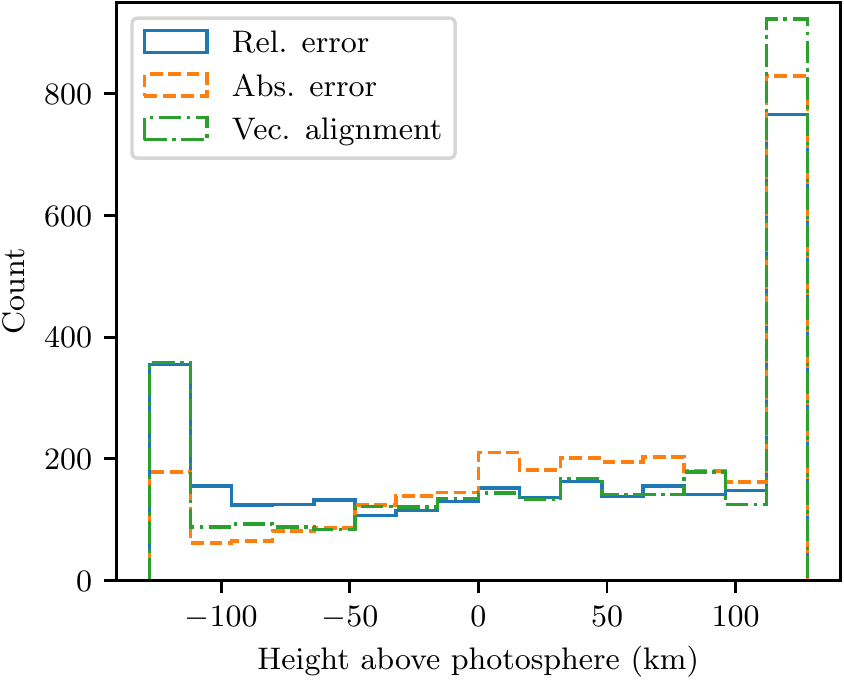}
	\caption[Height of best EMD--\muram\ velocity agreement for each \bp]{Height of best EMD--\muram\ velocity agreement for each \bp\ and each time step. This height is computed separately for each of the relative error, absolute error, and vector alignment metrics, and only heights within the range $\pm 100$~km from the photosphere are considered.}
	\label{fig:emd-comp-best-heights}
\end{figure}

In light of the agreement between our expected location where the EMD--\muram\ agreement is strongest and what we see in our data, and considering the physical justifications previously discussed for this expected range, we set a restricted height range of $\pm 125$~km from the photosphere as the range in which agreement between the EMD and \muram\ velocities can be considered credible, and we consider agreement outside this range to be largely spurious and by chance.
With this range set, we compute the height at which each \bp, at each time step, achieves best agreement between the two velocity fields, considering only those heights within our plausibility range.
These heights are shown in Figure~\ref{fig:emd-comp-best-heights}.
The heights are roughly uniformly-distributed across the range with a slight bias toward higher heights, but with a large excess of \bp s falling into the bin at each end of the distribution.
We interpret the membership of these edge bins as primarily \bp s for which the best EMD--\muram\ agreement occurs outside our plausible range, with the best agreement within our range then falling at the edge closest to the actual height of best agreement.
Since we consider agreements outside our range to lack physical plausibility, we label the velocity fields that fall into these edge bins for any of the three metrics as implausible (a total of 1,675 inferred velocity fields), and we focus the rest of our analysis on the remaining \bp s, the plausible sample (totaling 1,367 velocity fields).
(In Section~\ref{sec:emd-comp-but-should-we}, we discuss the many ways that a \bp 's evolution might violate the assumptions made by our EMD approach and thereby cause poor agreement within our targeted range.)

\begin{figure}[t]
	\centering
	\includegraphics{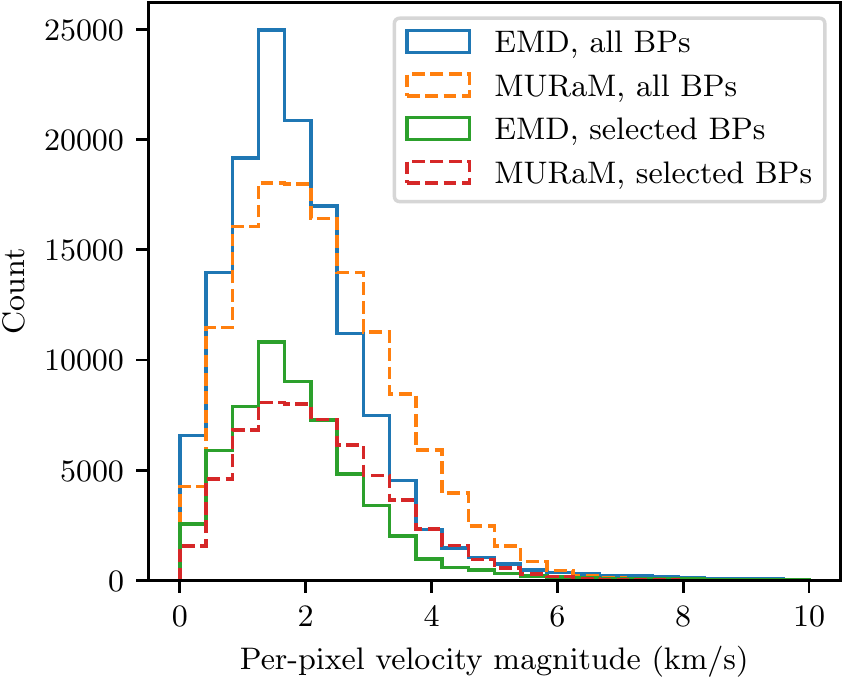}
	\caption[Per-pixel velocity magnitudes for \muram\ and EMD velocity fields]{Histogram of per-pixel velocity magnitudes for \muram\ and EMD velocity fields across all \bp s and time steps. The ``selected BPs'' distributions considers only those velocity fields in our ``plausible'' sample (see text). The \muram\ distributions show only those velocities at pixels within the boundary of the identified \bp\ (i.e., pixels where an EMD velocity has been computed) and only at the height (within our restricted range) at which the EMD--\muram\ agreement is best as measured by the absolute error metric.}
	\label{fig:emd-comp-vel-mags}
\end{figure}

In Figure~\ref{fig:emd-comp-vel-mags}, we show the distributions of per-pixel velocity magnitudes for both the EMD velocity fields and the \muram\ velocity fields within each feature and at that feature's height of lowest absolute error (within our restricted range).
It can be seen that the EMD and \muram\ distributions are very similar (with the EMD velocities overall being biased to slightly smaller values), indicating that our EMD approach is, in a sense, drawing from the correct distribution of velocity magnitudes.
It can also be seen that the sub-sample of plausible velocity fields and the corresponding sub-sample of \muram\ velocity fields are very similar, both to each other and to the corresponding distribution across all \bp s, suggesting that our concept of plausibility does not bias our sample in terms of velocity magnitudes.

We can now return to Figure~\ref{fig:emd-comp-line-plots}.
The left-hand column, as discussed previously, shows our error metrics across all \bp s as a function of the height at which the EMD--\muram\ comparison is made.
The center column shows the same for just our sub-sample of plausible velocity fields, and it can be seen that this selection process significantly improves the velocity field agreement near the photosphere while affecting the distribution of values very little at other heights.
Finally, the right-hand column provides a comparison where, for each velocity field, each per-pixel EMD velocity vector is replaced by a vector with a magnitude drawn randomly from the ``selected BPs'' distribution of EMD velocity magnitudes in Figure~\ref{fig:emd-comp-vel-mags}, and an orientation angle drawn randomly from a uniform distribution.
With these randomized replacement velocities filling the EMD velocity fields, the comparisons with the \muram\ velocity fields are repeated as before.
This shows that our EMD approach, both before and after removing our ``implausible'' subset, does much better than random chance at matching \muram\ velocities.

\subsection{Trends}
\label{sec:emd-comp-trends}

\begin{table}[t]
	\centering
	\begin{tabular}{l|ccc}
		& \multicolumn{3}{c}{Correlation coefficient $r$ with} \\
		Quantity & $E\abs$ & $E\rel$ & $A$ \\
		\hline
		BP Area & 0.29, 0.31 & 0.16, 0.21 & --0.22, --0.28 \\
		EMD velocity magnitudes & 0.45, 0.46 & 0.22, 0.24 & --0.01, --0.03 \\
		EMD velocity $\sigma$ & 0.56, 0.55 & 0.45, 0.45 & --0.23, --0.23 \\
		\muram\ velocity magnitudes & 0.41, 0.41 & 0.03, 0.07 & --0.08, --0.13 \\
		\muram\ velocity $\sigma$ & 0.67, 0.72 & 0.23, 0.29 & --0.36, --0.47 \\
		$\tau=1$ height $\sigma$ & 0.20, 0.22 & 0.09, 0.13 & --0.14, --0.16 \\
		Mean $B$ inclination & --0.37, --0.33 & --0.13, --0.09 & 0.06, 0.08 \\
	\end{tabular}
	\caption[Correlations between EMD--\muram\ error metrics and \bhp\ measurements]{Correlations between EMD--\muram\ error metrics and \bhp\ measurements. Each quantity and trend is discussed in the text. For each correlation, the first $r$ value given is computed with all EMD-inferred velocity fields, and the second is computed only with our ``plausible'' sub-sample. This choice of population causes only small changes in the $r$ values.}
	\label{table:emd-comp-correlations}
\end{table}

Having completed our EMD--\muram\ comparisons, we find some weak trends between our comparison metrics and various properties of the \bp\ or its surroundings.
We discuss these trends here, with values of the correlation coefficient $r$ presented in Table \ref{table:emd-comp-correlations}.
Velocity field agreement tends to improve for smaller \bp s, which might be expected as a smaller \bp\ does not have as much room for a complex flow field.
$E\abs$ is larger for \bp s with larger average velocity magnitudes, both \muram\ velocities and inferred EMD velocities.
$E\rel$ shows a weaker trend with EMD velocity magnitudes, and it shows no trend with \muram\ velocity magnitudes (which may be expected if the errors and the velocity magnitudes increase together, as suggested by the $E\abs$ trend).
$A$ shows no trend with velocity magnitudes, which is unsurprising since it considers only the velocity directions.
$E\abs$ increases strongly with the standard deviations of both the EMD velocities and the \muram\ velocities used in the comparison, with $E\rel$ and $A$ showing similar but weaker trends, indicating that our EMD approach struggles more with more spatially-variable velocities.
Variation of the height of the $\tau=1$ surface along the edge of the \bp\ might serve as a metric for both the accuracy of our identified boundary and (assuming the boundary is correct) the complexity of the plasma arrangement producing the \bp.
We find the error metrics worsen only slightly with increases in the standard deviation of the $\tau=1$ heights of the pixels immediately inside the identified \bhp\ boundary (with similar results when considering the pixels immediately \textit{outside} the boundary).
Errors increase with a decrease in the average inclination angle of the magnetic field relative to the horizontal (that is, errors are larger when the magnetic field is less vertical), with this trend much stronger in $E\abs$ than $E\rel$ or $A$.
This trend makes sense, as our EMD approach conceptually assumes a vertical flux tube.

We do not find trends between our error metrics and the amount of the \bp 's size change between frames, the divergence or curl of the \muram\ velocities in and immediately surrounding the \bp, the time variability of the \muram\ velocity field (measured through comparison to the velocities three frames or 6~s prior), the vertical derivative of the horizontal \muram\ velocities, the intensity contrast across the edge of the identified \bp, the average intensity throughout the \bp, the mean $\tau=1$ height along the edge of the \bp\ or throughout the entire \bp, or the magnetic field strength.

\subsection{On the Merits of Such Comparisons}
\label{sec:emd-comp-but-should-we}

Making these comparisons implicitly assumes that changes in the \bhp\ boundary are solely due to advection of the plasma at the boundary, and that the horizontal plasma velocities at one fixed height are solely responsible for that assumed advection---these are not trivial assumptions to make!
We discuss here a sample of the ways that the evolution of a \bhp\ boundary could violate these assumptions.

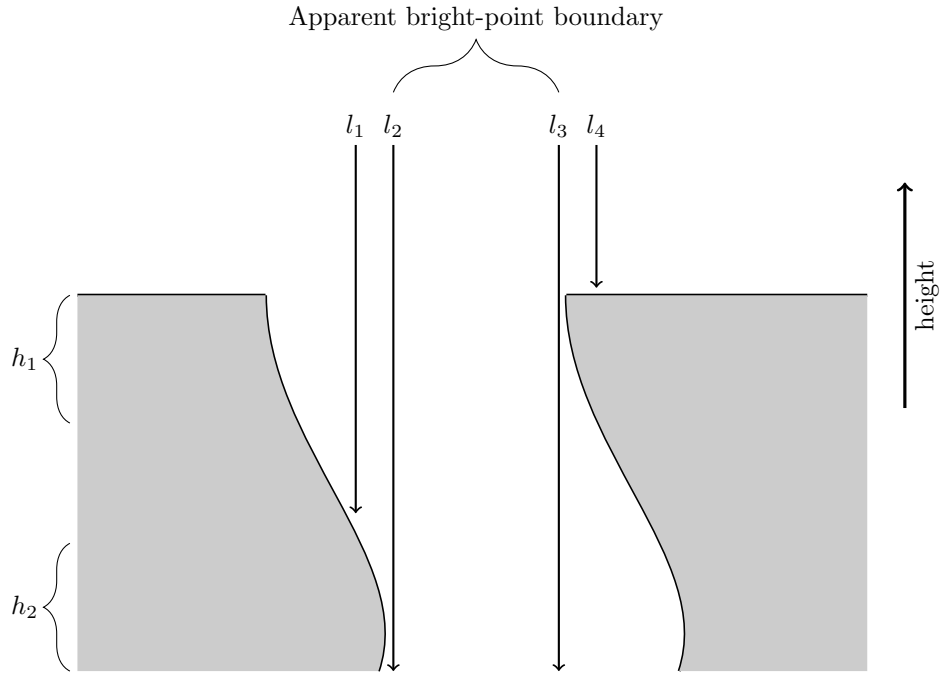
\begin{figure}[t]
	\centering
	\begin{tikzpicture}
		\coordinate (topleft) at (-4.5,5);
		\coordinate (bottomleft) at (-4.5,0);
		\coordinate (topright) at (6,5);
		\coordinate (bottomright) at (6,0);
		\coordinate (tubetopleft) at (-2,5);
		\coordinate (tubetopright) at (2,5);
		\coordinate (tubebottomleft) at (-.5,0);
		\coordinate (tubebottomright) at (3.5,0);
		\coordinate (cleft1) at (0,1.5);
		\coordinate (cleft2) at (-2,3);
		\coordinate (cright1) at (4,1.5);
		\coordinate (cright2) at (2,3);
		
		\draw[very thick] (tubebottomleft) .. controls (cleft1) and (cleft2)
			.. (tubetopleft);
		\draw[very thick] (tubebottomright) .. controls (cright1) and (cright2)
			.. (tubetopright);
		
		\draw[very thick] (topleft) -- (tubetopleft);
		\draw[very thick] (tubetopright) -- (topright);
		
		\fill[fill=gray!40] (tubebottomright) .. controls (cright1) and (cright2)
			.. (tubetopright) -- (topright) -- (bottomright) -- cycle;
		\fill[fill=gray!40] (tubebottomleft) .. controls (cleft1) and (cleft2)
			.. (tubetopleft) -- (topleft) -- (bottomleft) -- cycle;
		
		\draw[very thick,->] (6.5,3.5) -- (6.5,6.5)
			node[midway,rotate=90,below] {height};
		
		\draw[thick,->] (-.8,7) -- (-.8,2.1) node[pos=0,yshift=.25cm] {$l_1$};
		\draw[thick,->]  (-.3,7) -- (-.3,0) node[pos=0,yshift=.25cm] {$l_2$};
		\draw[thick,->] (1.9,7) -- (1.9,0) node[pos=0,yshift=.25cm] {$l_3$};
		\draw[thick,->] (2.4,7) -- (2.4,5.1) node[pos=0,yshift=.25cm] {$l_4$};
		
		\draw[decorate,decoration={brace, amplitude=20pt}] (-.3,7.7) -- (1.9,7.7)
		node [black,midway,above=20pt] {Apparent \bhp\ boundary};
		
		\draw[decorate,decoration={brace,amplitude=10pt}] (-4.6,3.3) -- (-4.6,5)
		node [black,midway,xshift=-0.6cm] {$h_1$};
		\draw[decorate,decoration={brace,amplitude=10pt}] (-4.6,0) -- (-4.6,1.7)
		node [black,midway,xshift=-0.6cm] {$h_2$};
	\end{tikzpicture}
	\caption[Cartoon depicting the formation of a \bhp\ boundary at varying heights]{Cartoon depicting the formation of a \bhp\ boundary at varying heights. This is a vertical slice through the very upper layers of a \bp. The shaded area represents the cool, dark plasma of the downflow lane surrounding the \bp\ and is assumed to be fully opaque. The central unshaded area is the upper portion of the flux tube, drawn with exaggerated curvature, and the overlying unshaded region lies above the average optical surface. All unshaded areas are assumed to be fully transparent. Below the bottom of the figure is the deeper, hotter plasma responsible for the bright appearance of the \bp. Marked are four lines of sight of an observer, $l_1$ through $l_4$, as well as two height ranges, $h_1$ and $h_2$. In the plane of the image, the observed \bp\ lies between $l_2$ and $l_3$.}
	\label{fig:emd-bp-sketch}
\end{figure}

The boundary of a \bp\ is not simply a disk of bright plasma embedded in the surrounding dark plasma, but rather it is a three-dimensional flux tube containing a region of low-density plasma that allows lines of sight to reach deeper, hotter, brighter layers of plasma.
The \bhp\ boundary is set by the transition between lines of sight reaching these deeper layers and lines of sight terminating at higher layers. 
To illustrate this, we model the region of low-density plasma as a cylinder which is pulled in different directions at different heights by external plasma flows, illustrated in Figure~\ref{fig:emd-bp-sketch}.
The boundary of the observed bright point along any one side (or portion of a side) is set at the height where the warped flux tube has the least outward extent.
For instance, along the left-hand side of the this \bp, the \bhp\ boundary is immediately to the left of line of sight $l_2$, as all lines of sight further to the left, such as $l_1$, terminate at cold, dark plasma.
Small, horizontal motions of the flux-tube boundary driven by horizontal plasma flows in the height range $h_1$ will not affect the apparent boundary of the \bp, since this boundary is determined within $h_2$ where the tube is narrowest.
Such motions will result in plasma flows without any corresponding motion in the apparent \bhp\ boundary.
On the other hand, small motions in $h_2$ can shift the left-hand boundary of the \bp, either by obstructing $l_2$ or by allowing $l_1$ to reach deeper layers.
This will result in \bhp\ boundary motion matching the plasma flow velocities.
However, on the right-hand side of the \bp, $h_1$ and $h_2$ switch roles, with the \bhp\ boundary sensitive to $h_1$ but not $h_2$.
Thus, even if the evolution of the \bhp\ boundary is driven purely by horizontal advection of the flux tube, different sides of the \bhp\ boundary could be affected by plasma flows at different heights, and advection of the flux tube will not always correspond with changes to the apparent boundary of the \bp.

Another possible violation of our assumptions is a case in which, instead of plasma flow driving the evolution of the flux-tube shape, the flux tube could drive the motion of the plasma.
Consider, for example, if the flux tube in Figure~\ref{fig:emd-bp-sketch} were to move vertically a distance comparable to the size of the figure while maintaining its shape.
The right-hand edge of the \bhp\ boundary would appear to move further right as the underlying flux-tube volume displaces the dense, dark plasma above it and allows $l_4$ to reach deeper layers.
If the displaced plasma flows to the right, then the plasma flow in $h_1$ would align with the apparent motion of the \bhp\ boundary.
However, if the displaced plasma flows vertically or into or out of the plane of the figure, the plasma flow would be transverse to the apparent boundary motion.
It is perhaps most likely that the plasma would flow in all of these directions, producing a diverging flow field with only some similarity to the linear boundary motion.

A modified version of this scenario may occur if a strong flow deforms the flux tube in between $h_1$ and $h_2$.
This may produce a strong magnetic tension force which causes similar motion of the flux-tube boundary within $h_1$ or $h_2$.
In this case, a diverging (or converging) flow similar to the one discussed previously might occur within $h_1$ or $h_2$, resulting in plasma flows just outside the flux tube which only partially resemble the motion of the \bhp\ boundary itself, while linear plasma motions more similar to the boundary motion occur at a different height and may not be fully aligned horizontally with the location of the observed \bhp\ boundary.
Building on this scenario, the observed \bhp\ boundary might be affected both by plasma flows at the height of the boundary and flows at other heights whose influence propagates along the flux tube to the height of boundary formation.

In all these scenarios, it must also be mentioned that when plasma motion is driven by motions of the flux tube (including, for example, compression of the tube), that motion may be vertical as well as or instead of horizontal, and any such vertical motions cannot be inferred by our EMD approach.
Vertical plasma flows are the primary plasma velocities associated with sausage-mode waves (as shown in Section~\ref{sec:sausage-modes}), and so these wave modes are largely undetectable by our EMD approach.

In addition to these geometric scenarios, a \bp\ might change shape if it merges with a smaller magnetic element, or a smaller element splits out of the main \bp\ \citep[such events are frequently observed; see, e.g.,][]{Berger1998,Iida2012,Liu2018}.
This smaller element might not form an associated \bp\ when it is an isolated feature, causing the smaller feature and the merging/splitting event to not be detected by our \bhp\ tracking.
The smaller feature also might form a \bp\ that is too small to be reliably detected by our tracking algorithm, or in observations it might form a \bp\ too small to be resolved by the telescope.
In a merging, for example, any one of these scenarios could lead to a growth in the apparent \bhp\ boundary as new flux is added, but any associated plasma flows would be external to the \bp\ and directed inward, toward the \bhp\ boundary, as they carry in the new flux.

More complex scenarios might involve changes to the properties governing radiative transfer through the \bp.
For example, evolution of the plasma and the convective flows outside the flux tube might change the radiative flux through the walls of the flux tube, increasing or decreasing the temperature of the plasma within the tube and altering the source function along the lines of sight reaching into the tube.
This may not be likely to rapidly turn a \bp\ dark, but it might cause a pixel at the edge of a \bp\ to cross a relevant threshold in our tracking algorithm and enter or exit the identified area of the feature, thereby moving the \bhp\ boundary.

Finally, in the present case of analyzing a \muram\ simulation at the pixel scale, we cannot discount the possibility that simulation inaccuracies and numerical and resolution effects might be disturbing the evolution of both the finest details of the velocity field and the exact location of the apparent boundary of the \bp, and this may introduce very difficult-to-predict variations between the plasma flows in the simulation and our EMD-inferred flows, as well as variations between the \bhp\ behavior seen in these simulated observations and what will be seen in DKIST observations.
The pixel scale in this simulation is very close to the expected diffraction limit of DKIST, and so we expect our analysis techniques to translate well to DKIST observations, where we expect that, under the telescope's keen eye, the Sun will allow the physics of \bp s to play out with full physical fidelity at all scales, free of any numerical effects.
Thus, final results of any EMD-based analysis must await appropriate observations, and we remind the reader that this study of simulated \bp s is meant as a proof-of-concept only.

\subsection{Estimated Energy Fluxes}
\label{sec:emd-energy-fluxes}

In Chapter~\ref{chap:ellipse-fitting}, we derived equations allowing us to move directly from time-series of moments calculated over \bhp\ shapes to an upward energy flux carried along the flux tubes by the waves excited by the motions in question.
(These fluxes were intended to account for the initial energy carried by these waves, before considering how each wave mode propagates or what fraction of flux tubes reach the corona.)
We seek to perform a similar calculation with our EMD velocity maps.
However, doing so will be hampered by the fact that our EMD approach cannot estimate vertical plasma velocities, which are the dominant component associated with sausage-mode waves, and it cannot constrain the plasma velocities outside the \bp, which are non-negligible for wave modes with small but non-zero values of $n$.
Additionally, while we still do not address the question of what fraction of the flux eventually reaches the corona, any future effort to do so will be hampered by the fact that our EMD velocities are expressed as raw and arbitrary velocity maps, rather than the analytical expressions associated with each wave mode in Chapter~\ref{chap:ellipse-fitting}.
Further, we do not even know for sure that every aspect of our velocity maps is associated with a propagating wave.

Nevertheless, we can compute an estimated energy flux as a benchmark value, knowing that it will not account for sausage-mode waves and that the degree of significant, upward propagation will remain ambiguous.
Such a benchmark value will serve as a metric by which these results can be compared to those of Chapter~\ref{chap:ellipse-fitting} and to observational estimates of kink-mode energy flux from \bp s, and as a way to assess the importance of further investigations of wave excitation by the shape changes of resolved \bhp\ observations.

In Chapter~\ref{chap:ellipse-fitting}, calculating the energy flux began at Equation~\eqref{eqn:epsilon} with
\begin{equation}
	\varepsilon(\vec{r}) = \frac{1}{4} \rho \left( \vec{v}(\vec{r}) \cdot \vec{v}^*(\vec{r}) \right),
\end{equation}
where $\varepsilon(\vec{r})$ is the instantaneous kinetic energy density carried by a flux-tube wave at a location $\vec{r}$, $\rho$ is the zeroth-order plasma density, and $\vec{v}(\vec{r})$ is the plasma velocity at $\vec{r}$.
Taking the same values for $\rho$ within and outside the flux tube as in Chapter~\ref{chap:ellipse-fitting}, this allows the direct calculation of the $\varepsilon$ (associated with a superposition of multiple wave modes) corresponding to the EMD velocity vector at each pixel.
As before, multiplying $\varepsilon$ by the group velocity of the waves produces an upward energy flux $F$.
Assuming that the complex waves excited by our EMD velocities are superpositions of the $n \ge 1$ thin-tube wave modes analyzed in Chapter~\ref{chap:ellipse-fitting}, which are dispersionless, the group velocity is the kink speed $c_k$, which is a constant defined in terms of the plasma density and \Al\ speed both within and outside the flux tube (values for which are specified in Chapter~\ref{chap:ellipse-fitting}).
As noted in Chapter~\ref{chap:ellipse-fitting}, for $n\ge1$ modes, the total integrated flux receives equal contributions from the $r<r_0$ and $r>r_0$ regimes (save for a varying factor of $\rho$).
Additionally, these expressions deal with kinetic energy, which is half of the total energy carried by linear MHD waves.
Accounting for these factors, we arrive at an expression for the instantaneous rate of transfer of energy (kinetic and otherwise) of
\begin{equation}
	\dot{E} = \int F \, dA = \frac{1}{2} (\rho_0 + \rho_e) V_{\rm gr} \int \left| \vec{v} \right|^2 dA \, ,
	\label{eqn:emd-flux}
\end{equation}
where the integral is taken over the area of the \bp.

For each identified \bp, and for each 10~s time step in our reduced-cadence sequence for that \bp, we use the per-pixel velocities to evaluate Equation~\eqref{eqn:emd-flux}, integrating over all pixels with an EMD-inferred velocity (that is, each pixel which is contained within at least one of the ``before'' or ``after'' shapes for that time step).
We take the resulting $\dot{E}$ values to be typical over the area of that \bp\ for the 10~s duration of that time step.
We integrate this transfer rate over all time steps and all \bp s, and then divide by the area and total duration of the simulation to produce a spatially- and temporally-averaged energy flux over the entire simulation domain.

To ensure the computed flux is directly comparable to those of Chapter~\ref{chap:ellipse-fitting}, we take two additional steps related to the temporal down-sampling in this chapter.
Each \bhp\ shape produced by this down-sampling is derived from five measured outlines, and it is associated with the time stamp of the central outline.
Additionally, after every block of five outlines is downsampled, if the total number of outlines is not divisible by five, the remaining outlines are discarded.
Integrating $\dot{E}$ from the first to the last time steps therefore involves a shorter integration range than integrating from the first to the last of the original sequence of \bhp\ outlines.
We therefore integrate the first $\dot{E}$ value of each \bp\ for an additional 4~s, and the last value between 4 and 14~s as required so that the total integrated lifetime is equal to that of the \bp\ before down-sampling.
Second, since our down-sampling factor of 5 requires each \bp\ to have a lifetime of at lest 10~time steps (as opposed to the 5~time steps listed with our tracking parameters in Table \ref{table:tracking_params}), we are analyzing a smaller ensemble of \bp s with the EMD technique (665 \bp s, versus 1,064 without the increased lifetime threshold).
To account for this in our averaged energy flux across the simulation domain, we calculate the sum of the lifetimes of the 1,064 \bp s in the original sample and the sum of the lifetimes of the 665 \bp s meeting the higher lifetime threshold, and we scale up our spatially-averaged flux by the ratio of these two sums.

The resulting energy flux is 24.2~kW~m$^{-2}$.
This value is comparable to the fluxes computed in Chapter~\ref{chap:ellipse-fitting}, and we will compare these values further in Chapter~\ref{chap:method-comp-and-conclusions}.
In calculating this number, we use all EMD velocity fields, not just our plausible sub-sample.
The average value of $\dot{E}$ across all \bp s and time-steps is $6.9 \times 10^{16}$~W for the plausible sub-sample (totaling 1,367 velocity fields) and $7.7 \times 10^{16}$~W for the velocity fields not in the sub-sample (totaling 1,675 fields).
This lack of a strong difference aligns well with the similarity of the velocity distributions in Figure~\ref{fig:emd-comp-vel-mags}.
Given the similarity between the two populations, as well as the inherent imprecision in the endeavor of computing energy fluxes for these EMD velocity fields, we do not attempt to account for the plausibility of velocity fields in computing our total energy flux.
Were we to do so, it would not be correct to simply ignore the implausible velocity fields, as the implausibility simply suggests that the \bhp\ evolution is more complex than the very simply model assumed by our EMD approach, and it does not suggest that there are no wave-driving motions present in the \bp 's evolution.
Instead, a possible approach might be to substitute the average energy flux of the plausible velocity fields for each implausible field (perhaps scaled by \bhp\ area), which would produce only a small effect in the total energy flux.

Our computed energy flux will be discussed further in Chapter~\ref{chap:method-comp-and-conclusions} and compared with the fluxes calculated in Chapter~\ref{chap:ellipse-fitting}.

\section{Summary}
\label{sec:emd-summary}

In this chapter, we described the earth mover's distance, a metric which compares two distributions by imagining them as describing piles of earth and considering how much effort would be required in optimally moving earth to rearrange one pile into the shape described by the other.
Viewing this motion of earth along optimal paths as a whole, it morphs the one distribution into the other as if it were a physical object, with the rearrangement more akin to the molding of clay than the shoveling of earth.
This rearrangement or morphing can serve as a first model of a vertically-homogeneous \bp\ evolving from one shape to another through advection.
We therefore apply the machinery of the earth mover's distance in calculating the optimal rearrangement strategy to our sample of \bp s in a \muram\ simulation as a proof-of-concept presentation.
We treat the shape of a \bp\ in two subsequent frames as two distributions, and we use off-the-shelf EMD algorithms to compute a mapping describing how algorithmic earth can be moved to morph the \bp\ from its shape in the first frame to the second, and from these mappings we produce velocity vectors over a regular grid.
This approach is meant to meet the goals and address the concerns discussed in Section~\ref{sec:need-new-techniques}.

Within the \muram\ simulation, to the extent that \bp s behave in the way this approach assumes, we expect to see plasma velocities similar to those produced by our EMD approach.
We do indeed see a level of agreement between EMD and \muram\ velocities.
The typical EMD velocity vector is within $45^\circ$ of the corresponding \muram\ velocity vector with a relative error in magnitude near unity, indicating the velocity amplitude is matched within a factor of two.
While this is not excellent agreement, we show that it is a good deal better than chance.
The relative error, for instance, is about half as large as would be obtained by chance.
This rough agreement appears impressive when considering the almost overwhelming degree to which we have simplified the processes by which observed \bp s can evolve.

We believe that the relative success we have obtained with our approach is acceptable as a first attempt at analyzing \bhp\ evolution at DKIST-like resolution.
We hope that this proof-of-concept will demonstrate the value of resolved \bhp\ evolution as a worthy target for future, observational study, and we hope that the method presented will inspire further development, both of this and other techniques.

\chapter{Method Comparison, Discussion, and Conclusions}
\label{chap:method-comp-and-conclusions}

In this chapter, we present cross-comparisons between our moment-fitting and EMD approaches to estimating horizontal velocities and the resulting wave-energy fluxes inside \bp s.
We compare both the methods themselves, as well as the results from applying these techniques to simulated images from the \muram\ simulation discussed in Section~\ref{sec:2s-muram}.

\section{Method Comparisons}

The moment and EMD methods take very different approaches to handling \bhp\ shape changes, each with its own strengths and weaknesses.
Using moments divides the inferred velocities into multiple components, each associated with a different wave mode.
Each of these wave modes is treatable analytically at and immediately above the photosphere, though modeling their propagation further into the solar atmosphere may require numerical integration.
Each wave mode will have different levels of height-dependent reflection and dissipation, and so calculating energy fluxes separately for each mode can directly facilitate modeling of this propagation.
However, limiting one's analysis to a fixed number of modes limits the analysis to working best when the actual \bhp\ motion is typical of those modes, and the results may become less connected to reality when other, more complex \bhp\ motions occur.
Likewise, the approach will struggle to interpret \bp s with more complex shapes that cannot be well-approximated as circular or elliptical (or otherwise, if higher-$n$ modes are added to the analysis).

The EMD approach is more capable of handling both shapes and shape-changes of arbitrary complexity.
However, this expanded capability comes at the cost of an algorithm connected less directly to the relevant wave physics, meaning the required assumptions are somewhat stronger.
Additionally, the output of the EMD approach is an arbitrary velocity field corresponding to some superposition of a wide range of wave modes, as well as some components that may not correspond to any propagating wave mode.
While estimating an upward energy flux at the level of the photosphere is possible (and we did so in Section~\ref{sec:emd-energy-fluxes}), estimating the degree of upward propagation is a much greater challenge.
Doing this for the arbitrary velocity fields produced by the EMD approach may require 3D, full-MHD simulations, which are much more demanding of the time of both the computer and the heliophysicist than single-wave-mode models and may have less generalizability from one set of observations to the next.
Additionally, were such simulations to be attempted, our approach does not provide any direct constraints on the plasma flows outside the \bp, meaning the EMD-inferred velocities provide only a partial boundary condition for the simulations.
Given these limitations, the EMD approach may find its place as a comparison point for the moment-based approach, as a separate type of measurement that can lend support to the approximate magnitude of the fluxes drawn from the moment analysis.
Since the EMD approach also facilitates direct comparisons with the \muram\ plasma velocities, and since the two approaches share a core assumption in how changes in the \bhp\ shape are connected to horizontal plasma velocities, the EMD approach's apparent level of validity may lend credence to the use of that same assumption by the moment-based approach.

\section{Flux Comparisons}

In Chapter~\ref{chap:ellipse-fitting}, we reported energy fluxes from our moment-based approach, averaged over the full spatial and temporal extent of the \muram\ simulation, of 10, 30, and 22~kW~m$^{-2}$ for the $n=0$, 1, and 2 modes, respectively, and a total flux 62~kW~m$^{-2}$.
The $n=1$ mode is clearly dominant, but adding the $n=0$ and 2 modes approximately doubles the total upward flux, and it is possible that this increase will be larger if additional wave modes were added to the analysis.

In Chapter~\ref{chap:emd}, we reported an estimated energy flux from our EMD approach of 24~kW~m$^{-2}$, which was meant to account for wave modes with $n\ge1$, as $n=0$ modes could not be estimated.
The EMD flux is very similar to the $n=2$ flux from the moment-fitting approach, and it is about half of the combined $n=1$ and $n=2$ moment-based fluxes\footnote{As discussed in Section~\ref{sec:emd-energy-fluxes}, the EMD approach cannot include $n=0$ waves in its estimates, so the appropriate comparison is to the $n=1,2$ fluxes only.} of 52~kW~m$^{-2}$, suggesting that the two approaches are making comparable (i.e., within a factor of two) estimates of the same physical process, despite the great differences between the two methods and their assumptions.
We believe this agreement allows each approach to lend some credence to the other, in that these two different attempts to account for wave modes other than and in addition to the $n=1$ mode produce estimates within a factor of 2.

These flux values can be put in context by comparing them to the upward Poynting flux in the \muram\ simulation.
We calculated this flux, averaged horizontally and across 19 data cubes evenly spaced through the simulation.
We found that this flux is negative immediately below the average $\tau=1$ surface, rises rapidly to a peak value of 33~kW~m$^{-2}$ at a height of 200~km above the photosphere, and drops to a value of zero at the simulation's upper boundary.
We interpret the negative fluxes below the photosphere as being associated with the strong convective downflows in the lanes where \bp s exist, as well as downward-propagating waves excited near the photosphere.
We interpret the drop-off toward the top of the simulation as damping by the upper boundary condition.
We therefore take the peak value of 33~kW~m$^{-2}$ as the most proper comparison point, where the influence of both convective downflows and boundary-condition damping is minimized.
When comparing the Poynting flux to the wave energy fluxes above, it is important to remember that the wave fluxes were arrived at by doubling the kinetic energy flux, since MHD-wave energy transport is half kinetic energy and half ``other'' or potential energy---the latter half includes magnetic and thermal energy \citep[see][]{Kulsrud2005}.
Appropriate values for the magnetic energy flux that can be compared with the Poynting flux are therefore at most half of the wave fluxes above (i.e. 5, 15, and 11~kW~m$^{-2}$ for the moment-based modes, and 12~kW~m$^{-2}$ for the EMD flux).
These values all fit within the Poynting flux budget (which is not expected to consist solely of flux tube waves above \bp s), indicating both that the order of magnitude of our fluxes is reasonable, and that the Poynting flux does not show our fluxes to be over-estimates.
Adding additional wave modes to the mode-based analysis might easily increase the total flux across wave modes to exceed the Poynting flux; however, the comparison values for the wave modes are upper-bound estimates of the magnetic portion of those waves---the true magnetic flux is likely less than the values presented here.

Our fluxes can also be contextualized by comparison to the energy flux required to maintain the coronal temperature.
These requirements are approximately 0.3~kW~m$^{-2}$ in quiet-Sun regions, 0.8~kW~m$^{-2}$ in coronal holes, and 10~kW~m$^{-2}$ in active regions \citep{Withbroe1977}.
For a proper comparison to these required fluxes, we must have a way to account for expected flux losses due to reflections as the waves propagate to the corona, as well as an estimate of the fraction of the energy flux expected to dissipate within the corona (versus continue out to the heliosphere).
As we discuss further in the next section, \citet{Cranmer2005} estimated that approximately 95\% of the $n=1$ mode flux does not reach the corona.
Even considering only 5\% of our reported values, our $n=1$ flux alone is more than sufficient to balance heat loss in the quiet-Sun corona, and the other wave modes only increase this ample source of energy.

In Chapter~\ref{chap:bp-centroids}, we computed an $n=1$ flux value from the observational \bhp\ centroid velocities of \citet{Chitta2012}.
Disregarding for now the factor used in that chapter to account for expected wave reflection, and using the $\rho$ and $B$ values of Chapter~\ref{chap:ellipse-fitting}, these observational velocities produce a comparison flux of 10.7~kW~m$^{-2}$ for the $n=1$ mode.
This value is comparable to, but about a third of, our estimated $n=1$ flux.
As discussed in Chapter~\ref{chap:bp-centroids}, possible explanations for this difference include the limited resolution of current observations of \bp s, inaccuracies in the \muram\ simulation, and differences in tracking algorithms, and these possible explanations can only be fully explored once DKIST observations become available.

Despite the similarity of the estimated total fluxes from the moment and EMD approaches, the factor-of-two difference between them remains, and it is difficult to say a priori which is more likely to be correct.
For small, localized changes to a \bp\ shape, the EMD approach may be more free to infer small and spatially-localized flows, whereas the moment-based approach can only produce plasma flows over the entire \bp, and so it may risk inferring a global oscillation from a local movement.
And for complex changes, the EMD approach is more capable of producing appropriate, complex velocity fields, whereas the moment approach is constrained to matching the complex shape-change with only a handful of wave modes relative to an assumed-circular flux tube, and the result may be somewhat detached from reality.
On the other hand, when multiple wave modes and frequencies are being driven simultaneously, the EMD approach might risk underestimating the energy transfer due to the constant, partial cancellation of different waves at any given pixel, whereas the moment-based approach may do better at accounting for all of the waves.
Given that the Poynting flux exceeds the upper bound for the magnetic component of the fluxes from both methods, it is even possible that both methods are under-estimates to different degrees.
A possible resolution to this question may come from comparing the Poynting flux above specific \bp s to the estimated wave fluxes of that \bp, as an attempt to isolate the Poynting flux specific to those wave motions, but we leave this project for future work.

\section{Comparison with Flux Elements}
As a final comparison, we also calculated wave-mode fluxes from the magnetic flux-element tracking of Section~\ref{sec:fe-tracking}.
From this data set, we compute an $n=0$ flux of 11.4~kW~m$^{-2}$, an $n=1$ flux of 46.8~kW~m$^{-2}$, and an $n=2$ flux of 53.8~kW~m$^{-2}$.
These values are 113\%, 155\%, and 236\% of their respective \bhp\ fluxes.
If each flux value is divided by the number of pixels contributing to that flux, to control for the differing numbers and sizes of identified features (i.e., dividing the \bhp\ fluxes by the total number of \bp\ pixels, and likewise for the flux elements), the ratios of these normalized fluxes are 89\%, 122\%, and 187\%, respectively.
From these numbers, one can see that the $n=0$ flux is comparable (within nearly 10\%) across these two data sets.
The $n=1$ flux is 55\% higher in the flux-element data, and just over half of this increase is accounted for by larger number of tracked pixels.
The $n=2$ flux varies most strongly between the two data sets, by nearly a factor of two even after accounting for the differing tracked area.
This may suggest that the $n=2$ mode, and potentially $n>2$ modes as well, are most sensitive to the variation between the two data sets (perhaps indicating that the identified flux elements and \bp s vary more significantly in shape than in size or location).
However, it is important to note that our flux-element tracking received less focus and scrutiny than the \bhp\ tracking that has been more central to this work, and so this higher flux may be due to some amount of ``jitter'' or other effects that can be mitigated by further refinement of the flux-element tracking.

\section{Effect of Resolution and Blurring the Data}
\label{sec:blurring-data}

A relevant question in this ``preparing for DKIST'' context is how our analyses change if the \muram\ images are degraded to a resolution more comparable to pre-DKIST observations.
We investigate this by convolving the \muram\ images with an Airy disk of radius 100~km (or 6.25 pixels).
We run our \bhp\ tracking on these blurred images, finding through manual inspection of a sample of \bp s that no modifications to our algorithm or its parameters are required for satisfactory tracking of the blurred \bp s.
The number of tracked and accepted \bp s is reduced by a factor of 3.3, from 1,064 to 324 \bp s.
The \bp s that are not detected in the blurred data tend to be difficult or impossible to detect by eye in the blurred data.
The average \bhp\ lifetime increases by a factor of 1.2, from 42~s to 52~s, and the average \bhp\ area increases by a factor of 1.6, from 36~px to 57~px, suggesting that it is predominantly small and short-lived \bp s that are lost.
This interpretation is confirmed by the size ratios of individual \bp s.
The 324 tracked \bp s in the blurred data account for 8,261 individual, single-frame outlines.
Of these outlines, 7,246 can be unambiguously linked to a single outline in the unblurred tracked results---that is, the blurred feature overlaps exactly one unblurred feature, and that unblurred feature overlaps exactly one blurred feature.
Among these one-to-one matches, the geometric mean of the ratios of the size of a \bp\ as detected in the blurred and unblurred tracking is 1.03, indicating that blurring does not typically affect the identified size of a \bp.
The increase in average \bhp\ size can therefore be interpreted as a loss of small \bp s.
(Only a third of the 324 \bp s in the blurred data show a one-to-one match that is consistent across their entire lifetime, so we do not attempt a similar analysis for the changes in lifetimes.)

\begin{figure}[t]
	\centering
	\includegraphics{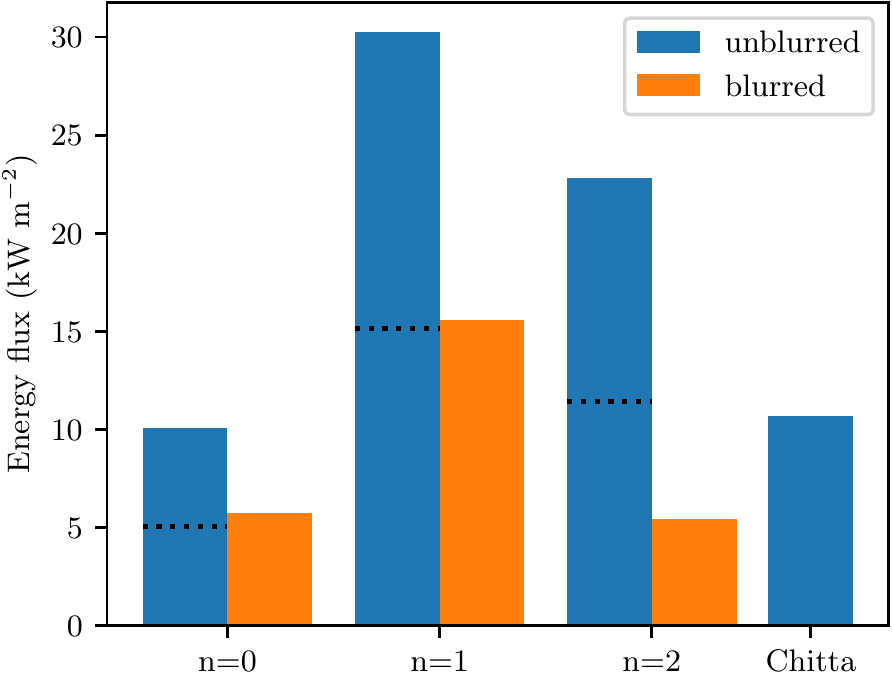}
	\caption[Energy fluxes for the $n=0-2$ wave modes computed from blurred and unblurred images]{Energy fluxes for the $n=0-2$ wave modes computed from blurred and unblurred images. Also shown is the comparison observational flux of \citet{Chitta2012}. As a guide to the eye, dotted lines indicate half of each of the unblurred flux values.}
	\label{fig:blurring-fluxes}
\end{figure}

Following the methods described earlier, we compute energy fluxes for the $n=0-2$ wave modes using our blurred tracking results.
This reveals a mode-dependent reduction in flux, shown in Figure~\ref{fig:blurring-fluxes}.
The $n=0$ flux is reduced from 11 to 5.8 kW~m$^{-2}$, the $n=1$ flux from 30 to 16 kW~m$^{-2}$, and the $n=2$ flux from 22 to 5.4 kW~m$^{-2}$.
This blurring produces an $n=1$ flux much more comparable to the value of 10.7 kW~m$^{-2}$ we compute from the observations of \citet{Chitta2012}, suggesting that much of the increased $n=1$ flux we detect in the \muram\ simulations is due to the presence of smaller \bp s that are only detectable in the higher-resolution data.
The flux reduction is approximately 50\% for the $n=0,1$ modes, but it is approximately 75\% for the $n=2$ mode, suggesting that this latter mode is most sensitive to the finer details of the \bhp\ shape.
This may be expected, as this mode is more dependent on the shape of the \bp\ than the other two modes.

\section{Next Steps}

At least three factors must be considered before moving forward with energy fluxes estimated by our techniques applied to DKIST data.

First is the ``true'' filling factor of \bp s.
Our tracking algorithm does not capture every \bp, and not every strong-field region corresponds to a visible \bp---this can be seen, for example, in Figure~\ref{fig:tracking-bp-sample} or \ref{fig:tracking-fe-sample}.
This means that we may be underestimating the total wave energy flux.
This can be resolved, in part, by further work refining the \bp\ tracking, but that still will not account for all vertical magnetic flux.
Our identified \bp s account for 42\% of pixels with $|B_z| > 1000$~G, 14\% of pixels with $|B_z| > 500$~G, and 4\% of all unsigned vertical flux.
(This distribution is shown in Figure~\ref{fig:tracking-pixel-distributions}.)
Thus, an ad-hoc correction factor of $1/0.42$ might be considered to approximately account for the energy flux in strong-field regions not included in the identified \bhp\ ensemble---regions which might be expected to experience wave driving very similar to the \bp s we measure.
An alternative way to approach this problem is to consider the differences in the fluxes computed from \bhp\ tracking and flux-element tracking as a rough estimate of the methodological uncertainty present in each flux value, including, in part, the variability due to the difficulty in identifying all flux tubes.

Second, one must assess the degree to which flux tubes reach the corona, as opposed to bending over and re-entering the photosphere (e.g., through an opposite-polarity \bp).
In regions with mixed magnetic polarity, the latter scenario may be expected to occur with some frequency.
Flux tubes with a large vertical extent may therefore be more commonly rooted in large, unipolar network regions.
(The \muram\ simulation we use does not include the supergranular network, and so this is an effect we are unable to test.)
An estimate of the fraction of \bp s reaching the corona might be attained by using a potential field extrapolation from measurements of the average magnetic field within individual \bp s \citep[see, e.g.,][]{Close2003,VanBallegooijen2003,Wiegelmann2004}.

Third, the degree of propagation of these waves (accounting for both reflection and damping) must be understood before these estimated wave fluxes can be considered in a coronal context.
For $n=1$ flux-tube waves, \citet{Cranmer2005} modeled the wave propagation from the photosphere, through the heights where the flux tubes are expected to merge into a ``canopy,'' through the rapidly-changing plasma properties of the chromosphere and transition region, and into the corona and heliosphere.
The degree of wave reflection can be treated by a reflection coefficient $\mathcal{R}$, with the net upward energy flux $F$ multiplied by a factor of $(1-\mathcal{R}) / (1+\mathcal{R})$.
\citet{Cranmer2005} found the value of $\mathcal{R}$ to be approximately 0.9 for propagation up to the transition region, indicating that the $n=1$ wave flux above the transition region is reduced to about 0.05 times its value at the photosphere.
(\citealp{Soler2019} find a similarly strong reduction in flux for the torsional \Al\ wave, a mode we do not consider in the present work.)
This factor cannot be directly transferred to the $n=0$ or $n=2$ modes (nor to the mixed-mode flux estimate from our EMD approach), but it may be expected that these other modes will also have a reduced fraction of their flux succeed in reaching the corona.
The modeling work required to address this question is beyond the scope of this thesis.

\section{Conclusions}
\label{sec:shapes-conclusions}

In this thesis so far, we have focused on preparing for DKIST observations of \bp s, which promise to resolve the shapes of \bp s at a level never before achieved.
These \bp s play a key role in one common model of coronal heating, which is MHD waves which are excited in flux tubes by convective buffeting and which are expected to propagate to and partially dissipate in the corona, delivering heat.
These resolved shapes can provide new insight on this wave driving---particularly the driving of waves other than $n=1$ or kink mode waves.
However, doing so requires analysis techniques beyond the centroid tracking that has traditionally been applied to unresolved \bp s to study $n=1$ waves.

Chapter~\ref{chap:bp_tracking} presented a revision of the \bhp\ tracking algorithm of Chapter~\ref{chap:bp-centroids}.
These revisions were focused on more robustly identifying the location of the \bhp\ edge, and we showed that these revisions largely eliminated the centroid jitter concerns raised in Section~\ref{sec:jitter}.
We also motivated the broad approach of the following chapters, which is to estimate the driving of waves in \bhp-associated flux tubes strictly from changes to the identified outline of the \bp.
This focus on the outline is due to the fact that the intensity structure within a \bp\ is not guaranteed to correspond to anything relevant to wave driving.
The \bhp\ edge, meanwhile, can be taken as a proxy for the boundary of the flux tube, and so changes in the \bhp\ shape (as defined by the location of its boundary) can be taken as a proxy for the evolving cross-sectional shape of the flux tube, which is associated with wave motion.
This is the core assumption that drives our approach.
Observational magnetic field information is a compelling alternative method for identifying the cross-sections of flux tubes, though in DKIST observations it will come with small trade-offs in resolution or cadence.
Nevertheless, it deserves attention in future work.

In Chapters \ref{chap:ellipse-fitting} and \ref{chap:emd}, we described our moment-based and EMD-based approaches, two distinct techniques for analyzing \bhp\ shape changes, and we demonstrated them on simulated observations from a \muram\ simulation of DKIST-like resolution.
While beginning with the same key assumptions, these techniques take very different approaches to connecting changing \bhp\ shapes to energy fluxes.
Despite these differences, they produced energy flux estimates within a factor of two of each other.
Taking these as quasi-independent estimates of the energy flux, this similarity allows each approach to lend support to the other.
Despite our estimated fluxes being notably larger than observational estimates of $n=1$ fluxes derived from centroid tracking, the magnetic component of our fluxes lies comfortably below the Poynting flux computed in the simulation, indicating a level of plausibility, and the velocity fields produced as an intermediary step by our EMD approach have distributions of magnitudes very similar to those seen in the simulation's plasma velocities.

The energy flux estimated by our moment-based approach is broken down by wave mode, with the $n=1$ mode being the only mode to receive significant study in observations to date.
Our results show that this mode is clearly dominant over the $n=0$ and $n=2$ modes, but together these latter modes equal the energy flux of the $n=1$ mode.
This indicates that these $n\ne1$ modes (potentially including $n>2$ modes not studied here) may make significant contributions to the wave-energy heating budget of the corona, and that the full energy budget of this heating mechanism may have been underestimated, or at least not fully observationally-supported, in studies to-date.

These claims must be taken with caution, of course, since they derive from analysis of simulations.
However, as we soon enter the new world of DKIST observations, our proposed techniques stand ready to be applied to these new data.
Future analyses of wave excitation in \bp s as seen by DKIST may lend credence to the claims we make here.
Alternatively, if observational fluxes are found to be lower than those presented here, they may motivate further development of photospheric simulations.
Either way, a new world of resolved \bhp\ observations awaits!

\paragraph{Acknowledgements}
We thank Matthias Rempel for making available the results of his \muram\ runs, Piyush Agrawal for assisting in acquiring those data files, and colleagues for fruitful discussions of this work in its early stages.
The work in Chapters \ref{chap:bp_tracking}--\ref{chap:method-comp-and-conclusions} was funded by NASA FINESST grant number 80NSSC20K1503 and National Science Foundation (NSF) grant number 1613207.

\biblio

\chapter{Granular Flicker in Stellar Observations}
\label{chap:flicker}

The following is the full text of my manuscript published in the Astrophysical Journal \citep{VanKooten2021} as it was when accepted for publication, with light adaptations for the formatting requirements of this thesis.

\section{Introduction}
\label{sec:kep-intro}

While primarily intended for exoplanet discovery, the \kep\ mission's long-duration light curves with high photometric precision have proven extremely valuable for investigation of the mission's target stars as well.
These stellar studies are important in their own right, but they also provide a benefit back to exoplanet science, as a better understanding of the stellar flux allows the exoplanetary signal to be more precisely separated from the stellar signal and allows the uncertainty in the process to be better described.

One topic of interest for many investigators has been short-period variation (called \textit{flicker}) in the \kep\ light curves of F, G, and K stars with convective surface-layers.
\citet{Bastien2013,Bastien2016} defined the quantity \fe, the root mean square (RMS) amplitude of the flicker that occurs on $<8$~hr timescales, and showed that it displays a very strong dependence on stellar surface gravity.
The evolving pattern of granulation, the convectively-driven warm and cool regions at the photosphere that is well-known in solar observations, has been shown to be a very plausible and likely driver of \fe\ \citep{Cranmer2014}, with the gravity dependence explained by the very strong dependence of granular size- and timescales on surface gravity.
Indeed, granulation and magnetic activity taken together have been shown to be sufficient to very closely reproduce the entirety of solar photometric variability (aside from the characteristic 5~min signal of acoustic oscillations) over timescales from minutes to decades \citep{Shapiro2017}.

One application of \fe\ measurements or related analyses is as a photometric proxy for surface gravity \citep{Bastien2013,Bastien2016,Pande2018} or stellar density \citep{Kipping2014}, especially where spectroscopic or asteroseismic measurements may not be available.
A related approach is to measure the \textit{timescales} of flicker, rather than the amplitude, as a proxy for surface gravity \citep{Kallinger2016}.
This approach is more resilient to noise and, when used to measure surface gravity, has the advantage that flicker timescales are more directly tied to surface gravity than are flicker amplitudes.
However, the approach requires that flicker timescales be temporally resolved, which is not always the case for \kep\ long-cadence data (the data available for the vast majority of \kep\ targets), where the 30~min cadence fails to resolve, for instance, G-dwarf granular timescales on the order of 10~min.
An additional approach that has been demonstrated is to use machine learning to extract stellar parameters such as surface gravity from the power spectra of stellar variability \citep{Sayeed2020}.

Another motivation for understanding stellar flicker is to constrain the noise present in planetary transit measurements.
Studies have shown that stellar granulation produces a non-negligible effect in the noise present during observations of planetary transits, introducing, e.g., uncertainty in planetary radius measurements of a few percent \citep{Chiavassa2017,Morris2020}, or even up to 10\% \citep{Sulis2020}, depending on the observation being modeled.
One cause of this noise is that the granulation provides one of many deviations from a perfect black body, affecting signals especially during spectroscopic observations.
Another is that the granular pattern, and thus the integrated flux of the star, varies during a transit (compare transit timescales of ~hours to granular timescales of $\sim10$~min for Sun-like stars).
A third is that the transit depth varies because the planet occults only small portions of the stellar disk, which consist of a mixture of granules and intergranular lanes in a ratio that varies from one local patch to another.
From a related, stellar perspective, preliminary simulation work from \citet{Bonifacio2018} has shown that neglecting the effect of granulation can introduce errors when inferring effective temperatures from photometric colors, as much as 200~K in the most extreme cases.
When mitigating these sorts of flicker-induced uncertainties, one option is to use techniques such as Gaussian Process regressions to fit and account for the stochastic flicker signal on a star-by-star basis \citep{Pereira2019}.

Efforts have been made to model the properties of granular flicker empirically \citep[e.g.][]{Corsaro2017,Tayar2018}.
\citet{Samadi2013a,Samadi2013b} derived an analytic model for flicker, and past work has carefully compared this model with observations and sought to reconcile the differences \citep{Cranmer2014}.
In this work we improve upon this modeling effort, and we compare the model with a much larger sample of stars with measured flicker (approximately 30 times as many stars).
We describe this larger sample in Section~\ref{sec:data}.
In Section~\ref{sec:model} we present the complete \fe\ model, including our additions.
In Section~\ref{sec:kep-results} we compare the model predictions to the observations, noting improved agreement with our model corrections.
We also pay special attention to the solar value of \fe, and we discuss the correlation, or lack thereof, of \fe\ with other observables.
In Section~\ref{sec:discussion} we discuss the remaining difference between observations and model predictions, and we finally summarize our work in Section~\ref{sec:conclusions}.

\section{Observational Data}
\label{sec:data}

\begin{figure}[tp]
	\centering
	\includegraphics[width=\textwidth]{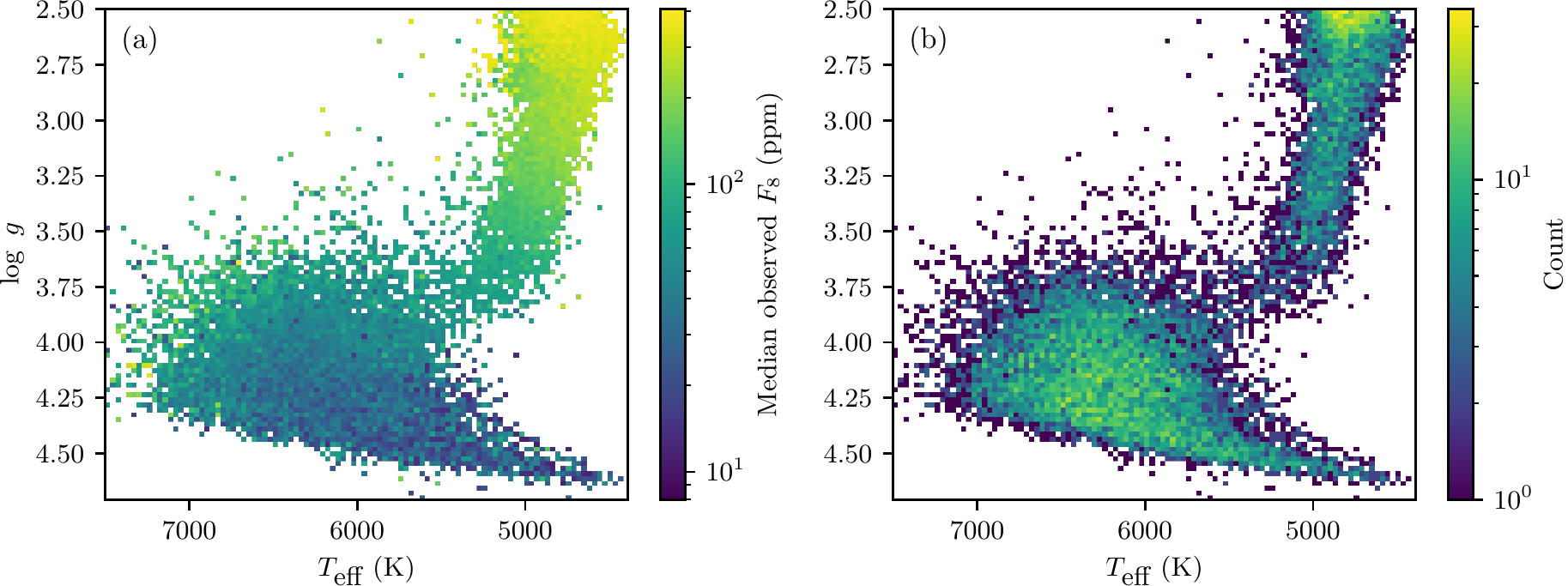}
	\caption[Our sample of 16,992 stars with both \fe\ measurements and \feh\ measurements]{Our sample of 16,992 stars with both \fe\ measurements and \feh\ measurements as described in Section~\ref{sec:data}. Both panels are 2D histograms showing, in panel (a), the median observed \fe\ per bin, and in panel (b), the number of stars per bin. Note that \logg\ is strongly influenced by a star's radius and therefore luminosity, and so the \logg\ axis is effectively a skewed luminosity axis and is plotted to preserve the traditional H-R diagram orientation. The main sequence runs along the bottom edge of the data cloud, and the red giant branch ascends toward and beyond the upper-right corner.}
	\label{fig:flicker_data}
\end{figure}

For this analysis we assembled a data catalog from a variety of sources, totaling 16,992 F, G, and K stars from the \kep\ catalog, shown in Figure~\ref{fig:flicker_data}.
This improves upon the past work of \citet{Cranmer2014}, which used a smaller sample of 508 stars with very few K dwarfs.
These stars have effective temperatures \Teff\ that range from 4390 to 7500 K, with \logg\ values ranging from 2.5 to 4.7 and typical field-star values of \feh\ ranging from --1 to +0.5.

We assembled our catalog by beginning with the 27,628 \fe\ values of \citet{Bastien2016}.
Since their work focuses on using \fe\ as a proxy for \logg, they report \logg\ values derived from their \fe\ measurements.
We therefore recover the original, observed (and corrected and calibrated) \fe\ values from their published \logg\ values by inverting their (monotonic, one-to-one) \fe --\logg\ relation\footnote{We note that, while the 16-pt RMS values used to calculate \fe\ are also reported for each star by \citet{Bastien2016} we have been informed via private communication that errors were inadvertently introduced in those values when preparing their table, with an erratum forthcoming. This motivates our more indirect route to recover the observed \fe\ values.}.
These \fe\ measurements, which measure the RMS amplitude of the portion of a star's variability that occurs on sub-eight-hour timescales, were extracted from \kep 's long-cadence, PDC-MAP light curves by subtracting from each light curve a smoothed version of itself, using an 8 hr smoothing window \citep[following][]{Basri2011,Bastien2013}.
Additional steps in their pipeline remove transient events such as flares and transits, subtract out the portion of the \fe\ measurement attributable to shot noise, and account for \kep\ pointing offsets and flux contamination from neighboring stars.

We merge this \fe\ catalog with the \Teff, \logg, and mass values of \citet{Berger2020}, which cover all but 848 of the stars with measured \fe\ values (a total of 26,780).
These values, part of a catalog intended to provide a comprehensive source of stellar parameters for \kep\ targets, are drawn from spectroscopic observations combined with \textit{Gaia} parallaxes and modeled isochrones.
The use of these \logg\ values from an independent source, rather than using those of \citet{Bastien2016}, ensures some independence between the \logg\ and \fe\ values we use.

Next, we merge into our catalog \feh\ metallicity data from the LAMOST-Kepler project \citep{Zong2018}, an effort to use the LAMOST telescope for spectroscopic follow-up observations of \kep\ targets.
This catalog covers an evenly-distributed 18,773 of the 26,780 \kep\ stars in both the \citet{Bastien2016} and \citet{Berger2020} catalogs, and we limit our analysis to this smaller sample.
While \feh\ values are reported by \citet{Berger2020}, their values are derived from multiple sources and they describe pipeline-to-pipeline variation as their dominant source of uncertainty for \feh.
We thus restrict ourselves to this one source to ensure a higher level of consistency.
We convert from \feh\ to heavy-element mass fraction $Z$ with a reference solar value of $Z\sun = 0.01696$ \citep{Grevesse1998}.

Finally, we remove all stars with $\logg < 2.5$, as the granular timescales for these large giants extend well beyond the 8-hour window that determines \fe\ \citep{Bastien2016}, and a handful of outlier stars with $\Teff > 7500$~K.
This removes 1,701 stars, yielding the final sample of 16,992 stars which is shown in Figure~\ref{fig:flicker_data}.
This final catalog is included in our code and data archive \citep{VanKooten2021_Zenodo}.

\section{Granulation Model}
\label{sec:model}

\citet{Samadi2013a,Samadi2013b} derived a theoretical model predicting the RMS amplitude $\sigma$ of granular flicker as a function of a star's effective temperature \Teff, surface gravity \logg, and mass $M$.
The model combines first-principles geometrical arguments, analytic derivations, and scaling relations, with some components further fit to numerical simulations.
Our use of this model closely follows that described by \citet{Cranmer2014}, including the conversion factor between the model-predicted $\sigma$ and the observational value \fe\ (i.e. the total RMS granular flicker amplitude versus the RMS amplitude of the flicker component occurring over $<8$ hour scales).
However, in this work we show that the model can be cast in more absolute terms, rather than as a scaling relation relative to the solar $\sigma$.
We also present updated functions for predicting: (1) the Mach number of near-surface, vertical plasma flows, (2) the relative temperature contrast between granular centers and lanes, and (3) the characteristic size of granules.
We also add a correction factor for the influence of \kep 's bandpass on observed \fe\ values.

Here we present in full the version of the model used in this work.
Our Python code implementing this model and producing our plots is included in our code and data archive \citep{VanKooten2021_Zenodo}.

\subsection{Bolometric Flicker}
\label{sec:bolo-amplitude}

\citet{Samadi2013a} derive an expression for $\sigma_\tau$, the RMS amplitude of the bolometric intensity variation in a stellar light curve due to granulation seen at optical depth $\tau$, of
\begin{equation}
	\sigma_\tau = \frac{12}{\sqrt{2}} \sqrt{\frac{\tau_g}{N_g}}\;\Theta\RMS^2,
	\label{eqn:sigma_tau}
\end{equation}
where $\tau_g$ is the characteristic optical thickness of granules (which is very nearly constant across the F, G and K dwarfs and giants used in this work), $N_g$ the average number of granules covering the visible half of the star, $\Theta\RMS$ is the RMS of the instantaneous temperature contrast $\Theta \equiv \Delta T / \left< T \right>$, $\left< T \right>$ is the average photospheric temperature, and $\Delta T \equiv T - \left< T \right>$ is the difference from the mean of the photospheric temperature at any one location.
\citet{Samadi2013a} also provides the expressions
\begin{align}
	N_g &= \frac{2\pi R_s^2}{\Lambda^2} \label{eqn:N_g} \\
	\tau_g &= \kappa\rho\Lambda \label{eqn:tau_g},
\end{align}
where $R_s$ is the radius of the star, $\kappa$ is the Rosseland mean absorption coefficient, $\rho$ is the mean photospheric density determined as in Section~\ref{sec:mach}, and $\Lambda$ is a characteristic granular size\footnote{This characteristic size is used as a horizontal size in Equation \eqref{eqn:N_g} and as a vertical size in Equation \eqref{eqn:tau_g}. This is because both sizes are of the order of the pressure scale height. The horizontal case is discussed in Section \ref{sec:Lambda}; for further discussion of the vertical case, see \citet{Trampedach2011}, among others.} described in Section~\ref{sec:Lambda}.

With Equations \eqref{eqn:N_g} and \eqref{eqn:tau_g} along with the expression $R_s^2 = GM/g$, Equation~\eqref{eqn:sigma_tau} can be written as
\begin{align}
	\sigma &= 6 \sqrt{\frac{\kappa\rho\Lambda^3}{\pi R_s^2}}\;\Theta\RMS^2 \\
	&= \frac{6}{\sqrt{\pi}} \sqrt{\frac{\kappa\rho g}{GM}} \;\Lambda ^{3/2} \;\Theta\RMS^2. \label{eqn:sigma}
\end{align}
The subscript $\tau$ has been removed, as granulation is seen in a small region around a single optical depth of $\tau \sim 1$, and so the observed, total fluctuation amplitude $\sigma$ can be taken to be $\sigma_{\tau=1}$.
In past work, the expression for $\sigma$ was written in terms of the observable $\nu\max$, the peak frequency of p-mode oscillations assumed to scale as $\nu\max \propto g / \sqrt{T}$; however, we omit this step in the present work.
Additionally, \citet{Samadi2013b} compared their version of this expression to $\sigma$ values measured in numerical models and found the fit could be slightly improved by raising the expression to the power 1.10.
However, we omit this step to to remain closer to a model derived from first principles.
The present expression for $\sigma$ differs from past work most notably in that it is presented in absolute form, rather than as a scaling relation normalized to a solar $\sigma$ value.

\subsection{Granular Size}
\label{sec:Lambda}

\citet{Samadi2013a,Samadi2013b} define the characteristic granular size $\Lambda \equiv \beta H_p$ as proportional to the pressure scale height $H_p$, which is itself a function of \Teff\ and \logg.
The proportionality constant $\beta$ is a free parameter of the model.
By comparison of the modeled and observed power spectra of granular flicker, appropriate values of $\beta$ are shown to fall approximately in the range 3--15, depending on how the modeled spectra is constructed.
(This range is in agreement with the grid of simulations of \citet{Magic2013}, which finds this proportionality constant to be $\sim 5$.)

In this work, we instead use the $\Lambda(\Teff,\logg)$ scaling of \citet{Trampedach2013a},
\begin{equation}
	\log \frac{\Lambda}{[\text{Mm}]} \simeq 1.3210\;\log \Teff - 1.0970\;\logg + 0.0306,
	\label{eqn:Lambda}
\end{equation}
which those authors produced by fitting granular size measurements taken from a grid of numerical simulations.
They note that this functional form provides a better fit than a simple proportionality with $H_p$, as they find differing best-fit values of $\beta$ to be appropriate across the H-R diagram, with values ranging from 9 to 13 and generally increasing from dwarf stars to more evolved stars.
We note that a similar trend can be found in the \kep\ \fe\ measurements: if we take the final and completed model for \fe\ described through this section but use $\Lambda = \beta H_p$, and if we divide the \kep\ sample into 2D bins in \Teff--\logg\ space and determine the value of $\beta$ that minimizes the RMS error of model-predicted \fe\ values within each bin, then the resulting best-fit $\beta$ values are near 8--10 for main-sequence stars and rise to 12--20 for giants, echoing the trend in $\beta$ values observed in the simulations of \citet{Trampedach2013a} and supporting the use of the alternative functional form of Equation~\eqref{eqn:Lambda}.

Some solar observations have found two populations of granules \citep[e.g.][]{Abramenko2012}.
The granule sizes of \citet{Trampedach2013a} are measured by finding the peak of the 2D spatial power spectra of granular images, and so their analysis includes both granular populations.
These two populations are divided in size: the large granules have diameters in a Gaussian distribution of approximately $1.2\pm0.5$~Mm and are believed to be traditional convective cells, whereas the smaller granules (or granule-like visible features) follow a decreasing power-law distribution in diameter, with typical sizes under 0.5~Mm.
This population of smaller features may have its origin in turbulent processes at the solar surface rather than convection \citep{VanKooten2017}, meaning they may follow different distributions in timescale $\tau_c$ and temperature contrast $\Theta\RMS$ than those assumed in the present work's model.
\citet{Abramenko2012} quantify the importance of a given size of granule to the overall appearance of the photosphere with the area contribution function (ACF), defined as the ratio of the total area of granules of a given size to the total area available, and this should serve as a good proxy of the influence of granules (or granule-like features) of a given size on \fe.
While those authors report a bimodal ACF distribution, with a peak in equivalent diameter near the typical granular size of 1.2~Mm and a second peak near 0.5~Mm, we note that when the ACF is computed using their fitted size distribution, which does not deviate significantly from the observed distribution, it produces only a single peak near 1.2~Mm.
We are thus unable to make a clear determination on the degree, if any, to which this population of small, granule-like features contributes to \fe.
In this work we do not attempt to account for these smaller features.

We also note that the scaling of \citet{Trampedach2013a} produces a value for solar parameters $\left(\Teff=5770~\textrm{K}, \logg=4.438\right)$ of 1.35~Mm, which is in good agreement with observational values of the typical solar granular size (e.g. the range around 1.2 Mm of \citet{Abramenko2012}).

\subsection{Determining the Temperature Contrast}
\label{sec:theta}

An expression for $\Theta\RMS$ must be determined before this model can be used.
Mixing length theory (MLT) predicts that $\Theta\RMS$ is proportional to the square of the Mach number \Ma.
Motivated by this, \citet{Samadi2013b} fit a quadratic polynomial to the $\Theta\RMS$ and \Ma\ values measured in a grid of numerical simulations.
We note, however, that their fitted polynomial is a concave-down parabola (i.e. $\Theta\RMS \propto -\Ma^2$), as opposed to the concave-up parabola (i.e. $\Theta\RMS \propto \Ma^2$) expected by MLT.
Additionally, the simulation measurements themselves that were used for this fit appear to show a possible trend of flattening off at larger \Ma\ values, despite a general adherence to an approximate $\Theta\RMS \propto \Ma^2$ scaling for low \Ma.

\begin{figure}[t]
	\centering
	\includegraphics[width=0.5\linewidth]{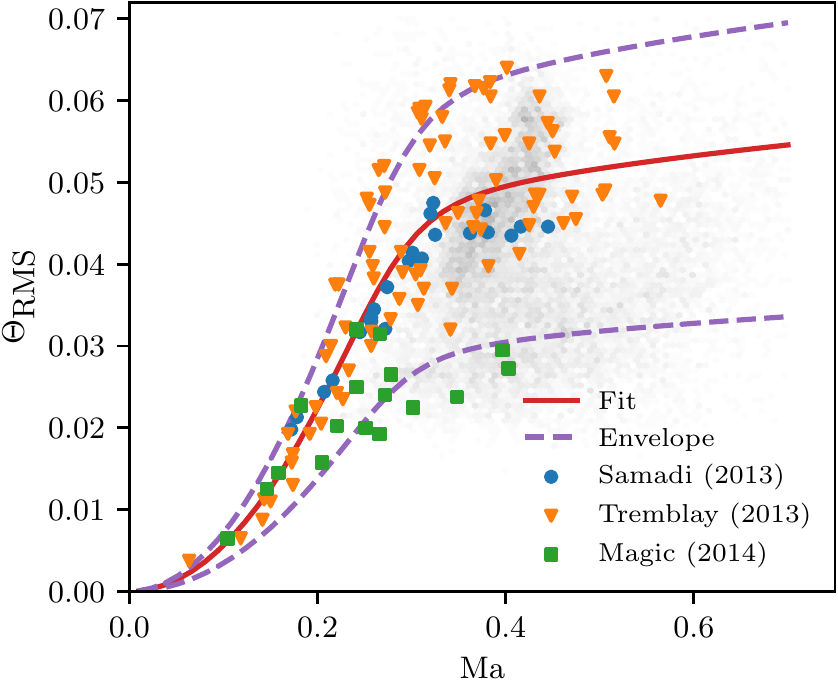}
	\caption[$\Theta\RMS$ as a function of Mach number \Ma]{$\Theta\RMS$ as a function of Mach number \Ma. Dots mark measurements from the three sources of numerical simulations (see Section~\ref{sec:theta}). Lines mark our central fit to all points as well as our envelope fit to the upper and lower bounds of the point cloud. The faint gray background is a 2D histogram of the Kepler stars (of Section~\ref{sec:data}) provided for reference, plotting the value of $\Theta\RMS$ required for our model to reproduce the observed \fe, and \Ma\ as computed from Equation~\eqref{eqn:ma}.}
	\label{fig:ma-theta}
\end{figure}

For this work, we use an expanded data set, with $\Theta\RMS$ and \Ma\ measurements from additional simulations \citep{Tremblay2013,Magic2014}, shown in Figure~\ref{fig:ma-theta}.
The expanded set of simulations notably provides a wider range of $\Theta\RMS$ values that are consistent with any given \Ma\ value.
To account for this, we produce both a central fit to the data set as well as a fitted envelope encapsulating this spread.
The envelope allows the model to produce a range of possible $\sigma$ values for any given star, all consistent with the range of $\Theta\RMS$ values produced by the numerical simulations.
We use the functional form
\begin{equation}
	\Theta\RMS = \frac{1}{(A_1 \Ma ^ {-2c} + A_2 \Ma ^ d)^{1/c}},
	\label{eqn:theta-ma-fit}
\end{equation}
which is able to act as the $\Theta\RMS \propto \Ma^2$ predicted by MLT for small values of \Ma\ while more freely fitting the data points at higher \Ma, which may be beyond the applicability of ideal MLT.
We produce a central-fit curve $\Theta_\text{central}(\Ma)$, shown in Figure~\ref{fig:ma-theta}, by fitting the complete set of data points from all three numerical experiments, producing the coefficients $A_1=21.0$, $A_2=3.54\times10^6$, $c=5.29$, and $d=-0.842$.
To produce the upper bound of our envelope we identify by hand a set of points representing the largest values of $\Theta\RMS$ predicted for any given $\Ma$, and we find a constant scaling factor which, multiplying $\Theta_\text{central}$, best fits that subset of points.
This produces $\Theta_\text{upper}(\Ma)=1.27\;\Theta_\text{central}(\Ma)$.
Determination of the lower bound of the envelope is less clear, since there is a sharp transition in the lowest-reported $\Theta\RMS$ values near $\Ma=0.4$.
Following the same method as for the upper bound, while focusing on matching the low-$\Theta\RMS$ points for $\Ma>0.4$, produces $\Theta_\text{lower}(\Ma)=0.82\;\Theta_\text{central}(\Ma)$; focusing on the low-$\Theta\RMS$ points for $\Ma<0.4$ produces $\Theta_\text{lower}(\Ma)=0.62\;\Theta_\text{central}(\Ma)$.
We choose the latter option, which produces a larger envelope and is more inclusive of the range of $\Theta\RMS$ values seen in simulations.

While use of this envelope fit may seem ad-hoc, we believe it has some justification.
In Figure~\ref{fig:ma-theta}, the faint gray background represents an attempt to position our \kep\ star sample in the plot.
For these stars, the \Ma\ is that computed from Equation~\eqref{eqn:ma}, and the $\Theta\RMS$ value is that which would be required for our model to reproduce the star's observed \fe.
The spread seen in these empirical $\Theta\RMS$ values is very comparable to that seen in the simulations and which we capture with our envelope fit, meaning that this level of spread appears plausible (though its origin is unclear---see Section~\ref{sec:discussion}).

When viewing the \kep\ sample in Figure~\ref{fig:ma-theta}, it can be seen that the simulations do not span the full range of \Ma\ values that we produce for the \kep\ stars.
This mismatch is explainable by the fact that these simulation grids were not designed with our particular sample in mind.
The highest \Ma\ values for our sample correspond to the hottest stars, near 6500--7000~K, while the simulation grids stop at slightly cooler stars.
Despite this, the bulk of our \kep\ sample is well-covered by these simulations.

Using Mach numbers computed as described in the following section, our computed $\Theta\RMS$ values increase toward higher temperatures along the main sequence, in agreement with the numerical results of \citet{Salhab2018}.
As an additional test, in Appendix~\ref{appendix:hinode} we describe our own determination of a solar value for $\Theta\RMS$ from \textit{Hinode}/SOT observations.
We find a value of $\Theta\RMS=0.0457$, which exceeds some other observational values but does have some support in existing literature.
This value compares well with the value predicted by our central-fit model for solar parameters, which is $\Theta\RMS=0.0455$.

\subsection{Determining the Mach Number}
\label{sec:mach}

\begin{figure}[t]
	\centering
	\includegraphics[width=0.5\linewidth]{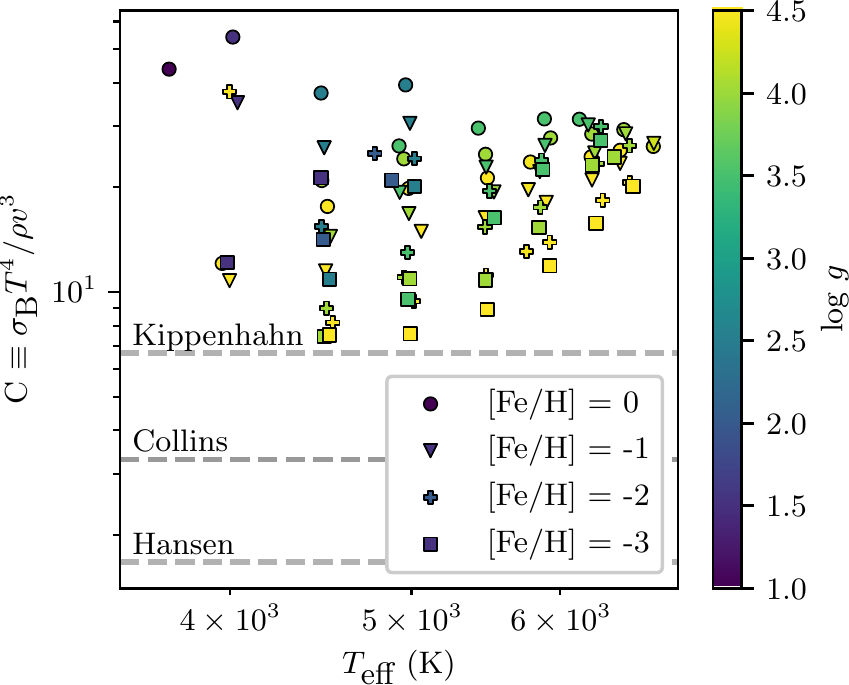}
	\caption[Computed $C$ values for the \citet{Tremblay2013} simulation grid]{Computed $C$ values for the \citet{Tremblay2013} simulation grid. The grid stars have one of four discrete values of \feh. Also indicated are three $C$ values predicted by various formulations of mixing-length theory (see text).}
	\label{fig:C-cf-ml}
\end{figure}

\begin{figure*}[t]
	\centering
	\includegraphics[width=\linewidth]{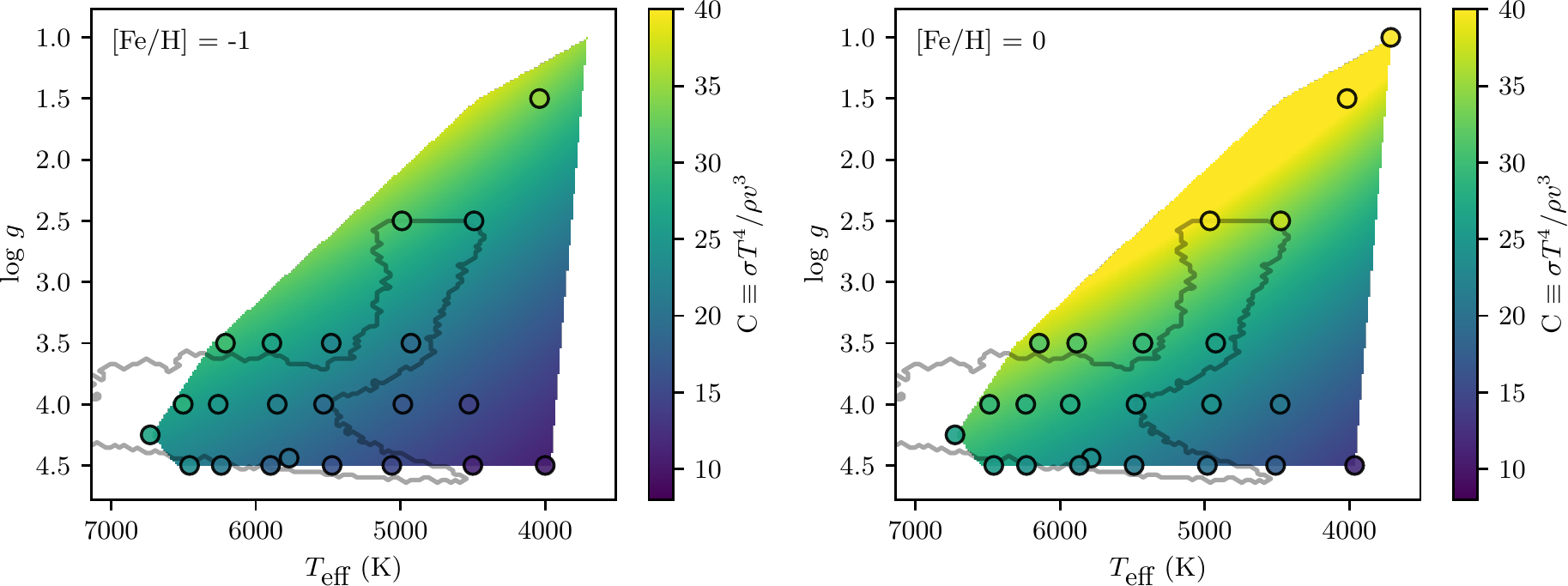}
	\caption[Our fitted function for $C$ as a function of \Teff, \logg\, and \feh]{Our fitted function (Equation~\eqref{eqn:fitted-C}) for $C$ as a function of \Teff, \logg\, and \feh\ (see Section~\ref{sec:mach}). The colored, circular points represent the $C$ values calculated for simulations in the grid of \citet{Tremblay2013}. These simulations have one of only four discrete values of \feh\ (-3, -2, -1, and 0); we show here simulations and our fit for the two \feh\ values within the range of our dataset ($\feh=-1$ on left and $\feh=0$ on right), whereas all four are used when producing our fit. Interpolation and extrapolation through our fit produces $C$ values for our \kep\ stars, which have typical \feh\ values in the range $[-0.75, +0.5]$. The gray outline in these figures indicates the location of our \kep\ data set, provided as a quick reference.}
	\label{fig:C-fit}
\end{figure*}

A value for the Mach number \Ma\ (which we define in terms of the RMS vertical velocity $v$) must now be determined.
We start by assuming that some fraction of the total stellar flux is carried by convection, and we adopt a standard mixing-length theory expression for the convective flux:
\begin{equation}
	\sigma_{\rm B} \Teff^4 = C \rho v^3,
\end{equation}
where $\sigma_{\rm B}$ is the Stefan-Boltzmann constant, $\rho$ and $v$ are the plasma density and RMS vertical plasma velocity, and $C$ is a constant of proportionality.
Density $\rho$ is a function of \Teff, \logg, and metallicity $Z$, and the dependence on metallicity is an addition over the model as employed by \citet{Cranmer2014}.
We determine density \citep[following][]{Cranmer2011} by finding an interpolated value for the Rosseland mean opacity $\kappa_\textrm{R}$ from an AESOPUS data grid \citep[see][]{Marigo2009}, computing a photospheric value for the density scale height $H_\rho$ as in \citet{Cranmer2011}, and solving for $\rho$ after setting the photospheric optical depth $\tau = \kappa_\textrm{R}\rho H_\rho = 2/3$.

The Mach number \Ma\ is then
\begin{equation}
	\Ma \equiv v / c_s = \frac{\left(\sigma_{\rm B} \Teff^4 / C \rho\right)^{1/3}}{c_s},
	\label{eqn:ma}
\end{equation}
where $c_s$ is the stellar surface sound speed defined as $c_s^2 = 5 k_B T / 3 m_H \mu(T)$, where $k_B$ is the Boltzmann constant, $m_H$ is the mass of a hydrogen atom, and $\mu(T)$ is the mean atomic weight for which we use the fitted expression
\begin{equation}
	\mu \approx \frac{7}{4} + \frac{1}{2} \tanh \left( \frac{3500 - \Teff}{600} \right)
	\label{eqn:mu}
\end{equation}
of \citet{Cranmer2011}.

We require \Ma\ near the stellar surface ($\tau=2/3$) where the MLT expression may not be fully valid, and we account for this through the $C$ parameter.
To determine this proportionality constant, we turn to the simulation grid of \citet{Tremblay2013}, for which the surface values for the plasma density and Mach number are reported.
This allows a value of $C$ to be inferred for each simulated star in the grid, which we show in Figure~\ref{fig:C-cf-ml}.
(We note that the reported Mach numbers are the RMS velocity amplitude.
We rescale these values by $1/\sqrt{3}$ to produce vertical-component Mach numbers.
This scaling is supported by our own analysis of $\tau=1$ slices from a \muram\ solar-surface simulation \citep[see][]{Rempel2014}, in which the space- and time-averaged velocity components $v_x$, $v_y$, and $v_z$ are in near-exact equipartition.)

Of note is that all of these computed $C$ values exceed those predicted by multiple MLT formulations.
In Figure~\ref{fig:C-cf-ml}, we mark three predicted $C$ values.
The first, derived from the formulation of \citet{Kippenhahn2012}, is $C=4/(\alpha \nabla_\textrm{ad})$, using the standard MLT quantities the mixing-length parameter $\alpha$ and the dimensionless adiabatic temperature gradient $\nabla_\textrm{ad}$.
With $\alpha=1.5$ and $\nabla_\textrm{ad} = 2/5$, this gives $C=6.67$.
The second, of \citet{Collins1989}, is $C=2/(\alpha \nabla_\textrm{ad})$, yielding $C=3.3$ using the values above.
The third, due to \citet{Hansen2004}, is $C=1/(\alpha \nabla_\textrm{ad})$, giving $C=1.67$.
The fact that all these MLT predictions fall short of our computed $C$ values may be due to the breakdown of MLT assumptions at the stellar surface, where convective flows must come to a halt.

We fit these $C$ values with the function
\begin{equation}
	C = 6.086\times10^{-4} \; \Teff^{1.406} \; g^{-0.157} \; \left( Z / Z_\odot \right)^{0.0975},
	\label{eqn:fitted-C}
\end{equation}
where \Teff\ is expressed in K and $g$ is expressed in cm~s$^{-2}$.
Figure~\ref{fig:C-fit} shows how this function compares with the simulation grid, and we go on to use this function to calculate $C$ values for all stars when computing Mach numbers.

As a point of reference, for solar parameters ($T=5770$~K, $\logg=4.438$) this fit produces $C=23.648$.
This corresponds to a solar \Ma\ of 0.32 (or 2.6~km~s$^{-1}$).
Observational values of vertical velocities at the solar photosphere range from 1 to 3~km~s$^{-1}$ \citep[][and references therein]{Oba2017a}, with the range in values possibly due in part to observational limitations, and possibly due in part to a strong gradient in the vertical velocity seen in numerical simulations near the photosphere \citep{Fleck2020}, raising the possibility that different observations sample different portions of this stratification.
Nonetheless, the predicted solar \Ma, while high, is consistent with at least some observations.
Since the prediction that $\Theta\RMS$ scales with $\Ma$ is based on mixing-length theory, which does not hold at the surface of the convection zone, it may be appropriate that our model predictions for $\Theta\RMS$ are driven by relatively higher \Ma\ values that correspond to layers slightly deeper than the photosphere where MLT holds more strongly.

We note that this \Ma\ calculation includes the effect of metallicity at two points, in the calculation of $\rho$ and $C$.
Over the range of \feh\ values in our catalog (approximately $-1 < \feh < 0.5$) and for otherwise solar parameters, increasing metallicity causes a decrease in $\rho$ by a factor of approximately 3 over the full range of metallicity values.
The Mach number, however, has a minimum near $\feh=-0.75$, with an increase of approximately 20\% up to $\feh=0.5$ but a much slower increase toward lower metallicity.
The corresponding effect on \fe\ is very comparable to that in \Ma.

\subsection[Limiting to $<$8 hr Timescales]{Limiting to $<8$~hr Timescales}
\label{sec:sigma-to-f8}

Once a value of $\sigma$ (the RMS flicker amplitude over all timescales) is calculated, it must be converted to the observational \fe\ (the RMS flicker amplitude over the $<8$ hr timescales dominated by granulation).
We follow \citet{Cranmer2014}, which assumed a Lorentzian function for the granular power spectrum and derived the relation
\begin{align}
	\label{eqn:fe-sigma}
	\frac{\fe}{\sigma} &= \CBP \sqrt{1 - \frac{2}{\pi} \tan^{-1}\left(4\;\tau\eff\;\nu_8\right)},
\end{align}
where $\nu_8 \equiv \left( 8\;\;\text{hr} \right)^{-1}$.
The factor \CBP\ has been inserted in this work to represent our bandpass correction term (see Section~\ref{sec:bandpass}).
We follow \citet{Samadi2013b} in defining $\tau\eff$, the characteristic timescale of granulation, as $\tau\eff = \Lambda / v$, where $\Lambda$ is the characteristic granular size of Section~\ref{sec:Lambda} and $v = \Ma c_s$ is the characteristic vertical plasma velocity.
In the present work, $\Lambda$ and $\tau\eff$ can be computed directly, and so we are not required to compute $\tau\eff$ as a scaling relation relative to a solar value, as was done in past work.
Nevertheless, the model-predicted $\tau\eff$ for solar parameters is 8.8~min, in line with observational values for granular lifetimes in the range 7--9~min \citep{Nesis2002}.

Excluding \CBP, the conversion factor tends to be very close to 1 for dwarf stars, and tends toward 0.5 for giants.

\subsection{Bandpass Correction}
\label{sec:bandpass}

We add to our version of this model a \kep\ bandpass correction factor, given that our observational \fe\ values are \kep-derived.
At the conceptual level, this accounts for the fact that the blackbody spectra of different portions of a stellar photosphere (i.e. granules versus lanes) have different amounts of overlap with the \kep\ bandpass.
For cooler K-type stars with spectra nearer to the long-wavelength end of the bandpass, the bandpass will pass a smaller fraction of the light from the cooler lanes than from the relatively warmer granules, dimming the lanes relative to the granule centers and exaggerating the effective temperature contrast when viewed in terms of measured intensity.
Conversely, for warmer F-type stars nearer to the short-wavelength end of the bandpass, the warmer granules will be dimmed relative to the lanes, reducing the effective temperature contrast.
(We find the effect to be much more pronounced for K stars than F stars.)
Analytically, this effect alters the $B = \sigma_{\rm B} T^4$ scaling inherent in \citet{Samadi2013a}'s derivation leading to our Equation~\eqref{eqn:sigma_tau}, producing a much stronger dependence on temperature when a star's peak wavelength of emission is near either end of the bandpass.
In Appendix~\ref{appendix:bandpass} we derive the correction factor \CBP\, which takes values of $\sim1$ (for F-type stars) to $\sim2$ (for K-type stars) and is nearly monotonic in effective temperature (see Figure~\ref{fig:sigma_correction_factor}). 

\section{Results}
\label{sec:kep-results}

In this section we compare the observed and model-predicted \fe\ values, describe how the changes we made to the model affect that comparison, and discuss possible ways to interpret the final ratio of these two values.

\subsection{Initial Comparisons}
\label{sec:initial-comp}

First, as a simple benchmark, we compute the quantity $\Delta$, defined as
\begin{equation}
	\log_{10} \Delta \, \equiv \, \rm{RMS} \left[ \log_{10} \left( \frac{\feobs}{\femod} \right) \right],
\end{equation}
which measures the typical multiplicative factor separating each star's observed and model-predicted \fe\ values.
(For example, a value of $2$ would indicate the typical star is either over- or under-predicted by a factor of $2$, with mispredictions in either direction contributing equally to the overall score.)
Calculated across all stars in our catalog, we find a value of $2.50$ when using the predictions of the model of \citet{Cranmer2014} (without their empirical correction factor), and a decreased value of $2.02$ when using the model presented in the present work.
Perfect model--observation agreement would produce a value of $1$, so this decrease represents a 32\% reduction in the typical, relative prediction error.
As an additional metric, if an ad-hoc correction factor were to be introduced that multiplies all model predictions, $\Delta$ is minimized when that factor is $0.66$, with $\Delta=1.75$.
(To be clear, we do not propose such a correction factor in this work.)

\begin{figure}[tp]
	\centering
	\includegraphics[width=0.5\linewidth]{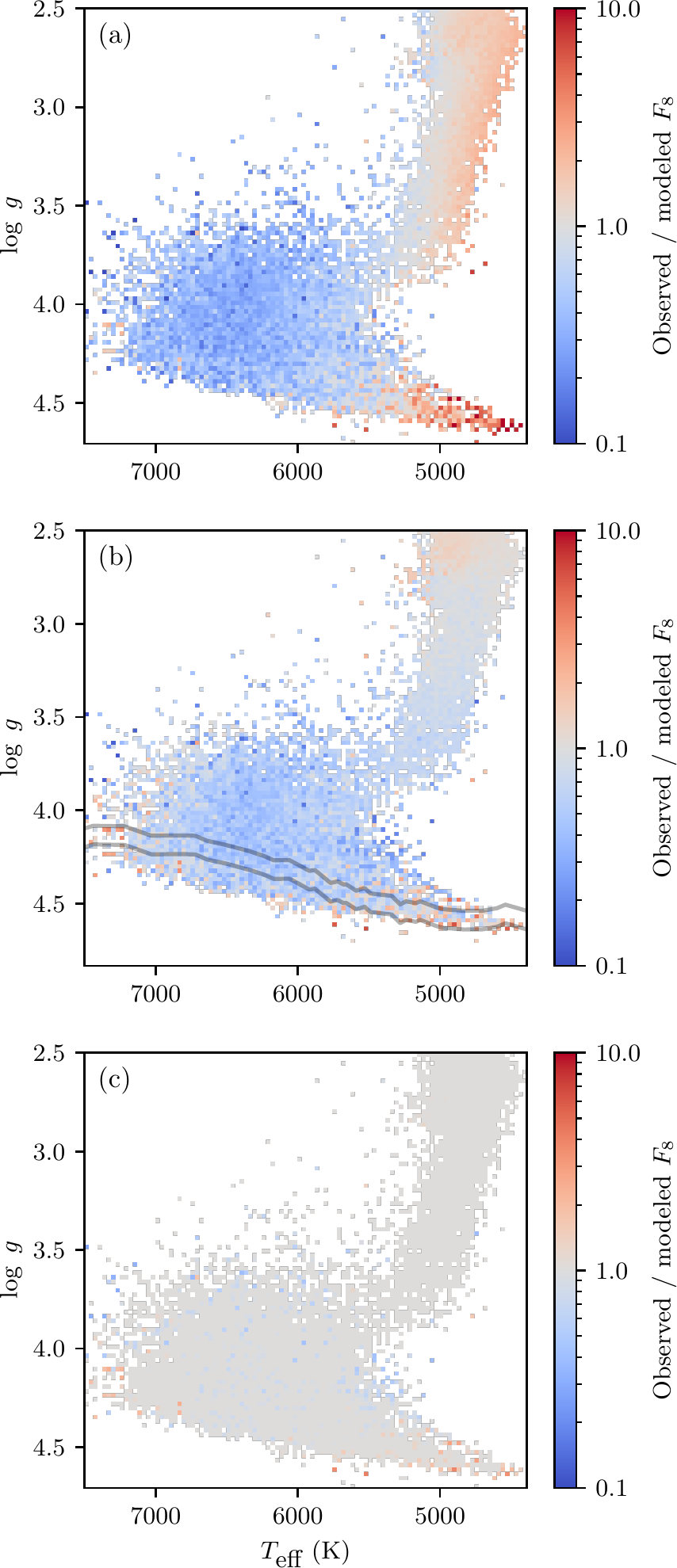}
	\caption[Two-dimensional histograms showing the median ratio of observed to modeled \fe]{Two-dimensional histograms showing the median ratio of observed to modeled \fe, (a) using the model as in \citet{Cranmer2014} (without their empirical correction factor), (b) using our model, and (c) using our model with the full envelope of possible $\Theta\RMS$ values for each predicted \Ma. That is, the ratio is shown as 1 if any of the possible $\Theta\RMS$ values reproduce the median observed \fe\ in a histogram bin, and otherwise the ratio is between the observed \fe\ and the nearest of the possible model predictions. The dark lines in (b) mark the main-sequence slice we use in Figure~\ref{fig:ms-slice}.}
	\label{fig:before-and-after}
\end{figure}

\begin{figure*}[t]
	\centering
	\includegraphics[width=\linewidth]{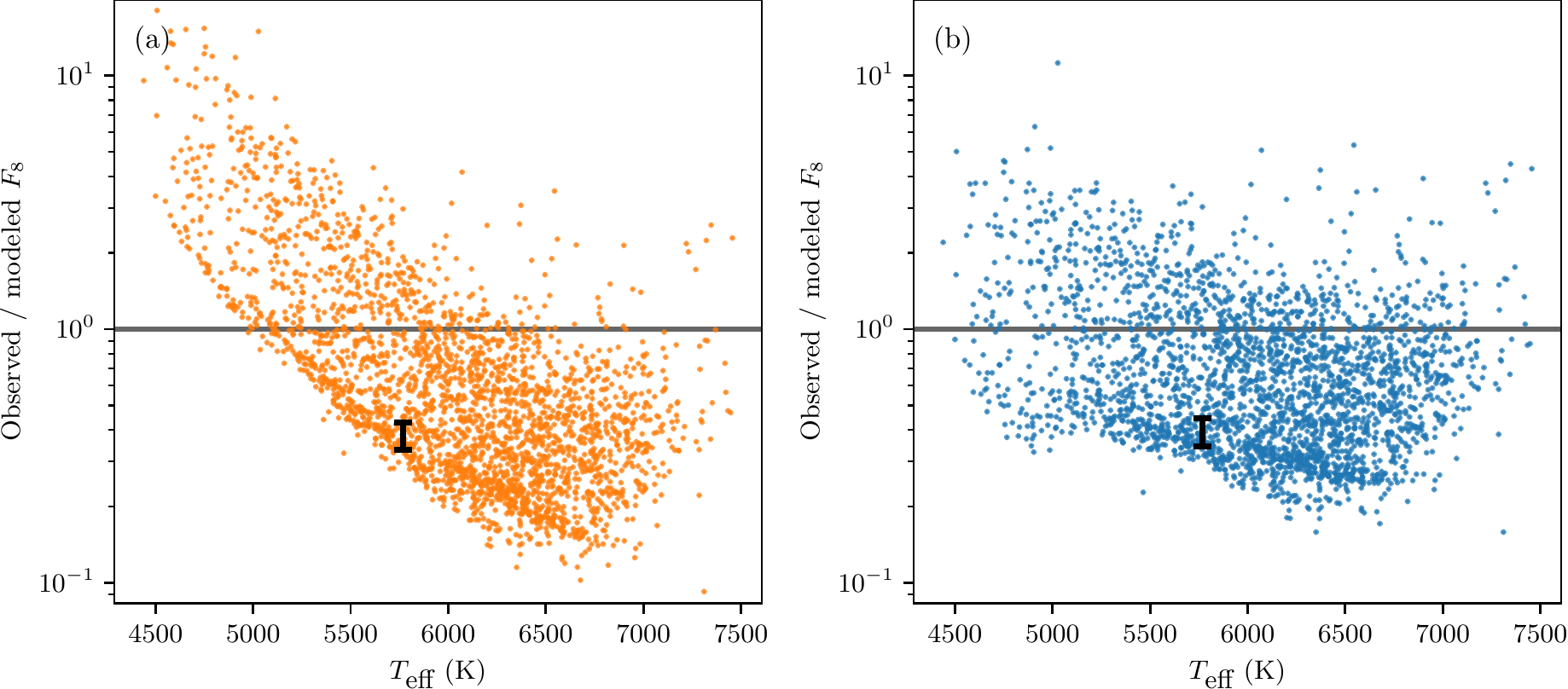}
	\caption[Ratio of observed to modeled \fe\ values along the main-sequence slice]{The ratio of observed to modeled \fe\ values along the main-sequence slice shown in Figure~\ref{fig:before-and-after}b. In (a) we use the model as in \citet{Cranmer2014} (without their empirical correction factor), which shows a strong trend with temperature, and in (b) we use our model (without the envelope fit for $\Theta\RMS$), showing that the temperature trend has been significantly mitigated. The black bars mark the Sun and its range of observed \fe\ values (discussed further in Section~\ref{sec:solar_value}).}
	\label{fig:ms-slice}
\end{figure*}

In Figure~\ref{fig:before-and-after} we show the ratio of observed to modeled \fe\ across stellar types, both before and after the modifications we have made to the model in the present work.
We also show the expanded agreement afforded by the ``envelope'' in the $\Theta\RMS(\Ma)$ fit (see Section~\ref{sec:theta}).
Considering first the central-fit model (comparing Figures \ref{fig:before-and-after}a and \ref{fig:before-and-after}b), the model agreement has improved almost universally, with significant improvement among the K-type dwarfs and the cool edge of the giant branch.
However, along the main sequence, the appearance of a small band of maximal model agreement near 5500~K has been reduced.
(This feature's appearance in the model of prior work may have been due to the handful of areas in which the model was calibrated relative to solar parameters, whereas our current model removes these close ties to the Sun.)

We turn now to the envelope fit in Figure~\ref{fig:before-and-after}(c), which accounts for the fact that simulations predict a range of possible $\Theta\RMS$ values for a given \Ma\ (producing an envelope around the $\Theta\RMS(\Ma)$ fit line; see Figure~\ref{fig:ma-theta}), leading to a range of possible \fe\ values.
The median observed \fe\ values for nearly all bins fall within the \fe\ range predicted by this envelope model, and so these bins appear in the plot with a ratio of 1.
On a star-by-star basis, 78\% of stars have an observed \fe\ falling within the appropriate envelope range.
Given the spread in predicted $\Theta\RMS$ values from numerical simulations, this envelope fit may be the more reasonable representation of the level of certainty available in modeling \fe.

Past work \citep{Cranmer2014} noted that F-type stars display less flicker than predicted in a manner that appeared temperature-dependent, and the prior model applied to our current, expanded sample of measured \fe\ values shows the opposite effect in the K-type dwarfs, yielding a strong, temperature-dependent trend along the main sequence.
We illustrate this more clearly in Figure~\ref{fig:ms-slice}, in which we plot $\feobs / \femod$ for the main-sequence slice indicated in Figure~\ref{fig:before-and-after}b.
(This slice is those stars within $0.05$~dex in \logg\ of the dwarf sequence values\footnote{Some of these values were originally presented in Table 5 of \citet{Pecaut2013}; we obtain updated values from version 2019.3.22 of \url{http://www.pas.rochester.edu/~emamajek/EEM_dwarf_UBVIJHK_colors_Teff.txt}} of \citet{Pecaut2013}.) 
Our updated model significantly mitigates that apparent trend, producing a more uniform model discrepancy along the main sequence.
Indeed, the remaining residual for our updated model trends more strongly (if at all) with \logg\ instead of \Teff.
\fe\ itself has a strong dependence on \logg\ (varying by an order of magnitude across $2.5 < \logg < 4.5$, see Figure~\ref{fig:flicker_data}) which is generally accounted for by the model.
It may be encouraging to see that the remaining discrepancy relates more closely to the primary variable \logg, and that the \Teff\ dependence present in past work is now resolved.

\begin{figure*}[t]
	\centering
	\includegraphics[width=\linewidth]{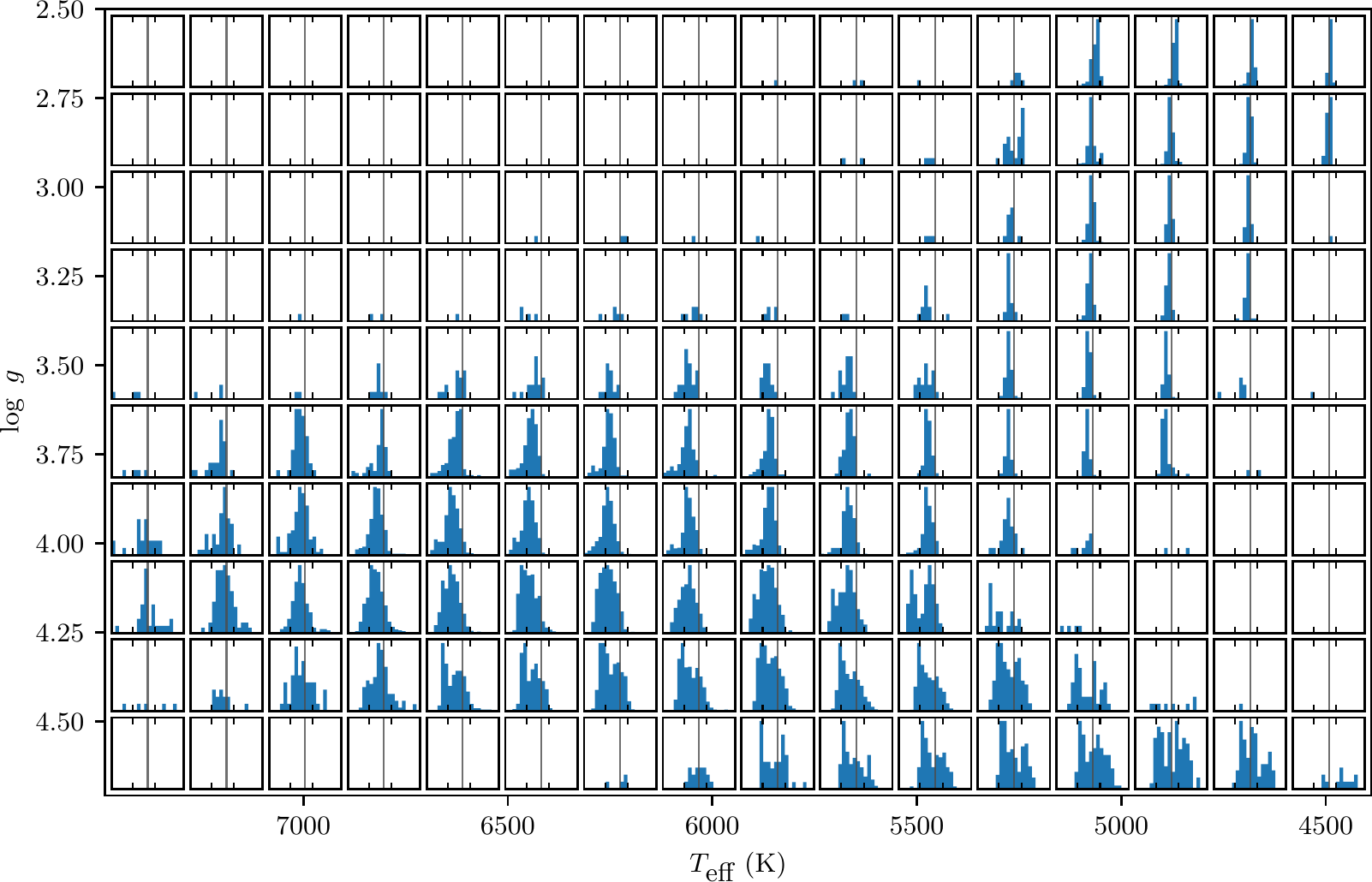}
	\caption[Distribution of $\feobs/\femod$ values]{The distribution of $\feobs/\femod$ values. We have taken the $\Teff-\logg$ domain of Figure~\ref{fig:before-and-after} and divided it into a $10\times16$ grid. Within each grid cell, we plot a histogram of $\feobs/\femod$ for the stars within that cell. Each histogram's horizontal axis is logarithmically-scaled and runs from 0.1 to 10. A vertical, gray line marks 1, and small ticks mark the range afforded by our envelope fit (see text). Each cell's vertical axis is scaled independently to match the histogram's maximum value, but each vertical axis is required to span at least 0 to 10 so that cells with few stars can be identified visually.}
	\label{fig:mini-dists}
\end{figure*}

Figure~\ref{fig:before-and-after} compares only the median observed \fe\ value in each bin to model predictions.
In Figure~\ref{fig:mini-dists}, we display the distribution of $\feobs/\femod$ values (using our central-fit model) across the sample of stars.
When discussing these ratios, the ideal value is $1$ for every star, and since our envelope fit produces a constant multiplier for $\Theta$, with $\fe \propto \Theta^2$, the envelope produces a fixed range of explainable ratio values (from 0.38 to 1.61), indicated by the small tick marks in each sub-plot.
It can be seen that there exists a spread of observed \fe\ values relative to model-predicted values, especially nearer to the main sequence.
The typical spread of values is approximately symmetric (in logarithmic space) around some central value, and that central value is often near 1 and in almost all cases is within the envelope range.
However, in most of the cells in our plot, the central value is slightly below 1, in correspondence with the typical ratio values seen in Figure \ref{fig:before-and-after}b.
Visualizing these distributions provides an important reminder that the model predictions are often the  most accurate for the average star within any one cell.
While some of this spread is certainly due to measurement uncertainty, some of it seems to be real variation from star to star (as we illustrate for the Sun in Section~\ref{sec:solar_value}).
Of note is that the distribution of $\feobs/\femod$ values is more narrow for giant stars.
Recalling that \fe\ is larger by an order of magnitude for giant stars as compared to dwarfs, this might be explained by some fixed amount of variation in \feobs\ (whether arising from the star itself or from observational noise) causing a much smaller relative variation in \feobs\ for giants.

\subsection{On the Solar Value}
\label{sec:solar_value}

Of interest is a careful look at the solar value of \fe, since portions of our model can be independently compared with solar observables.

\begin{figure}[t]
	\centering
	\includegraphics[width=0.5\linewidth]{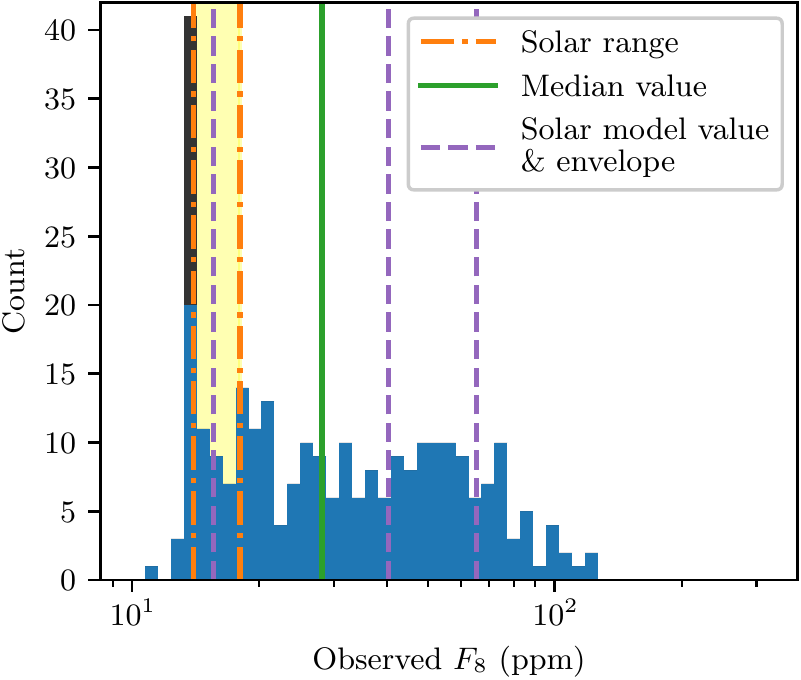}
	\caption[Histogram of 273 \kep\ stars in the ``solar neighborhood'']{Histogram of 273 \kep\ stars in the ``solar neighborhood'' (5620~K $<\Teff<$ 5920~K, 4.363 $<\logg<$4.513). Vertical lines mark the median \fe\ of the \kep\ stars, the central-fit model-predicted \fe\ and the bounds of the envelope prediction for solar parameters, and the range of the solar \fe\ over its 11-year magnetic cycle. The excess of stars in the largest bin is due in part to a clamping applied by \citet{Bastien2016} during shot-noise removal when computing the measured \fe\ values, with the affected stars indicated by the darker portion of the histogram bar---many of those stars likely should be spread over even lower \fe\ values. (Such a clamping is present only in a narrow band of $T-\logg$ bins along the main sequence.)}
	\label{fig:sun_hist}
\end{figure}

First, we look at the solar value of \fe\ itself.
The Sun's \fe\ can be computed from SOHO/VIRGO light curves \citep{Basri2013}, and it ranges from 14 to 18~ppm, a range that includes the variation due to the 11~yr solar cycle \citep{Bastien2013}.
(These values are very similar to the \fe\ values computed by \citealp{Sulis2020}.)
In Figure~\ref{fig:sun_hist} we show this range in a histogram of \fe\ values for \kep\ stars within 150~K in \Teff\ and 0.075~dex in \logg\ of the Sun.
Of the 273 stars in the histogram, 71\% have a measured \fe\ greater than the upper end of the solar range (i.e. 18~ppm), and 85\% exceed the lower end of the solar range (i.e. 14~ppm).
The solar \fe\ also straddles the edge of the range allowed by our envelope fit.

A similar relationship between the Sun and the \kep\ sample was shown by \citet{Gilliland2011} using the Combined Differential Photometric Precision (CDPP) metric, which measures the total noise budget in \kep\ observations, including the ``intrinsic'' noise of the star itself over 6.5~hr timescales (i.e., flicker in the star's light curve).
Those authors find that only 23\% of solar-type \kep\ stars have a CDPP as low as the Sun, giving the Sun a position in the CDPP distribution comparable to its position in the \fe\ distribution for Sun-like stars.
In an alternative measure of photometric variability, the total range of variability (which is tied to magnetic activity but is not strongly influenced by granulation), the Sun is more typical compared to Sun-like stars \citep{Basri2013} though less active than those with detectable, Sun-like periodicity \citep{Reinhold2020}.

This difference between the Sun and most of the sample of Sun-like stars suggests that the Sun's granulation pattern, and perhaps the convection driving it, may vary to some degree from that of a typical Sun-like star.
Within this sub-sample, neither the observed \fe\ values nor the ratio of observed and model-predicted \fe\ values correlates strongly with \Teff, \logg, \feh, rotation rate (for the 72 stars with measured periods) or magnetic activity indices (for the 53 stars with measured indices).
(Rotation rates and magnetic indices are discussed further in Section~\ref{sec:addl-quantities}.)
Additionally, model values such as the characteristic size ($\Lambda$), temperature contrast ($\Theta\RMS$), and timescale ($\tau\eff$) of granulation for solar parameters compare well to observed solar values (as shown in Sections \ref{sec:Lambda}, \ref{sec:theta}, and \ref{sec:sigma-to-f8}).
Despite this, the model-predicted \fe\ for solar parameters agrees well with the population median of Sun-like stars but not the Sun itself.
We discuss this disagreement further in Section~\ref{sec:discussion}.

This variation also enhances the value of the model presented in this paper, which is provided in more absolute terms, whereas past iterations included some scaling relations relative to solar values.
It should be noted, though, that past versions normalized their $\sigma$ (and therefore \fe) predictions not to the observed solar $\sigma$, but to the $\sigma$ of the solar-like simulation in the grid of \citet{Samadi2013b}.
The latter value, converted to \fe\, is in fact very close to the average \fe\ observed in the \kep\ sample for Sun-like stars.

\subsection{Additional Observables Considered}
\label{sec:addl-quantities}

Here we describe a few observables which we considered as factors which might explain the remaining discrepancy between model predictions and observations (i.e. the quantity $\feobs/\femod$), but which we did not find to be useful.

\subsubsection{Rotation Rate}

\begin{figure*}[t]
	\centering
	\includegraphics[width=\linewidth]{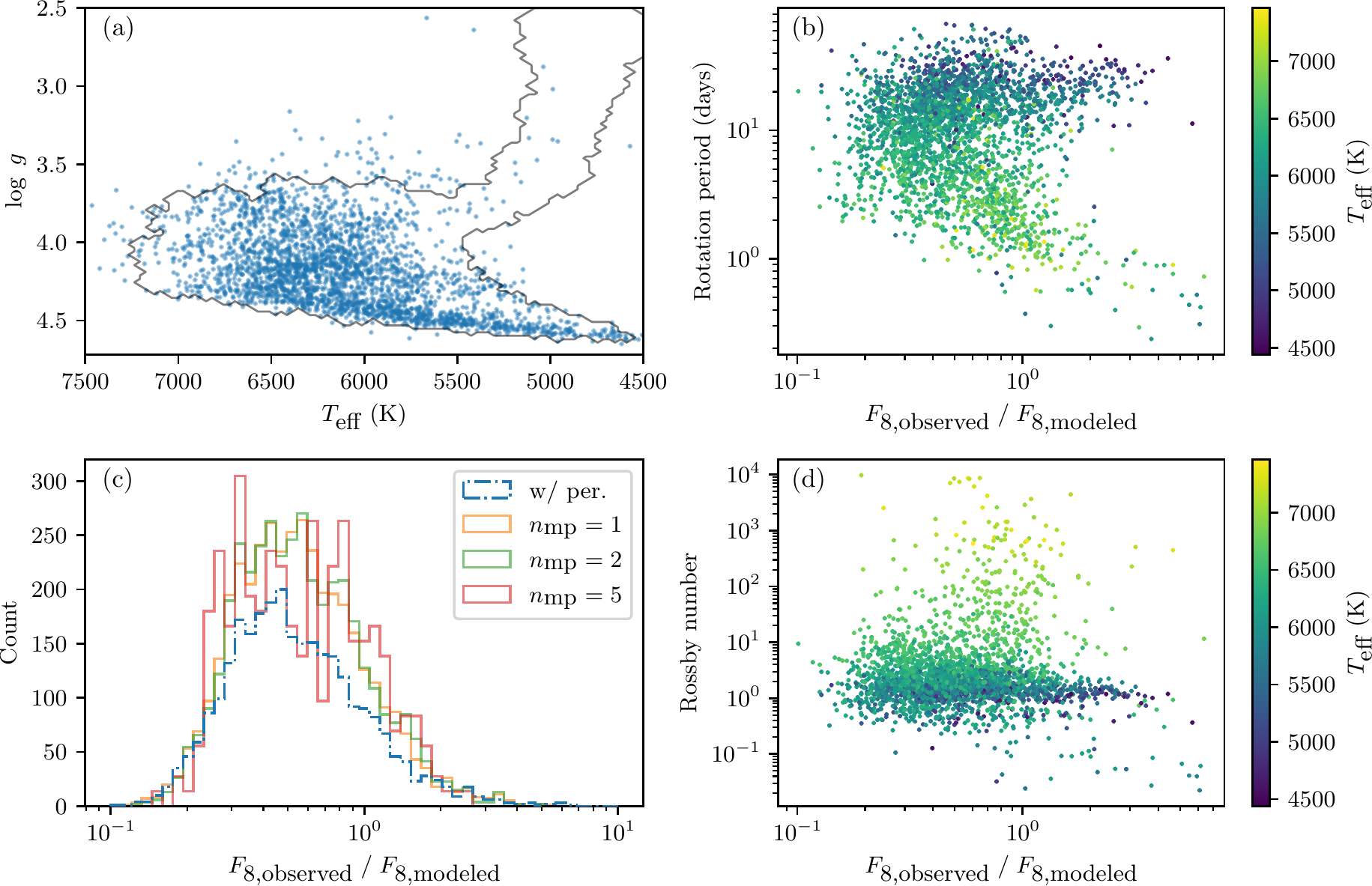}
	\caption[Analysis of rotation rate with respect to \fe]{(a) Coverage of the \citet{McQuillan2014} sample. Blue dots mark stars included in both the rotation rate sample and our joint flicker-metallicity sample. The gray line marks the boundary of the bulk of our complete flicker-metallicity sample. (b,d) Distribution of periods or Rossby number with \fe\ model discrepancy. No clear correlation is seen, though we note the fastest-rotating stars tend toward increased \fe\ as a portion of the rotational signal leaks into the \fe\ window. (c) Histograms of the distribution of model discrepancies for stars with detectable periods (a marginalization of (b)) and similar stars, under varying criteria, without detectable periods (see text). The $n_\text{mp}=2,5$ histograms are scaled vertically to align with the $n_\text{mp}=1$ curve.}
	\label{fig:rot_rate}
\end{figure*}

\citet{McQuillan2014} provide rotation rate measurements of \kep\ main-sequence stars derived via the auto-correlation method.
These data cover 3,954 of the 27,628 stars in the flicker sample and 2,820 of the 16,992 stars in our joint flicker-metallicity sample (shown in Figure~\ref{fig:rot_rate}a).
While this sample does not include any giant stars, it covers most of the $\logg > 3.5$ portion of our \fe\ sample, which contains the bulk of the remaining model disagreement (as seen in Figure~\ref{fig:before-and-after}b).

$\feobs/\femod$ shows no clear trend with rotation period or Rossby number (calculated using Equation (36) of \citet{Cranmer2011}), suggesting that the remaining discrepancy in the model predictions is not due to rotation rate effects (Figure~\ref{fig:rot_rate}b,d).
The correlation coefficient of $\log\left(\feobs/\femod\right)$ with the logarithm of the rotation period is $r=-0.11$, and with $\log\left(\text{Rossby number}\right)$ is 0.025.
Notably, though, the fastest-rotating stars (periods $\lesssim 1$~day) do show an enhanced \fe, as the rotational power spectrum begins to cross into the 8~hr window of \fe.

Additionally, the model discrepancy does not vary strongly between stars with and without detected rotation periods.
(Such a trend could occur if stellar properties that affect the detectability of a rotation period correlate with flicker.)
To determine this, we divide the ($\Teff$, \logg) parameter space plotted in Figure~\ref{fig:rot_rate}a into a $100\times100$ grid (the same grid used, e.g., for binning in Figures~\ref{fig:flicker_data} and~\ref{fig:before-and-after}) and identify those stars without measured periods which fall within a grid cell containing at least $n_\text{mp}$ stars with a measured period.
This is to ensure similar sample populations for stars with and without measured periods, since stars with measured periods span a smaller range on the H-R diagram than our full \fe\ sample.
We repeat this for $n_\text{mp}=1,2,5$, which strikes different balances between covering the full extent of the stars with measured periods and covering only the core of this population, and we see no difference in the results.
Shown in Figure~\ref{fig:rot_rate}c, the population-matched sample of stars without measured periods shows a slightly flatter distribution, but the distribution of ratios is otherwise similar in location and extent for stars with and without measured periods.

Within the ``solar neighborhood'' of Section~\ref{sec:solar_value}, these results also hold.
Little correlation is seen between either rotation period or Rossby number and $\feobs/\femod$ (with Pearson $r=-0.080$ and $-0.068$, respectively), and no clear difference is seen between the distributions of $\feobs/\femod$ for stars with and without measured periods.

\subsubsection{Magnetic Activity}

\begin{figure*}[t]
	\centering
	\includegraphics[width=\linewidth]{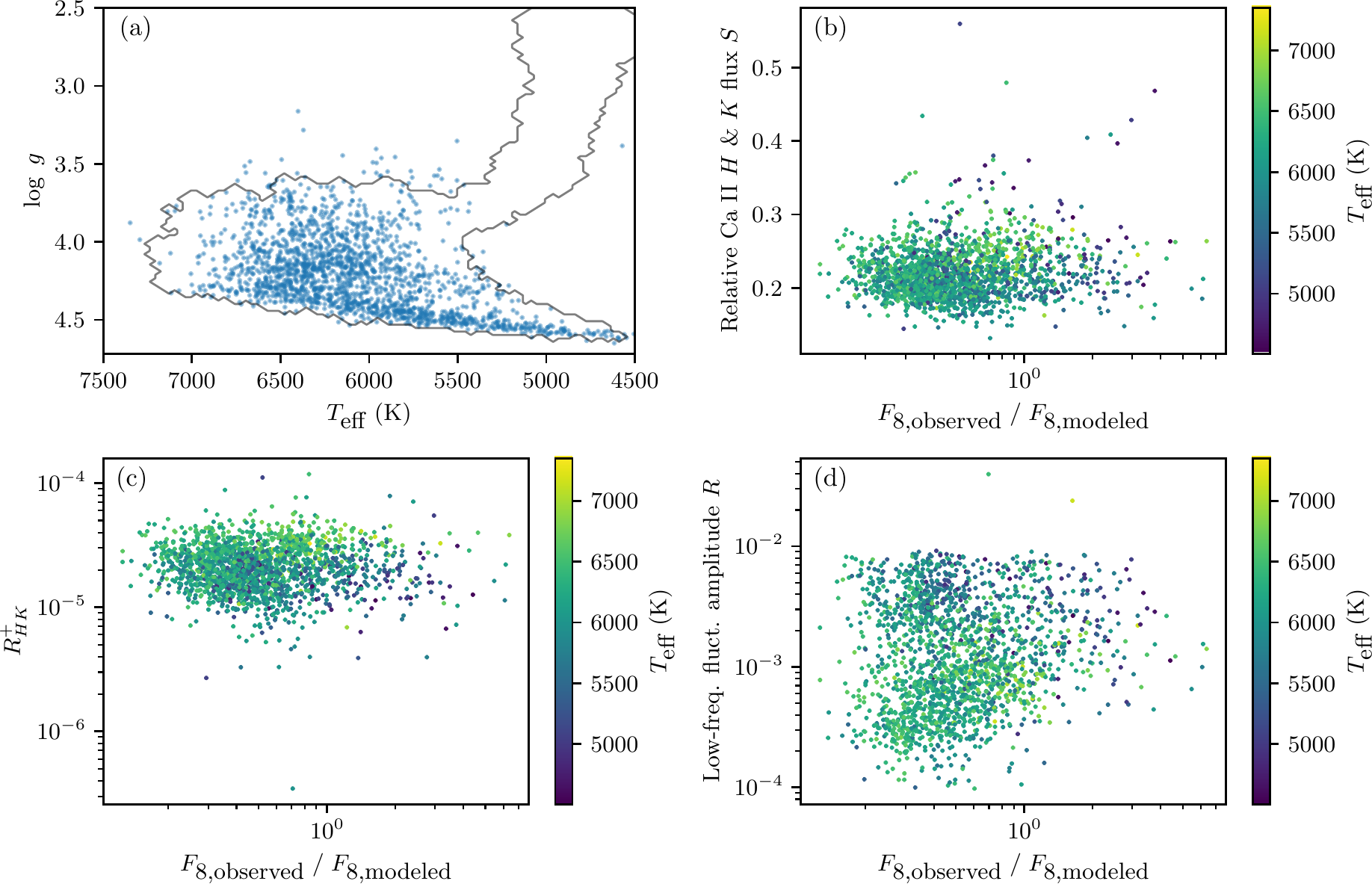}
	\caption[Analysis of magnetic activity with respect to \fe]{(a) Coverage of the \citet{Zhang2020} sample. Blue dots mark stars included in both the magnetic indices sample and our joint flicker-metallicity sample. The gray line marks the boundary of the bulk of our complete flicker-metallicity sample. There is no strong correlation between the ratio of observed to modeled \fe\ values and (b) relative Ca II $H$ and $K$ flux $S$, (c) $R^+_{HK}$, or (d) the relative low-frequency fluctuation amplitude $R$. The correlation coefficients, respectively, are $r=0.12, -0.011$, and $0.115$ (calculated using the logarithm of the latter two quantities).}
	\label{fig:mag_indices}
\end{figure*}

\citet{Zhang2020} provide measurements of the magnetic activity indices $S$, the ratio of emission in the Ca~II $H$ and $K$ lines to the continuum; $R^+_{HK}$, a proxy derived from $S$ by eliminating the contributions to the Ca~II $H$ and $K$ flux of the photosphere and the basal chromospheric flux; and $R_\text{eff}$, the range of low-frequency fluctuation in the \kep\ light curves of stars (i.e. the range of fluctuation due to magnetic activity, not the fluctuation due to granular activity considered in this paper).
$S$ and $R^+_{HK}$ are measured from LAMOST spectra.

The sample covers 2,021 stars with measured \fe\ and 1,895 stars in our joint flicker-metallicity sample.
As with the rotation rate sample, this sample covers most of the $\logg > 3.5$ portion of our \fe\ sample.
As shown in Figure~\ref{fig:mag_indices}, none of the three magnetic activity indices show any strong correlation with the ratio $\feobs/\femod$, indicating that the remaining discrepancy in the model predictions is not due to the types of magnetic activity measured by these indices (i.e., stellar magnetic flux in the chromosphere, and starspots and faculae on the photosphere).
This is in line with the findings of \citet{Meunier2017} that, while stellar magnetic activity does correlate with reduced convective signals, that effect is uniform across G and K dwarfs (whereas our model--observation discrepancy varies from early-G to K dwarfs).

In the solar neighborhood of Section~\ref{sec:solar_value}, only small, insignificant correlations are seen with these quantities.

\subsubsection{Binary Stars}
\label{sec:binaries}

\begin{figure}[t]
	\centering
	\includegraphics[width=0.5\linewidth]{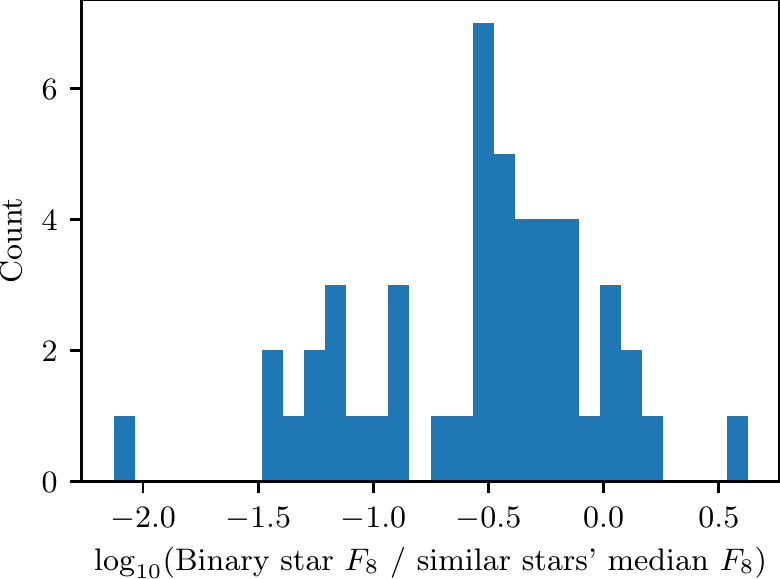}
	\caption[Histogram of the ratio of a binary star's observed \fe\ to the median of stars with similar \Teff\ and \logg]{Histogram of the ratio of a binary star's observed \fe\ to the median of stars with similar \Teff\ and \logg, for each of 48 known binary stars.}
	\label{fig:binaries}
\end{figure}

\citet{Kirk2016} provide a catalog of 2,878 identified eclipsing and ellipsoidal binary stars in the \kep\ field (though we use the third revision of their online catalog,\footnote{\url{http://keplerebs.villanova.edu/}} which is expanded to 2,922 binaries).
This catalog includes 73 known binaries included in the flicker sample, and 48 known binaries in the joint flicker-metallicity sample.
This can by no means be considered a complete catalog of the binary stars in our sample, but it does allow the properties of known binaries to be compared to those of a mixed binary/non-binary population.
For each of the 48 known binaries in our sample, we identified all stars with a \Teff\ within 100~K and a \logg\ within 0.05~dex of the binary star (which will include a mixture of unknown binary stars and non-binary stars).
This selection produces of order 100 ``neighbor'' stars for each known binary.
We computed the ratio of each binary star's observed \fe\ to the median \fe\ of its neighbors, and these ratios are shown in Figure~\ref{fig:binaries}.
The geometric mean of this ratio across all 48 known binaries is 0.30, with nearly all values falling below one.
While the sample size is low, this seems to suggest that binary stars (or at least those binary stars that are easiest to detect) display lower \fe\ than non-binaries (since the ``similar stars'' comparison is a mixture of unknown binaries and non-binaries).

This reduction in mean \fe\ aligns with the fact that, in the case of an unresolved binary system, the presence of a binary companion star will cause the observed light curve to be the sum of the light curves of the two individual stars. Both stars' granules will therefore contribute to the variability of the combined light curve (along with a more difficult-to-quantify change to the effective values of the timescale and temperature contrast of granulation when viewing these commingled granular patterns), and $\fe \propto 1 / \sqrt{N_g}$.
This decrease in \fe\ may be able to explain some of the over-prediction of \fe\ by our model for most non-giant stars.
While our sample of binaries is small, the estimated multiplicity fraction for FGK dwarfs is of order 50\% \citep{Raghavan2010} (though we note the \kep\ sample of stars may be biased against binary stars; see \citet{Wolniewicz2021}), meaning binarity could be a factor for many stars.
In addition, we showed in Section~\ref{sec:initial-comp} that multiplying all model-predicted \fe\ values by 0.66 minimizes one metric of overall observation-model disagreement.
This factor is intriguingly close to the factor of $1/\sqrt{2}=0.71$ that would arise under the very simple assumption that the presence of a binary companion simply doubles the number of visible granules, with no other changes.

However, we emphasize that the very small number of identified binaries in our flicker-metallicity sample makes it difficult to draw any strong conclusions about the impact of binarity on \fe.
We feel that this may be an interesting topic for additional study with a larger sample of known binaries, but we do not attempt to incorporate binarity into the present work.

\section{Discussion}
\label{sec:discussion}

We have improved the ability of the theoretical model to predict \fe, but some discrepancies with the \kep\ data remain.
An important fact to note is that our model--observation comparisons are based on the median observed \fe\ within a \Teff--\logg\ bin, and that the population of stars within each bin shows a non-trivial degree of spread (as illustrated in Figure~\ref{fig:mini-dists}, and for Sun-like stars in Figure~\ref{fig:sun_hist}).
To some degree this is certainly due to measurement uncertainty in \fe.
However, the observed solar \fe\ is relatively well-constrained and shows relatively strong deviation from the typical \fe\ for Sun-like stars, suggesting a level of intrinsic variation in \fe\ between stars.
Thus, eliminating the model error seen in Figure~\ref{fig:before-and-after} is only the first step in fully modeling stellar flicker.

A number of possible avenues might resolve the remaining model errors.

The \fe\ model includes a handful of quantities (such as the characteristic size $\Lambda$, temperature contrast $\Theta\RMS$, plasma flow velocity \Ma, and timescale of granulation $\tau\eff$) which certainly vary across the H-R diagram, and this variation is characterized through numerical simulations.
Further numerical work to develop increasingly accurate scaling relations for these quantities will directly improve the \fe\ model.
In particular, an improved scaling for the granular temperature contrast $\Theta\RMS$, as well as for the Mach number on which $\Theta\RMS$ is expected to depend, may help explain the spread in $\Theta\RMS$ values seen in Section~\ref{sec:theta} and may eliminate the need for our envelope fit, or at least provide a more theoretically-grounded replacement for it.
Conversely, to the degree that any one of these quantities' scaling behavior is less well understood than the others, our \fe\ model may provide a route for observationally constraining that quantity for future simulations.

Beyond better constraints for factors already incorporated in the model, additional factors might need to be included.
One such factor is the presence of binary companions, discussed in Section~\ref{sec:binaries}.

Another possible factor is the effect of star spots.
While the 8~hr filtering applied to calculate \fe\ removes the direct influence of starspots on flicker (as their evolution and their rotation across the disk of the star both occur on longer timescales for all but the fastest-rotating stars), it is possible that spots still produce a small effect in \fe.
One mechanism might be the suppression of granulation within the boundaries of a spot, which reduces the number of granules contributing to \fe\ and thereby increases \fe.
The solar \fe\ has been shown to be independent of the solar cycle \citep{Bastien2013,Sulis2020}, suggesting that activity levels up to solar levels do not affect \fe.
We have not attempted to model this possibility for greater activity levels in the present work.

Another possibility lies in the influence of smaller-scale magnetic fields, including weaker, background fields, on convection.
\citet{Cranmer2014} presented an ad-hoc, empirical factor premised on the idea that F-type stars, with shallower convective layers, may be more susceptible to magnetic fields that constrain plasma and reduce convective velocities (an idea supported by some of the results of \citet{Bhatia2020}).
While the specific trend motivating that explanation (the model rather uniformly over-predicting \fe\ for F-type stars but not G-type stars) is no longer as clear in the present work, the idea may still warrant exploration.
In Appendix~\ref{appendix:mag-sup} we present a simple, theoretical treatment of convection in the presence of a magnetic field, in which we derive a magnetic suppression factor that allows interpretation of model over-predictions of \fe\ in terms of an ambient magnetic field strength, for which we find values on the order of 10--40~G.
We note that this treatment is very simplified and speculative, and it is only able to treat cases where the model over-predicts \fe.
However, we hope that it inspires further analysis.

An additional path to comparing observed and modeled flicker may be to replace the \fe\ metric with some other quantity or quantities.
A number of statistical regression techniques attempt to fit the variability of stochastic processes with multiple parameters, and this may provide a richer characterization of stellar flicker, which may in turn provide clearer insights on trends in the data.
Examples include ARIMA models \citep[applied to \kep\ data by, e.g.,][]{Feigelson2018,Caceres2019,Caceres2019a}, which fit a set number of parameters describing how each point in the light curve relates to the preceding $n$ points, and Gaussian process regression \citep[applied to \kep\ data by, e.g.][]{Pereira2019}, in which the model is described by a number of specified kernels which are fit to the data, each intended to describe different sources of stochasticity.
Using these sorts of approaches to characterize granular flicker in \kep\ stars, and producing model-predicted values of these metrics as a function of stellar parameters (perhaps by building on the model of \citealp{Samadi2013a,Samadi2013b}) may open a new arena in which granular models and observations can be compared.

\section{Conclusions}
\label{sec:conclusions}

Building on the modeling work of \citet{Samadi2013a,Samadi2013b} and \citet{Cranmer2014}, we have confronted model predictions of \fe\ (the RMS amplitude of $<8$~hr stellar variability) with a larger sample of observational measurements by \citet{Bastien2016}, allowing the discrepancy between observed and modeled \fe\ to be analyzed over a wider range of stellar types.
We have also refined that model.
In every aspect in which the model previously was cast as a theoretical scaling relation relative to a solar value, our model instead draws on scaling relations from grids of numerical simulations covering a range of stellar types, and it directly calculates \fe\ from the other modeled quantities.
This includes Mach numbers calculated in a way that incorporates the stellar metallicity and allows for divergences from ideal theory near the surface of the photosphere through reference to the numerical simulations of \citet{Tremblay2013}, and granular sizes calculated via the scaling relation of \citet{Trampedach2013a} (which removes the free parameter $\beta$ in the original derivation of \citet{Samadi2013a,Samadi2013b}).
We also consider multiple numerical experiments when determining the scaling of the temperature contrast $\Theta\RMS$ with respect to the Mach number, and we attempt to account for the spread in these simulated values through an ``envelope fit'', which produces a range of plausible \fe\ values for a given set of stellar parameters.
Additionally, we have included a term to correct for the influence of \kep 's bandpass on \fe.

These changes to the model have improved its agreement with the observations (reducing the typical misprediction from a factor of 2.5 to a factor of 2), especially when using the envelope fit (which can explain the observed \fe\ of 78\% of our \kep\ sample), though some disagreement remains.
With respect to this disagreement, we have ruled out rotation period and the signatures of large-scale magnetic activity as possible explanations for most convecting dwarf stars, and we have shown that the status of a star as a binary may have an influence on observed \fe, though a larger sample of known binaries is required to confirm this.
We have also discussed other possible influences not considered in detail in the present work.

Given the reliance of the model on results from numerical simulations, the model will be enhanced by further constraints from simulations---particularly improved constraints on the dependence of granular temperature contrast on stellar parameters.
Additionally, it may provide a route for using \fe\ to place constraints on some granular properties for convective simulations.

With interest in exoplanet discovery and characterization as high as ever, stellar flicker \citep[as well as the closely-related radial-velocity jitter, e.g.][]{Hojjatpanah2020,Luhn2020} remains relevant as a source of noise for these observations, and this work has made progress toward fully modeling this noise source from a theoretical perspective.

\paragraph{Acknowledgements}
The authors thank Fabienne Bastien and Keivan Stassun for discussions of their past work on the topic, and the anonymous reviewer and the AAS statistics editor for their comments which have strengthened this article.

This work was supported by start-up funds from the Department of Astrophysical and Planetary Sciences at the University of Colorado Boulder, and by the National Science Foundation (NSF) under grant 1613207.
This research has made use of NASA’s Astrophysics Data System Bibliographic Services.

Hinode is a Japanese mission developed and launched by ISAS/JAXA, collaborating with NAOJ as a domestic partner, NASA and STFC (UK) as international partners.  Scientific operation of the Hinode mission is conducted by the Hinode science team organized at ISAS/JAXA.  This team mainly consists of scientists from institutes in the partner countries.  Support for the post-launch operation is provided by JAXA and NAOJ(Japan), STFC (U.K.), NASA, ESA, and NSC (Norway).

This paper draws upon measurements from data collected by the Kepler mission and obtained from the MAST data archive at the Space Telescope Science Institute (STScI). Funding for the Kepler mission is provided by the NASA Science Mission Directorate. STScI is operated by the Association of Universities for Research in Astronomy, Inc., under NASA contract NAS 5–26555.

Guoshoujing Telescope (the Large Sky Area Multi-Object Fiber Spectroscopic Telescope LAMOST) is a National Major Scientific Project built by the Chinese Academy of Sciences. Funding for the project has been provided by the National Development and Reform Commission. LAMOST is operated and managed by the National Astronomical Observatories, Chinese Academy of Sciences.

\paragraph{Software used}
Astropy version 4.2 \citep{astropy1,astropy2}, Matplotlib version 3.3.3 \citep{matplotlib,matplotlib3.3.3}, NumPy version 1.19.5 \citep{numpy}, SciPy version 1.6.0 \citep{scipy,scipy1.5.3}

\section{Appendix A: Derivation of \kep\ Bandpass Correction Term}
\label{appendix:bandpass}

The \kep\ bandpass correction term, described in Section~\ref{sec:bandpass}, accounts for the fact that at different temperatures, different fractions of a stellar spectrum will fall within the \kep\ bandpass, meaning observed variations in flux have a complex dependence on stellar temperature.
The derivation of our correction factor involves a modification of Equations (A.16) through (A.29) of \citet{Samadi2013a}, and here we provide an outline of how our correction modifies key quantities in the original derivation.

We begin by generalizing the bolometric intensity $B = \sigma_{\rm B} T^4/\pi$ to $B \propto T^m$.
This yields
\begin{equation}
	\Delta B = \left(\left(1 + \Theta\right)^m - 1 \right) \left<B\right>_t
\end{equation}
and a Taylor expansion of
\begin{equation}
	\Delta B = \left( m\Theta + \frac{m(m-1)}{2} \Theta^2 \right) \left<B\right>_t.
\end{equation}
This provides an expression for $\left<\Delta B_1 \Delta B_2\right>$, the correlation product of the instantaneous intensity variation of the granules at two points in space and time, of
\begin{equation}
	\left<\Delta B_1 \Delta B_2\right> / \left<B\right>_t = m^2\left< \Theta_1 \Theta_2 \right> \;+\; \frac{m^2(m-1)}{2}\left< \Theta_1 \Theta_2^2 \right> \;+\; \frac{m^2(m-1)}{2} \left< \Theta_1^2 \Theta_2 \right> \;+\; \frac{m^2(m-1)^2}{4}\left< \Theta_1^2 \Theta_2^2 \right>.
\end{equation}
As in the original derivation, the $\left< \Theta_1 \Theta_2^2 \right>$ and $\left< \Theta_1^2 \Theta_2 \right>$ terms are assumed to be zero, and a quasi-normal approximation allows the $\left<\Theta_1^2 \Theta_2^2\right>$ term to be expanded to $2\left<\Theta_1 \Theta_2\right>^2$, yielding
\begin{equation}
	\left<\Delta B_1 \Delta B_2\right> / \left<B\right>_t = m^2\left< \Theta_1 \Theta_2 \right> \;+\; \frac{m^2(m-1)^2}{2}\left< \Theta_1 \Theta_2 \right>^2
\end{equation}
and eventually producing
\begin{equation}
	\mathcal{F}_\tau(\tau,\nu) = \frac{(2\pi)^2\kappa\rho}{R_s^2}\left[ m^2 \left<\widetilde{\Theta_1\Theta_2}\right> + \frac{m^2(m-1)^2}{2} \tilde{\mathcal{B}}_\Theta\right],
\end{equation}
wherein the $\left<\widetilde{\Theta_1\Theta_2}\right>$ term (derived from the $\left<\Theta_1\Theta_2\right>$ term) was found to be negligibly small and taken to be zero by \citet{Samadi2013a}, and so we do likewise.
This final expression for $\mathcal{F}_\tau$ then differs from that in the original derivation by a factor of $m^2(m-1)^2/144$ (which is equal to 1 when $m=4$).
This factor can be carried through Equations (7) and (13) of \citet{Samadi2013a} to find a bandpass correction factor of
\begin{equation}
	\CBP \equiv \frac{\sigma_\text{corrected}}{\sigma_\text{original}} = \sqrt{\frac{m^2(m-1)^2}{144}}.
	\label{eqn:sigma_corrected}
\end{equation}

\begin{figure}[t]
	\centering
	\includegraphics[width=0.5\linewidth]{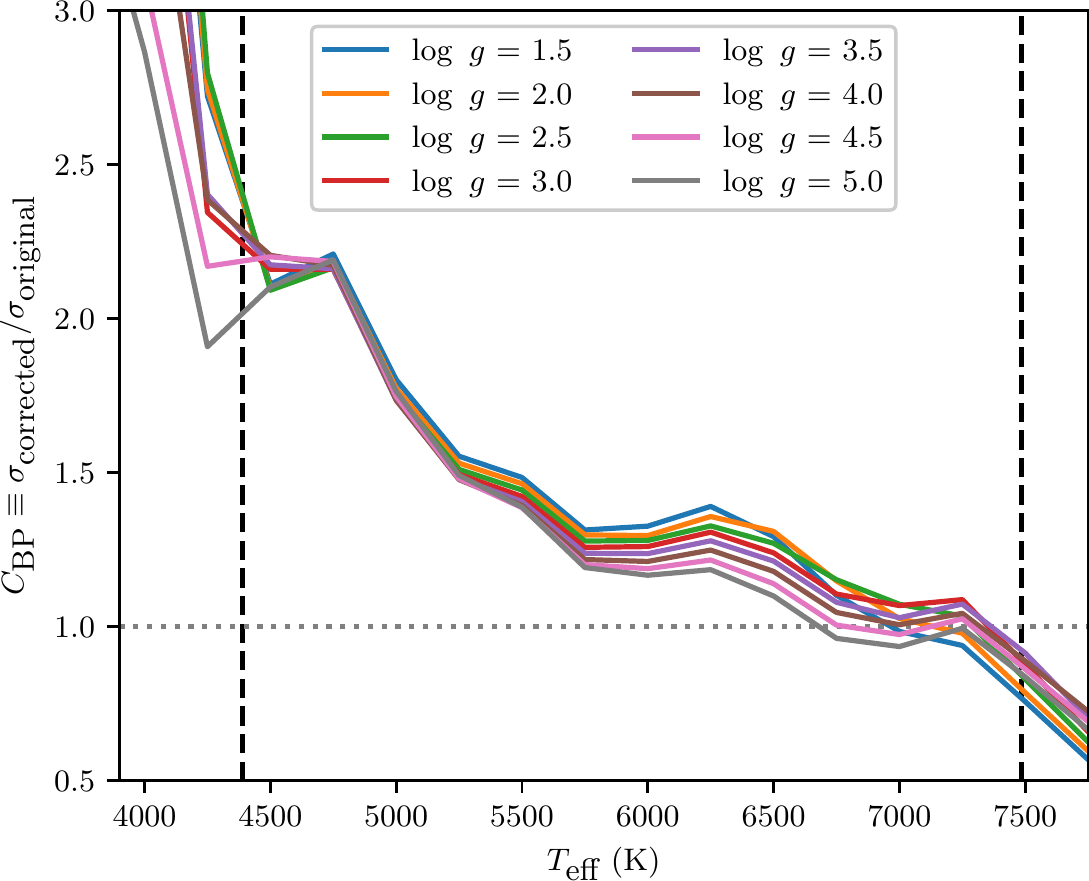}
	\caption[Factor by which our \kep\ bandpass correction factor adjusts the model-predicted $\sigma$ values]{The factor by which our \kep\ bandpass correction factor adjusts the model-predicted $\sigma$ values. Dashed vertical lines indicate the minimum and maximum temperatures within our data catalog.}
	\label{fig:sigma_correction_factor}
\end{figure}

What remains is to determine the values of $m$ to use.
For this task we used the synthetic spectral library of \citet{Lejeune1997}.
For simplicity, we used only their solar-metallicity grid ([M/H] = 0), which contained a collection of 467 stellar spectra with values of \Teff\ between 2,000 and 50,000~K, and \logg\ between $-1$ and 5.5.
For the wavelengths that overlap with the \kep\ passband (i.e., 400--900~nm), the modeled spectra were provided on a grid with a wavelength spacing of 2~nm.
Each spectrum was integrated in two ways: once over all wavelengths to obtain the bolometric flux $F_{\rm bol}$ (which we verified to be equal to $\sigma_{\rm B} \Teff^4$), and once weighted by the
\kep\ spectral response function \citep{Koch2010} to obtain a bandpass-limited flux $F_{\rm Kep}$.
For each subset of models at a fixed value of \logg, we found the logarithmic slope of $F_{\rm Kep} \propto \Teff^m$ by computing
\begin{equation}
	m(T) \, = \, \frac{d \ln F_{\rm Kep}(T)}{d \ln T_{\rm eff}}.
\end{equation}
In concert with Equation~\eqref{eqn:sigma_corrected}, these fitted values produce the scaling factors plotted in Figure~\ref{fig:sigma_correction_factor} and which we apply (interpolating to actual stellar values of (\Teff, \logg)) in Section~\ref{sec:bandpass}.
Our tabulated values, and code using them, are included in our code and data archive \citep{VanKooten2021_Zenodo}.

\section{Appendix B: Calculation of Temperature Contrast from Hinode/SOT data}
\label{appendix:hinode}

\begin{table}[t]
	\centering
	\begin{tabular}{cccccc}
		& Date &&& Time (UTC) & \\
		\hline
		\rule{0pt}{1\normalbaselineskip}	
		& 2008-01-04 &&& 11:06:26 & \\
		& 2008-01-05 &&& 00:00:27 & \\
		& 2008-01-08 &&& 06:13:25 & \\
		& 2008-02-02 &&& 18:03:55 & \\
		& 2008-02-03 &&& 18:09:03 & \\
		& 2008-02-04 &&& 18:27:57 & \\
		& 2008-02-05 &&& 05:58:57 & \\
		& 2008-02-05 &&& 10:37:57 & \\
		& 2008-02-06 &&& 00:02:27 & \\
		& 2008-02-06 &&& 10:58:27 & \\
	\end{tabular}
	\caption[\textit{Hinode}/SOT Observations used]{\textit{Hinode}/SOT Observations used. Note: All times given are for the first of the three continuum filters to be imaged. The remaining two images follow within the subsequent 20--40~s.}
	\label{table:hinode-dates}
\end{table}

To compute a solar value of $\Theta\RMS$, the RMS of $\left( T - \left<T\right>\right) / \left<T\right>$, we turn to the \textit{Solar Optical Telescope} (SOT) on \textit{Hinode} \citep{Kosugi2007,Tsuneta2008}.
Its Broadband Filter Imager (BFI) includes three ``continuum'' filters (in the blue at 450.45~nm, green at 555.05~nm, and red at 668.40~nm, each with a band width of 0.4~nm), and it images with a field of view of 218\arcsec$\times$109\arcsec\ ($\sim$160$\times$80~Mm on the solar surface, containing of order 10,000 granules), a pixel scale of 0.054\arcsec\ ($\sim$40~km on the solar surface), and a diffraction-limited resolution of 0.25\arcsec\ at 500~nm ($\sim$180~km on the solar surface).
We obtained BFI data from the SOT archive for a selection of dates (listed in Table~\ref{table:hinode-dates}), where for each date we used one image from each of the three continuum filters.
Each of the three images were taken in rapid succession ($\sim$10~s cadence) with the same pointing, so that they image very nearly identical conditions (compare the 10~s cadence to typical granular lifetimes of order 10~min).
Each set of images is at or near disk-center, and is a full-resolution, unbinned image (though we note that applying $2\times2$ binning to the level 1 images changes our final results no more than about $5\%$).
Within the range of relevant temperatures, the ratios of fluxes between any two of the three bands maps uniquely to a temperature when assuming blackbody emission, allowing photospheric temperatures to be determined \citep[see, e.g.,][for similar analyses]{Choudhary2013,Watanabe2013,Goodarzi2016}.

Before computing these ratios for any given triplet of images, we first calibrate and align the images.
To calibrate level 1 data units to relative flux, we assume that the mean pixel value in each image corresponds to the mean solar photospheric temperature of 5770~K.
We uniformly scale the pixel values of each image so that the ratios of the means of the three images are equal to the ratios of the fluxes of a 5770~K black body at the three filters' central wavelengths.
We next correct for small, fixed, filter-dependent variations in translation and pixel scale between images \citep[see][]{Shimizu2007}, as well as any small offsets in pointing that may occur between images.
Within each image triplet, we find the translations and scalings that maximize the sum of products of the corresponding pixels for each pair of images.
These values are consistent with those of \citet{Shimizu2007}.
For our final, aligned images we use the average of the fitted scaling factors across all image triplets (calculated per wavelength pair) in light of the fixed nature of the varying plate scales, but we use directly the best-fit translational factors for each image pair to guard against any slight variations in telescope pointing.

With these calibrated and aligned images, we compute flux ratios and map them to temperatures.
We find that the green-to-red ratio produces maps that are significantly more noisy than the blue-to-red and blue-to-green ratios, and so we use only the latter two ratios, and we use in the rest of our analysis the mean of the two temperature maps from those two ratios.

We find that $\Theta\RMS$, computed from individual temperature maps, ranges from 0.044 to 0.049.
Calculated across all maps, $\Theta\RMS=0.0457$ (or $\pm260$~K from the mean in absolute terms).
Our value is slightly higher than some literature values \citep[see][]{Puschmann2005,Baran2015,Gray2018}, which typically note $150-200$~K variations from the mean in photospheric temperatures.
However, we note that \citet{Beck2013} present measurements of temperature fluctuations in solar simulations which are comparable to our values, as well as measurements of observational temperature fluctuations more comparable to the literature values noted previously, and they show that the simulations, when degraded with an appropriate PSF and modeled stray-light contribution, closely match the observational values.
They interpret this as indicating that the higher temperature fluctuations in the simulations can be reasonably taken as representative of the actual Sun, while the typically-lower observational values are due to spatial and spectral degradation and stray light.
This suggests that our higher observational value may also be reasonable, especially since our use of un-binned BFI images affords us higher spatial resolution than the SP data used by \citet{Beck2013}, and our broadband technique limits the effect of spectral degradation.

\section{Appendix C: Avenues Toward Understanding Magnetic Suppression of Flicker}
\label{appendix:mag-sup}

Here we present a simple discussion of small-scale magnetic fields in the presence of convection and speculate on how those fields could affect flicker measurements.
Numerous, more advanced approaches exist in the literature \citep[e.g.,][and references therein]{Gough1966, Knoelker1988,Cattaneo2003,MacDonald2014}; however, our discussion is based on first-principles arguments of convective force balances, which have very successfully described convection in the presence of both strong and weak global rotation \citep[reviewed and applied to the Sun respectively in][]{aurnou_etal_2020, vasil_etal_2020}.

We assume an ideal gas with an equation of state of $p = R \rho T$ and write the fully compressible, magnetohydrodynamic (MHD) momentum equation,
\begin{equation}
	\partial_t \bm{u} + \bm{u}\cdot\grad\bm{u} + \frac{1}{\rho}\bm{J}\times\bm{B} = -\frac{1}{\rho}\grad p + \bm{g},
\end{equation}
where we assume ideal MHD so that $\bm{J} = (\grad\times\bm{B})/4\pi$.
We next decompose thermodynamics into background (subscript 0) and fluctuating (subscript 1) pieces, and we assume that the fluctuations are relatively small compared to the background (which may not be a good assumption for all of the stars in this work).
We assume that the background pressure gradient is in hydrostatic equilibrium, and subtract these background terms from the above equation to find
\begin{equation}
	\partial_t \bm{u} + \bm{u}\cdot\grad\bm{u} + \frac{1}{\rho_0}\bm{J}\times\bm{B} = -\frac{1}{\rho_0}\grad p_1,
	\label{eqn:momentum_fluctuations}
\end{equation}
where we have assumed that $\rho^{-1} \approx \rho_0^{-1}$ for small thermodynamic perturbations.

In the absence of magnetism and in a statistically stationary state, the advective forces should roughly balance the pressure gradient,
\begin{equation}
	\bm{u}\cdot\grad\bm{u} \sim -\frac{1}{\rho_0}\grad p_1.
\end{equation}
Expressing the gradient as an inverse characteristic length scale ($\grad \approx L^{-1}$) and dividing this balance by the square sound speed ($c_s^2 = \partial p/\partial \rho = R T_0$), we find that
\begin{equation}
	\frac{|u|^2}{L c_s^2} \sim \frac{p_1}{L p_0}
	\qquad\Rightarrow\qquad
	\mathcal{M}^2 \sim \frac{T_1}{T_0}.
	\label{eqn:hydro_balance}
\end{equation}
The latter expression relies on the linearized equation of state \citep[per, e.g., Equation~(11) of][]{brown_etal_2012} to assume that $p_1/p_0 \sim T_1/T_0$, and retrieves the classical expression that the squared Mach number scales like the temperature perturbations \citep[discused in Section~\ref{sec:theta} and verified in][]{anders_brown_2017}.

Most careful treatments of magnetoconvection have focused on the case where there is a strong background magnetic field \citep[for a brief review, see][]{plumley_julien_2019}.
However, to understand stellar magnetism outside of starspots we are interested in the less-understood case where there is no defined, strong, background magnetic field and fluctuations dominate.
We now \emph{speculate} that this magnetoconvection at the stellar surface exhibits a triple force balance between nonlinear magnetic forces, nonlinear inertial forces, and the pressure gradient \citep[inspired by the Coriolis-inertial-Archimedean (or CIA) balance of rapidly rotating convection, see Equation~(24) of][]{aurnou_etal_2020}.
Put differently, we assume that induction generates magnetic fields whose amplitudes saturate once they are strong enough to feed back on the convection which generates them.
Assuming this balance and following the same arguments as the non-magnetized balance, we retrieve
\begin{equation}
	\frac{1}{\rho_0}\bm{J}\times\bm{B} \sim \bm{u}\cdot\grad\bm{u} \sim -\frac{1}{\rho_0}\grad p_1
	\qquad\Rightarrow\qquad
	\frac{|B^2|}{4\pi \rho_0 c_s^2} \sim \mathcal{M}^2 \sim \frac{T_1}{T_0}.
	\label{eqn:B_triple_balance}
\end{equation}
This balance allows for an immediate estimation of the magnitude of small-scale magnetic fields at the stellar surface in cgs units,
\begin{equation}
	|B| \sim \sqrt{4\pi\rho_0} c_s \mathcal{M}.
	\label{eqn:B_est}
\end{equation}
At the Sun's surface, $\rho_0 \approx 2 \times 10^{-7}$ g/cm$^{-3}$ and $c_s \approx 10^6$ cm s$^{-1}$ \citep{avrett_loeser_2008}, for an approximate magnitude of $|B| \sim 10^{3} \mathcal{M}$, which suggests magnetic field strengths of order 100~G for Mach numbers of order 0.1 at the solar surface.
This is quite a reasonable estimate; for one of many observational estimates of typical, quiet-sun magnetic field strengths, see \citet{OrozcoSuarez2012}, who find a distribution peaking near 100~G.
This suggests that the triple-balance of Equation~\eqref{eqn:B_triple_balance} is a plausible starting point.

Armed with this assumption, we return to Equation~\eqref{eqn:momentum_fluctuations} and take a time- and volume- average (represented as an overbar).
Continuing to assume a statistically-stationary flow, the mean forces must satisfy
\begin{equation}
	\overline{\bm{u}\cdot\grad\bm{u}} = -\overline{\frac{1}{\rho_0}\left(\grad p_1 + \bm{J}\times\bm{B} \right)} \approx -f\overline{\frac{1}{\rho_0}\grad p_1}.
	\label{eqn:average_forces}
\end{equation}
Rearranging Equation~\eqref{eqn:average_forces} and applying the procedure used to derive Equation~\eqref{eqn:hydro_balance}, we find
\begin{equation}
\frac{T_1}{T_0} \sim \frac{\mathcal{M}^2}{f}.
\label{eqn:mag_suppression}
\end{equation}
Here we have assumed that the Lorentz force and pressure gradient have the same magnitude, but we have left $f$ as a free parameter which describes how these vectors are on average oriented with respect to one another ($f = 2$ for uniformly parallel vectors and $f = 0$ for antiparallel).
A value of $f > 1$ corresponds to magnetic suppression of convection in Equation~\eqref{eqn:mag_suppression} (i.e. a reduction in the effective \Ma\ that is controlling the temperature fluctuations); it has been known for decades that strong magnetic fields suppress convection \citep{chandrasekhar_1961}, and it would make sense for weak fields to have a similar (but smaller) effect.
If the influence of magnetism as considered here is the cause for our remaining model--observation discrepancy (or some portion thereof), values of $f$ greater than one are the values that will reduce that remaining discrepancy.
That discrepancy is most typically a model overprediction by a factor of about 2 (see Section \ref{sec:initial-comp}), and since $\fe \propto \left(T_1 / T_0 \right)^2$, this misprediction would be resolved by a factor of $f \sim \sqrt{2}$ (neglecting for now any differences in the amplitudes of the forces in the triple balance).

A determination of the validity of this treatment is beyond the scope of this work.
We leave the reader with a few questions which could be answered by targeted magnetohydrodynamic simulations of stellar surface convection:
\begin{enumerate}
	\item Is the triple-balance described in Equation~\eqref{eqn:B_triple_balance} seen in evolved simulations of nonlinear MHD convection?
	If so, Equation~\eqref{eqn:B_est} provides a straightforward way of estimating the magnitude of surface magnetism for stars with convective envelopes.
	\item Regardless of whether the triple-balance is achieved, is the assumption that introduces the magnetic suppression factor $f$ in Equation~\eqref{eqn:average_forces} valid?
	If so, what is the magnitude of $f$?
\end{enumerate}

\biblio

\chapter{Conclusions and Future Work}
\label{chap:conclusions}

\section{Implications of this Work}

\subsection{For \BP s and Energy Fluxes}

As summarized in Section~\ref{sec:shapes-conclusions}, I developed and demonstrated two very different techniques for analyzing \bhp\ shape changes.
These techniques produced estimated upward wave energy fluxes within a factor of two of each other---a similarity that allows these different approaches to mutually support each other.
The moment-based approach of Chapter~\ref{chap:ellipse-fitting} produces estimated fluxes separated by wave modes, facilitating future work to estimate the degree of upward propagation of these waves and the location of their eventual dissipation.
The earth mover's distance (EMD) approach of Chapter~\ref{chap:emd}, meanwhile, produces estimated horizontal velocities for each pixel within the \bp, facilitating a direct comparison to the plasma velocities in the \muram\ simulations.
This comparison showed that EMD and \muram\ velocity vectors are typically within $45^\circ$ degrees of each other and agree in magnitude within a factor of two, which can be viewed as rather good agreement considering the many ways that the assumptions inherent in the EMD approach (as well as the moment-based approach) can over-simplify the complex, 3D structure of \bp s and their flux tubes (Section~\ref{sec:emd-comp-but-should-we}).
The results of this EMD-\muram\ comparison support our use of the assumptions I make as a first approach at understanding \bhp\ shape changes.

The energy fluxes I estimated significantly exceed some observation-based estimates, and it falls to DKIST observations to determine whether this difference will exist in real data.
But my estimates also predict that there is a significant amount of energy flux present in modes other than the kink modes that have long been studied (a factor of two increase over just the kink-mode flux).
I hope this will motivate careful investigations of \bp s with DKIST to verify this claim observationally, and modeling work to understand if these modes can make significant contributions to wave-based models of coronal heating.
If so, this may significantly increase the estimated energy budget of this commonly-discussed model.
In addition, DKIST observers will certainly be hoping to find direct evidence of small-scale wave propagation and dissipation in the chromosphere and corona, and I hope these results provide additional motivation for these searches.

I used a custom-developed algorithm to identify and track \bp s.
First described in Chapter~\ref{chap:bp-centroids}, the algorithm often did a poor job of identifying an appropriate edge for each \bp, but its results were only intended to be used for intensity-weighted centroid tracking, in which the darker edge pixels are less important that the bright core of each \bp.
My shape-change work, however, depends entirely on the outline of each identified \bp, and so I made significant algorithmic improvements in Chapter~\ref{chap:bp_tracking}.
These changes incidentally (and happily) resolved much of the concern over centroid jitter discussed in Section~\ref{sec:jitter} by drawing \bhp\ boundaries more consistently from frame to frame, which emphasizes the importance of careful algorithm design in this and any automated analysis.

From a practical perspective, my algorithm improvements were approximately equally split between changes to the algorithm itself and changes to the various tunable parameters of the algorithm.
This parameter tuning was fueled by the elimination of a computational bottleneck to reduce my tracking code's runtime from hours to minutes, enabling rapid iteration and a finer search for optimal parameter values, and emphasizing the value of time spent developing good code.
My parameter tuning also benefited greatly from developing a more capable set of visualization tools for both accepted and rejected \bp s, making it easier to visualize and consistently track the effect of each parameter change.
I have often found myself having to decide whether to invest time up front in developing and refining code for plotting or other tasks, or whether to push forward quickly on the science and hope I will not need complex tooling.
While this sort of question never has one universal answer, I now tend to feel it is often better to invest the time up front to prepare well whatever tools and systems will be needed (to the extent that these needs can be predicted), with these late changes to my tracking algorithm as one case study.
This allows the science, once it begins, to be less frequently interrupted by the need to build out more programmatic features, and it allows the science to benefit from the outset from carefully-developed tools rather than relying on ad-hoc and quickly-produced scripts that all too often grow slowly to an unmanageable size.

\subsection{For Convection}

In Chapter~\ref{chap:bp-centroids}, I investigated the motion of passive tracers in the \muram\ horizontal velocity field and in the simplified, turbulence-free granular pattern of the \RF\ model.
I found that the two models produce very similar low-frequency passive-tracer motion, but that the motions diverge sharply for high-frequency motion.
Since \RF\ was designed to contain only laminar flows but was otherwise tuned to match the granular characteristics of the \muram\ simulation, this suggests that horizontal plasma flows driven by convection at the photosphere might be thought of broadly as two separate components: large-scale, largely laminar flows responsible for the overall structure of the granular pattern; and small-scale, more turbulent flows that superpose more rapidly-varying details on the overall granular structure.
Given that observed granules show an apparently two-component size distribution \citep{Abramenko2012}, for which \muram\ matched both components but \RF\ matched only the large-feature component, and given that different size-lifetime correlations have been found for granules persisting for over versus under 8~minutes \citep{DelMoro2004}, there may indeed be two distributions of granules (or granule-like features) on the Sun.
I speculated that the larger granules, which appeared in both \muram\ and \RF, are true granules---convective cells consisting of (relatively) large-scale and quasi-laminar flows---while the smaller features which appeared only in \muram\ may not be convective cells, but rather small features of turbulent origin.

In my work with granulation in \kep\ stars, I advanced efforts to compare these stellar signals to predictions made by a physical model of granular flicker.
This model, while physical, is not a from-first-principles model, instead depending on certain scaling relations for granular properties that are derived from numerical simulations.
My work, therefore, can be seen as advancing an effort to confront those simulations of granulation with the wealth of stellar data in the \kep\ database.
In the future, these comparisons, from me and others, may provide some small, additional motivation for advancing the state of understanding and simulating stellar convective processes.
Perhaps more importantly, though, it may provide a way to use \kep\ data to provide valuable constraints for convective simulations across a range of stellar parameters.
While understanding stellar convection is valuable in its own right, improvements brought on by new constraints may, in turn, fuel more accurate simulations of solar convection as well.

\subsection{For the Solar-Stellar Connection}

The solar-stellar connection was implicit throughout Chapter~\ref{chap:flicker}, even despite one focus of my work being to ``de-Sun-ify'' the flicker model by removing its scalings relative to solar values.
I still compared model values, both flicker itself and intermediary quantities such as granular sizes and temperature contrasts, with solar values---even to the point of determining my own solar value for the granular temperature contrast, the only solar observations used directly in this thesis!
This highlights the extreme utility of the Sun as the only star we can study up close, the only star for which we can make detailed observations of subtle features, and the blueprint for our understanding of every other star.

These results may some day be relevant to work along the solar-stellar connection.
Interest has been shown by many \citep[e.g.,][]{Schrijver1989,Ciaravella1996,Peres2004,Cranmer2011,Cranmer2013} in understanding the properties of stellar coronae---how they are structured, whether they are hot, and if so, how they are heated.
Studies of stellar coronae do not have the wealth of observational data that solar studies do, and so they are presently restricted to relatively broad-brush approaches.
But if, some day in the future, the focus turns to the precise nature of stellar coronal heating, solar models such as wave-driven heating will become relevant.
The wave-driven heating discussed through much of this thesis is dependent on the convective flows at the solar surface to dictate the amount of wave energy as well as the modes and frequencies of waves that determine the degree of propagation to the corona.
My work with \kep\ stars sought to improve understanding of how surface convection varies between stars of different types, and further work in this direction might some day become relevant to characterizing and constraining how those stars' coronae are heated.

\section{Future Directions}

\subsection{Solar \BP s}

I am a co-investigator on a proposal that is planned to be submitted to the next DKIST proposal call, which aims to collect data suitable for \bhp\ tracking as well as some spectropolarimetric data that can be used for initial probes into the magnetic field of \bp s.
Unfortunately, DKIST construction and testing have been delayed by the on-going pandemic and initial science operations have not yet begun, and so it may be some time yet before the techniques proposed in this thesis can be applied to real data.
However, I join many others in looking forward to the beginning of DKIST observations, and I hope to be involved in the new frontier of \bhp\ observations that will be opened.

Despite my statements that developing my shape-change analysis techniques now will leave me well-positioned to analyze \bp s in DKIST data, I anticipate that that DKIST analysis will be a full project unto itself.
Calibration steps will not be well-established in the early days of DKIST, and so any early work will require careful thought about how to approach the observations.
(My focus solely on the borders of \bp s, however, may side-step many data-calibration concerns.)
Next, the parameters of my tracking algorithm may need to be adjusted to better match the properties of real observations.
Some parameters, such as the number of expansion iterations, are closely connected to the pixel resolution of the images, while others, such as the threshold for initial seed-pixel selection, depend on the distribution of intensity values within the image.
Next, it will be prudent to inspect the EMD-inferred velocity fields to verify that real \bp s do not display behavior that produces velocity fields that are contrary to physical intuition (or if they do, whether that defiance of intuition should be disqualifying).
Additionally, observational effects must be carefully considered.
The amount of noise and its effects, if any, on inferred velocities and fluxes should be considered.
Any ``shaking'' of the image (due to atmospheric effects, for instance) must also be carefully addressed, lest it be interpreted as \bhp\ motion.

This all assumes that \bp s in DKIST observations appear similar to those in \muram.
In the DKIST first-light images released last year, the \bp s appear to show structure in their intensity profiles, whereas \bp s in the \muram\ simulation have a much smoother appearance.
(Compare the DKIST image in Figure~\ref{fig:intro-dkist-first-light-full} to the \muram\ \bp s in Figures \ref{fig:tracking-bp-sample}, \ref{fig:moments-example-fits}, or \ref{fig:emd-real-samples}.)
If \bp s in calibrated DKIST images show similar intensity structure, the tracking algorithm may require significant adjustments to handle these different feature appearances.

Finally, my proposed techniques depend entirely on the assumption that \bp s serve as proxies for the location and cross-sectional shape of flux tubes, which has never been tested at DKIST's expected resolution.
It is therefore important to see whether this expected correlation holds at the smallest scales, and this may be a separate project entirely (dependent, as it is, on extracting precise magnetic field information at the highest possible resolution).

\subsection{Stellar Granulation}

I laid out several ideas in Chapter~\ref{chap:flicker} for investigations that may better constrain the dependence of stellar granulation of stellar parameters.
This included a better understanding of how the granular temperature contrast varies with stellar parameters, and an understanding of how stellar magnetism might affect convection.
Pursuing this further is not a short-term focus of mine, but I hope to collaborate with others to perform these investigations and see how they inform my revised flicker model.
One interesting direction might be to explore whether JWST can provide observational constraints on stellar convection---perhaps by constraining the granular temperature contrast.

\section{Final Thoughts}

One unifying theme of this thesis has been how new telescopes can enable new classes of observations and promote new developments in theory and modeling.
My work with \kep\ was a part of that new frontier opened by the mission, but my work with \bp s attempts to anticipate what the new DKIST frontier will be and where it will lead, and such predictions are simultaneously exciting and fraught with peril.
It may be that DKIST observations lead to conclusions opposite those in this thesis, or even conclusions that are orthogonal to mine, revealing an entirely new direction of more fruitful inquiry.
It really is not possible to predict what the state of knowledge will be for \bp s in a decade, and that is just a small part of what makes this current era of heliophysics so exciting (filled, as it is, with big names like DKIST, Parker Solar Probe, and Solar Orbiter, and smaller yet still frontier-opening missions such as PUNCH).
As a key observational signature of one mechanism of coronal heating, \bp s, despite being one of the smallest observed features on the Sun, may well play a large role in the big questions that will hopefully be answered in the coming decades.

\biblio



\end{document}